\documentclass[11pt]{book}
\usepackage{enumerate,amssymb}
\usepackage{cite}
\usepackage{latexsym}
\usepackage{amsmath}
\usepackage{amsfonts}
\usepackage[dvips]{graphicx}

\newcommand{\beq}{\begin{equation}}
\newcommand{\eeq}{\end{equation}}
\newcommand{\bea}{\begin{eqnarray}}
\newcommand{\eea}{\end{eqnarray}}
\newcommand{\ba}{\begin{array}}
\newcommand{\ea}{\end{array}}
\newcommand{\bt}{\begin{tabular}}
\newcommand{\et}{\end{tabular}}
\newcommand{\bc}{\begin{center}}
\newcommand{\ec}{\end{center}}

\newcommand{\ax}{\alpha}
\newcommand{\bx}{\beta}
\newcommand{\cx}{\gamma}
\newcommand{\dx}{\delta}
\newcommand{\ox}{\omega}
\newcommand{\lx}{\lambda}
\newcommand{\ab}{\bar\alpha}
\newcommand{\bb}{\bar\beta}
\newcommand{\cb}{\bar\gamma}
\newcommand{\db}{\bar\delta}

\newcommand{\Ox}{\Omega}

\newcommand{\Gx}{\Gamma}

\newcommand{\cC}{\mathcal{C}}
\newcommand{\cD}{\mathcal{D}}
\newcommand{\cL}{\mathcal{L}}
\newcommand{\cK}{\mathcal{K}}
\newcommand{\cN}{\mathcal{N}}

\newcommand{\cF}{\mathcal{F}}
\newcommand{\cI}{\mathcal{I}}
\newcommand{\cR}{\mathcal{R}}
\newcommand{\cM}{\mathcal M}
\newcommand{\W}{\mathcal W}

\newcommand{\bi}{{\bar i}}
\newcommand{\bj}{{\bar j}}

\newcommand{\wg}{\wedge}

\newcommand{\del}{\partial}

\DeclareMathOperator{\SU}{\mathit{SU}}
\DeclareMathOperator{\SO}{\mathit{SO}}

\DeclareMathOperator{\Spin}{\mathit{Spin}}
\DeclareMathOperator{\so}{\mathit{so}}
\DeclareMathOperator{\su}{\mathit{su}}

\newcommand{\rep}[1]{\mathbf{#1}}

\newcommand{\dd}{\mathrm{d}}

\newcommand{\ii}{\mathrm{i}}

\newcommand{\bbZ}{\mathbb{Z}}
\newcommand{\bbR}{\mathbb{R}}
\newcommand{\bbC}{\mathbb{C}}

\newcommand{\CY}{Calabi-Yau\hspace{0.2cm}}

\newcommand{\nn}{\nonumber}

\newcommand{\dg}{\delta g}

\newcommand{\tox}{\tilde\omega}
\newcommand{\txi}{\tilde\xi}
\newcommand{\IM}{\textrm{Im} \,}
\newcommand{\RE}{\textrm{Re} \,}

\newcommand{\CT}{\kappa} 
\newcommand{\p}{\tilde m}
\newcommand{\q}{\tilde e}

\newcommand{\hg}{{\hat{g}}}
\newcommand{\LCY}{\tilde{L}}
\newcommand{\ff}{{\zeta}}
\newcommand{\moo}{\kappa_0}

\def\rme{{\rm e}}

\newcommand{\rd}{\textrm {d}}

\renewcommand{\a}{\alpha}
\renewcommand{\b}{\beta}
\newcommand{\g}{\gamma}           
\renewcommand{\d}{\delta}         

\newcommand{\ki}{\chi}
\newcommand{\la}{\lambda}

\newcommand{\s}{\sigma}           \renewcommand{\S}{\Sigma}
\newcommand{\tet}{\theta}         
\newcommand{\f}{{\phi}}           \newcommand{\F}{{\Phi}}

\newcommand{\eps}{{\epsilon}}

\newcommand{\ca}{{\cal A}}
\newcommand{\cc}{{\cal C}}
\newcommand{\cd}{{\cal D}}

\newcommand{\cf}{{\cal F}}
\newcommand{\cg}{{\cal G}}

\newcommand{\cm}{{\cal M}}

\newcommand{\cw}{{\cal W}}

\newcommand{\be}{\begin{equation}}
\newcommand{\eqn}[1]{\label{#1}\end{equation}}
\newcommand{\equ}[1]{(\ref{#1})}

\newcommand{\eqan}[1]{\label{#1}\end{eqnarray}}

\newcommand{\loco}{{\mathop{ \, \rule[-.06in]{.2mm}{3.8mm}\,}}}
\newcommand{\doubar}{{{\loco}\!{\loco}}}

\newcommand{\az}{{\bf{z}}}
\newcommand{\au}{{\bf{u}}}
\newcommand{\av}{{\bf{v}}}
\newcommand{\aw}{{\bf{w}}}

\newcommand{\da}{{\dot{\alpha}}}

\newcommand{\ta}{{\mbox{\tiny{A}}}}
\newcommand{\tb}{{\mbox{\tiny{B}}}}
\newcommand{\tc}{{\mbox{\tiny{C}}}}
\newcommand{\td}{{\mbox{\tiny{D}}}}
\newcommand{\te}{{\mbox{\tiny{E}}}}
\newcommand{\tf}{{\mbox{\tiny{F}}}}
\newcommand{\tg}{{\mbox{\tiny{G}}}}

\renewcommand{\gg}{\mathfrak{g}}
\newcommand{\hh}{{\mathfrak{h}}}

\newcommand{\vd}{\textrm{d}}
\newcommand{\bP}{\bar{P}}
\newcommand{\bs}{\bar{\s}}
\newcommand{\bp}{\bar{\psi}}
\newcommand{\bla}{\bar{\la}}

\begin{document}

\strut\thispagestyle{empty} \vspace{-2cm}
\begin{center}

\begin{tabular}{cp{8.3cm}c}
\includegraphics[height=3cm]{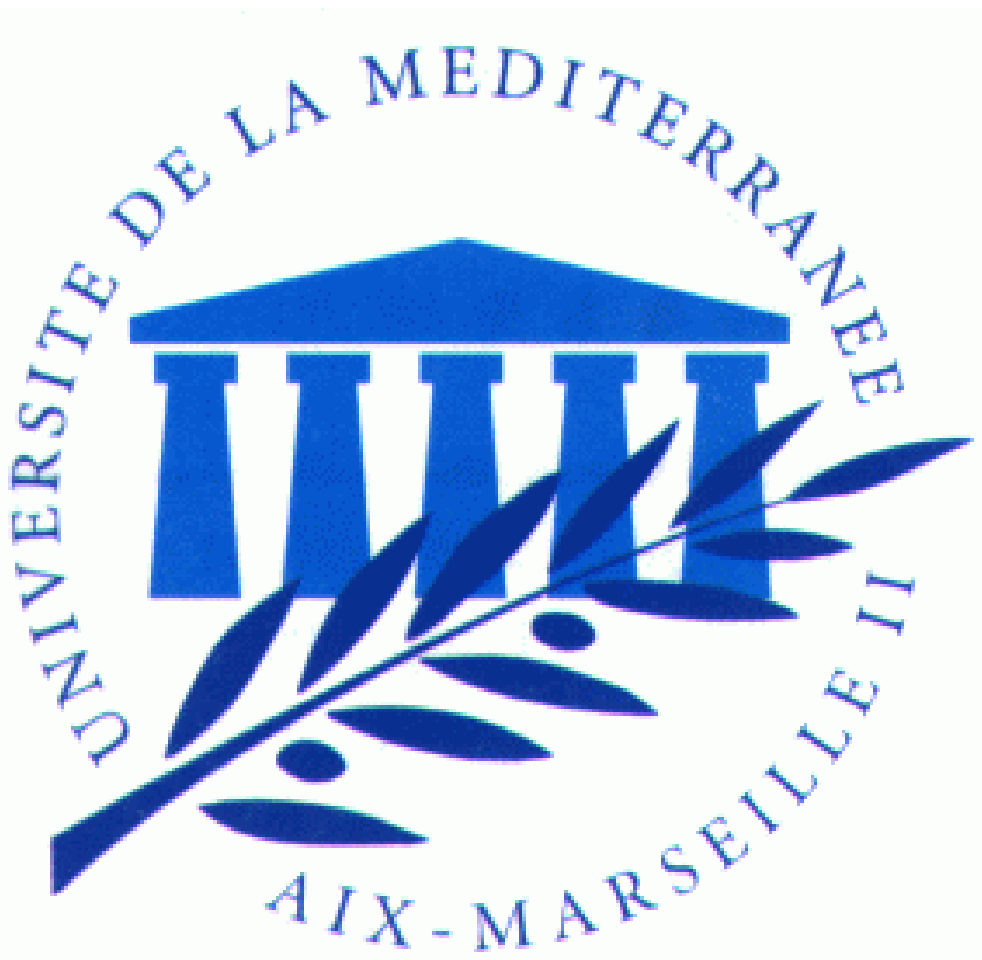}
&
\vspace*{-1.7cm}
\begin{tabular}{c}
\hline
\hspace{8cm}
\end{tabular}
&
\includegraphics[height=3cm]{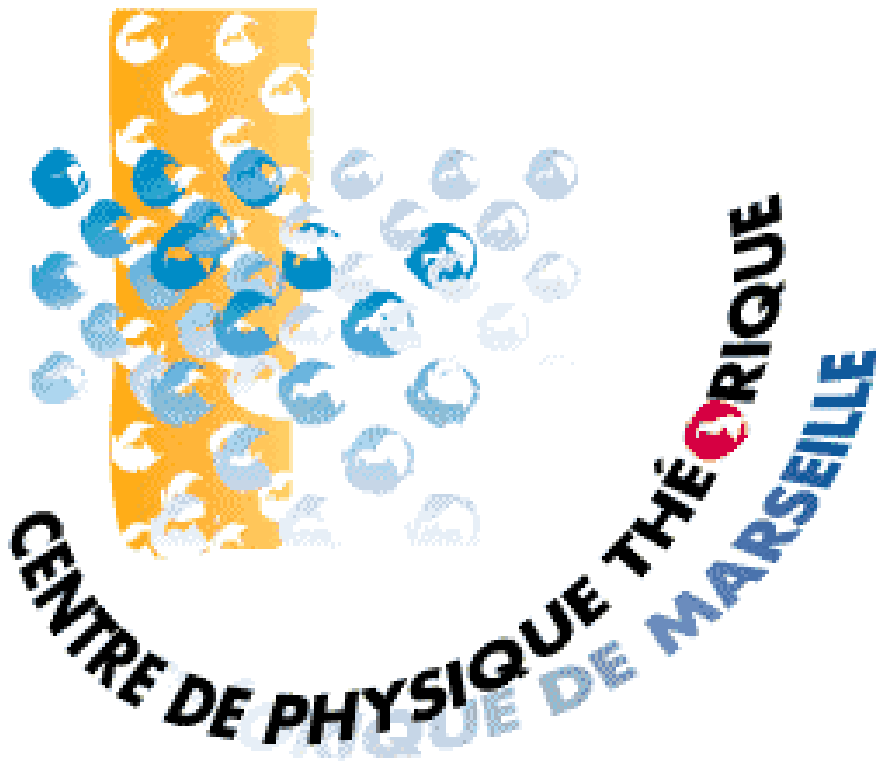}
\end{tabular}

\vspace{1cm}

\Large
\textsc{
Universite de la Mediterranee -
Aix-Marseille II
\\
Centre de Physique Theorique
}

\vspace{0.4cm}


\vspace{1.2cm}

\textsc{These}
\vspace{0.5cm}

\large
pr\'esent\'ee par
\vspace{0.5cm}

\Large
\textsc{Sebastien GURRIERI}
\vspace{0.5cm}

\large
en vue d'obtenir le grade de
\vspace{0.5cm}

\Large
Docteur de l'Universit\'e de la M\'editerran\'ee
\vspace{0.5cm}

\large
Sp\'ecialit\'e : Physique des particules, physique math\'ematique
et mod\'elisation
\vspace{1.2cm}

\LARGE \textsc{$N=2$ and $N=4$ supergravities as compactifications
from string theories in 10 dimensions.}

\vspace{1cm}
\large
Soutenue le 13 juin 2003 devant le jury compos\'e de :
\vspace{0.5cm}

\begin{tabular}{llcl}
MM.& Grimm & R. & {\it Directeur de th\`ese} \\
 & Klimcik & C. & {\it Pr\'esident}\\
 & Louis & J. & \\
 & Van Proeyen & A. & \\

\end{tabular}

\vspace{.5cm}
Apr\`es les rapports de :
\vspace{0.5cm}

\begin{tabular}{lll}
MM. & Louis & J.  \\
 & Van Proeyen & A.  \\

\end{tabular}

\end{center}
\vfill
\pagebreak

\newpage
\emph{}
\newpage

\vspace*{10cm}

\hspace{5cm}\emph{A mon p\`ere et mon grand-p\`ere qui auraient aim\'e \^etre pr\'esents}

\hspace{5cm}\emph{lors de l'ach\`evement de ce travail.}

\newpage
\emph{}

\newpage

\tableofcontents

\newpage

\numberwithin{equation}{chapter}

\chapter{Introduction}

The early excitements about string theory came from its possible
ability to reconcile General Relativity with Quantum Mechanics
\cite{GSW,LT,JP}. On the one hand, General Relativity explains the
behaviour of gravity at macroscopic scales. Among its main
predictions one can cite the deviation of light near a matter
source, or the relativity and local dependence of time. On the
other hand Quantum Mechanics brought new insights in the
structure of matter at microscopic scales, and introduced the
ideas of quantization and uncertainty. It explained the discrete
spectrum of hydrogenoid atoms, or the black body radiation.
Quantum Mechanics was extended to Quantum Field Theory to describe
the interactions of particles when creation and annihilation
processes take place. Three of the four fundamental forces,
electromagnetism, weak and strong interactions, could be unified
in the same Standard Model, a quantum field theory based on the
gauge group $SU(3)\times SU(2)\times U(1)$. The Standard Model
has given the most accurate results ever predicted by a physical
theory. However, it suffers from important drawbacks. The gauge
group, as well as the masses and coupling constants, are put ``by
hand''. Their values are measured, and no internal principle can
guide us to their origin. Moreover, the Standard Model is only
defined in flat space-time, which means that it says nothing about
the other fundamental force, gravity. General relativity describes
with great accuracy the behaviour of gravity, but it does not
include any possible quantum effects of this interaction. It has
become clearer and clearer, for example regarding space-time
singularities in black holes, that a quantum formulation of
gravity has to be discovered. Besides this, the constant progress
of Physics towards unification of all interactions is an
indication that a formulation of the Standard Model where all
forces are treated on the same footing may exist. This is where
string theory enters the game.

The fundamental object, the string, has dimension 1. The
corresponding action is a Quantum Field Theory on the world sheet
swept by the strings as it moves in a $D$-dimensional space-time.
It has invariance under Poincar\'e transformations, world sheet
diffeomorphisms and Weyl symmetry.  If one considers
superstrings, the world sheet action is extended to have
invariance under supersymmetry transformations. The mass of the
string is determined in terms of its internal oscillation degrees
of freedom according to

\beq M^2 \sim \frac{1}{\ax'}(N -A)\eeq where $A$ is a constant of
zero-energy and $N$ is the number of oscillations. For the open
string, $N$ is a general integer and $A=1$, for the closed string
$N$ is even and $A=2$. The ground state ($N=0$) is a tachyon and
is projected out by supersymmetry. The first excited level has
zero-mass, and all other states have masses quantized in units of
the Planck mass $\frac{1}{\sqrt{\ax'}}\sim 10^{19}$ GeV. Of
course, the massive states have no physical interest, and the
particles of the Standard Model are to be found in the massless
sector.

One should stress that, in string theory, the dimension
of  space-time is not set by hand, but is determined by
consistency considerations. The Fock space contains states which,
due to the negative signature of space-time, have negative norms.
Only in the particular case of $D=10$ are these states absent.
The massless physical states thus correspond to representations of
the transverse group $SO(8)$. It turns out that one can
distinguish 5 different string theories. Here we give their main
characteristics and their bosonic massless
spectra\footnote{Supersymmetry insures that there is an equal
number of bosonic and fermionic degrees of freedom. The fermionic
spectra are generically composed of $N$ gravitini and a number of
spin $1/2$ fermions such their degrees of freedom match the
bosonic ones.}.

\vspace{0.5cm}{\bf Type I} It is composed of open and unoriented
closed strings, has $N=1$ supersymmetry, and contains the metric
$g_{MN}$, an antisymmetric tensor $B_{MN}$, the dilaton $\phi$,
and 496 vectors $A_M^a$ with gauge group $SO(32)$.

\vspace{0.5cm}{\bf Heterotic $SO(32)$ or $E_8\times E_8$}
Heterotic theories are hybrids of closed strings and
superstrings. They have $N=1$ supersymmetry, and contain the
metric $g_{MN}$, an antisymmetric tensor $B_{MN}$, the dilaton
$\phi$, and 496 vectors $A_M^a$ with gauge group $SO(32)$ or
$E_8\times E_8$.

\vspace{0.5cm}{\bf Type IIA} It is made of closed strings, has
$N=2$ supersymmetry, contains the metric $g_{MN}$, an
antisymmetric tensor $B_{MN}$, the dilaton $\phi$, a vector $A_M$
and a 3-form $C_{MNP}$.

\vspace{0.5cm}{\bf Type IIB} It is made of closed strings, has
$N=2$ supersymmetry, contains the metric $g_{MN}$, an
antisymmetric tensor $B_{MN}$, the dilaton $\phi$, a scalar $l$,
a 2-form $C_{MN}$ and a 4-form $A_{MNPQ}$ with self-dual field
strength.

\vspace{0.5cm}At first the fact that several consistent string
theories could exist seemed unappealing, because string theory
was supposed to unify all forces in a single framework, and this
lack of unicity was a serious drawback. But it was realized in
the mid 90ies that several relations, the dualities, hold between
these 5 theories. In this thesis we will concentrate on one of these dualities,
mirror symmetry. Before describing it in more detail, we need to introduce
the notion of phenomenology and compactification in string theory.

If string theory is to be viable, then, in some limit to define,
it should be able to reduce to the Standard Model, which has given
extremely accurate predictions for numerous experiments up to
now. The mass scale of string theory is of Planck
order, and the masses that today's experiments can probe are of
order 1 Tev, so it is obvious that only a low-energy limit of
string theory should give the Standard Model. As a first step,
one restricts the spectrum to the massless particles, which are
the only plausible candidates for the known particles. In a
second step one needs to find a space-time action for these
fields. To do so, one uses the constraints implied by Weyl
symmetry of the string action. The massless fields obtained in
the various spectra of the 5 string theories can serve as
coupling functions in the world sheet action. However, not all
field configurations preserve Weyl symmetry at the quantum level.
There is a Weyl anomaly which lies in the trace of the
energy-momentum tensor and is absent only if the $\beta$-functions
governing the behaviour of each coupling vanish. This leads to a
set of equations that take the form of equations of motion. These
equations in turn can be obtained from a variational
principle applied to a space-time action, the supergravity action.
This new theory is the low-energy limit of the corresponding
string theory.

A fundamental issue is then the search for consistent solutions to
these equations of motion, called backgrounds or vacua. Finding a
consistent background means giving vacuum expectation values to
all fields of the theory in such a way that the equations of
motion are satisfied. A very general supergravity describes the
dynamics of a graviton, some p-forms, and a set of scalars (or
0-forms) in the bosonic sector, as well as their supersymmetry
partners in the fermionic sector. If one's interest is restricted
to massless fields without potential, flat Minkowski space-time
with vanishing values for all fields (other than the metric) is
always a solution. However one may be interested in more
sophisticated backgrounds for which some fields acquire non
vanishing values and/or the metric is no longer flat.

Finding all solutions is of course an extremely difficult problem;
one starts instead from simplifying ans\"atze. In the
Kaluza-Klein (KK) framework, the geometry of the $D$-dimensional
space-time is that of a product of the space-time in $d<D$
dimensions times an internal manifold. Since in this thesis we
want to consider theories in $d=4$, the KK ansatz reads

\beq \cM_{10} = \cM_4\times\cI_6 \eeq where $\cM_{10}$ is the
product of the 4-dimensional space-time $\cM_4$ and an internal
compact\footnote{Recently it has been proposed that some non
compact spaces may play the role of internal manifold, leading to
large extra dimensions of millimeter size. The mass scale of the
string theory is no longer the Planck mass, but a mass of order 1
Tev, which solves the hierarchy problem. We will not
consider this formalism in the following, and we refer the reader
for example to refs \cite{AHDD,AAHDD}.} manifold $\cI_6$. On the
fields, it amounts to a separation of space-time and internal
variables. A 10-dimensional field $\hat\phi$ is expanded
according to

\beq \hat\phi (\hat x) = \phi_n(x)h_n(y) \eeq where $x$ and $y$
are the space-time and internal coordinates, and $h_n$ are
harmonic functions on the internal manifold.

 The geometry of the internal space has to be
consistent with Einstein's equation

\beq R_{MN} = F_{M..}^{(i)}F_{N}^{(i)..} \eeq where $R_{MN}$ is
the Ricci tensor and $F_{M..}^{(i)}$ is an i-form field strength.
If one assumes that the field strengths have no purely internal
components, then the internal manifold has to be Ricci-flat.

Further information can be obtained by imposing a specific number
of conserved supercharges in 4 dimensions. Such a condition
highly constrains the internal manifold. Take for instance type
II supergravities in 10 dimensions. These have invariance under
the maximal number of supercharges, that is, 32. If one wants to
obtain an $N=2$ supergravity in 4 dimensions, the internal
manifold has to preserve $1/4$th of the supercharges. These
manifolds are known as Calabi-Yau spaces. Calabi-Yau
compactifications do not produce any potential for the moduli,
the scalars that parameterize degenerate vacua. Consequently, the
vacua correspond to arbitrary constant values of the moduli and
are degenerate. Moreover, $N=2$ supersymmetry remains completely
unbroken.

However, for phenomenological reasons, it is interesting to look
for vacua where $N=2$ supersymmetry is partially or completely
broken. To this end, one can relax the above constraints for the
field strengths and allow for some purely internal components. For
consistency with Bianchi Identities, the field strengths have to
be expanded on the harmonic forms on the internal \CY space, and
the number of flux parameters is determined by the Betti number
of the cohomology class which is expanded on. The main features
of the introduction of such fluxes are the gauging of some
isometries of the scalar manifold and the appearance of a scalar
potential. The minimum of this potential is generally obtained
for non-vanishing values of the scalars, and $N=2$ supersymmetry
is broken\footnote{At generic points in the field space $N=2$
supersymmetry is completely broken, but at some particular points
it can be either partially broken to $N=1$ or completely unbroken
\cite{CKLT}.}.

Under the relaxed constraints including fluxes, it is interesting
to study the fate of mirror symmetry. Mirror symmetry is one of
the dualities relating the 5 string theories in 10 dimensions. In
the case of type II theories, it states that type IIA supergravity
on a \CY 3-fold is the same as type IIB on the mirror \CY
manifold, defined by the exchange of odd and even cohomologies.
This symmetry still holds when Ramond-Ramond (R-R) fluxes are
turned on. Recall that type II spectra are divided into two
sectors. The Neveu-Schwarz (NS-NS) sector, common to both type II
theories, contains the metric, the dilaton, and an antisymmetric
tensor (or 2-form), while the R-R spectra contains only forms of
different degrees, depending on the type of the supergravity. In
type IIA, the R-R sector contains a 4-form and a 2-form field
 strengths. This means that the R-R fluxes lie in the even cohomologies. On the other
side, in type IIB, the R-R field strengths are a 1-form, a
3-form, and a 5-form, and the corresponding fluxes are found in
the odd cohomologies. It can be checked by a KK reduction that
the fluxes are correctly mapped \cite{LM2}.

The case of the NS-NS fluxes is less clear. Since the only NS-NS
form is a 3-form field strength in both type IIA and IIB, the
form fluxes should be found in the odd cohomologies in both
cases, and there is no possible mirror map. However, the mirror
fluxes may lie in the other fields of the NS-NS sector, the
metric and the dilaton. This has led Vafa to conjecture that the
mirror manifold should no longer be Calabi-Yau, but should instead
have a non-integrable complex structure, and the fluxes would be
expected to lead to this deformation of the geometry \cite{Vafa}. In this
thesis we will introduce new manifolds, called half-flat spaces,
which we conjecture to be the mirror image of a Calabi-Yau
manifold when NS-NS fluxes are turned on. We will display several
checks for this conjecture, which are based on the papers
\cite{GLMW} and \cite{GM}.

Compactifications of $N=2$ $d=10$ supergravities will be the
subject of the first two chapters. These supergravities bear
invariance under 32 supercharges.
We will consider compactifications on general manifolds with
$SU(3)$-structure (which includes Calabi-Yau), and consequently we will obtain (gauged or
ungauged) $N=2$ supergravities in 4 dimensions. Such theories
contain the gravity multiplet, as well as vector multiplets and
hypermultiplets in the matter sector. The couplings of these
multiplets are characterized by two holomorphic homogeneous
functions of degree 2, the prepotentials $\cF(X)$ and $\cF(z)$,
one describing each sector. The dynamics of these theories is
highly constrained by supersymmetry, but there is still room for
the choice of a prepotential. Supergravities with $N>2$ are of
course more constrained, and the maximal $N$ for which matter
multiplets exist is $N=4$. Thus $N=4$ supergravities are of
particular interest. Their dynamics allows for matter multiplets,
but their structure is quite simple since entirely determined by
supersymmetry. Although the action for this multiplet is known
for quite long \cite{Cha81a,NT81}, until recently a completely
satisfactory formulation in superspace was still missing. In
\cite{GHK01}, a set of constraints on the torsion was proposed,
and it was argued that the resulting multiplet should be
equivalent to the one of \cite{NT81}. The particular features of
this formulation were that it was making use of central charge
superspace, and the vectors (graviphotons) were identified in the
components of the vielbein carrying central charge indices.
However, a complete proof of the equivalence between the two
formalisms, in components and in superspace, was not yet worked
out. In this thesis, we will fill in this gap. We will compute the
equations of motion for all members of the multiplet in the
superspace formalism, and we will show that they are exactly the
same as the one arising from the Lagrangian of \cite{NT81}. These
results were obtained in \cite{GK02}.

In the second chapter, we describe in more detail the procedure
of compactification, in general, and then in the particular case
of a Calabi-Yau manifold. The main features of these
compactifications in the case of type II theories are reviewed,
the stress is put on the notion of moduli space, and as a
conclusion we display the map between type IIA and type IIB
supergravities, illustrating mirror symmetry.

In the third chapter, we introduce the half-flat manifolds. We
recall their properties relevant to our purpose, and we motivate
our conjecture by compactifying type IIA supergravity on such
manifolds and showing that it is equivalent to type IIB
supergravity on a Calabi-Yau manifold with NS-NS fluxes turned
on. We then check the converse of the above procedure.
We compactify type IIB theory on a half-flat manifold, and we
show that it is equivalent to type IIA on a Calabi-Yau manifold
with NS-NS fluxes turned on. We conclude with some conjectures
about the moduli space of half-flat manifolds and we try to give
hints about possible ans\"atze for a description of the magnetic
fluxes.

In chapter 4, we present our work on $N=4$ supergravity in 4
dimensions. We recall how the use of central charge superspace
made it possible to identify the gravity multiplet in the
components of the vielbein and the torsion, and we show the
equivalence with the formulation in components by deriving the
equations of motions for all members of the multiplet and
identifying them with the ones obtained from the Lagrangian of
\cite{NT81}.

Various elementary notions of differential geometry can be found
in appendix \ref{AppA}. Appendix \ref{AppB} contains basic
properties of Calabi-Yau manifolds, and many formulae useful for
the compactification are gathered. In appendix \ref{typeIINS} we
briefly recall the compactification of type II theories on \CY
manifolds with NS fluxes. In appendix \ref{acs} we review a few
facts about G-structures from the mathematical point of view. We
present the computation of the Ricci scalar of half-flat
manifolds in appendix \ref{Rhf}. Finally, appendix \ref{torcurv}
displays the components of the torsion and curvature for $N=4$
supergravity in central charge superspace.

\chapter{Calabi-Yau compactifications}\label{CYcomp}

This section is a review of the compactification of type II
supergravities on Calabi-Yau manifolds. Such compactifications were
first considered in \cite{BCF} for type IIA and in
\cite{Bodner:1990ca} for type IIB, see \cite{BGHL} for a review in
more recent notations.

In the first part we recall briefly the basics of the theory of
reduction \`a la Kaluza-Klein, first on the circle, and then on a
general n-dimensional manifold. We emphasize the main features of
such reductions, and the relations between the topology of the
internal manifold and the properties of the 4-dimensional theory.
 Then we deal with the issue of supersymmetry
conservation during compactification which leads to the emergence
of Calabi-Yau manifolds. We describe in details the
reduction of the Ricci scalar and we introduce the notion
of moduli space. In the second and the third parts, we turn to the
compactification of the bosonic actions for type IIA and IIB
supergravities. We show how the fields arrange in supergravity
multiplets and we give the 4-dimensional action. We conclude by
displaying the mirror map between the two theories.

\section{Kaluza-Klein compactification}

\subsection{Reduction on a circle}\label{cir}

In the case of the compactification on a circle, the $d$
space-time coordinates $\hat{x}^M$, split into one set of $d-1$
space-time coordinate $x^{\mu}$ and one internal coordinate $y$,
subject to the periodicity condition $y\sim y+R$ where $R$ is the
radius of the circle. It is well known that with such a
periodicity property, any quantity $\hat{\Phi}$ can be expanded on
a basis of periodic functions

\begin{equation} \hat{\Phi}(\hat{x}) = \sum_n
\tilde\Phi_n(x)\rme^{in\frac{y}{R}}\label{cir1}\end{equation}
where $\tilde\Phi_n(x)$ is the Fourrier transform of
$\hat\Phi(\hat x)$. We note that the $\rme^{in\frac{y}{R}}$ are
solutions to Laplace equation

\beq \Delta \rme^{in\frac{y}{R}} = \frac{\del^2}{\del y^2}
\rme^{in\frac{y}{R}} =
-\frac{n^2}{R^2}\rme^{in\frac{y}{R}}\label{cir2}\eeq with "mass"
$n/R$. In all that follows, we will only consider the low energy
limits of string theory, supergravities. Thus we will always
truncate the summation at the massless level, which means that we
only keep the term with $n=0$ in (\ref{cir2}), called massless
mode. For more details about the consistency of this procedure,
see \cite{Haack:2001ha} p.85 and references therein. As an
example, we display the massless reduction of the metric

\beq \hat g_{MN} \longrightarrow \left(\begin{array}{cc} g_{\mu\nu} & V_{\mu}\\[2mm]
V_{\mu} & \phi\end{array}\right). \label{cir3}\eeq When the two
indices are external, the $d$-dimensional metric becomes the
$(d-1)$-dimensional one, when one index is external and one is
internal, it has the index structure of a space-time vector
$V_{\mu}$, and when the two indices are internal, it is a scalar
$\phi$. None of these new fields depends on the internal
coordinate $y$, which corresponds to a massless expansion. This leads, up to field
redefinitions, to the following expansion for the Ricci scalar

\beq \hat\cR = \cR + F_{\mu\nu}F^{\mu\nu} +
\del_{\mu}\phi\del^{\mu}\phi\label{cir4}\eeq where $F_{\mu\nu}$
is the field strength of the vector $V_{\mu}$. This is the action
for gravity coupled to electro-magnetism and an uncharged scalar.

\subsection{Reduction on a compact manifold of dimension n}\label{gen}

Kaluza-Klein reductions applied to supergravity have been
described in details in \cite{Duff:1986np,Nieu:1985ni}. Here we
only give the features that will be relevant to the next
sections. The $d$ space-time coordinates $\hat{x}^M$, split into
one set of $d-n$ space-time coordinates $x^{\mu}$ and one set of
$n$ internal coordinates $y^m$. Let us take the example of a
scalar field $\hat\Phi(\hat x)$. Suppose that the metric is
block-diagonal, which will always be the case from now on. Then
the $d$-dimensional equation of motion can be written as

\beq \hat\Delta\hat\Phi = m^2_d\hat\Phi(\hat x) =
\Delta_{d-n}\hat\Phi + \Delta_n\hat\Phi.\label{gen1}\eeq where
$\Delta_{d-n}$ and $\Delta_n$ are Laplacians in lower dimensions.
The Kaluza-Klein ansatz on $\hat\Phi$ reads

\beq \hat\Phi(\hat x^M) =
\phi^i(x^{\mu})\ox_i(y^m)\label{gen2}\eeq where $\ox_i$ is a set
of a priori unknown functions, counted by the index $i$. We
assume that the $(d-n)$-dimensional scalars $\phi^i$ also obey
their usual field equation, which leads to

\beq m^2_d \phi^i(x^{\mu})\ox_i(y^m) =
m^2_{d-n}\phi^i(x^{\mu})\ox_i(y^m) +
\phi^i(x^{\mu})\Delta_n[\ox_i(y^m)]. \label{gen3}\eeq This means
that the functions $\ox_i$ have to satisfy Laplace equation

\beq \Delta_n\ox_i = (m^2_d-m^2_{d-n}) \ox_i.\label{gen4}\eeq A
result of differential analysis states that on a compact
manifold, the eigenvalues of Laplace operator $\Delta$ are of the
form $n/S$ where $S$ corresponds to the size of the manifold, and
$n$ is an integer. Again, we will only consider massless
expansions, for which $n=0$. The unknown functions are thus
harmonic

\beq \Delta \ox_i = 0. \label{gen5}\eeq

The bosonic fields of supergravities are either the metric or
forms. The case of the metric is described in section \ref{ms}.
For the forms, the above result generalizes in the following way.
A p-form $\hat B_p$ is expanded on all harmonic q-forms for $0\leq
q\leq p$

\beq \hat B_p = B_p^{i_0}\ox_{i_0} + B_{p-1}^{i_1}\ox_{i_1} + ...
+ B_0^{i_p}\ox_{i_p}\label{gen6}\eeq where $\ox_{i_0}$ is a basis
for the harmonic 0-forms and so on. For the definition of
harmonicity on forms, see appendix \ref{forms}.

\subsection{Calabi-Yau requirement}\label{cyr}

Consider now a supergravity theory in 10 dimensions, compactified
on a 6-dimensional compact space. On the fermionic side, there is
always a gravitino, whose supersymmetry transformation is related
to the covariant derivative of the parameter. Suppose we are
looking for a supersymmetric background, then all vacuum values of
transformations of fermionic fields must vanish. For the
gravitino, we obtain

\beq <\dx \hat\Psi_M> = <\cD_M \hat\epsilon> = 0,
\label{cyr1}\eeq where $\hat\epsilon$ is the supersymmetry parameter.
We use the Kaluza-Klein ansatz for this spinor

\beq \hat\epsilon(\hat x) = \epsilon(x)\eta(y)\label{cyr2}\eeq
where $\epsilon$ is a spinor in 4 dimensions, and $\eta$ is a
spinor on the internal space. If we take an internal component in
(\ref{cyr1}), we obtain that $\eta$ must be covariantly constant

\beq \nabla_m\eta = 0. \label{cyr3}\eeq This is a very strong
statement. For each covariantly constant spinor on the internal
space, there is one conserved supersymmetry in 4
dimensions. For type II theories, the
Kaluza-Klein ansatz reads

\beq \hat\epsilon^A(\hat x) =
\epsilon^A(x)\eta(y)\label{cyr4}\eeq where $A = 1,2$ counts the
supersymmetry operators. Since we are interested in $N=2$
supergravities in 4 dimensions,
 we must compactify on a space possessing one covariantly constant spinor: such
 spaces are called Calabi-Yau manifolds. These manifolds have numerous
interesting properties. The fact that they admit one covariantly constant spinor
restricts their holonomy group to $SU(3)$. They are complex, K\"ahler and
Ricci-flat. This is consistent with Einstein's equation which relates the Ricci tensor to
squares of field strengths of forms appearing in the spectra of type II supergravities

\beq \hat R_{MN} \sim \hat F_{MPQ...}\hat F_N{}^{PQ...}.
\label{ein1}\eeq In the case of KK
compactifications, the field strengths have no purely internal components, which
means that to be a solution to the equations of motions,
the internal space must be Ricci-flat. Among other properties, \CY manifolds have one and
only one covariantly constant harmonic (3,0)-form, and their Hodge diamond
has the form

\beq\begin{array}{cccccccl}&&&1&&&&\qquad b^0=1\\
&&0&&0&&&\qquad b^1=0\\
&0&&h^{(1,1)}&&0&&\qquad b^2=h^{(1,1)}\\
1&&h^{(2,1)}&&h^{(2,1)}&&1&\qquad b^3=2+2h^{(2,1)}\\
&0&&h^{(1,1)}&&0&&\qquad b^4=h^{(1,1)}\\
&&0&&0&&&\qquad b^5=0\\
&&&1&&&&\qquad b^6=1.\end{array}\label{cyr5}\eeq

\subsection{Moduli space}\label{ms}

We have seen in section \ref{gen} that the general ansatz for the
compactification of forms requires expansion on harmonic forms on
the Calabi-Yau manifold. This is in some sense also true for the
metric. When the two indices are external, we obtain the 4-dimensional metric.
There will be no components with one external and one internal index,
because there are no
1-forms on the Calabi-Yau (\ref{cyr5}). Now we are left with purely internal
components. Solutions to supergravities are generally not isolated, but come in
continuous families, parameterized by the moduli. The metric
moduli are thus infinitesimal deformations of the internal metric that conserve
the Calabi-Yau conditions. Let us start from a background with
hermitian metric $g^0_{mn}$ and define the metric on a Calabi-Yau
manifold infinitesimally close to the first one by $g_{mn} =
g^0_{mn} +\dg_{mn}$. The
constraint that the new manifold is Calabi-Yau can be expressed
by imposing Ricci-flatness. This gives Lichnerowicz
equation

\beq \nabla^l\nabla_l\dg_{mn}-[\nabla^l\, ,\,
\nabla_m]\dg_{ln}-[\nabla^l\, ,\, \nabla_n]\dg_{lm} = 0.
\label{ms1}\eeq Considering the properties of the Riemann tensor
of a K\"ahler space \ref{ack}, we can see that, in complex indices, this equation
splits into one on the mixed part of the metric, and one on
the pure part. In the coordinates (\ref{ack2}), the K\"ahler class is
directly related to the metric

\beq J_{\ax\ab} = ig_{\ax\ab},\label{ka1}\eeq so the mixed
variations of the metric are K\"ahler class deformations. Suppose
we apply to the metric a pure variation followed by a small change
of coordinates $f^m$ in such a way that the first variation is annihilated. Then we
have

\beq \dx g_{\ax\bx} = \frac{\del\bar f^{\ab}}{\del z^{\ax}}
g_{\ab\bx}+\frac{\del\bar f^{\ab}}{\del z^{\bx}}
g_{\ab\ax}.\label{com1}\eeq This means that $f$ cannot be
holomorphic and the manifold obtained after a pure variation $\dx
g_{\ax\bx}$ has a different complex structure. This is why such
deformations are complex structure deformations.

Obviously, (\ref{ms1}) is Laplace equation (\ref{forms12}), except
that $\dg_{mn}$ is not a form. Following this idea, we expand the
mixed component of $\dg$ on the (1,1)-forms on the Calabi-Yau

\beq \dg_{\ax\bb} = -iv^i(\ox_i)_{\ax\bb} \label{ms2}\eeq where
$v^i$ are $h^{(1,1)}$ real scalars. $\ox_i$ is harmonic, so it
satisfies Laplace equation, and (\ref{ms2}) solves the mixed part
of Lichnerowicz equation. Since there are no (2,0)-forms, and
anyway $\dg_{\ax\bx}$ is symmetric, it is not possible to expand
it directly. We take instead

\beq \dg_{\ax\bx} = \frac{i}{||\Ox||^2}\bar
z^a(\bar\eta_a)_{\ax\bb\cb}\Ox^{\bb\cb}{}_{\bx}, \label{ms3}\eeq
where $\Ox$ is the (3,0)-form and we have expanded on the
$(1,2)$-forms $\bar\eta_a$, with $h^{(2,1)}$ scalar complex coefficients
$\bar z^a$. It is
shown in appendix \ref{lse} that this is symmetric and solution to
(\ref{ms1}). The metric moduli space is thus generated by
$h^{(1,1)}+2h^{(2,1)}$ real parameters. For further information about
its structure, see appendix \ref{msa}.

We also need to compute the Ricci scalar in the
10-dimensional action. Since the moduli are infinitesimal
parameters, we make a perturbation expansion, and we keep all
terms up to order 2. The detailed calculation can be found in
appendix \ref{crs}. Here we only display the result, which, as
might be expected from (\ref{cir4}), shows the emergence of
kinetic terms for the scalars $v^i\, ,\, z^a$

\bea \int\rd^{10}\hat x\sqrt{-\hat g}\hat R  = &
\int\rd^4x\sqrt{-g_4}& \left(\cK R_4+
P_{ij}\del_{\mu} v^i\del^{\mu} v^j+Q_{ab}\del_{\mu} z^a\del^{\mu}\bar z^b\right)
 \label{ms4}\eea where $\cK$ is the volume of the Calabi-Yau manifold and the couplings
 are defined in (\ref{crs16}) and (\ref{crs0017}).
 The scalars are organized in two non-linear sigma models, whose metrics
  are both Special K\"ahler with K\"ahler potentials (\ref{msa5}) and
  (\ref{msa8}). The whole moduli
 space has the structure of a product of
 two Special K\"ahler manifolds, one corresponding to the K\"ahler
 class deformations $\cM_{1,1}$, of complex\footnote{Once the $B_2$ moduli
 are taken into
 account
 as in (\ref{msa2}).} dimension
 $h^{(1,1)}$, and one corresponding to the complex
 structure deformations $\cM_{2,1}$ of complex dimension $h^{(2,1)}$

 \beq
\cM = \cM_{1,1}\times\cM_{2,1}.\label{pro1}
 \eeq

%
%
\section{Compactification of type IIA supergravity}
%
%

%
\subsection{Reduction of the Ricci scalar and the
dilaton}\label{crd}
%

As we will see in the next sections, there is a part of the
Lagrangian which is common to type IIA and type IIB supergravities.
This comes from the fact that the spectrum of both theories is
composed of two sets of fields, the R-R and the NS-NS fields.
Type IIA and type IIB supergravities have the same NS-NS
spectrum, but differ in the R-R sector. Thus their NS-NS
Lagrangian is identical. It contains the graviton $\hat g_{MN}$ ,
the dilaton $\hat \phi$, and the NS 2-form $\hat B_2$. Later on in
this thesis, we will be interested in turning on fluxes for some
forms, including $\hat B_2$, such that the part of the Lagrangian
which contains only the graviton and the dilaton will always be
compactified in the same way. This procedure is described below.
The action we will study is

\bea
  S & = & \int \, e^{-2\hat\phi} \left( -\frac12 \hat R *\! {\bf 1} + 2
  d \hat\phi \wg * d \hat\phi\right).\label{crd1}
\eea Remark that the kinetic term for the scalars has a wrong
sign. This is because this action is written in the string frame.
We will now perform a Weyl rescaling on (\ref{crd1}) to go to
Einstein's frame. Recall also that formula (\ref{crs15}) is only
true up to total derivatives, which means that it is not possible
to use it directly as it stands in the string frame.

\subsubsection{$1^{\rm{st}}$ step : going to Einstein's frame}

We perform the Weyl rescaling (\ref{eg6}) with $\Ox =
\rme^{-\hat\phi/4}$. We obtain

\bea
  S & = & \int \, -\frac12 \hat R *\! {\bf 1} -\frac14
  d \hat\phi \wg * d \hat\phi.\label{crd3}\eea Here we have to be careful about the
  fact that under this rescaling, the determinant of the metric and the inverse
  metrics used to contract indices are not written explicitly, but should be
  transformed. Remark that now the kinetic term for the dilaton
  has a correct sign.

\subsubsection{$2^{\rm{nd}}$ step : compactifying}

We use formula (\ref{crs15}). This leads to

\bea
  S & = & \int \, -\frac12\cK R *\! {\bf 1} -\cK\frac14
  d \hat\phi \wg * d \hat\phi -\frac12 P_{ij}d v^i\wg * d v^j\nn\\[2mm]
&&-\frac{1}{2}Q_{ab}d z^a\wg * d\bar z^b,\label{crd4}\eea
where the integral is now only on the 4-dimensional space-time.

\subsubsection{$3^{\rm{rd}}$ step : Weyl rescaling for the volume $\cK$}

In order to have the usual normalization for the Ricci scalar, we
perform the Weyl rescaling (\ref{eg6}) with $\Ox = \cK^{\frac12}$. We obtain

\bea
  S & = & \int \, -\frac12 R *\! {\bf 1} -\frac34
  d \ln\cK \wg * d \ln\cK-\frac14
  d \hat\phi \wg * d \hat\phi -\frac{1}{2\cK} P_{ij}d v^i\wg * d v^j\nn\\[2mm]
&&-\frac{1}{2\cK}Q_{ab}d z^a\wg * d\bar z^b.\label{crd5}\eea

\subsubsection{$4^{\rm{th}}$ step : rotation of the $v^i$}

We realize now a rotation of the $v^i$. The purpose is to
eliminate the term $d \ln\cK \wg * d \ln\cK$ which is not a
new scalar, but depends on the $v^i$. We define

\beq v^i = \rme^{-\frac12\hat\phi}\tilde v^i.\label{crd6}\eeq
Considering that the basis forms are independent of $v^i$, we
find the following transformation rules for the integrals defined
in appendix \ref{icy}

\bea \cK_{ijk} = \tilde\cK_{ijk}\quad & ; & \quad \cK_{ij} =
\rme^{-\frac12\hat\phi}\tilde\cK_{ij}\nn\\[2mm]
P_{ij} = \rme^{-\frac12\hat\phi}\tilde P_{ij}\quad & ; & \quad
\cK_{i} =
\rme^{-\hat\phi}\tilde\cK_{i}\label{crd7}\\[2mm]
\cK = \rme^{-\frac32\hat\phi}\tilde\cK\quad & ; & \quad g_{\ax\bb} =
\rme^{-\frac12\hat\phi}\tilde g_{\ax\bb},\nn
 \eea the last equation holding because the volume is an integral on $\sqrt{g_6}$
 which is of order 3 in the metric with lower indices. Performing a careful counting of
 the number of lower metrics in $Q_{ab}$, we find

 \beq
Q_{ab} = \rme^{-\frac32\hat\phi}\tilde Q_{ab}.
 \label{crd8}\eeq The transformation of the kinetic terms for the $v^i$ is

 \bea
\frac{1}{2\cK}P_{ij}\del v^i\del v^j & = & \tilde g_{ij}\del
\tilde v^i\del \tilde v^j
-\frac12\del\ln\tilde\cK\del\ln\tilde\cK+\frac54\del\ln\tilde\cK\del\hat\phi-\frac{15}{16}
\del\hat\phi\del\hat\phi \label{crd9} \eea where the metric
$g_{ij}$ is given in (\ref{gH11}) and we have used

\beq \tilde\cK_i\del \tilde v^i =
2\del\tilde\cK.\label{crd10}\eeq Finally we obtain the action

\bea
  S & = & \int \, -\frac12 R *\! {\bf 1} -d \phi \wg * d\phi -g_{ij}
  d v^i\wg * d v^j\nn\\[2mm]
&&-g_{ab}d z^a\wg * d\bar z^b.\label{crd11}\eea Here we
defined the 4-dimensional dilaton by

\beq \phi = \hat\phi-\frac12\ln\tilde\cK, \label{crd12}\eeq we
dropped the tildes, and the metric for the scalars $z^a$ is

\beq g_{ab} = \frac{1}{2\cK}Q_{ab}.\label{crd13}\eeq The metrics
$g_{ij}$ and $g_{ab}$ exhibit properties detailed in appendix
\ref{msa}.

%
\subsection{Matter part of type IIA supergravity}\label{cA}
%

In this section we recall the known results of type IIA
supergravity compactified on a Calabi-Yau threefold $Y$. The NS-NS
spectrum consists of the graviton, the
dilaton, a 2-form $\hat B_2$, and the R-R spectrum contains a
1-form $\hat A_1$ and a 3-form $\hat C_3$. We start from the
following action in 10 dimensions

\begin{eqnarray}
  \label{SIIA10}
  S & = & \int \, e^{-2\hat\phi} \left( -\frac12 \hat R *\! {\bf 1} + 2
  d \hat\phi \wg * d \hat\phi - \frac14  \hat H_3\wg * \hat H_3 \right) \nn \\
  & & - \frac12  \int \, \left(\hat  F_2 \wg * \hat F_2 + \hat F_4 \wg * \hat F_4
  \right) + \frac12 \int \hat H_3 \wedge \hat C_3 \wedge d \hat C_3 \, ,
\end{eqnarray} where

\beq\label{HFdef} \hat H_3 = d\hat B_2\ ,\qquad \hat F_2 = d\hat
A_1\ ,\qquad \hat F_4 = d \hat C_3 - \hat A_1 \wedge\hat H_3,
\eeq and we will follow the above procedure step by step. From now on
we will not display the part with the graviton and the dilaton whose
behaviour has been studied in the previous section.

\subsubsection{$1^{\rm{st}}$ step : going to the Einstein's frame}

We perform the Weyl rescaling (\ref{eg6}) with $\Ox =
\rme^{-\hat\phi/4}$. We obtain

\begin{eqnarray}
  \label{cA1}
  S & = & - \frac14\int\, \rme^{-\hat\phi}  \hat H_3\wg * \hat H_3 - \frac12  \int \,
  \rme^{\frac32\hat\phi}\hat  F_2 \wg * \hat F_2 -\frac12\int\,\rme^{\frac12\hat\phi}
   \hat F_4 \wg * \hat F_4 \nn \\
  & &  + \frac12 \int \hat H_3 \wedge \hat C_3 \wedge d \hat C_3 \, ,
\end{eqnarray} where the Chern-Simons term remains unchanged because it does not
contain any metric.

\subsubsection{$2^{\rm{nd}}$ step : expanding}

In the KK reduction we expand the ten-dimensional fields in terms
of harmonic forms on $Y$ \begin{eqnarray}
  \label{fexpA}
  \hat A_1 & = & A^0 \ ,\nn \\
  \hat C_3 & = & C_3 + A^i \wg \ox_i + \xi^A \ax_A + \txi_A \bx^A \ ,\\
  \hat B_2 & = & B_2 + b^i \ox_i \ ,\nn
\end{eqnarray} where $C_3$ is a 3-form, $B_2$ a 2-form, $(A^0,A^i)$ are
1-forms and $b^i, \xi^A, \txi_A$ are scalar fields in $D=4$. The
$\ox_i$ are a basis for the harmonic $(1,1)$-forms and $(\ax_A\,
,\, \bx^A)$ a basis for the harmonic 3-forms, see appendix
\ref{icy}. All these 4-dimensional fields are organized in
supergravity multiplets. The gravity multiplet contains the metric
$g_{\mu\nu}$ and the graviphoton $A^0$. The $h^{(1,1)}$ vectors
$A^i$, together with the $2h^{(1,1)}$ scalars $v^i,b^i$ belong to
$h^{(1,1)}$ vector multiplets. The rest of the fields only consists
of scalars. Therefore all these scalars belong to hypermultiplets.
Indeed, $C_3$ is non dynamical in 4 dimensions, and $B_2$ is
dual to a scalar $a$. Collecting the remaining $4h^{(2,1)}+4$
scalars $\phi$, $z^a$, $a$, $\xi^A$ and $\tilde\xi_A$, we obtain
$h^{(2,1)}+1$ hypermultiplets. Since the harmonic forms are closed,
the differential operator $d$ acts only on the space-time forms

\bea
d \hat A_1 & = & d A^0 \label{cA001} \\[2mm]
d \hat C_3 & = & d C_3 + d A^i \wg \ox_i + d\xi^A\wg \ax_A + d\txi_A\wg \bx^A
\label{cA002}\\[2mm]
\hat H_3 & = & H_3 + d b^i\wg \ox_i.\label{cA003}\eea The terms of (\ref{cA1})
are thus
expanded according to

\begin{eqnarray}
   - \frac14\rme^{-\hat\phi} \int_Y\hat H_3 \wg * \hat H_3
  & = & - \frac{\cK}{4}\rme^{-\hat\phi} \, H_3 \wg * H_3 - \cK\rme^{-\hat\phi}
  g_{ij} db^i \wg * db^j \,\label{cA2} \\ [5mm]
  - \frac12\rme^{\frac32\hat\phi} \int_Y \hat F_2 \wg * \hat F_2 & = &
  - \frac{\cK}{2}\rme^{\frac32\hat\phi} \, d A^0 \wg * dA^0 \ ,\label{cA3}\\ [5mm]
  - \frac12 \rme^{\frac12\hat\phi}\int_Y {\hat {F}_4} \wg * {\hat {F}_4} & = & -
  \frac{\cK}{2}\rme^{\frac12\hat\phi}\, (d C_3 - A^0\wedge H_3) \wg * (d C_3 - A^0\wg H_3)
  \nn \\
  & & - 2 \cK\rme^{\frac12\hat\phi} g_{ij} (d A^i - A^0 d b^i) \wg * (d A^j - A^0 d b^j)
 \nn \\
  & & + \frac{1}{2}\rme^{\frac12\hat\phi}\left(\IM \cM ^{-1} \right)^{AB}
  \Big[ d\tilde\xi_A +  \cM_{AC} d\xi^C \Big] \wg * \Big[ d\tilde\xi_B +
  \bar \cM_{BD} d\xi^D \Big] \ , \nn \\[5mm]
  \frac12 \int_Y \hat H_3 \wedge \hat C_3 \wedge d \hat C_3
  & = &  - \frac12 H_3 \wg (\xi^A d \txi_A - \txi_A d \xi^A) +\frac12 d b^i
  \wg A^j \wg d A^k \cK_{ijk}  \, . \label{cA5}
\end{eqnarray} where the integration on the internal manifold has been carried
out
and the matrix $\cM$ is defined in appendix \ref{msa}.

\subsubsection{$3^{\rm{rd}}$ step : Weyl rescaling for the volume $\cK$}

In order to recover a standard kinetic term for gravity, we need to perform
a Weyl rescaling with Weyl factor $\cK^{\frac12}$

\begin{eqnarray}
   - \frac14\rme^{-\hat\phi} \int_Y\hat H_3 \wg * \hat H_3
  & = & - \frac{\cK^2}{4}\rme^{-\hat\phi} \, H_3 \wg * H_3 - \rme^{-\hat\phi}
  g_{ij} db^i \wg * db^j \,\label{cA6} \\ [5mm]
  - \frac12\rme^{\frac32\hat\phi} \int_Y \hat F_2 \wg * \hat F_2 & = &
  - \frac{\cK}{2}\rme^{\frac32\hat\phi} \, d A^0 \wg * dA^0 \ ,\label{cA7}\\ [5mm]
  - \frac12 \rme^{\frac12\hat\phi}\int_Y {\hat {F}_4} \wg * {\hat {F}_4} & = & -
  \frac{\cK^3}{2}\rme^{\frac12\hat\phi}\, (d C_3 - A^0\wedge H_3) \wg * (d C_3 - A^0\wg
 H_3)
  \nn \\
  & & - 2 \cK\rme^{\frac12\hat\phi} g_{ij} (d A^i - A^0 d b^i) \wg * (d A^j - A^0 d b^j)
 \nn \\
  & & + \frac{1}{2\cK}\rme^{\frac12\hat\phi}\left(\IM \cM ^{-1} \right)^{AB}
  \Big[ d\tilde\xi_A +  \cM_{AC} d\xi^C \Big] \wg * \Big[ d\tilde\xi_B +
  \bar \cM_{BD} d\xi^D \Big] \ , \nn \\[5mm]
  \frac12 \int_Y \hat H_3 \wedge \hat C_3 \wedge d \hat C_3
  & = &  - \frac12 H_3 \wg (\xi^A d \txi_A - \txi_A d \xi^A) +\frac12 d b^i
  \wg A^j \wg d A^k \cK_{ijk}  \, . \label{cA9}
\end{eqnarray}

\subsubsection{$4^{\rm{th}}$ step : rotation of the $v^i$}

Counting the powers of the metric in the definition of $\cM$ as an integral, we can
deduce that $\cM$ is invariant under this rotation. Once everything is written in
terms of the 4-dimensional dilaton, we obtain

\begin{eqnarray}
   - \frac14\rme^{-\hat\phi} \int_Y\hat H_3 \wg * \hat H_3
  & = & - \frac{1}{4}\rme^{-4\phi} \, H_3 \wg * H_3 -
  g_{ij} db^i \wg * db^j \,\label{cA10} \\ [5mm]
  - \frac12\rme^{\frac32\hat\phi} \int_Y \hat F_2 \wg * \hat F_2 & = &
  - \frac{\cK}{2}\, d A^0 \wg * dA^0 \ ,\label{cA11}\\ [5mm]
  - \frac12 \rme^{\frac12\hat\phi}\int_Y {\hat {F}_4} \wg * {\hat {F}_4} & = & -
  \frac{\cK}{2}\rme^{-4\phi}\, (d C_3 - A^0\wedge H_3) \wg * (d C_3 - A^0\wg H_3)
  \nn \\
  & & - 2 \cK g_{ij} (d A^i - A^0 d b^i) \wg * (d A^j - A^0 d b^j) \nn \\
  & & + \frac{1}{2}\rme^{2\phi}\left(\IM \cM ^{-1} \right)^{AB}
  \Big[ d\tilde\xi_A +  \cM_{AC} d\xi^C \Big] \wg * \Big[ d\tilde\xi_B +
  \bar \cM_{BD} d\xi^D \Big] \ , \nn \\[5mm]
  \frac12 \int_Y \hat H_3 \wedge \hat C_3 \wedge d \hat C_3
  & = &  - \frac12 H_3 \wg (\xi^A d \txi_A - \txi_A d \xi^A) +\frac12 d b^i
  \wg A^j \wg d A^k \cK_{ijk}  \, . \label{cA13}
\end{eqnarray}
With these expressions we can now combine the different terms
appearing in the action (\ref{SIIA10}) The dualization of the
3-form $C_3$ in 4 dimensions produces a contribution to the
cosmological constant. As shown in \cite{Louis:2002am} this
constant can be viewed as a specific RR-flux. Since we are not
interested in RR-fluxes here we choose it to be zero and hence
discard the contribution of $C_3$ in 4 dimensions. Thus the only
thing we still need to do in order to recover the standard
spectrum of $N=2$ supergravity in 4 dimensions is to dualize the
2-form $B_2$ to the axion $a$. The action for $B_2$ is

\begin{equation}
  \label{cA14}
  \cL_{H_3} = - \frac{1}{4} e^{-4\phi}H_3 \wg * H_3  + \frac12 H_3 \wg
  \left(\tilde\xi_A d \xi^A -\xi^A d \tilde\xi_A \right).
\end{equation} Counting the degrees of freedom of $B_2$, we know that it should be
dual to a scalar $a$, called the axion. We add to (\ref{cA14})
the term

\beq +\frac12 H_3\wg d a \label{cA15}\eeq which realizes the
Bianchi identity of $B_2$ as an equation of motion for $a$.
$H_3$ can consequently be considered as a fundamental field, and eliminated
through its equation of motion

\begin{equation}
  \label{cA16}
  - \frac{1}{2} e^{-4\phi}* H_3  + \frac12 \left( d a + \tilde\xi_A d \xi^A
  -\xi^A d \tilde\xi_A \right) = 0.
\end{equation} The Lagrangian for $a$ becomes

\beq  \cL_a = - \frac{1}{4} e^{+4\phi}\left( d a +
\tilde\xi_A d \xi^A -\xi^A d \tilde\xi_A \right) \wg *
\left( d a + \tilde\xi_A d \xi^A -\xi^A d \tilde\xi_A
\right). \label{cA17}\eeq The usual $N=2$ supergravity couplings
can be read off after redefining the gauge fields $A^i \to A^i -
b^i A^0$ and introducing the collective notation $A^I = (A^0,
A^i)$ where $I=(0,i) = 0, \ldots, h^{(1,1)}$.

Collecting all terms from (\ref{cA10})-(\ref{cA13}) and taking
into account the gravity sector \ref{crd11}, we obtain

 \begin{eqnarray}
  \label{S4A0}
  S_{IIA} & = & \int \Big[ -\frac12 R ^* {\bf 1} - g_{ij} dt^i \wg * d
  {\bar t}^j - h_{uv} d q^u \wg * d q ^v \nn \\
  & & \qquad + \frac{1}{2}\, \IM \cN_{IJ} F^I\wg * F^{J}
  + \frac{1}{2} \, \RE \cN_{IJ} F^I \wg F^J \Big] ,
\end{eqnarray} where the gauge coupling matrix $\cN$ is defined in appendix
\ref{msa}, the scalars $b^i$ coming from the NS 2-form and $v^i$
from the K\"ahler class deformations are complexified into $t^i =
b^i + iv^i$, and $h_{uv}$ is the $\sigma$-model metric for the
scalars in the hypermultiplets \cite{FeS}

\begin{eqnarray}
  \label{qktNS}
  h_{uv} dq^u \wg * dq ^v& = &  d\phi \wg * d\phi + g_{ab} dz^a \wg * d \bar z^b
  \\
  & & + \frac{e^{4\phi}}{4} \, \Big[ da +
  (\tilde\xi_A d \xi^A-\xi^A d\tilde\xi_A) \Big] \wg \rule{0pt}{12pt}^*
  \Big[da +
  (\tilde\xi_A d \xi^A-\xi^A d \tilde\xi_A) \Big] \nn \\
  & & - \frac{e^{2\phi}}{2}\left(\IM \cM^{-1} \right)^{AB}
  \Big[ d\tilde\xi_A + \cM_{AC} d\xi^C \Big]
   \wg \rule{0pt}{12pt}^* \Big[ d\tilde\xi_B + \bar \cM_{BD} d\xi^D \Big]  \,
  .\nn
\end{eqnarray}

\section{Compactification of type IIB supergravity}\label{cB}

In this section we recall the KK compactification of type IIB
supergravity on a \CY{} 3-fold $\tilde Y$. The 10 dimensional bosonic
spectrum of type IIB supergravity consists of the metric
$\hat{g}$, the antisymmetric tensor field $\hat{B_2}$ and the
dilaton $\hat{\phi}$ in the NS-NS sector, an axion $\hat{l}$,
a 2-form $\hat{C_2}$ and a 4-form $\hat{A_4}$ with self-dual
field strength $* \hat F_5 = \hat F_5$ in the R-R sector. No
local covariant action can be written for this theory in 10
dimensions due to the self-duality of $\hat F_5$. Instead we use
the action \cite{JP}
\begin{eqnarray}
  \label{cB1}
  S_{IIB}^{(10)} &=& \int e^{-2\hat\phi}\left(-\frac{1}{2}\hat R *\! {\bf 1} +
2d\hat\phi\wg     *d\hat\phi-\frac{1}{4}d\hat B_2\wg * d\hat B_2\right)\nonumber\\[2mm]
  &&-\frac{1}{2}\int \left( d\hat l\wg *d\hat l+ \hat F_3\wg
    * \hat F_3+\frac12\hat F_5\wg * \hat F_5\right)\nonumber\\[2mm]
  &&-\frac12\int\hat A_4\wg d\hat B_2\wg d\hat C_2 ,
\end{eqnarray} where

\begin{eqnarray}
\hat F_3 & = & d \hat C_2 -\hat l \hat H_3\label{CB02}\\[2mm]
 \hat F_5 & = & d \hat A_4 - d \hat B_2 \wg \hat C_2\ ,
  \label{cB2}
\end{eqnarray} and impose the self-duality of $\hat F_5$  separately, at the
level of the equations of motion. The compactification proceeds
as usual, by following the steps of the above sections. However, we will skip the
steps that are not essential to our purpose.

We expand the 10-dimensional quantities in terms of
harmonic forms on the \CY{} manifold as

\begin{eqnarray}
  \label{cB4}
  \hat B_2 & = &  B_2 +  b^i \wg \ox_i \ ,\qquad i = 1,\ldots,h^{(1,1)}\ , \\
  \hat C_2 & = & C_2 + c^i \wg \ox_i\ , \nn \\
  \hat A_4 & = & D_2^i \wg \ox_i + \rho_i \wg \tilde \ox^i + V^A \wg \ax_A
  - U_A \wg \bx^A\ ,\qquad A = 1,\ldots,h^{(2,1)} \ , \nn
\end{eqnarray} where $B_2,C_2,D_2^i$ are two-forms, $V^A,U_A$ are one-forms and
$b^i,c^i,\rho_i$ are scalar fields in $D=4$. Only half of the
fields in the expansion of $\hat A_4$ are independent due to the
self-duality of $\hat F_5$. We choose to keep $\rho_i$ and $V^A$
as independent fields. The 4-dimensional spectrum arranges as
follows. The gravity multiplet contains the metric $g_{\mu\nu}$
and the graviphoton $V^0$. The $h^{(2,1)}$ vectors $V^a$, together
with the $2h^{(2,1)}$ scalars $z^a$ belong to $h^{(2,1)}$ vector
multiplets. Again, the rest of the fields only consists of
scalars which go to hypermultiplets. Indeed, $C_2$ and $B_2$ are
dual to two scalars $h_1,h_2$. Collecting the remaining
$4h^{(1,1)}+4$ scalars $\phi , b^i , v^i , c^i , h_1 , h_2 , l ,
\rho^i$, we obtain $h^{(1,1)}+1$ hypermultiplets. For the field
strengths, this gives

\bea
   \hat H_3 & = & H_3 + d b^i \wg \ox_i \\[2mm]
  d \hat C_2 & = & d C_2 + d c^i \wg \ox_i  \label{cB5} \\[2mm]
  d \hat A_4 & = & d D_2^i \wg \ox_i + d\rho_i \wg \tilde \ox^i + F^A \wg \ax_A
  - G_A \wg \bx^A. \nn
\eea where $F^A = d V^A$ and $G_A = d U_A$. For $\hat F_5$,
this leads to

\bea \hat F_5 & = &  F^A\wg\ax_A- G_A\bx^A + \left(
d D_2^i - d b^i\wg C_2 - c^iH_3\right)\wg\ox_i\nn\\[2mm]
&& +d\rho_i\wg\tox^i-c^i d b^j\wg\ox_i\wg\ox_j.\label{cB6} \eea
The straightforward expansion reads

\begin{eqnarray}
   - \frac14 \int_Y\hat H_3 \wg * \hat H_3
  & = & - \frac{\cK}{4} \, H_3 \wg * H_3 - \cK
  g_{ij} db^i \wg * db^j \label{cB8}\\[5mm]
 - \frac12 \int_Y  d \hat l \wg
* d \hat l & = &
  - \frac{\cK}{2} \, d l \wg * d l \ ,\label{cB9}\\ [5mm]
  - \frac12 \int_Y {\hat {F}_3} \wg * {\hat {F}_3} & = & -
  \frac{\cK}{2}\, (d C_2 - l H_3) \wg * (d C_2 - l H_3)
  \nn \\
  & & - 2 \cK g_{ij} (d c^i - l d b^i) \wg * (d c^j - l d b^j)
  \label{cB10} \\[5mm]
 -\frac{1}{4}\int_Y \hat F_5\wg * \hat F_5
  & = &  + \frac14 Im\cM^{-1}\left(\tilde G-\cM \tilde F
\right)\wg *\left(\tilde G-\bar\cM \tilde F\right) \nn\\
  && -\cK g_{ij}d \tilde D_2^i \wg *d\tilde D_2^j-\frac{1}{16\cK} g^{ij}
  d \tilde \rho_i \wg *d\tilde\rho_j\label{cB11}\\[5mm]
  -\frac12\int_Y\hat A_4\wg \hat H_3\wg d\hat C_2 & = &
  -\frac12\cK_{ijk}D_2^i\wg d b^j\wg d c^k-\frac12\rho_i\left( d
  B_2\wg d c^i + d b^i\wg d C_2\right)\label{cB012}
\end{eqnarray} with

\bea d\tilde D_2^i & = & d D_2^i -d b^i\wg C_2-c^i d
B_2\label{cB12}\\[2mm]
d\tilde\rho_i & = & d\rho_i -\cK_{ikl}c^k d b^l.\label{cB13}
 \eea

The self-duality of  $\hat F_5$ implies that only half of the
fields appearing in the expansion of $\hat A_4$ in (\ref{cB4})
are independent. Thus the expansion above cannot be used
directly. On the other hand, we cannot impose the self-duality

\beq \hat F_5 = *\hat F_5\label{cB014}\eeq in the action because
the kinetic term $\hat F_5\wg *\hat F_5$ would vanish. In this
thesis we want to show two different but equivalent strategies to
discard half of the fields in $\hat A_4$. The first one starts by
writing a general Lagrangian involving the remaining terms, and
then identifying the reduced 10-dimensional equations of motion
with the 4-dimensional ones calculated from the general
Lagrangian. To make this more precise, we decide to discard first
$D_2^i$, using the self-duality

\beq d\tilde D_2^i = \frac{1}{4\cK}g^{ij}* d\tilde\rho_j.
\label{cB013}\eeq From looking at all possible terms, we can infer the following
Lagrangian

\bea \cL_{inf} & = &  k_1g^{ij}( d\rho_i-\cK_{ikl}c^k d b^l)\wg
* ( d\rho_j-\cK_{jpq}c^p d b^q)\nn\\[2mm]
&& +k_3 d\rho_i\wg(c^i d B_2+d b^i\wg
C_2)+k_4\cK_{ijk}c^ic^j d B_2\wg d b^k.\label{cB14} \eea From
the action (\ref{cB1}), we derive the 10-dimensional equation of
 motion for $\hat C_2$, in the limit where $\hat l=0$, $\hat\phi$
 is constant and the metric is flat,

\beq d *d\hat C_2 = \hat F_5\wg d\hat B_2
\label{cB15}\eeq where we have used (\ref{cB014}). This equation
has two components $(4,4)$ and $(2,6)$, where $(e,i)$ means
order $e$ in space-time indices and order $i$ in internal
indices. The component $(4,4)$ reads (after integration over the internal manifold)

\bea d *d c^i & = &
-\frac{1}{16\cK^2}g^{il}g^{pk}\cK_{lpj}d b^j\wg
*d\tilde\rho_k+\frac{1}{4\cK}g^{ij}d\tilde\rho_j\wg d
B_2.\label{cB16}\eea The kinetic term for $c^i$ can be
read off in (\ref{cB10}), and the identification with the equation
calculated from $\cL_{inf}$  gives the values

\beq k_1 = -\frac{1}{8\cK}\quad ;\quad k_3 = -1\quad ;\quad k_4 =
-\frac{1}{2}. \label{cB17}\eeq After integration over the
internal manifold, the component $(2,6)$ reads

\bea \cK d * d C_2 & = & d\tilde\rho_i\wg d
b^i.\label{cB18}\eea The identification with the equation
calculated from $\cL_{inf}$ gives again $k_3 = -1$. To deal with
the vectors, we choose to display an other strategy. We take the Lagrangian
from the expansions (\ref{cB11}) and (\ref{cB012})

\bea \cL_{F^A} & = & + \frac14 Im\cM^{-1}\left( G-\cM  F
\right)\wg *\left( G-\bar\cM  F\right).\label{cB19} \eea For the
same reason as above, it is not
  possible to impose the self-duality of $\hat F_5$

\bea * G & = & Re\cM * F - Im\cM  F
\label{cB20}\\[2mm]
 G & = & Re\cM  F + Im\cM  *F \label{cB21}\eea
  directly. First we add the total derivative

  \beq +\frac12 F^A\wg G_A \, ,\label{cB22}\eeq and we remark that
  the equation of motion of $G_A$ is exactly (\ref{cB20}), so the
  self-duality will be taken into account in a non-trivial way
  once $G_A$ is eliminated with its equation of motion. We find

\bea \cL_{F^A} & = & + \frac12 Im\cM_{AB} F^A \wg * F^B +\frac12
Re\cM_{AB}  F^A\wg  F^B.\label{cB23} \eea Finally, after the Weyl
rescaling of the volume and the rotation of $v^i$, the whole
action is

\begin{eqnarray}
  S_{IIB}^{(4)} &=& \int - \frac{1}{2} R *\! {\bf 1} - g_{ab} dz^a \wg
  *d\bar{z}^{b} - g_{ij} dt^i \wg *d\bar{t}^j - d\phi \wg *d \phi
  \nonumber\\[2mm]
  && -\frac{1}{4} e^{-4\phi} dB_2 \wg * dB_2 - \frac{1}{2} e^{-2\phi} \cK
  \left( dC_2 - l dB_2 \right) \wg *\left( dC_2 - l dB_2 \right)
  \nonumber\\[2mm]
  && - \frac{1}{2} \cK e^{2\phi} dl \wg * dl - 2 \cK e^{2\phi} g_{ij} \left(
    dc^i - l db^i \right)\wg * \left( dc^j - l db^j \right)\nonumber\\[2mm]
  && - \frac{e^{2\phi}}{8 \cK} g^{-1\,ij} \left( d\rho_i -
    \cK_{ikl} c^k db^l \right) \wg *\left( d\rho_j -
    \cK_{jmn} c^m db^n \right) \nonumber \\[2mm]
  && +  \left( db^i \wg C_2 + c^i dB_2 \right)\wg
  \left( d\rho_i -  \cK_{ijk} c^j db^k \right) + \frac{1}{2}
  \cK_{ijk} c^i c^j dB_2 \wg db^k \nonumber \\[2mm]
  && + \frac{1}{2} Re\cM_{AB} F^A \wg F^B + \frac{1}{2} Im
  \cM_{AB} F^A \wg * F^B \ .\label{cB24}
\end{eqnarray} Now we want to dualize the 2-forms $C_2$ and $B_2$ with scalar
duals $h_1$ and $h_2$. We add first

\beq +d C_2\wg d h_1 \label{cB25}\eeq and the Lagrangian of
interest for $C_2$ is

\bea \cL_{C_2} & = & - \frac{1}{2} e^{-2\phi} \cK
  \left( dC_2 - l dB_2 \right) \wg *\left( dC_2 - l dB_2 \right)\nn\\[2mm]
  && -  b^i d C_2 \wg d\rho_i + d C_2\wg d h_1.\label{cB26}
    \eea We eliminate $d C_2$ with its equation of motion and we
    find

\bea \cL_{h_1} & = & - \frac{1}{2\cK} e^{2\phi}
  \left( d h_1 - b^i d\rho_i \right) \wg *\left( d h_1 - b^j d\rho_j \right)\nn\\[2mm]
  && +l d B_2 \wg\left( d h_1-b^i d\rho_i\right).\label{cB27}
    \eea Repeating the same procedure with $B_2$, we obtain the action for
    type IIB supergravity on a Calabi-Yau manifold

\begin{eqnarray}
  S_{IIB}^{(4)} &=& \int - \frac{1}{2} R *\! {\bf 1} - g_{ab} dz^a \wg
  *d\bar{z}^{b} - g_{ij} dt^i \wg *d\bar{t}^j - d\phi \wg *d \phi
  \nonumber\\[2mm]
  && - \frac{e^{2\phi}}{8 \cK} g^{-1\,ij} \left( d\rho_i -
    \cK_{ikl} c^k db^l \right) \wg *\left( d\rho_j -
    \cK_{jmn} c^m db^n \right) \nonumber \\[2mm]
  &&  - 2 \cK e^{2\phi} g_{ij} \left( dc^i - l db^i \right)\wg * \left( dc^j - l db^j \right)
  - \frac{1}{2} \cK e^{2\phi} dl \wg * dl\nonumber\\[2mm]
  &&  - \frac{1}{2\cK} e^{2\phi}\left( d h_1 - b^i d\rho_i \right)
  \wg *\left( d h_1 - b^j d\rho_j \right) \nonumber \\[2mm]
  && -\rme^{4\phi}D\tilde h\wg *D\tilde h\nn\\[2mm]
  && + \frac{1}{2} Re\cM_{AB} F^A \wg F^B + \frac{1}{2} Im
  \cM_{AB} F^A \wg * F^B \label{cB28}
\end{eqnarray} with

\beq D\tilde h = d h_2 +l d h_1 +
  (c^i-lb^i)d\rho_i-\frac12\cK_{ijk}c^ic^j d
  b^k.\label{cB29}\eeq

\section{Mirror symmetry}\label{msy}

Comparing the field content for the reduced type IIA and type IIB,
comes the striking fact that the spectra are almost identical.
The only difference lies in the number of vector and
hypermultiplets. For type IIA, we found $h^{(1,1)}$ vector
multiplets and $h^{(2,1)}+1$ hypermultiplets. For type IIB, it is
exactly the other way around : $h^{(2,1)}$ vector multiplets and
$h^{(1,1)}+1$ hypermultiplets. This strongly suggests that, defining
the "mirror" manifold $\tilde Y$ of $Y$ by

\beq \tilde Y\quad : \quad
\bigg\{ \begin{array}{lll} \tilde h^{(1,1)} & = & h^{(2,1)}\\
\tilde h^{(2,1)} & = & h^{(1,1)},\end{array}\label{msy0}\eeq type IIA
compactified on Y would be identical to type IIB on $\tilde Y$.
Moreover, when we compare (\ref{cB28}) and (\ref{S4A0}), we can
see that in the sector of the vectors the identification of the
two matrices $\cN$ and $\cM$ is required. In order to prove that
type IIA and type IIB are mirror image of each other, we need to
show that this identification still holds for the hypermultiplets.
Thus we expect to find, up to some field redefinitions, the
metric for the hypermultiplets in type IIB

\bea \tilde h_{uv} d q^u\wg * d q^v & = &  g_{ij} dt^i \wg
*d\bar{t}^j + d\phi \wg *d \phi\nn\\[2mm]
&& + \frac{e^{4\phi}}{4} \, \Big[ da +
  (\tilde\xi_I d \xi^I-\xi^I d\tilde\xi_I) \Big] \wg \rule{0pt}{12pt}^*
  \Big[da +
  (\tilde\xi_J d \xi^J-\xi^J d \tilde\xi_J) \Big] \nn \\
  && - \frac{e^{2\phi}}{2}\left(\IM \cN^{-1} \right)^{IJ}
  \Big[ d\tilde\xi_I + \cN_{IK} d\xi^K \Big]
   \wg \rule{0pt}{12pt}^* \Big[ d\tilde\xi_J + \bar \cN_{JL} d\xi^L \Big]  \,
  .\label{cB30} \eea Guided by the powers of the dilaton and
  the explicit expressions for the matrix $\cN$ (\ref{eq:N}) and (\ref{ImN-1e}),
we decompose the last line into

\bea &&\left(\IM \cN^{-1} \right)^{IJ}
  \Big[ d\tilde\xi_I + \RE\cN_{IK} d\xi^K \Big]
   \wg \rule{0pt}{12pt}^* \Big[ d\tilde\xi_J + \RE\cN_{JL} d\xi^L
   \Big]\label{cB31}\\
   & + &\IM\cN_{IJ} d\xi^I\wg * d\xi^J\label{cB32}
\eea and we expand

\bea \IM\cN_{IJ} d\xi^I\wg * d\xi^J & = & -4\cK
g_{ij}\left( d\xi^i-b^i d\xi^0\right)\wg
*\left( d\xi^j-b^j d\xi^0\right) -\cK d\xi^0\wg
* d\xi^0\label{cB33}\eea This suggests to map

\bea l & \longleftrightarrow & \xi^0\label{cB34}\\[2mm]
lb^i - c^i & \longleftrightarrow & \xi^i.\label{cB35}
 \eea In (\ref{cB31}), the part of the component of $\left(\IM \cN^{-1} \right)^{ij}$
 which only contains the inverse metric $g^{ij}$ reads

\beq -\frac{1}{4\cK}g^{ij}\Big[ d\tilde\xi_i + \RE\cN_{iK} d\xi^K
\Big]
   \wg \rule{0pt}{12pt}^* \Big[ d\tilde\xi_j + \RE\cN_{jL} d\xi^L
   \Big]. \label{cB36}\eeq This leads to the identification

\beq d\rho_i -
    \cK_{ikl} c^k db^l \longleftrightarrow d\tilde\xi_i + \RE\cN_{iK} d\xi^K \label{cB37}\eeq
which yields, with (\ref{cB34}) and (\ref{cB35}),

\beq \rho_i +\frac12\cK_{ikl}b^kb^l l-\cK_{ikl} c^k b^l
\longleftrightarrow \tilde\xi_i. \label{cB38}\eeq The rest of
(\ref{cB31}) can be written

\beq -\frac{1}{\cK}\Big[ d\tilde\xi_0+\RE\cN_{0K} d\xi^K + b^i(
d\tilde\xi_i + \RE\cN_{iK} d\xi^K) \Big]
   \wg \rule{0pt}{12pt}^* \Big[ d\tilde\xi_0+\RE\cN_{0L} d\xi^L + b^j(
d\tilde\xi_j + \RE\cN_{jL} d\xi^L) \Big], \nn\eeq where $\tilde\xi_0$
is identified as

\beq -h_1 +\frac12\cK_{ikl}b^ib^kc^l - \frac16\cK_{ikl}b^ib^kb^l l
 \longleftrightarrow \tilde\xi_0. \label{cB40}\eeq The last of
these redefinitions corresponds to the term

\beq - \frac{e^{4\phi}}{4} \, \Big[ da +
  (\tilde\xi_I d \xi^I-\xi^I d\tilde\xi_I) \Big] \wg \rule{0pt}{12pt}^*
  \Big[da +
  (\tilde\xi_J d \xi^J-\xi^J d \tilde\xi_J) \Big] \label{cB040}\eeq and
  gives us $a$

\beq 2h_2+lh_1+\rho_i(c^i-lb^i) \longleftrightarrow a .
\label{cB41}\eeq Expressing the action (\ref{cB28}) in this new
set of fields, we finally find the mirror of (\ref{S4A0})

 \begin{eqnarray}
  \label{cB42}
  S_{IIA} & = & \int \Big[ -\frac12 R ^* {\bf 1} - g_{ab} dz^a \wg * d
  {\bar z}^b - \tilde h_{uv} d q^u \wg * d q ^v \nn \\
  & & \qquad + \frac{1}{2}\, \IM \cM_{AB} F^A\wg * F^{B}
  + \frac{1}{2} \, \RE \cM_{AB} F^A \wg F^B \Big]
\end{eqnarray} with $\tilde h_{uv}$ given in (\ref{cB30}). The
matrices $\cM$ and $\cN$ have the explicit expressions
(\ref{msa0021}) and (\ref{msa21}) which we recall here

\bea \cN_{IJ} & = & \bar\cF_{IJ}+\frac{2i}{X^M\IM\cF_{MN} X^N
}\IM\cF_{IK}X^K\IM\cF_{JL}X^L\nn\\[2mm]
\cM_{AB} & = & \bar\cF_{AB}+\frac{2i}{z^E\IM\cF_{EG}
z^G}\IM\cF_{AC}z^C\IM\cF_{BD}z^D. \eea in terms of prepotentials
$\cF$ depending holomorphically on the coordinates $z^A =
(1,z^a)$ and $X^I = (1,t^i)$. The K\"ahler potentials can also be written using the
prepotentials according to

\bea \rme^{-K_A} & = & i\left(\bar X^I\cF_I-
X^I\bar\cF_I\right)\nn\\[2mm]
\rme^{-K_B} & = & i\left(\bar z^A\cF_A-
z^A\bar\cF_A\right),\nn\eea and are obviously mapped. To sum up, mirror symmetry states
that type IIA on some \CY manifold $Y$ is the same as type IIB on the mirror manifold
$\tilde Y$

\beq IIA/Y \longleftrightarrow IIB/\tilde Y \eeq with $\tilde Y$ obtained from $Y$
by exchanging the even and odd cohomology classes following (\ref{msy0}). This equivalence
can be checked at the level of the supergravity by performing KK expansions of the
type IIA and type IIB actions in 10 dimensions. The resulting actions are equivalent once
the gauge coupling matrices $\cN$ and $\cM$ are mapped.

\newpage

\chapter{Generalized Calabi-Yau compactifications}

Let us review the main results of last chapter. In ten space-time
dimensions there exist two inequivalent $N=2$ supergravities denoted type IIA
and type IIB. Both theories have the maximal amount
of 32 local supersymmetries but they differ
in their field content~\cite{GSW,LT,JP}.
{}Phenomenologically interesting backgrounds correspond to compactifications on
$\bbR^{1,3}\times Y$ where $\bbR^{1,3}$ is the four-dimensional Minkowski space
while $Y$ is a compact six-dimensional Euclidean manifold. The amount of
supersymmetry which is left unbroken by the background depends on the
holonomy group. The maximal holonomy group for a metric-compatible connection is $SO(6)$.
It breaks all 32 supercharges whereas only some of the supercharges are broken by any of
its subgroups. Calabi--Yau threefolds are a particularly interesting class of
compactification manifolds as their holonomy group
is $SU(3)$ and as a consequence they preserve only eight supercharges (\ref{cyr4}).

In a KK compactification on a Calabi--Yau threefold, the light modes
of the effective theory are the coefficients of an expansion on solutions to
Laplace equation with zero mass on $Y$. Such harmonic forms are the non-trivial elements
of the cohomology groups $H^{(p,q)}(Y)$. The interactions of the light modes are captured
by a low-energy effective Lagrangian ${\cal L}_{{\rm eff}}$ which can be computed via a KK

reduction of the ten-dimensional Lagrangian. The resulting theory is a four-dimensional
 $N=2$ supergravity
coupled to vector- and
hypermultiplets~\cite{deWit:2002am,Bagger:1983wi,deWit:1985lp,Dauria:2001fp,
N=2,BCF,BGHL,Bodner:1990ca}. Mirror symmetry relates the effective theories of
type IIA and type IIB in 4 dimensions. Type IIA compactified on
$Y$ is equivalent to type IIB compactified on the mirror manifold
$\tilde Y$~\cite{mirror} defined by the exchange of the even and odd cohomology groups
according to  $H^{(1,1)}(Y)\leftrightarrow H^{(2,1)}(\tilde Y)$ and vice-versa.

From the phenomenology point of view, one drawback of Calabi-Yau
compactifications is the absence of a scalar potential lifting
the vacuum degeneracy. One possible way to obtain a scalar
potential is to include background fluxes. Type II supergravities
contain several kinds of $(p-1)$-forms $C_{p-1}$ with $p$-form
field strengths $F_{p} = dC_{p-1}$. When the exterior derivative
$d$ is applied to such a form, expanded according to
(\ref{gen6}), it only acts on the space-time coefficients, due to
harmonicity of the internal forms. Hence it is impossible to have
a term with a space-time 0-form coefficient in the expansion of a
field strength. However, remembering that a harmonic form is
\emph{locally} exact, one can consider a term of the form

\begin{equation}
F_{p} = e_i \omega_{p}^i\ ,
\end{equation} where $\ox^i = d\chi^i$ is only true locally. Since we want $F_p$
to be the exterior derivative of some form, $e_i$ must be a
constant, called a background flux. The name "background" comes
from the fact that such a term gives a background value to the
field strength\footnote{This is analogous to Gauss's theorem in electromagnetism
$\oint \overrightarrow{E}\cdot d\overrightarrow{S} = Q$ where $Q$ is the charge. Since the

fluxes indeed parameterize some gaugings, they are usually called electric and magnetic
fluxes.}

\begin{equation}\label{fluxdef}
  \int_{\gamma_p^i} F_p  = e_i \ .
\end{equation} where $\gamma_p^i$ is a $p$-cycle  in $Y$ Poincar\'{e}-dual to
 $\omega_p^i$. For consistency it should be required that $\chi^i$ never appear
explicitly in the action. This is easily achieved if the form
$C_{p-1}$ only participate to the action through its field
strength $F_p$. We will always consider this case in the
following.

Recently generalized Calabi--Yau compactifications of type II
string theories have been considered where background fluxes for
the field strengths $F_{p}$ are turned
on~\cite{PS,JM,TV,PM1,CKLT,GKP,CKKL,GD,LM2,CKL}.

Due to a Dirac condition, the fluxes $e_i$ are
quantized in string theory. They are thus
integers and their number is given by the Betti number $h^p$.
However, in the supergravity approximation the fluxes can be
considered as continuous parameters which represent a small
perturbation of the original Calabi-Yau compactification. The light modes are no longer
massless but acquire masses depending continuously on the fluxes.  Nevertheless their
induced masses are much smaller than the ones of the heavy KK states of order the
compactification scale. The field content is consequently unchanged, and the interactions
of the light modes continue to be captured by an effective Lagrangian ${\cal L}_{{\rm
eff}}$ which describes the dynamics of the fluctuations around
the background values of the theory in the absence of fluxes. The fluxes appear as gauge
or mass parameters and deform the original supergravity into a gauged or massive
supergravity. They introduce a non-trivial potential for some of the massless fields and
spontaneously break (part of) the supersymmetry.

${\cal L}_{{\rm eff}}$ has been computed in various situations.
In refs.~\cite{JM,TV,GD,LM2} type IIB compactified on Calabi-Yau
threefolds $\tilde{Y}$ in the presence of RR-three-form flux
$F_3$ and NS-three-form flux $H_3$ was derived. In
refs.~\cite{PS,CKKL,LM2} type IIA compactified on the mirror
manifold $Y$ with RR-fluxes $F_0$, $F_2$, $F_4$ and $F_6$ present
was considered. The resulting low-energy effective action was
equivalent to the type IIB action on the mirror manifold
$\tilde{Y}$ with $F_3$ non-zero, but $H_3=0$~\cite{LM2}. As
expected given the matching of odd and even cohomologies on
mirror pairs, the type IIB RR-fluxes $F_3$ in the third
cohomology group $H^3(\tilde{Y})$ are mapped to the type IIA
RR-fluxes in the even cohomology groups $H^0(Y)$, $H^2(Y)$,
$H^4(Y)$ and $H^6(Y)$~\cite{GVW,gukov}.

However, for non-vanishing NS-fluxes the situation is less clear
as no obvious mirror symmetric compactification is known. In both
type IIA and type IIB on $Y$ an NS three-form $H_3$ exists which
can give a non-trivial NS-flux in $H^3(Y)$. However, in neither
case is there an NS form field which can give fluxes in the mirror
symmetric even cohomologies $H^0(Y)$, $H^2(Y)$, $H^4(Y)$ and
$H^6(Y)$. Vafa~\cite{Vafa} suggested that the mirror symmetric
configuration is related to compactifying on a manifold $\hat{Y}$
which is not complex but only admits a non integrable almost
complex structure. The purpose of this chapter is to make this
proposal more precise. As a first step we demand that the $D=4$
effective action continues to have $N=2$ supersymmetry, that is,
eight local supersymmetries. According to (\ref{cyr2}), this
implies that there is a single globally defined spinor $\eta$ on
$\hat Y$ so that each of the $D=10$ supersymmetry parameters
gives a single local four-dimensional supersymmetry. As a result,
the structure group has to reduce from $\SO(6)$ to $SU(3)$ or one
of its subgroups. If we further demand that the two $D=4$
supersymmetries are unbroken in a Minkowskian ground state $\eta$
has to be covariantly constant with respect to the Levi-Civita
connection $\nabla$, see \eqref{cyr3}, or equivalently the
holonomy group has to be $SU(3)$. This second requirement
uniquely singles out Calabi--Yau threefolds as the correct
compactification manifolds, see section \ref{cyr}. However, in
this chapter we relax this second condition and only insist that
a globally defined $\SU(3)$-invariant spinor exists. Manifolds
with this property have been discussed in the mathematics and
physics literature and are known as manifolds with $SU(3)$
structure (see, for example,
refs.~\cite{FFS,salamonb,joyce,friedrich,salamon,
CS,AS1,Hull1,rocek,rocek2,papa,papa2,papa3,papa4,waldram,waldram2,KMPT}).
They admit an almost complex structure $J$, a metric $g$ which is
hermitian with respect to $J$ and a unique $(3,0)$-form $\Omega$.
Generically, since $\eta$ is no longer covariantly constant, the
Levi-Civita connection now fails to have $\SU(3)$-holonomy.
However one can always write $\nabla\eta$ in terms of a
three-index tensor, $T^0$, contracted with gamma matrices, acting
on $\eta$. In the same way $\nabla J$ and $\nabla\Omega$ can be
also written in terms of contractions of $T^0$ with $J$ and
$\Omega$ respectively. This tensor $T^0$, known as the intrinsic
torsion, is thus a measure of the obstruction to having $\SU(3)$
holonomy.

Different classes of manifolds with $\SU(3)$ structure exist and they
are classified by the different elements in the decomposition of the
intrinsic torsion into irreducible $\SU(3)$ representations.
We will mostly consider the slightly non-generic situation where only
``electric'' flux is present.
In this case, we find that mirror symmetry restricts us to a
particular class of manifolds with $SU(3)$ structure called
{\it half-flat} manifolds \cite{CS}.\footnote{%
Manifolds with torsion have also been considered in
refs.~\cite{AS1,Hull1,rocek,rocek2,papa,papa2,papa3,papa4,friedrich,BD,waldram,waldram2}.
However, in these papers the torsion is usually chosen to be
completely antisymmetric in its indices or in other words it is a
three-form. This turns out to be a different condition on the
torsion and these manifolds are not half-flat.} They are neither
complex, nor K\"ahler, nor Ricci-flat but they are characterized
by the conditions
\begin{equation}
d\Omega^- =\ 0 =\ d(J\wedge J) \ ,
\end{equation}
where $\Omega^-$ is the imaginary part of the $(3,0)$-form.
On the other hand the real part of $\Omega$ is not closed
and plays precisely the role of an NS four-form
$d\Omega^+ \sim F_4^{NS}$ corresponding to fluxes along
$H^4(Y)$~\cite{Vafa}. Thus the `missing' NS-fluxes are purely
geometrical and arise directly from  the change in the
compactification geometry.

This chapter is organized as follows. In section~\ref{mirrorreview} we recall in more
details mirror symmetry in Calabi-Yau compactifications with RR-flux. In
section~\ref{Yhat} we discuss properties of manifolds with $SU(3)$ structure and the way
they realize supersymmetry in the effective action. These manifolds are classified in
terms of irreducible representations of the structure group $SU(3)$ and in
section~\ref{sec:hf} we argue that the class of half-flat
manifolds are likely to be the mirror geometry of Calabi-Yau
manifolds with electric NS-fluxes.

In section~\ref{KK} we
perform the KK-reduction of type IIA compactified on $\hat Y$,
derive the low energy effective action and show that it is mirror
symmetric to type IIB compactified on threefolds $Y$ with
non-trivial electric NS-flux $H_3$. The effect of the altered
geometry is as expected. It turns an ordinary supergravity into a
gauged supergravity in that scalar fields become charged and a
potential is induced. This potential receives contributions from
different terms in the ten-dimensional effective action, one of which
arises from the non-vanishing Ricci-scalar. This contribution is
crucial to obtain the exact mirror symmetric form of the
potential. Of course, if $\hat Y$ is to be the mirror image of a Calabi-Yau manifold
when NS fluxes are present, this relation should not depend on which theory one
considers. Thus type IIB compactified on $\hat Y$  should also be equivalent to type IIA
compactified on a Calabi-Yau manifold with NS fluxes. This is showed in
section~\ref{IIBhalf}. Section \ref{conc} contains our conclusions.
Calabi-Yau compactifications of type II theories with NS form-fluxes are reviewed in
appendix \ref{typeIINS}. Some technical details about G-structure are gathered in appendix

\ref{acs} while in appendix \ref{Rhf} we compute the Ricci-scalar for half-flat manifolds.

\section{Mirror symmetry in CY compactifications with fluxes}
\label{mirrorreview}

 Let us begin by reviewing mirror symmetry for Calabi--Yau compactifications
 with non-trivial RR fluxes. Consider first type IIB. The only allowed RR flux on the
internal Calabi--Yau manifold $\tilde{Y}$ is the three-form $F_3=dC_2$.
The flux $F_3$ then defines $2(h^{(1,2)}+1)$ flux parameters
$(\q_A, \p^A)$ according to
\begin{equation}
F_3 = dC_2 + \p^A \ax_A - \q_A \bx^A   \ .
\end{equation}
The effective action of this compactification is worked out in refs.~\cite{JM,TV,GD,LM2}.
A KK reduction is performed on the original Calabi-Yau geometry
with the non-vanishing fluxes taken into account. This leads to a
potential which induces small masses for some of the scalar
fields and spontaneously breaks supersymmetry.

It was shown in~\cite{LM2} that this IIB effective action is
manifestly mirror symmetric to the one arising from the
compactification of massive type IIA supergravity \cite{Romans}
on $Y$ with RR-fluxes turned on in the even cohomology of $Y$.
More precisely, in IIA compactifications the RR two-form field
strength $F_2$ can have non-trivial flux in $H^2(Y)$ while
the four-form field strength $F_4$ has fluxes in $H^4(Y)$.
Then there are $2h^{(1,1)}$ IIA RR-flux parameters given by
\begin{equation}
F_2 =dA_1 +  m^i \omega_i \ , \qquad
F_4 = dC_3 -A_1\wedge H_3 + e_i \tilde\omega^i  \ .
\end{equation}
In addition there are the two extra parameters $m^0$ and $e_0$,
where $e_0$ is the dual of the space-time part of the four-form
$F_{4\,\mu\nu\rho\sigma}$ and $m^0$ is the mass parameter of the
original ten-dimensional massive type IIA theory \cite{LM2}.
Altogether there are $2(h^{(1,1)}+1)$ real RR-flux parameters
$(e_I,m^J), I,J=0,1,\ldots, h^{(1,1)}$ which precisely map to the
$2(h^{(1,2)}+1)$ type IIB RR-flux parameters under mirror
symmetry. This is confirmed by an explicit KK-reduction of the
respective effective actions and one
finds~\cite{LM2}\footnote{For $m^I = 0$ one finds a
  standard $N = 2$ gauged supergravity with a potential for the moduli
  scalars of the vector multiplets. For $m^I \neq 0$ a non-standard
  supergravity occurs where the two-form $B_2$ becomes massive. For a more
  detailed discussion and a derivation of the effective action we refer
  the reader to ref.\ \cite{LM2}.}
\begin{equation}
{\cal L}^{(IIA)}(Y, e_I, m^J)\ \equiv\ {\cal L}^{(IIB)}(\tilde Y,\q_A, \p^B)\ .
\end{equation}

We expect that mirror symmetry continues to hold when one considers
fluxes in the NS-sector. However, in this case, the situation is more
complicated. In both type IIA and type IIB there is a NS two-form
$B_2$ with a three-form field strength $H_3$, so one can consider
fluxes in $H^3(Y)$ in IIA and $H^3(\tilde{Y})$ in
IIB. However, these are clearly not mirror symmetric since
mirror symmetry exchanges the even and odd cohomologies. One appears
to be missing $2(h^{(1,1)}+1)$ NS-fluxes, lying along the even
cohomology of $Y$ and $\tilde Y$, respectively.
Since the NS fields include only the metric,
dilaton and two-form $B_2$, there is no candidate NS even-degree
form-field strength to provide the missing fluxes. Instead, they must
be generated by the metric and the dilaton. Thus we are led to
consider compactifications on a generalized class of manifolds
$\hat{Y}$ with a metric which is no longer Calabi--Yau, and perhaps
a non-trivial dilaton in order to find a mirror-symmetric effective
action. This necessity was anticipated by Vafa in
ref.~\cite{Vafa}.

We now turn to what characterizes this generalized class of
compactifications on $\hat{Y}$.  We choose to first present the
IIA compactification on a half-flat manifold $\hat Y$ compared to
the IIB compactification on a Calabi-Yau manifold $Y$ with NS flux
$H_3$. Since the NS sectors are identical this should be, of
course, equivalent to the problem with the roles of IIA and IIB
reversed. This one will be addressed in section~\ref{IIBhalf}.

\section{Half-flat spaces as mirror manifolds}

\subsection{Supersymmetry and manifolds with $\SU(3)$-structure}
\label{Yhat}

The low-energy effective action arising from IIB compactifications
with non-trivial $H_3$-flux describes a massive deformation of an
$N=2$ supergravity \cite{JM,TV,GD,LM2}. Compactification on the
conjectured generalized mirror IIA manifold $\hat{Y}$ should lead
to the same effective action. Thus the first constraint on
$\hat{Y}$ is that the resulting low-energy theory preserves $N=2$
supersymmetry.

Let us first briefly review how supersymmetry is realized in the
conventional Calabi--Yau compactification. Ten-dimensional type IIA
supergravity has two supersymmetry parameters $\epsilon^\pm$ of
opposite chirality each transforming in a real 16-dimensional spinor
representation of the Lorentz group $\Spin(1,9)$. In particular, the
variation of the two gravitini in type IIA is schematically given
by~\cite{GP}
\begin{equation}\label{gravitino}
\delta \psi_M^\pm =
   \left[\nabla_M + (\Gamma \cdot H_3)_M\right]\epsilon^\pm
   + \left[(\Gamma \cdot F_2)_M
      + (\Gamma \cdot F_4)_M\right]\epsilon^\mp + \ldots \ ,
\end{equation}
where the dots indicate further fermionic terms.
Next one dimensionally reduces on a six-dimensional manifold $Y$
and requires that the theory has a supersymmetric vacuum of the
form $\bbR^{1,3}\times Y$ with all other fields trivial.
Following the discussion in section \ref{cyr}, this implies that
there are particular spinors $\epsilon^\pm$ for which the
gravitino variations~\eqref{gravitino} vanish. On
$\bbR^{1,3}\times Y$ the Lorentz group $\Spin(1,9)$ decomposes
into $\Spin(1,3)\times\Spin(6)$ and we can write
$\epsilon^\pm=\theta^\pm\otimes\eta$. In the supersymmetric
vacuum, the vanishing of the gravitino variations imply the
$\theta^\pm$ are constant and $\eta$ is a solution of
\begin{equation}
\label{ccons}
\nabla_m \eta=0\ , \qquad m = 1,\dots,6 \ .
\end{equation}
If this equation has a single solution, each $\epsilon^\pm$ gives a
Killing spinor and we see that the background preserves $N=2$
supersymmetry in four dimensions as required. Equivalently, if we
compactify on $Y$, the low-energy effective action will have $N=2$
supersymmetry and admits a flat supersymmetric ground state
$\bbR^{1,3}$.

The condition~\eqref{ccons} really splits into two parts: first the
existence of a non-vanishing globally defined spinor $\eta$ on $Y$ and
second that $\eta$ is covariantly constant. The first condition
implies the existence of two four-dimensional supersymmetry parameters
and hence that the effective action has $N=2$ supersymmetry. The
second condition that $\eta$ is covariantly constant implies that the
effective action has a flat supersymmetric ground state.

The existence of $\eta$ is equivalent to the statement that the
structure group is reduced. To see what this means, recall that
the structure group  refers to the group of transformations of
orthonormal frames over the manifold. Thus on a space-time of the
form $\bbR^{1,3}\times Y$ the structure group reduces from
$\SO(1,9)$ to $\SO(1,3)\times\SO(6)$ and the spinor
representation decomposes accordingly as
$\rep{16}\to\rep{(2,4)}+\rep{(\bar 2,\bar 4)}$. Suppose now that
the structure group of $Y$ reduces further to
$\SU(3)\subset\SO(6)\cong\SU(4)$. The $\rep{4}$ then decomposes
as $\rep{3}+\rep{1}$ under the $\SU(3)$ subgroup. An invariant
spinor $\eta$ in the singlet representation of $\SU(3)$ thus
depends trivially on the tangent space of $Y$ and so is globally
defined and non-vanishing. Conversely, the existence of such a
globally defined spinor implies that the structure group of $Y$
is $\SU(3)$ (or a subgroup thereof). Mathematically, one says
that $Y$ has $\SU(3)$-structure. In appendix~\ref{acs} we review
some of the properties of such manifolds from a more mathematical
point of view and for a more detailed discussion we refer the
reader to the mathematics
literature~\cite{FFS,salamonb,joyce,friedrich,salamon,CS}. Here
we will concentrate on the physical implications.

The second condition that $\eta$ is covariantly constant has well
known consequences (as reviewed for instance
in~\cite{Green:1987mn}). It is equivalent to the statements that
the Levi--Civita connection has $\SU(3)$ holonomy or similarly
that $Y$ is Calabi--Yau. It implies that an integrable complex
structure exists and that the corresponding fundamental two-form
$J$ is closed. In addition, there is a unique closed holomorphic
three-form $\Omega$. Together these structure and integrability
conditions imply that Calabi-Yau manifolds are complex,
Ricci-flat and K\"ahler.

Symmetry with the low-energy IIB theory with $H_3$-flux implies
that compactification on generalized mirror manifold $\hat Y$
still leads to an effective action that is $N=2$ supersymmetric.
However, the IIB theory with flux in general no longer has a
flat-space ground state which preserves all
supercharges~\cite{TV,PM1,CKLT}. From the above discussion, we see
that this implies that we still have a globally defined
non-vanishing spinor $\eta$, but we no longer require that $\eta$
is covariantly constant, so $\nabla_m \eta\neq0$. In other words,
$\hat{Y}$ has $\SU(3)$-structure but generically the Levi--Civita
connection no longer has $\SU(3)$-holonomy, so in general,
$\hat{Y}$ is not Calabi--Yau. In particular, as discussed in
appendix~\ref{Rhf}, generic manifolds with $\SU(3)$-structure are
not Ricci-flat.

In analogy with Calabi-Yau manifolds
let us first use the existence of the globally defined
spinor $\eta$  to define other invariant
tensor fields.\footnote{For Calabi-Yau manifolds these constructions are
  reviewed, for example, in ref.~\cite{Green:1987mn}. For compactifications
  with torsion  they are generalized in ref. \cite{dWS,AS1,Hull1,KMPT} and
  here we closely follow these references.}
Specifically, one has a fundamental two-form
\begin{equation}
\label{defJ}
    J_{mn} = -i \eta^\dagger \Gamma_7 \Gx_{mn} \eta\ ,
\end{equation}
and a three-form
\begin{equation}\label{defO}
   \Omega =  \Omega^+ + i \Omega^-\ ,
\end{equation}
where
\begin{equation}
   \Omega^+_{mnp} = -i \eta^\dagger \Gx_{mnp} \eta \ ,\qquad
   \Omega^- _{mnp} = -i \eta^\dagger \Gx_7\Gx_{mnp} \eta \ .
\end{equation}
By applying Fierz identities one shows
\begin{equation}
\label{eq:JOmegaIds}
\begin{gathered}
   J\wedge J \wedge J
      = \frac{3\ii}{4}\, \Omega \wedge \bar{\Omega} \ , \\
   J \wedge \Omega = 0 \ ,
\end{gathered}
\end{equation}
exactly as for Calabi--Yau manifolds.
Similarly, raising an index on $J_{mn}$ and assuming a
normalization $\eta^\dag\eta=1$, one finds
\begin{equation}\label{Jac}
J_m{}^p J_p{}^n = - \dx_m{}^n\ , \qquad
J_m{}^p J_n{}^r g_{pr} = g_{mn}\ ,
\end{equation}
by virtue of the $\Gamma$-matrix algebra (\ref{cyp}). This implies that
$J_m{}^p$ defines an almost complex structure such that the
metric $g_{mn}$ is Hermitian with respect to $J_m{}^p$. The
existence of an almost complex structure is sufficient to define
$(p,q)$-forms as we review in  appendix~\ref{acs}. In particular,
one can see that $\Omega$ is a $(3,0)$-form.

Thus far we have used the existence of the $SU(3)$-invariant spinor $\eta$ to
construct $J$ and $\Omega$. One can equivalently characterize manifolds
with $SU(3)$-structure by the existence of a globally defined,
non-degenerate two-form $J$ and a globally defined non-vanishing complex
three-form $\Omega$ satisfying the conditions~\eqref{eq:JOmegaIds}.
Together these then define a metric \cite{joyce,H}.

The key difference from the Calabi--Yau case is that
a generic $\hat Y$ does not have $SU(3)$ holonomy
since $\nabla_m\eta\neq0$. Using  \eqref{defJ}  and \eqref{defO}
this immediately implies that
also $J$ and $\Omega$ are  generically no longer covariantly constant
$\nabla_m J_{np} \neq 0 \ ,  \nabla_m \Omega_{npq} \neq 0$.
In other words the deviation from being covariantly constant
is a measure of the deviation from $SU(3)$ holonomy
and thus a measure of the deviation from the Calabi--Yau condition.
This can be made more explicit by using the fact that
on $\hat Y$ there always exists another connection
$\nabla^{(T)}$, which is metric compatible
(implying $\nabla_m^{(T)}g_{np}=0$), and which does
satisfy $\nabla_m^{(T)} \eta = 0$
\cite{salamonb,joyce}.
The difference between any two metric-compatible
connections is a tensor, known as the contorsion $\CT_{mnp}$,
and thus we have explicitly
\begin{equation}
\label{torKS}
\nabla_m^{(T)} \eta = \nabla_m \eta-\frac14\CT_{mnp}\Gx^{np} \eta = 0\ ,
\end{equation}
where $\Gx^{np}$ is the antisymmetrized product of
$\Gamma$-matrices defined in appendix~\ref{cad} and $\CT_{mnp}$
takes values in $\Lambda^1\otimes\Lambda^2$ ($\Lambda^p$ being
the space of $p$-forms). We see that $\CT_{mnp}$ is the
obstruction to $\eta$ being covariantly constant with respect to
the Levi-Civita connection and thus for non-vanishing $\CT$ the
manifold $\hat Y$ can not be a Calabi-Yau manifold. Similarly,
using \eqref{defJ}, \eqref{defO} and \eqref{torKS} one shows that
also $J$ and $\Omega$ are generically no longer covariantly
constant but
 instead obey
\begin{equation}
\label{torJO}
\begin{aligned}
   \nabla_m^{(T)} J_{np} &=
      \nabla_m J_{np} - \CT_{mn}{}^r J_{rp} - \CT_{mp}{}^r J_{nr}
      = 0 \ , \\
   \nabla_m^{(T)} \Omega_{nmp} &=
      \nabla_m \Omega_{npq} - \CT_{mn}{}^r\Omega_{rpq}
         - \CT_{mp}{}^r\Omega_{nrq} - \CT_{mq}{}^r\Omega_{npr}
      = 0 \ ,
\end{aligned}
\end{equation}
where again $\CT$ is measuring the obstruction to $J$ and $\Omega$
being covariantly constant with respect to the Levi-Civita connection.
We see that the connection $\nabla^{(T)}$ preserves the $SU(3)$ structure
in that $\eta$ or equivalently $J$ and $\Omega$ are constant with
respect to $\nabla^{(T)}$.

Let us now analyze the contorsion $\CT\in\Lambda^1\otimes\Lambda^2$ in
a little more detail.  Recall that $\Lambda^2$ is isomorphic to the
Lie algebra $\so(6)$, which in turn decomposes into $\su(3)$ and
$\su(3)^\perp$, with the latter defined by
$\su(3)\oplus\su(3)^\perp\cong\so(6)$. Thus the contorsion actually
decomposes as $\CT^{\su(3)}+\CT^0$ where
$\CT^{\su(3)}\in\Lambda^1\otimes\su(3)$ and
$\CT^0\in\Lambda^1\otimes\su(3)^\perp$. Consider now the action of
$\CT$ on the spinor $\eta$. Since $\eta$ is an $\SU(3)$ singlet, the
action of $\su(3)$ on $\eta$ vanishes, and thus, from~\eqref{torKS},
we see that
\begin{equation}
\label{eq:CTdef}
   \nabla_m \eta = \frac14 \CT^0_{mnp} \Gamma^{np}\eta\ .
\end{equation}
{}From~\eqref{torJO}, one finds that analogous expressions hold for
$\nabla_mJ_{np}$ and $\nabla_m\Omega_{npq}$. We see that the
obstruction to having a covariantly constant spinor (or equivalently
$J$ and $\Omega$) is actually measured by not the full contorsion
$\CT$ but by the so-called ``intrinsic contorsion'' part $\CT^0$.
Eq.~\eqref{eq:CTdef} implies that $\CT^0$ is independent
of the choice of $\nabla^{(T)}$ satisfying~\eqref{torKS}, and thus
is a property only of the $\SU(3)$-structure. This fact is reviewed in more
detail in appendix~\ref{acs}.

Mathematically,
it is sometimes more conventional to use the torsion $T$ instead
of the contorsion $\CT$; the two are related via
$T_{mnp}=\frac12(\CT_{mnp}-\CT_{nmp})$ and  $T_{mnp}$
also satisfies \eqref{RT}.
Similarly, one usually refers to the corresponding ``intrinsic torsion''
$T^0_{mnp}=\frac12(\CT^0_{mnp}-\CT^0_{nmp})$ which also is an element
of $\Lambda^1\otimes\su(3)^\perp$ and is in one-to-one correspondence
with $\CT^0$.\footnote{%
Note that our terminology is not very precise in that whenever we use the
notion of torsion we in fact mean by this intrinsic torsion.}
If $\CT^0$ and hence $T^0$ vanishes, we say that the
$\SU(3)$ structure is torsion-free. This implies $\nabla_m\eta=0$ and
the manifold is Calabi--Yau.

Both $\CT^0$ and $T^0$  can be decomposed
in terms of irreducible $SU(3)$ representations and hence
different $\SU(3)$ structures
can be characterized by the non-trivial $SU(3)$ representations
$T^0$ carries. Adopting the
notation used in~\cite{salamon,CS} we denote this decomposition by
\begin{equation}
  \label{TinW}
  T^0 \in \W_1 \oplus \W_2  \oplus \W_3 \oplus \W_4 \oplus \W_5\ ,
\end{equation}
with the corresponding parts of $T^0$ labeled by $T_i$ with
$i=1,\dots,5$ and where the representations corresponding to the
different $\W_i$ are given in table \ref{tabW}.

\begin{table}[htbp]
  \begin{center}
    \begin{tabular}[h]{|c|c|c|}
      \hline
      component & interpretation & $SU(3)$-representation \\
      \hline
      $\W_1$ & $J\wg d\Ox$\ \ or\ \ $\Omega\wg d J$ &
           $\rep 1 \oplus \rep{1}$ \\
      \hline
      $\W_2$ & $( d \Ox)^{2,2}_0$ & $\rep 8 \oplus \rep{8}$ \\
      \hline
      $\W_3$ & $ ( d J)^{2,1}_0+( d J)^{1,2}_0$ &
         $\rep 6 \oplus \rep{\bar 6}$ \\
      \hline
      $\W_4$ & $J \wg  d J$ & $\rep 3 \oplus \rep{\bar 3}$ \\
      \hline
      $\W_5$ & $ d \Ox^{3,1}$
         & $\rep 3 \oplus \rep{\bar 3}$ \\
      \hline
    \end{tabular}
    \caption{The five classes of the intrinsic torsion of a space with
      $\SU(3)$ structure.}
    \label{tabW}
  \end{center}
\end{table}

The second column of table~\ref{tabW}, gives an interpretation of each
component of $T^0$ in terms of exterior derivatives of $J$ and
$\Omega$. The superscripts refer to projecting onto a particular
$(p,q)$-type, while the $0$ subscript  refers to the irreducible
$\SU(3)$ representation with any trace part proportional to $J^n$
removed (see appendix~\ref{app:IT}). This interpretation arises since,
from~\eqref{torJO}, we have
\begin{equation}
\label{dJO}
\begin{aligned}
   d J_{mnp} &= 6 T^0_{[mn}{}^r J_{r|p]} \ , \\
   d \Ox_{mnpq} &= 12 T^0_{[mn}{}^r \Ox_{r|pq]}\ .
\end{aligned}
\end{equation}
These can then be inverted to give an expression for each component
$T_i$ of $T^0$ in terms of $d J$ and $d\Omega$. This is discussed
in more detail from the point of view of $SU(3)$ representations in
appendix~\ref{app:IT}.

Manifolds with $SU(3)$ structure are in general not complex
manifolds. An almost complex structure $J$ (obeying \eqref{Jac})
necessarily exists but the integrability of $J$ is
determined by the vanishing of the Nijenhuis tensor $N_{mn}{}^p$.
{}From its definition~(\ref{Ntens}) we see that a covariantly constant
$J$ has a vanishing $N_{mn}{}^p$ and in this situation the manifold
is complex and K\"ahler (as is the case for Calabi--Yau manifolds).
However, for a generic $J$  the Nijenhuis tensor does not vanish
and is instead determined by the (con-) torsion using
(\ref{Ntens}) and (\ref{torJO}). Thus $T^0$ also is an obstruction to
$\hat Y$ being a complex manifold. However, one can show~\cite{salamon,CS}
that $N_{mn}{}^p$ does not depend on all torsion components but is
determined entirely by the component of the torsion
$T_{1\oplus2}\in\W_1\oplus\W_2$, through
\begin{equation}
\label{NIT}
   N_{mn}{}^p = 8 (T_{1\oplus2})_{mn}{}^p \ .
\end{equation}

Before we proceed let us summarize the story so far. The requirement
of an $N=2$ supersymmetric effective action led us to consider
manifolds $\hat{Y}$ with $SU(3)$-structure. Such manifolds admit a
globally defined $SU(3)$-invariant spinor $\eta$ but the holonomy group of
the Levi-Civita connection is no longer $SU(3)$. The deviation from
$SU(3)$ holonomy is measured by the intrinsic (con-)torsion, and
implies that generically the manifold is neither complex nor
K\"ahler. However, the fundamental two-form $J$ and the $(3,0)$-form
$\Omega$ can still be defined; in fact their existence is equivalent
to the requirement that $\hat{Y}$ has $\SU(3)$-structure.
Different classes of manifolds with $SU(3)$ structure are
labeled by the $SU(3)$-representations in which the intrinsic torsion
tensor resides. In terms of $J$ and $\Omega$ this is measured by which
components of the exterior derivatives $d J$ and $d\Omega$ are
non-vanishing.

\subsection{Half-flat manifolds}
\label{sec:hf}

In general, we might expect that there are further restrictions on
$\hat{Y}$ beyond the supersymmetry condition that it has
$\SU(3)$-structure. This would correspond to constraining the
intrinsic torsion so that only certain components in table
\ref{tabW} are non-vanishing. We provide evidence for  a particular
set of constraints in the following subsections.
Then, in section~\ref{KK}, we verify that these conditions do lead to
the required mirror symmetric type IIA effective action.

Before doing so, however, let us consider two arguments suggesting how
these constraints might appear.  First, recall that the K\"{a}hler
moduli on the Calabi--Yau manifold are paired with the $B_2$ moduli as
an element $B_2+\ii J$ of $H^2(Y,\bbC)$ where $J$ is the K\"{a}hler
form. Under mirror symmetry, these moduli map to the complex structure
moduli of $\tilde{Y}$ which are encoded in the closed holomorphic
$(3,0)$-form $\Omega$.
Turning on $H_3$ flux on the original Calabi--Yau manifold means that
the real part of the complex K\"{a}hler form $B_2+\ii J$ is no longer
closed. Under the mirror symmetry, this suggests that we now have a
manifold $\hat{Y}$ where half of $\Omega=\Omega^++\ii\Omega^-$, in
particular $\Omega^+$, is no longer closed.
From table~\ref{tabW}, we see that $ d\Omega^{2,2}$ is related to
the classes $\W_1$ and $\W_2$ which can be further decomposed into
$\W^+_1\oplus\W^-_1$ and $\W^+_2\oplus\W^-_2$ giving
\begin{equation}
\label{T+-}
\begin{aligned}
   T^+_{1\oplus2} &\text{ corresponding to }
      (d\Omega^+)^{2,2} \ , \\
   T^-_{1\oplus2} &\text{ corresponding to }
      (d\Omega^-)^{2,2} \ .
\end{aligned}
\end{equation}
Thus, the above result that only $\Ox^-$ remains closed suggests that,
\begin{equation}
\label{eq:T-}
   T_{1\oplus 2}^- = 0 \ .
\end{equation}
One might expect that it also implies that half of the $\W_5$
component vanishes. However, as discussed in~\cite{CS},
$(d\Omega^+)^{3,1}$ and $(d\Omega^-)^{3,1}$ are related, so, in
fact, all of the component in $\W_5$ vanishes and we have in addition
\begin{equation}
   T_5 = 0 \ .
\end{equation}

The second argument comes from the fact that the intrinsic torsion $T^0$
should be such that it supplies the missing $2(h^{(1,1)}+1)$ NS-fluxes.
In other words we need the new fluxes to be counted by the even
cohomology of the original Calabi-Yau manifold $Y$. This implies that
there should be some well-defined relation between $\hat Y$ and the
Calabi-Yau manifold $Y$. We return to this relation in more detail in
section~\ref{spectrum} but here let us simply make the very naive
assumption that we try to match the $\SU(3)$ representations of the
$H^{p,q}(Y)$ cohomology group with the $\SU(3)$ representations of
$T^0$. This suggests setting
\begin{equation}
   T_4 = T_5 = 0 \ .
\end{equation}
since the corresponding $H^{3,2}(Y)$ and $H^{3,1}(Y)$ groups vanish on
$Y$. On the other hand $T_{1,2,3}$ can be non-zero as the
corresponding cohomologies do exist on $Y$.

Taken together, these arguments suggest that the appropriate
conditions might be
\begin{equation}
   T_{1\oplus2}^- = T_4 = T_5 = 0 \ .
\end{equation}
This is in fact a known class of manifolds, denoted
\textit{half-flat}~\cite{CS}. From table~\ref{tabW} it is easy to see
that the necessary and sufficient conditions can be written as
\begin{equation}
\label{eq:hf}
\begin{aligned}
    d \Omega^- &= 0 , \\
    d \left( J \wedge J \right) &= 0 .
\end{aligned}
\end{equation}

It will be useful in the following to have explicit expressions for
the components of the intrinsic torsion $T_1$, $T_2$ and $T_3$ which
are non-vanishing when the manifold is half-flat. From table
\ref{tabW} we recall that $T_{1\oplus2}$ is in the same $\SU(3)$
representation as a complex four-form $F^{(2,2)}$ of type
$(2,2)$. Explicitly we have
\begin{equation}
  \label{cxT12}
  (T_{1\oplus 2})_{mn}{}^p =
      F_{mnrs} \Ox^{rsp} + \bar F_{mnrs} \bar \Ox^{rsp} \ .
\end{equation}
The half-flatness condition $T_{1\oplus 2}^- = 0$ just imposes that
$F$ is real ($F=\bar F$) so that
\begin{equation}
\label{su3Nt}
   (T_{1\oplus2})_{mn}{}^p\ = \ (T_{1\oplus2}^+)_{mn}{}^p\ =  \
   2F_{mnrs}^{(2,2)}\, \Ox^{+rsp} \ ,
\end{equation}
where we have used \eqref{defO}. Explicitly, from the
relations~\eqref{dJO} one has that $F$ is related to $ d\Omega$
by\footnote{%
  Note, that up to this point, the normalization~$\eta^\dag\eta=1$
  fixed the normalization of $J$ and $\Omega$. In the following it
  will be useful to allow an arbitrary normalization of $\Omega$, thus
  we have included in this expression the general factor
  $||\Ox||^2\equiv\frac{1}{3!}\Ox_{\ax\bx\cx}\bar\Ox^{\ax\bx\cx}$. }
\begin{equation}\label{F22}
 F_{mnrs}^{(2,2)}
     \equiv \frac{1}{4||\Omega||^2}\, ( d\Omega)^{2,2}_{mnrs}
     = \frac{1}{4||\Omega||^2}\, ( d\Omega^+)^{2,2}_{mnrs}\ .
\end{equation}
We will see in section~\ref{KK} that this plays the role of the NS
four-form which precisely complexifies the RR 4-form background
flux in the low-energy effective action.  This fact was
anticipated in~\cite{Vafa}. However, it will only generate the
electric fluxes defined in \eqref{cBNS5}, i.e.~half of the missing
NS-fluxes. As we said in the introduction, the treatment of the
magnetic fluxes, corresponding to the NS two-form flux is more
involved and will be discussed in a separate publication
\cite{GLMW2}.

Similarly, we see from table~\ref{tabW} that the component $T_{3}$ of
the torsion is in the same representation as a real traceless
three-form $A^{(2,1)}_0+\bar{A}^{(1,2)}_0$ of type $(2,1)+(1,2)$ (see
also appendix~\ref{app:IT}). From~\eqref{dJO} we see that this form is
proportional to $( d J)^{(2,0)}_0$. Explicitly we have
\begin{eqnarray}
  \label{T3}
  (T_3)_{mnp} = \frac14 \Big(\dx_m^{m'} \dx_n^{n'} - J_m{}^{m'}
J_n{}^{n'}
  \Big) J_p{}^{p'} ( d J)_{m'n'p'} - 2 F\, (\Ox^+)_{mnp} \ ,
\end{eqnarray}
where by $F$ we denoted the trace in complex indices
$F_{\ax\bx}{}^{\ax\bx}$.

The remainder of the section focuses on providing evidence that
equations \eqref{eq:hf} are indeed the correct conditions. Before
doing so, recall that compactifications on manifolds with torsion
have also been discussed in
refs.~\cite{dWS,AS1,Hull1,rocek,rocek2,friedrich,papa,papa2,papa3,papa4,waldram,waldram2}.
The philosophy of these papers was slightly different in that
they considered backgrounds where some of the $p$-form field
strength were chosen non-zero and in order to satisfy $\delta
\psi_m =0$. Here instead we want the torsion to generate terms
which mimic or rather are mirror symmetric to NS-flux
backgrounds. As a consequence, one finds rather different
conditions. Since in both cases one wants $N=2$ supersymmetry in
four dimensions, the class of manifolds discussed
in~\cite{dWS,AS1,Hull1,rocek,rocek2,friedrich,papa,papa2,papa3,papa4}
are also manifolds with $SU(3)$ structure. However, in these
cases the torsion is a traceless real three-form. This implies $T
\in \W_3 \oplus \W_4$ so that $T_1=T_2=T_5=0$. As a consequence
the Nijenhuis tensor vanishes (since it depends only on
$T_{1\oplus2}$) and the manifolds are complex but not K\"ahler.


\section{Type IIA on a half-flat manifold}
\label{KK}

Before we launch into the details of the dimensional reduction,
recall that we are aiming at the derivation of a type IIA
effective action which is mirror symmetric to the type IIB
effective action obtained from compactifications on Calabi-Yau
threefolds with (electric) NS 3-form flux $H_3$ turned on.  This
effective theory is reviewed in section \ref{cBNS} while the
Calabi-Yau compactification of type IIA without fluxes is
recalled in section \ref{cA}. As we have stressed throughout, the
central problem is that in IIA theory there is no NS form-field
which can reproduce the NS-fluxes which are the mirrors of $H_3$
in the type IIB theory. Vafa suggested that the type IIA mirror
symmetric configuration is a different geometry where the complex
structure is no longer integrable~\cite{Vafa}, so that the
compactification manifold $\hat{Y}$ is not Calabi--Yau. In the
previous section we have already collected evidence that
half-flat manifolds are promising candidates for $\hat{Y}$. The
additional flux was characterized by the four-form
$F^{(2,2)}\sim d\Omega^{2,2}$. The purpose of this section is to
calculate the effective action, in an appropriate limit, for type
IIA compactified on a half-flat $\hat{Y}$, and show that it is
exactly equivalent to the known effective theory for the mirror
type IIB compactification with electric flux.

The basic problem we are facing in this section is that so far we have no
mathematical procedure for constructing a half-flat manifold
$\hat{Y}$ from a given
Calabi-Yau manifold $Y$. Instead we will give a set of rules
for the structure of $\hat{Y}$ and the corresponding light spectrum
by using physical considerations and in particular using
mirror symmetry as a guiding principle.
Specifically, we will write a set of two-, three- and
four-forms on $\hat{Y}$ which are in some sense ``almost
harmonic''. By expanding the IIA fields in these forms, we can then
derive the four-dimensional effective action which is equivalent to
the known mirror type IIB action.

\subsection{The light spectrum and the moduli space of $\hat{Y}$}

\label{spectrum}

To derive the effective four-dimensional theory we first have to
identify the light modes in the compactification such as the
metric moduli. Unlike the case of a conventional reduction on a
Calabi--Yau manifold, from the IIB calculation we know that the
low-energy theory has a potential~\eqref{potIIB} and so not all the
light fields are massless. In any dimensional reduction there is
always an infinite tower of massive Kaluza--Klein states, thus we need
some criterion for determining which modes we keep in the effective
action.

Recall first how this worked in the type IIB case. One starts with a
background Calabi--Yau manifold $\tilde{Y}$ and makes a perturbative
expansion in the flux $H_3$. To linear order, $H_3$ only appears in
its own equation of motion, while it appears quadratically in the
other equations of motion, such as the Einstein and dilaton
equations, so, heuristically,
\begin{equation}
\label{eq:eom}
\begin{gathered}
   \nabla^m H_{mnp} = \dots \ , \\
   R_{mn} = H^2_{mn} + \dots \ .
\end{gathered}
\end{equation}
In the perturbation expansion we first solve the linear
equation on $\tilde{Y}$ which implies that $H_3$ is harmonic. We then
consider the quadratic backreaction on the geometry of $\tilde{Y}$ and
the dilaton. The backreaction will be small provided $H_3$ is small
compared to the curvature of the compactification, set by the inverse size
of the Calabi--Yau manifold $1/\LCY$. Recall, however, that in
string theory the flux $\int_{\gamma_3}H$, where $\gamma_3$ is any
three-cycle in $\tilde{Y}$ is quantized in units of
$\alpha'$. Consequently $H_3\sim\alpha'/\LCY^3$ and so for a small
backreaction we require $H_3/\LCY^{-1}\sim\alpha'/\LCY^2$ to be small. In
other words, we must be in the large volume limit where the
Calabi--Yau manifold is much larger than the string length, which
anyway is the region where supergravity is applicable. The
Kaluza--Klein masses will be of order $1/\LCY$. The mass correction
due to $H_3$ is proportional to $\alpha'/\LCY^3$ and so is comparatively
small in the large volume limit. Thus in the dimensional
reduction it is consistent to keep only the zero-modes on $\tilde{Y}$
which get small masses of order $\alpha'/\LCY^3$ and to drop all the
higher Kaluza--Klein modes with masses of order $1/\LCY$.

We would like to make the same kind of expansion in IIA and think
of the generalized mirror manifold $\hat{Y}$ as some small
perturbation of the original Calabi--Yau $Y$ mirror to
$\tilde{Y}$ without flux. The problem we will face throughout
this section is that we do not have, in general, an explicit
construction of $\hat{Y}$ from $Y$. Thus we can only give general
arguments about the meaning of such a limit. From the previous
discussion we saw that it is the intrinsic torsion $T^0$ which
measures the deviation of $\hat{Y}$ from a Calabi--Yau manifold.
Thus we would like to think that in the limit where $T^0$ is
small $\hat{Y}$ approaches $Y$. The problem is that in general $Y$
and $\hat{Y}$ have different topology. Thus, at the best, we can
only expect that $\hat{Y}$ approaches $Y$ locally in the limit of
small intrinsic torsion. Put another way, the torsion, like $H_3$
is really ``quantized''. Consequently, it cannot really be put to
zero, instead we can only try distorting the space to a limit
where locally $T^0$ is small and then locally the manifold looks
like $Y$.

This can be made slightly more formal in the following way. It is a
general result~\cite{FFS} that the Riemann tensor of any manifold with
$\SU(n)$ structure has a decomposition as
\begin{equation}
\label{Rdecomp}
   R = R_{\text{CY}} + R_\perp \ ,
\end{equation}
where the tensor $R_{\text{CY}}$ has the symmetry properties of the
curvature tensor of a true Calabi--Yau manifold, so that, for instance
the corresponding Ricci tensor vanishes. The orthogonal component
$R_\perp$ is completely determined in terms of $\nabla T^0$ and
$(T^0)^2$. (Note that the corresponding decomposition of the Ricci
scalar in the half-flat case is calculated explicitly in
appendix~\ref{Rhf}.) From this perspective, we can think of $R_\perp$
as a correction to the Einstein equation on a Calabi--Yau manifold,
analogous to the $H_3^2$ correction in the IIB theory. In particular,
if $\hat{Y}$ is to be locally like $Y$ in the limit of small torsion,
we require
\begin{equation}
   R_{\text{CY}}(\hat{Y}) = R(Y) \ .
\end{equation}

What, however, characterizes the limit where the intrinsic
torsion is small? Unlike the IIB case the string scale does not
appear in $T^0$. Typically both curvatures $R_{\text{CY}}$ and
$R_\perp$ are of order $1/\hat{L}^2$ where $\hat{L}$ is the size
of $\hat{Y}$. Thus making $\hat{Y}$ large will not help us.
Instead, we must consider some distortion of the manifold so that
$R_\perp\ll R_{\text{CY}}$. What this distortion might be is
suggested by mirror symmetry. We know that, without flux, a large
radius $\tilde{Y}$ is mapped to $Y$ with large complex structure.
Thus we might expect that we are interested in the large complex
structure limit of $\hat{Y}$.

In this limit, the conjecture is that $R_\perp(\hat{Y})$ becomes a
small perturbation, with a mass scale much smaller than the
Kaluza--Klein scale set by the average size of $\hat{Y}$. Thus,
as in the IIB case, at least locally, the original zero modes on
$Y$ become approximate massless modes on $\hat{Y}$ gaining a
small mass due to the non-trivial torsion. This suggests it is
again consistent in this limit to consider a dimensional
reduction keeping only the deformations of $\hat{Y}$ which
correspond locally to zero modes of $Y$. This holds both for the
ten-dimensional gauge potentials given in the case without flux
in~\eqref{fexpA} and the deformations of the metric as
in~\eqref{Jexp} and~\eqref{icy32}.

Having discussed the approximation, let us now turn to trying to
identify this light spectrum more precisely and characterizing how the
missing NS flux enters the problem. As discussed, it is the intrinsic
torsion of $\hat{Y}$ which characterizes the deviation of $\hat{Y}$
from a Calabi--Yau manifold therefore we expect that this encodes the
NS-flux parameters we are looking for. Mirror symmetry requires that
these new NS-fluxes are counted by the even cohomology of the ``limiting''
Calabi-Yau manifold $Y$.
As we saw above, in the case of half-flat
manifolds this suggests that the real $(2,2)$-form $F\sim d\Omega$ on
$\hat{Y}$, introduced in~\eqref{su3Nt} and discussed by
Vafa~\cite{Vafa}, can be viewed as specifying some ``extra data"
on $Y$ which is  a harmonic form
$\ff\in H^4(Y,\bbR)$ (or equivalently $H^2(Y,\bbR)$) measuring, at
least part of, the missing NS flux.

As mentioned above, the problem is that we have no explicit
construction of $\hat{Y}$ in terms of $Y$ and some given flux
$\ff$. 
Nonetheless, we expect, if mirror symmetry is
to hold, that for each pair $(Y,\ff)$ there is a unique half-flat
manifold $\hat{Y}_\ff$, so that there is a map
\begin{equation}
\label{eq:ident}
   (Y,\ff) \Leftrightarrow \hat{Y}_\ff \ ,
\end{equation}
where, in the limit of small torsion (large complex structure),
$Y$ and $\hat{Y}_\ff$ with the corresponding metrics
are locally diffeomorphic. In fact, we can argue two more conditions. First,
the identification~\eqref{eq:ident} can be applied at each point in
the moduli space of $Y$ giving us, assuming uniqueness, a
corresponding moduli space of $\hat{Y}_\ff$. Furthermore, it can be checked
, for the simple case of the torus, that the type IIB $H_3$-flux only
effect the
topology of $\hat{Y}$ in the sense that all points in the moduli space
of $\hat{Y}_\ff$ for given flux had the same topology. Thus we see
that, if mirror symmetry is to hold, the moduli space of metrics
$\mathcal{M}(Y)$ and $\mathcal{M}(\hat{Y}_\ff)$ of $Y$ and $\hat{Y}$
are the same
\begin{equation}
   \mathcal{M}(\hat{Y}_\ff) = \mathcal{M}(Y)
      \qquad \text{for any given $\ff$} \ ,
\end{equation}
where $\ff$ only effects the topology of $\hat{Y}$. This gives
the full moduli space of all $\hat{Y}_\ff$ the structure of an
infinite number of copies of $\mathcal{M}(Y)$ labeled by $\ff$.

More explicitly, the matching of moduli spaces means that for each
$(\Omega,J)$ on $Y$, since $\hat{Y}_\ff$ has $\SU(3)$ structure, we
have a unique corresponding $(\Omega,J)$ on $\hat{Y}$ and we must have
a corresponding expansion in terms of a basis of forms on $\hat{Y}$
\begin{equation}
\label{Oxz2}
\begin{aligned}
   \Ox &= z^A\, \ax_A - \cF_A\, \bx^A\ ,
     \qquad  A= 0,1,\ldots,h^{(1,2)}(Y)\ , \\
   J &= v^i\, \omega_i \ ,\qquad  i=1,\ldots, h^{(1,1)}(Y)\ ,
\end{aligned}
\end{equation}
where $z^A= (1,z^a)$ with $a=1,\ldots,h^{(1,2)}(Y)$ and the $z^a$
are the scalar fields corresponding to the deformations of the
complex structure ($\cF_A$ is  defined in section \ref{msa}),
while the $v^i$  are the scalar fields corresponding to the
K\"ahler deformations. The key point here is that although
$(\ax_A,\bx^A)$ form a basis for $\Omega$ and the $\omega_i$ form
a basis for $J$ they are not, in general, harmonic, and thus are
not bases for $H^3(\hat{Y})$ and $H^{(1,1)}(\hat{Y})$. Locally,
however, in the limit of small intrinsic torsion, they should
coincide with the harmonic basis of $H^3(Y)$ and $H^{(1,1)}(Y)$
on $Y$. For $*J$ one has an analogous expansion in terms of
four-forms on $\hat{Y}$ as in \eqref{oxstar}
\begin{equation}
\label{*J}
   *J = 4\mathcal{K} g_{ij}\nu^i\tilde{\omega}^j \ ,
      \qquad  i=1,\ldots, h^{(1,1)}(Y)\ ,
\end{equation}
where, again, there is no condition on $\tilde{\omega}^i$ being
harmonic on $\hat{Y}$, but in the small torsion limit they again
locally approach harmonic forms on $Y$.

The above expressions~\eqref{Oxz2} and~\eqref{*J} have been written in
terms of a prepotential $\mathcal{F}$ and a metric $g_{ij}$
which is the metric on the moduli space just as for
$Y$. If the
low-energy effective action is to be mirror symmetric we necessarily
have that the metrics on the moduli spaces
$\mathcal{M}(\hat{Y}_\ff)$ and $\mathcal{M}(Y)$ agree. This means
that the corresponding kinetic terms in the low-energy effective
action agree and implies the conditions
\begin{equation}
\label{normYhat}
   \int_{\hat Y} \ox_i \wg \tilde \ox^j =  \dx_i^j \  ,  \quad
   \int_{\hat Y} \ax_A \wg \bx^B =  \dx_A^B\ ,
      \quad
  \int_{\hat Y} \ax_A \wg \ax_B = \int_{\hat Y} \bx^A \wg \bx^B = 0\ ,
\end{equation}
exactly as on $Y$ in (\ref{normH2}) and (\ref{icy31}).

Now let us return to the flux and the restrictions implied
by $\hat{Y}_\ff$ being half-flat. Recall that we have argued that
the four-form $F^{(2,2)}\sim( d\Omega)^{2,2}$ corresponds to a
harmonic form $\ff\in H^4(Y,\bbZ)$ measuring the flux. Given the map
between harmonic four-forms on $Y$ and the basis $\tilde{\omega}^i$
introduced in~\eqref{*J}, we are naturally led to
rewrite \eqref{F22} as
\begin{equation}
\label{F4}
\begin{split}
  F_{mnpq}^{(2,2)} &\equiv
        \frac{1}{4||\Ox||^2} ( d\Omega)^{2,2}_{mnpq} \\
     &= \frac{1}{4||\Ox||^2}\, e_i\, \tox^i_{mnpq}\ ,
        \qquad i= 1, \ldots , h^{(1,1)}(Y) \ ,
\end{split}
\end{equation}
where the $e_i$ are
constants parameterizing the flux. Again, in the limit of small
torsion, locally $F$ is equivalent to a harmonic form on $Y$, namely
$\ff$.

Inserting (\ref{Oxz2}) into (\ref{F4}), we have
\begin{equation}
    d\Omega = z^A  d\alpha_A - \mathcal{F}_A  d\beta^A = e_i\tox^i \ .
\end{equation}
However, we argued that the flux only effects the topology of
$\hat{Y}$ and does not depend on the point in moduli space. Thus,
we require that this condition is satisfied independently of the
choice of moduli $z^A=(1,z^a)$. This is only possible if we have
\begin{equation}
  \label{dax}
   d \ax_0 = e_i\, \tox^i\ , \qquad  d \ax_a =  d \bx^A = 0\ ,
\end{equation}
where $\ax_0$ is singled out since it is the only direction in
$\Omega$ which is independent of $z^a$.\footnote{%
Of course this corresponds to a specific choice
of the symplectic basis of $H^3$. It is the same choice which
is conventionally used in establishing the mirror map without fluxes.}
Furthermore, inserting
(\ref{dax}) into (\ref{normYhat}) gives
\begin{equation}
  \label{dox1}
  e_i = \int \ox_i \wg d \ax_0 = - \int   d \ox_i \wg \ax_0 \ .
\end{equation}
Thus consistency requires
\begin{equation}
  \label{dox}
   d \ox_i = e_i \bx^0\  , \qquad  d \tilde\omega^i = 0\ ,
\end{equation}
where the second equation follows from (\ref{dax}).\footnote{%
Strictly speaking also  $d \ox_i =
  e_i \bx^0 + a^A \ax_A + b_a \bx^a$ for some yet undetermined
  coefficients $a^A, b_a$ solves (\ref{dox1}). However by a similar
  argument as presented for the exterior derivative of $\ox_i$ one can
  see that any non-vanishing such coefficient will produce a nonzero
  derivative of $\ax_a$ or/and $\bx^A$ contradicting (\ref{dax}). From
  this one concludes that the only solution of (\ref{dox1}) is
  (\ref{dox}).}

Eqs.~(\ref{dax}) and~(\ref{dox}) imply, just as we anticipated
above that neither $\omega_i$ nor $\tox^i$ are harmonic. In
particular, $\omega_i$ are no longer closed while the dual forms
$\tox^i$  are no longer co-closed, since at least one linear
combination $e_i\tox^i$ is exact. However, assuming for instance
that $e_1$ is non-zero, the linear combinations
\begin{equation}
  \label{ox'}
  \ox_i' = \ox_i - \frac{e_i}{e_1}\, \ox_1\ , \qquad i\ne 1\ ,
\end{equation}
are harmonic in that they satisfy
\begin{equation}
  d \ox_i' = d^\dagger \ox_i' = 0 \ ,
\end{equation}
where we used $ d^\dagger \ox_i' = *d* \ox_i' \sim *d \tilde\ox^{\prime i}$.
Thus there are still at least $h^{(1,1)}(Y)-1$ harmonic forms $\ox_i'$
on $\hat Y$.
The same argument can be repeated for $H^3$ where one
finds $2h^{(1,2)}$ harmonic forms or in other words the dimension
of $H^3$ has changed by two and we have together
\begin{equation}
h^{(2)}(\hat Y) = h^{(1,1)}(Y)-1\ , \qquad
h^{(3)}(\hat Y) = h^{(3)}(Y)-2\ .
\end{equation}
Physically this can be
understood from the fact that some of the scalar fields gain a mass
proportional to the flux parameters and no longer appear as zero modes
of the compactification. Similarly, from mirror symmetry
we do not expect the occurrence of new
zero modes on $\hat Y$ as these would correspond to additional new
massless fields in the effective action.
This is also consistent with our expectation
that $\hat Y$ is topologically  different from $Y$ which
stresses the
point that $Y$ and $\hat{Y}$ can only be locally close to each other in
the large complex structure limit.

Simply from the moduli space of $\SU(3)$-structure of $\hat{Y}_\ff$
and the relation~\eqref{F4} we have conjectured the existence of a set
of forms on $\hat{Y}_\ff$ satisfying the conditions~\eqref{dax}
and~\eqref{dox} which essentially encode information about the
topology of $\hat{Y}_\ff$. We should now see if this is compatible
with a half-flat structure. In particular we find, given~\eqref{Oxz2},
\begin{equation}
\label{eq:dJOe}
\begin{aligned}
    d J &= v^i e_i \beta^0 \ , \\
    d \Omega &= e_i \tox^i \ .
\end{aligned}
\end{equation}
From the standard $\SU(3)$ relation $J\wedge\Omega=0$ we have
that $\omega_i\wedge\alpha^A=\omega_i\wedge\beta^A=0$ for all $A$ and
hence in particular $J\wedge d J=0$. Furthermore, since the $e_i$ are
real, $ d\Omega^-=0$. Thus we see that
\eqref{dax} and~\eqref{dox} are consistent
with half-flat structure.\footnote{
  It would be interesting to calculate the moduli space of half-flat
  metrics on $\hat{Y}_\ff$ directly and see that it agreed with, or
  at least had a subspace, of the form given by~\eqref{Oxz2}
  and~\eqref{*J} together with~\eqref{dax} and~\eqref{dox}.}
Furthermore, since $ d J$ and $ d\Omega$
completely determine the intrinsic torsion $T^0$, we see that all the
components of $T^0$ are given in terms of the constants $e_i$ without
the need for any additional information.

Let us summarize. We proposed a set of rules for identifying the
light modes for compactification on $\hat{Y}$ compatible with
mirror symmetry and half-flatness. We first argued that in the
limit of large complex structure the torsion of $\hat{Y}$ is
small, and locally $\hat{Y}$ and $Y$ are diffeomorphic, even
though globally they have different topology. In this limit, the
light spectrum corresponds to modes on $\hat{Y}$ which locally
map to the zero modes of $Y$. This was made more precise by first
noting that mirror symmetry implies a one-to-one correspondence
between each pair of a Calabi--Yau manifold $Y$ and flux $\ff\in
H^4(Y,\bbZ)$ and a unique half-flat manifold $\hat{Y}_\ff$. As a
consequence the moduli space of half-flat metrics on
$\hat{Y}_\ff$  has to be identical with the moduli space of
Calabi--Yau metrics on $Y$. This in turn implies that the metrics
on these moduli spaces agree and a basis of forms for $J$ and
$\Omega$ exist on $\hat{Y}$ which coincides with the
corresponding basis of harmonic forms on $Y$ in the small torsion
limit. Identifying the missing NS flux $e_i$ as
$F\sim d\Omega^{2,2}\sim e_i \tilde\omega^i$ led to a set of
differential relations  among this basis of forms in terms of the
$h^{(1,1)}(Y)$ flux parameters $e_i$. We further showed that
these relations are compatible with the conditions of
half-flatness. As we will see more explicitly in the next section
these forms give the correct basis for expanding the
ten-dimensional fields on $\hat Y$ and obtaining a mirror
symmetric effective action. We will find that the masses of the
light modes are proportional to the fluxes and thus to the
intrinsic torsion of $\hat Y$.


\subsection{The effective action}
\label{LLEA}

In this section we present the derivation of the low energy
effective action of type IIA supergravity compactified on the
manifold $\hat Y$ described in sections \ref{sec:hf} and
\ref{spectrum}. As argued in the previous section we insist on
keeping the same light spectrum as for Calabi-Yau
compactifications and therefore the KK-reduction is closely
related to the reduction on Calabi-Yau manifolds which we recall
in appendix \ref{cA}. The difference is that the differential
forms we expand in are no longer harmonic but instead obey
\begin{equation}
  \label{dab}
  d \ax_0 = e_i \tox^i\ , \qquad d \ax_a = d \bx^A = 0 \ ,\qquad
 d \ox_i = e_i \bx^0\ ,  \qquad \qquad d \tox^i = 0 \ .
\end{equation}
However, we continue to demand that these forms have identical
intersection numbers as on the Calabi-Yau or in other words obey
unmodified (\ref{normYhat}). As we are going to see shortly the
relations (\ref{dab}) are responsible for generating mass terms
in the effective action
consistent with the discussion in the previous section.\footnote{%
  Note that we are not expanding in the harmonic forms $\ox'_i$
  defined in (\ref{ox'}) but continue to use the non-harmonic
  $\ox_i$. The reason is that in the  $\ox_i$-basis mirror symmetry
  will be manifest. An expansion in the $\ox'_i$-basis merely
  corresponds to field redefinition in the effective action as they
  are just linear combinations of the $\ox_i$.}

Let us start  from the type IIA action in $D=10$ \cite{JP}
\begin{eqnarray}
  \label{SIIA}
  S & = & \int \, e^{-2\hat\phi} \left( -\frac12 \hat R *\! {\bf 1} + 2
  d \hat\phi \wg * d \hat\phi - \frac14  \hat H_3\wg * \hat H_3 \right) \nn \\
  & & - \frac12  \int \left(\hat  F_2 \wg * \hat F_2 + \hat F_4 \wg * \hat F_4
  \right) + \frac12 \int \hat H_3 \wedge \hat C_3 \wedge d \hat C_3 \, ,
\end{eqnarray}
where the notations are explained in more detail in section
\ref{cA}. In the KK-reduction the ten-dimensional (hatted) fields
are expanded in terms of the forms $\ox_i, \ax_A,\bx^A$
introduced in (\ref{Oxz2})
\begin{eqnarray}
  \label{fexp}
 \hat \phi & = & \phi \ ,\qquad   \hat A_1  =  A^0 \ ,\qquad
\hat B_2 =  B_2 + b^i \ox_i      \nn \\
  \hat C_3 & = & C_3 + A^i \wg \ox_i + \xi^A \ax_A + \txi_A \bx^A \ ,
\end{eqnarray}
where $A^0,A^i$ are one-forms in $D=4$ (they will generate
$h^{(1,1)}$ vector multiplets and contribute the graviphoton to
the gravitational multiplet) while $\xi^A,\txi_A,b^i$ are scalar
fields in $D=4$. The $b^i$ combine with the K\"ahler deformations
$v^i$ of (\ref{Oxz2}) to form the complex scalars $t^i = b^i +
iv^i$ sitting in the $h^{(1,1)}$ vector multiplets. The
$\xi^a,\txi_a$ together with the complex structure deformations
$z^a$ of (\ref{Oxz2}) are members of $h^{(1,2)}$ hypermultiplets
while $\xi^0,\txi_0$ together with the dilaton $\phi$ and $B_2$
form the tensor multiplet.

The difference with Calabi-Yau compactifications results from the fact
that the derivatives of $\hat B_2, \hat C_3 $ in (\ref{fexp}) are modified
as a consequence of  (\ref{dab}) and we find
\begin{eqnarray}
  \label{dfexp}
  d \hat C_3 & = & d C_3 + (d A^i) \wg \ox_i
+ (d \xi^A) \ax_A + (d\txi_a) \bx^a + (d\txi_0 - e_i A^i)  \bx^0
  + \xi^0 e_i \tox^i \ ,\nn \\
  d \hat B_2 & = & d B_2 + (d b^i) \ox_i + e_i b^i \bx^0\ .
\end{eqnarray}
We already see that the scalar $\txi_0$ becomes charged precisely
due to (\ref{dab}) which is exactly what we expect from the type IIB
action. However, on the type
IIB side we have $(h^{(1,2)} +1)$ electric flux parameters while in
(\ref{dfexp}) only $h^{(1,1)}$ fluxes $e_i$ appear. The missing flux arises
from  the NS 3-form
field strength $\hat H_3 = d \hat B_2$ in the direction of $\bx^0$.
Turning on this additional NS flux
amounts to a shift
\begin{equation}
  \label{flux}
  \hat H_3 \to \hat H_3 + e_0 \bx^0 \, ,
\end{equation}
where $e_0$ is the additional mass parameter.
Using (\ref{dfexp}), (\ref{flux}) and
(\ref{HFdef}) we see
that the parameter $e_0$ introduced in this way naturally combines with the
other fluxes $e_i$ into
\begin{eqnarray}
  \label{modFS}
  \hat H_3 & = &  d B_2 + d b^i \ox_i + (e_i b^i +e_0) \bx^0\ ,  \\
  \hat F_4 & = & (d C_3 - A^0 \wg dB_2) + (d A^i - A^0 d b^i) \wg \ox_i + D
  \xi^A \ax_A + D \txi_A \bx^A  + \xi^0 e_i \tox^i\ ,\nn
\end{eqnarray}
where the covariant derivatives are given by
\begin{equation}
  \label{cd}
  D \txi_0 = d \txi_0 - e_i (A^i + b^i A^0) - e_0 A^0 , \qquad
  D \xi^A = d \xi^A , \qquad D \txi_a = d \txi_a .
\end{equation}
This formula is one of the major consequences of compactifying on $\hat Y$ (in
particular of expanding the 10 dimensional fields in forms which are not
harmonic) as one of the scalars, $\txi_0$, becomes charged.

{}From here  on the compactification proceeds as in the massless
case by inserting (\ref{modFS}) into the action (\ref{SIIA}).
Except for few differences which we point out, the calculation
continues as in section \ref{cA} and we are not going to repeat
this calculation here.
Using (\ref{fexpA}),  (\ref{dfexp}) and (\ref{modFS})
one can see that the parameters $e_0$ and $e_i$ give rise to new interactions
coming from the topological term in (\ref{SIIA})
\begin{eqnarray}
  \label{ltop}
\frac12 \int_{\hat Y} \hat H_3 \wedge \hat C_3 \wedge d \hat C_3
  & = &  \frac{\xi^0}{2} d B_2 \wg A^i e_i - \frac12 dB_2
  \wg \left(\xi^0 (d \txi_0 -e_i A^i) + \xi^a d \txi_a - \txi_A d \xi^A
  \right)  \nn \\
  & & + \frac{\xi^0}{2} e_i d b^i \wg C_3 +\frac12 d b^i \wg A^j \wg d A^k
  \cK_{ijk}  \\
  & & - \frac{\xi^0}{2} (e_i b^i + e_0) d C_3 -
  \frac12(e_i b^i + e_0) \wg C_3 \wg d \xi^0\ ,\quad \nn
\end{eqnarray}
where $\cK_{ijk} $ is defined in (\ref{K}).

The 3-form $C_3$ in 4 dimensions carries no physical degrees of freedom. Nevertheless
it can not be neglected as it may introduce a cosmological constant. Moreover
when such a form interacts non-trivially with the other fields present in the
theory as in (\ref{ltop}) its dualization to a constant requires more
care. Collecting all terms which contain $C_3$ we find
\begin{equation}
  \label{SC3}
  S_{C_3} = - \frac{\cK}{2} (d C_3 - A^0 \wg dB_2) \wg * (d C_3 - A^0 \wg d B_2)
  - \xi^0 \, (e_i b^i + e_0) dC_3 \, .
\end{equation}
As shown in \cite{BGGF,BGG} the proper way of performing this
dualization is by adding a Lagrange multiplier $\lx dC_3$. The
3-form $C_3$ is dual to the constant $\lx$ which was shown to be
mirror symmetric to a RR-flux in ref.\ \cite{LM2}   and
consequently plays no role in the analysis here. Solving for
$dC_3$, inserting the result back into (\ref{SC3}) and in the end
setting $\lx= 0$ we obtain the action dual to (\ref{SC3})
\begin{equation}
  \label{Cdual}
  S_{dual} = - \frac{(\xi^0)^2}{2 \cK} (e_i b^i + e_0)^2 - \xi^0 \, (e_i b^i +
  e_0)  A^0 \wg dB_2 \, .
\end{equation}

Finally, in order to obtain the usual $N=2$ spectrum we dualize $B_2$ to a
scalar field denoted by $a$.
Due to the Green-Schwarz type interaction of $B_2$ (the first term in
(\ref{ltop}) and the second term in (\ref{Cdual})) $a$ is charged, but beside
that the dualization proceeds as usually.
Putting together all the pieces and after going to the Einstein frame one can
write the compactified action in the standard $N=2$ form
\begin{eqnarray}
  \label{S4A}
  S_{IIA} & = & \int \Big[ -\frac12 R *\! {\bf 1} - g_{ij} dt^i \wg * d
  {\bar t}^j - h_{uv} Dq^u \wg * Dq ^v \nn \\
  & & \qquad + \frac{1}{2}\, \IM \cN_{IJ} F^I\wg * F^{J}
  + \frac{1}{2} \, \RE \cN_{IJ} F^I \wg F^J - V_{IIA} *\! {\bf 1} \Big] ,
\end{eqnarray}
where the gauge coupling matrix $\cN_{IJ}$ and the metrics
$g_{ij}, h_{uv}$ are given in (\ref{eq:N}), (\ref{gH11}) and
(\ref{qktNS}) respectively. As explained in section \ref{cA} the
gauge couplings can be properly identified after redefining the
gauge fields $A^i \to A^i - b^i A^0$. We have also introduced the
notation $I=(0,i) = 0, \ldots, h^{(1,1)}$ and so $A^I = (A^0,
A^i)$. Among the covariant derivatives of the hypermultiplet
scalars $Dq^u$ the only non-trivial ones are\footnote{Up to a redefinition of the sign of
the fluxes.}

\begin{equation}
  \label{cds}
  D a = d a - \xi^0 e_I A^I \, ; \qquad D \txi_0 = d \txi_0 + e_I A^I \, .
\end{equation}
We see that two scalars are charged under a Peccei-Quinn symmetry
as a consequence of the non-zero $e_I$.

Before discussing the potential  $V_{IIA}$ let us note that the
action (\ref{S4A}) already has the form expected from the mirror
symmetric action given in section \ref{cBNS}. In particular the
forms $\ax_0$ and $\bx^0$ in (\ref{dab}) single out the two
scalars $\xi^0,\txi_0$ from the expansion of $\hat C_3$. $\xi^0$
maps under mirror symmetry to the RR scalar $l$ which is already
present in the $D=10$ type IIB theory while  $\txi_0$ maps to the
charged RR scalar in type IIB. Moreover, using these
identifications one observes that the gauging (\ref{cds})  is
precisely what one obtains in the type IIB case with NS electric
fluxes turned on (\ref{action5}).

Finally, we  need to check that the potential from (\ref{S4A})
coincides with the one
obtained in the type IIB case (\ref{potIIB}).
In the case of type IIA compactified on $\hat Y$ one can identify four
distinct contributions to the potential: from the kinetic terms of $\hat B_2$
and $\hat C_3$, from the dualization of $C_3$ in 4 dimensions and
from the Ricci scalar of $\hat Y$. We study these contributions in turn.
We go directly to the four-dimensional Einstein frame which amounts to
multiplying every term in the potential by a factor $e^{4 \phi}$ coming from
the rescaling of $\sqrt {-g}$, $\phi$ being the four-dimensional
dilaton which is related to the ten-dimensional dilaton
$\hat \phi$ by $e^{-2 \phi} = e^{-2 \hat \phi} \cK $.

Using (\ref{modFS}) we see that the kinetic
term of $\hat B_2$ in (\ref{SIIA})  contributes to the potential
\begin{equation}
  \label{V1}
  V_1 = \frac{e^{2 \phi}}{4 \cK} (e_i b^i + e_0)^2 \int_{\hat Y} \bx^0 \wg *
  \bx^0 = - \frac{e^{- 2 \phi}}{4 \cK} (e_i b^i + e_0)^2 \left[(\IM
    \cM)^{-1}\right]^{00} \, ,
\end{equation}
where the integral over $\hat Y$ was performed using (\ref{star}), (\ref{A-N})
and (\ref{oxstar}).
Similarly, the kinetic term of $\hat C_3$ produces the following piece in the
potential
\begin{equation}
  \label{V2}
  V_2 = e^{4 \phi} \, \frac{(\xi^0)^2}{8 \cK}\, e_i e_j g^{ij}\ ,
\end{equation}
where $g^{ij}$ arises after integrating over $\hat Y$ using (\ref{oxstar}).
Furthermore, (\ref{Cdual}) contributes
\begin{equation}
  \label{V3}
  V_3 = e^{4 \phi} \, \frac{(\xi^0)^2}{2 \cK} (e_i b^i + e_0)^2 \ .
\end{equation}
Combining (\ref{V1}), (\ref{V2}) and (\ref{V3}) we arrive at
\begin{eqnarray}
  \label{potA}
    V_{IIA} & = &V_g + V_1 + V_2 + V_3 \\
    & =& V_g - \frac{e^{2 \phi}}{4 \cK} (e_i b^i + e_0)^2
    \left[(\IM \cM)^{-1}\right]^{00} - e^{4 \phi} \, \frac{(\xi^0)^2}{2}
    e_I e_J \left[(\IM \cN)^{-1} \right]^{IJ}\ ,\nn
\end{eqnarray}
where we used
the form of the matrix $(\IM \cN)^{-1}$ given in (\ref{ImN-1e}).
$V_g$ is a further contribution to the potential which arises from
the Ricci scalar. Since $\hat Y$ is no longer Ricci-flat $R$
contributes to the potential and in this way provides
 another sensitive test of the half-flat geometry.

In appendix \ref{Rhf} we show that for half-flat manifolds
the Ricci scalar  can be written in terms of the contorsion as
\begin{equation}
  \label{R}
  R = - \CT_{mnp} \CT^{npm} - \frac{1}{2} \epsilon^{mnpqrs} (\nabla_m \CT_{npq} -
  \CT_{mp}{}^l \CT_{nlq}) J_{rs} \ ,
\end{equation}
which, as expected, vanishes for $\CT=0$.
In order to evaluate the above expression we first we have to give a
prescription about how to compute $\nabla_m \CT_{npq}$. Taking into
account that at in the end the potential in the four-dimensional theory appears
after integrating over the internal manifold $\hat Y$
we can integrate by parts and `move' the covariant derivative to act
on $J$. This in turn can be computed by using the fact that $J$ is covariantly
constant with respect to the connection with torsion (\ref{torJO}).
Replacing the contorsion $\CT$ from (\ref{T-K}), going to complex indices and
using the defining relations for the torsion (\ref{su3Nt}), (\ref{T3}) and
(\ref{F4}) one can find after some straightforward but tedious algebra
the expression for the Ricci scalar.
The calculation is presented in appendix \ref{Rhf} and here we only record the
final result
\begin{equation}
  \label{Rf}
  R = - \frac{1}{8} \, e_i e_j g^{ij} \left[(\IM \cM)^{-1}
  \right]^{00} \ .
\end{equation}
Taking into account the factor $\frac{e^{-2\hat \phi}}{2}$ which
multiplies the Ricci scalar in the 10 dimensional action (\ref{SIIA}) and the
factor $e^{4 \phi}$ coming from the four-dimensional Weyl rescaling one obtains
the contribution to the potential coming from the gravity sector to be
\begin{equation}
  \label{Vg}
  V_g = - \frac{e^{2 \phi}}{16 \cK} \, e_i e_j g^{ij} \left[(\IM
  \cM)^{-1}\right]^{00}\ .
\end{equation}
Inserted into (\ref{potA}) and using again (\ref{ImN-1e}) we can finally
write the entire potential
which appears in the compactification of type IIA supergravity on $\hat Y$
\begin{equation}
  \label{potAfin}
  V_{IIA} = - \frac{e^{4 \phi}}{2} \left( (\xi^0)^2 - \frac{e^{-2 \phi}}{2}
  \left[(\IM \cM)^{-1}\right]^{00}  \right) e_I e_J
  \left[(\IM \cN)^{-1} \right]^{IJ} .
\end{equation}

In order to compare this potential to the one obtained in type IIB case
(\ref{potIIB}) we should first see how the formula (\ref{potAfin}) changes under
the mirror map. We know that under mirror symmetry the gauge coupling matrices
$\cM$ and $\cN$ are mapped into one another. In particular this means
that\footnote{In order to avoid confusions we have added the label $A/B$ to
  specify the fact that the corresponding quantity appears in type IIA/IIB
  theory.}
\begin{equation}
  \left[(\IM \cM_A)^{-1}\right]^{00} \leftrightarrow \left[(\IM
  \cN_B)^{-1} \right]^{00} = - \frac{1}{\cK_B} \ .
\end{equation}
where we used the expression for $(Im \cN)^{-1}$ from (\ref{ImN-1e}).
With this observation it can be easily seen that the type IIA potential
(\ref{potAfin}) is precisely mapped into the type IIB one (\ref{potIIB})
provided one identifies the electric flux parameters
$e_I \leftrightarrow \q_A$
and the four-dimensional dilatons on the two sides.

To summarize the results obtained in this section, we have seen
that the low energy effective action of type IIA theory
compactified on $\hat Y$ is precisely the mirror of the effective
action obtained in section \ref{cBNS} for type IIB theory
compactified on $Y$ in the presence of NS electric fluxes. This
is our final argument that the half-flat manifold $\hat Y$ is the
right compactification manifold for obtaining the mirror partners
of the NS electric fluxes of type IIB theory. In particular the
interplay between the gravity and the matter sector which
resulted in the potential (\ref{potAfin}) provided  a highly
nontrivial check on this assumption.

\section{Type IIB on a half-flat manifold}\label{IIBhalf}

In the previous section, Vafa's proposal that the mirror of the
NS fluxes should come from the geometry of the internal manifold
was made more concrete : it was conjectured that when NS fluxes
are turned on in type IIB theory, mirror symmetry requires the
presence of a new class of manifolds, known as half-flat
manifolds with $\SU(3)$ structure on type IIA side. The main
argument supporting this proposal was provided by showing that
the low-energy effective actions for the type IIB compactified on
a \CY three-fold in the presence of electric NS three-form flux
and type IIA compactified on a half-flat space are equivalent.
The purpose of this section is to test this conjecture in the
reversed situation. We want to show that compactifying type IIB
theory on half-flat manifolds produces an effective action which
is mirror equivalent to type IIA theory compactified on \CY
three-folds with NS three-form flux turned on, whose action is
reviewed in appendix \ref{IIANS}.

Following the previous sections, we will now perform the
compactification of type IIB on a manifold $\hat Y$ obeying
\eqref{dax} and \eqref{dox}, which again will turn out to be
responsible for generating mass terms and gaugings in the
lower-dimensional action.

Let us start by shortly recording type IIB supergravity in ten
dimensions. The NS-NS sector of the bosonic spectrum consists of
the metric $\hat g_{MN}$, an antisymmetric tensor field $\hat
B_2$ and the dilaton $\hat \phi$. In the RR sector one finds the
0-, 2-, and 4-form potentials $\hat l, \ \hat C_2, \ \hat A_4$.
The four-form potential satisfies a further constraint in that its
field strength $\hat F_5$ is self-dual.
The interactions of the above fields are described by the
ten-dimensional action \cite{JP}
\begin{eqnarray}
  S_{IIB}^{(10)} &=& \int e^{-2\hat\phi} \left(- \frac{1}{2}\hat R *\! {\bf 1} +
    2  d  \hat \phi \wg *  d  \hat\phi - \frac{1}{4}  d  \hat B_2 \wg *  d
    \hat B_2\right) \nn \\[2mm]
  &&-\frac{1}{2}\int \left(  d\hat l\wg *  d\hat l+ \hat F_3 \wg * \hat F_3 +
    \frac12 \hat F_5\wg * \hat F_5\right)  \label{actionIIB}\\[2mm]
  && - \frac12 \int \hat A_4 \wg  d \hat B_2\wg  d \hat C_2 \nn\ ,
  \end{eqnarray}
where the field strengths $\hat F_3$ and $\hat F_5$ are defined as
\begin{eqnarray}
  \hat F_3 & = &  d  \hat C_2 - \hat l  d \hat B_2 \ , \label{fsIIB}\\[2mm]
  \hat F_5 & = &  d  \hat A_4 -  d  \hat B_2 \wg \hat C_2 \nn \ .
\end{eqnarray}
As it is well known the action \eqref{actionIIB} does not
reproduce the correct dynamics of type IIB supergravity as the
self-duality condition of $\hat F_5$ can not be derived from a
variational principle. Rather this should be imposed by hand in
order to obtain the correct equations of motion and we will come
back to this constraint later as it plays a major role in the
following analysis.

In order to compactify the action \eqref{actionIIB} on a half flat
manifold we proceed as in section \ref{LLEA} and continue to
expand the ten dimensional fields in the forms which appear in
\eqref{dax} and \eqref{dox} even though they are not harmonic.
The 4-dimensional spectrum is not modified by the introduction of
the fluxes, and is still obtained by a regular KK expansion
\begin{eqnarray}
  \hat B_2 & = &  B_2 +  b^i \wg \ox_i \ ,\qquad i = 1,\ldots,h^{(1,1)}\ , \\
  \hat C_2 & = & C_2 + c^i \wg \ox_i\ , \label{expanIIB}\\
  \hat A_4 & = & D_2^i \wg \ox_i + \rho_i \wg \tilde \ox^i + V^A \wg \ax_A
  - U_A \wg \bx^A\ ,\qquad A = 0,\ldots,h^{(1,2)} \ , \nn
\end{eqnarray}
and thus one finds the two forms $B_2,C_2,D_2^i$, the vector
fields $V^A,U_A$, and the scalars $b^i,c^i,\rho_i$. Additionally,
from the metric fluctuations on the internal space one obtains
the scalar fields $z^a$ and $v^i$ \eqref{Oxz2}, which correspond
to the \CY complex structure  and K\"ahler class deformations
respectively. Due to the self-duality condition which one has to
impose on $\hat F_5$, not all the fields listed above describe
physically independent degrees of freedom. Thus as four
dimensional gauge fields one only encounters either $V^A$ or
$U_A$. In the same way, the scalars $\rho_i$ and the two forms
$D_2^i$ are related by Hodge duality and one can eliminate either
of the two in the four dimensional action. In the end one obtains
an $N=2$ supersymmetric spectrum consisting of a gravity multiplet
$(g_{\mu\nu},V^0)$, $h^{(2,1)}$ vector multiplets $(V^a, z^a)$
and $4(h^{(1,1)} +1)$ scalars $\phi, ~h_1,~ h_2,~ l,~ b^i,~ c^i,
~v^i, ~\rho^i$ which form $h^{(1,1)} +1$
hypermultiplets.\footnote{
  We have implicitly assumed that the two-forms $C_2$ and $B_2$ remain massless
  in four dimensions and they can be Hodge dualized to scalars  which we have
  denoted $h_1$ and $h_2$ respectively.}

Up to this point everything looks like the ordinary \CY
compactification reviewed in section \ref{cB}. The difference
comes when one inserts the above expansion back into the action
\eqref{actionIIB}. Due to \eqref{dax} and \eqref{dox}, the
exterior derivatives of the fields \eqref{expanIIB} are going to
differ from the standard case
\begin{eqnarray}
  \label{cBhf2}
  d \hat B_2 & = &  d  B_2 +  d  b^i \wg \ox_i +e_i b^i \bx^0 + e_0 \bx^0
  \ , \nn \\[2mm]
  d  \hat C_2 & = &  d  C_2 +  d  c^i\wg \ox_i + e_i c^i \bx^0 \ , \\[2mm]
  d  \hat A_4 & = &  d   D_2^i \wg \ox_i + e_i D_2^i \wg \bx^0 +  d  V^A \wg
  \ax_A -  d  U_A \wg \bx^A + ( d  \rho_i - e_i V^0 )\wg \tox^i \ .\nn
\end{eqnarray}
In analogy with type IIA case, we have also allowed for a normal
$H_3$ flux proportional to $\bx^0$. This naturally combines with
the other fluxes parameters $e_i$ defined in \eqref{dax} to
provide all the $h^{(1,1)}+1$  electric fluxes. With these
expressions one can immediately write the field strengths $F_3$
and $F_5$ from \eqref{fsIIB}
\begin{eqnarray}
  \label{F35}
  \hat F_3 & = & ( d  C_2 - l  d  B_2) + ( d  c^i - l  d  b^i) \wg \ox_i
  + e_i (c^i - l b^i) \bx^0 - l e_0 \bx^0 \ , \\[2mm]
  \hat F_5 & = & ( d  D_2^i - d  b^i \wg C_2 - c^i  d  B_2) \wg \ox_i +
  (D \rho_i -\cK_{ijk}c^j d  b^k) \wg \tox^i + F^A \wg \ax_A - \tilde G_A \wg
  \bx^A \ , \nn
\end{eqnarray} where we have defined
\begin{eqnarray}
  \label{cBhf5}
  D \rho_i & = &  d  \rho_i - e_i V^0 \ , \nn \\[2mm]
  F^A & = &  d  V^A \ , \quad G_A =  d  U_A \ , \\[2mm]
  \tilde G_0 & = & G_0 - e_i(D_2^i - b^i C_2) + e_0 C_2 \ ; \quad
  \tilde G_a = G_a \ . \nn
\end{eqnarray}

In order to derive the lower-dimensional action we adopt the
following strategy \cite{GD}. In the first stage we are going to
ignore the self-duality condition which should be imposed on
$\hat F_5$ and treat the fields coming from the expansion of
$\hat A_4$ as independent. Thus, initially we naively insert the
expansions \eqref{cBhf2} into \eqref{F35} and perform the
integrals over the internal space. To obtain the correct action
we will further add suitable total derivative terms so that the
self-duality conditions appear from a variational principle. At
this point one can eliminate the redundant fields and in this way
obtain the four-dimensional effective action and no other
constraint has to be imposed. It can be checked that the result
obtained in this way is compatible with the ten dimensional
equations of motion.

Let us apply this procedure step by step. First one inserts the
expansions \eqref{cBhf2} and \eqref{F35} into the ten-dimensional
action \eqref{actionIIB}. The various terms of this action take
the form
\begin{eqnarray}
  \label{cBhf10}
   - \frac14 \int_Y  d  \hat B_2 \wg *  d  \hat B_2
  & = & - \frac{\cK}{4} \,  d  B_2 \wg *  d  B_2 - \cK g_{ij}  d  b^i \wg *
   d  b^j +\frac14(e_ib^i+e_0)^2 \moo * {\bf 1} \ , \nn \\[5mm]
  - \frac12 \int_Y {\hat {F}_3} \wg * {\hat {F}_3} & = & -
  \frac{\cK}{2}\, ( d  C_2 - l  d  B_2) \wg * ( d  C_2 - l  d  B_2) \nn \\
  & & - 2 \cK g_{ij} ( d  c^i - l  d  b^i) \wg * ( d  c^j - l  d  b^j)
  +\frac12 \Big[e_i(c^i-lb^i)-le_0\Big]^2 \moo * {\bf 1} \ , \nn
\end{eqnarray}
\begin{eqnarray}
  \label{cBhf11}
  -\frac{1}{4}\int \hat F_5\wg * \hat F_5
  & = &  + \frac14 \IM \cM^{-1} \left(\tilde G - \cM F \right)
  \wg * \left(\tilde G - \bar \cM  F\right) \\*
  && -\cK g_{ij} ( d  D_2^i - d  b^i \wg C_2 - c^i  d  B_2) \wg
   *( d  D_2^j - d  b^j \wg C_2 - c^j  d  B_2) \nn \\*
  & & - \frac{1}{16\cK} g^{ij}  (D \rho_i -\cK_{ilm}c^l d  b^m) \wg
   *(D \rho_i -\cK_{jnp}c^n d  b^p)  \ , \nn \\
  & & \nn \\
  - \frac12 \int \hat A_4 \wg d \hat B_2 \wg d \hat C_2 & = &
  - \frac12\cK_{ijk}D_2^i\wg d  b^j\wg d  c^k-\frac12\rho_i\left( d
  B_2 \wg d  c^i +  d  b^i\wg  d  C_2 \right) \ , \nn\\*
  && +\frac12e_iV^0\wg\left( c^i d  B_2-b^i d  C_2\right)-\frac12e_0V^0\wg  d
  C_2 \ . \nn
\end{eqnarray}
In order to write the above formulae we have defined $\moo =
\left(\IM \cM^{-1} \right)^{00}$. In the gravitational sector,
beyond the usual part containing the kinetic terms for the moduli
of $\hat Y$ there will be a further contribution coming entirely
from the internal manifold which is due to the fact that $\hat Y$
is not Ricci-flat and which will generate a piece of potential in
four dimensions. The Ricci scalar for half-flat manifolds was
computed in appendix \ref{Rhf} and here we will just record the
effective potential generated in this way
\begin{equation}
  \label{pot1}
  V_g =  -\frac{\moo}{16\cK}\rme^{2\phi}e_ie_jg^{ij}  \ .
\end{equation}

At this point we have to impose the self-duality condition for
$\hat F_5$ which translates into the following constraints on the
four dimensional fields
\begin{eqnarray}
  \label{cBhf13}
   d  D_2^i - d  b^i \wg C_2 - c^i  d  B_2 & = & \frac{1}{4 \cK} g^{ij}
  * (D \rho_i -\cK_{ijk}c^j d  b^k) \ , \nn\\[2mm]
  *\tilde G_{A} & = & \RE\cM_{AC} * F^C - \IM \cM_{AC}  F^C \ ,
\end{eqnarray}
with $D \rho_i$ and $\tilde G_A$ defined in \eqref{cBhf5}. By
adding the following total derivative term to the action
\begin{eqnarray}
  \label{cBhf15}
  \cL_{\rm{td}} & = & + \frac12  d  D_2^i \wg  d  \rho_i + \frac12
  F^A \wg G_A \nn\\[2mm]
  & = & + \frac12  d  D_2^i \wg D \rho_i + \frac12 F^A \wg \tilde
  G_A - \frac12 (e_i b^i + e_0) F^0 \wg C_2
\end{eqnarray}
the constraints \eqref{cBhf13} can be found upon variation with
respect to $d  D_2^i$ and $G_A$ respectively. This allows us to
eliminate the fields $ d  D_2^i$ and $G_A$ using their equations
of motion and consequently the effective action obtained in this
way describes the correct dynamics for the remaining fields which
now do not have to satisfy any further constraint.

After the dualization of the 2-forms $C_2$ and $B_2$ to the
scalars $h_1$ and $h_2$ one obtains the effective action for type
IIB supergravity compactified to four dimensions on a half-flat
manifold
\begin{eqnarray}
  \label{cBhf28}
  S_{IIB}^{(4)} &=& \int - \frac{1}{2} R *\! {\bf 1} - g_{ab} dz^a \wg
  *d\bar{z}^{b} - g_{ij} dt^i \wg *d\bar{t}^j - d\phi \wg *d \phi
  \nonumber\\[2mm]
  && - \frac{e^{2\phi}}{8 \cK} g^{-1\,ij} \left( D\rho_i -
    \cK_{ikl} c^k db^l \right) \wg *\left( D\rho_j -
    \cK_{jmn} c^m db^n \right) \nonumber \\[2mm]
  &&  - 2 \cK e^{2\phi} g_{ij} \left( dc^i - l db^i \right)\wg * \left( dc^j - l db^j \right)
  - \frac{1}{2} \cK e^{2\phi} dl \wg * dl \\[2mm]
  &&  - \frac{1}{2\cK} e^{2\phi}\left(  d  h_1 - b^i D\rho_i +e_0V^0\right)
  \wg *\left(  d  h_1 - b^j D\rho_j +e_0V^0\right) -\rme^{4\phi}D\tilde h\wg *D\tilde h\nonumber \\[2mm]
  && + \frac{1}{2} \RE \cM_{AB} F^A \wg F^B + \frac{1}{2} \IM
  \cM_{AB} F^A \wg * F^B - V_{IIB} * {\bf 1} \ , \nn
\end{eqnarray}
where
\begin{eqnarray}
  \label{cBhf29}
  D \tilde h =  d  h_2 +l  d  h_1 + (c^i-l b^i) D \rho_i + l e_0 V^0 -
  \frac12 \cK_{ijk} c^i c^j  d  b^k \ .
\end{eqnarray}
Performing the field redefinitions \cite{BGHL}
\begin{eqnarray}
  \label{mmap}
  a = 2 h_2 + l h_1 + \rho_i (c^i - l b^i) \ , & & \xi^0 = l \ , \qquad  \xi^i
  = l b^i - c^i \ ,  \\
  \txi_i = \rho_i + \frac{l}{2} \cK_{ijk} b^j b^k - \cK_{ijk} b^j c^k \ ,
  & &
  \txi_0 = - h_1 -\frac{l}{6} \cK_{ijk} b^i b^j b^k + \frac12 \cK_{ijk}
  b^i b^j c^k \ ,\nn
\end{eqnarray}
the metric for the hyperscalars takes the standard quaternionic
form of \cite{FeS} which is now exactly the mirror image of
\eqref{cAns21} with the gauge coupling matrices $\cN$ and $\cM$
exchanged as prescribed by the mirror map. Introducing the
collective notation $q^u = (\phi, a, \xi^I, \txi_I)$ we can write
the final form of the four dimensional action
\begin{eqnarray}
  \label{cBhf33}
  S_{IIA} & = & \int \Big[ -\frac12 R ^* {\bf 1} - g_{ab} dz^a \wg * d
  {\bar z}^b - \tilde h_{uv} D q^u \wg * D q ^v  - V_{IIB}*{\bf 1}\nn \\
  & & \qquad + \frac{1}{2}\, \IM \cM_{AB} F^A\wg * F^B
  + \frac{1}{2} \, \RE \cM_{AB} F^A \wg F^B \Big] \ ,
\end{eqnarray}
where the scalar potential has the form
\begin{eqnarray}
  \label{cBhf34}
  V_{IIB} =  \frac{\moo}{4} \rme^{+ 2 \phi}
  e_I e_J \left(\IM \cN^{-1} \right)^{IJ} - \frac{\moo}{2} \rme^{4 \phi}(e_I
  \xi^I)^2 .
\end{eqnarray} The non-trivial covariant derivatives are
\begin{equation}
  \label{cBhf30}
  D\tilde\xi_I =  d \tilde\xi_I-e_IV^0 \ ; \qquad
  Da =  d  a + e_IV^0\xi^I \ ,
\end{equation}
while all the other fields remain neutral.

This ends the derivation of the effective action of type IIB
theory compactified to four dimensions on half-flat manifolds.
One can immediately notice that the gaugings \eqref{cBhf30} are
precisely the same as in the case of type IIA theory
\eqref{cAns07} and \eqref{cAns18} when all the magnetic fluxes
$p^A$ are set to zero. It is not difficult to see that in this
case also the potentials \eqref{cBhf34} and \eqref{cAns20}
coincide. For this one should just note that under mirror
symmetry $\moo=(\IM\cM_B^{-1})^{00}$ is mapped to
$-\frac{1}{\cK_A}$, $\cK_A$ being the volume of the \CY manifold
on which type IIA is compactified.


\section{Conclusions}
\label{conc}\setcounter{equation}{0}

In this chapter we proposed that type IIB (respectively IIA)
compactified on a Calabi-Yau threefold $\tilde{Y}$ with electric
NS three-form flux
is mirror symmetric to type IIA (respectively IIB) compactified
on a half-flat manifold $\hat{Y}$ with $SU(3)$ structure. The
manifold $\hat{Y}$ is neither complex nor is it Ricci-flat.
Nonetheless, though topologically distinct, it is closely related
to the ordinary Calabi--Yau mirror partner $Y$ of the original
threefold $\tilde{Y}$. In particular, we argued that the moduli
space of half-flat metrics on $\hat{Y}$ must be the same as the
moduli space of Calabi--Yau metrics on $Y$. Furthermore, it is
the topology of $\hat{Y}$ that encodes the even-dimensional
NS-flux mirror to the original $H_3$-flux on $\tilde{Y}$.

We further strengthened this proposal by deriving the low-energy
type IIA (IIB) effective action in the supergravity limit
and showing that it is exactly equivalent to the appropriate type IIB
(IIA) effective action. In particular, the resulting potential delicately
depends on the non-vanishing Ricci scalar of the half-flat geometry
and thus provided a highly non-trivial check on our proposal.

It is interesting to note that one particular NS flux $e_0$ played a
special role in that it did not arise from the half-flat geometry but
 appeared as a NS three-form flux $H_3\in
H^{(3,0)}(\tilde{Y})$. In this context, it appears that mirror
symmetry only acts on the `interior' of the Hodge diamond in that it
exchanges $H^{(1,1)} \leftrightarrow H^{(1,2)}$ but
leaves $H^{(3,3)}\oplus H^{(0,0)}$ and $H^{(3,0)}\oplus H^{(0,3)}$
untouched. Put another way, it appears that it is the same single NS
electric flux which is associated to both  $H^{(3,0)}\oplus
H^{(0,3)}$ and $H^{(3,3)}\oplus H^{(0,0)}$ on a given Calabi--Yau
manifold.

We found that requirements of mirror symmetry provided a number of
conjectures about the geometry of the half-flat manifold
$\hat{Y}$. For instance the cohomology groups of $\hat Y$ shrink
compared to those of $Y$ in that the Hodge numbers $h^{(1,1)}$ and
$h^{(1,2)}$ are reduced by one.
In addition, a non-standard KK reduction had to be performed in order
to obtain masses for some of the scalar fields. This in turn led us to
make a number of assumptions which need
to be better understood from a mathematical point of view.
One particular conjecture is the following. In general, the electric
NS $H_3$-flux maps under mirror symmetry to some element $\zeta\in
H^4(Y)$. Mirror symmetry would appear to imply that
\begin{quote}
   for all integer fluxes $\zeta\in H^4(Y)$ there should be
   a unique manifold $\hat{Y}_\zeta$ admitting a family of half-flat
   metrics such that the moduli space of such metrics $\mathcal{M}(Y_\zeta)$
   is equal to the moduli space $\mathcal{M}(Y)$ of Calabi--Yau metrics on $Y$.
\end{quote}
We note that it should be possible to determine this moduli space
of half-flat geometries directly from its definition and without
relying on the physical relation with Calabi-Yau threefold
compactification.\footnote{In this respect a generalization of
ref.~\cite{H2} might be useful.} Moreover, a more precise
mathematical statement about the relationship between a given
Calabi-Yau threefold $Y$ and its `cousin' half-flat geometry on
$\hat{Y}$ should also be possible.

Finally, our analysis only treated electric NS fluxes. The
discussion of the magnetic ones is technically more involved.
Indeed, when type IIB is compactified on a Calabi-Yau manifold with magnetic NS
form fluxes, a massive RR two-form appears which has no obvious counterpart
on the type IIA side. In the second approach where type IIA is
compactified on a Calabi-Yau manifold with magnetic NS three-form fluxes, no
 massive forms are present. Thus, it appears that in this picture
it would be easier to look for the magnetic fluxes. However, in this case we
encounter an other puzzle. Recall that when type IIA is compactified on a
\CY 3-fold with NS electric and magnetic fluxes (\ref{cAns07}), all the scalars
$\xi^A$ and $\txi_A$ are gauged, with respect to the same vector $A^0$. The
corresponding vector in type IIB is $V^0$ which only appears in the expansion of
the self-dual field strength $\hat F_5$. With a quick look at the action (\ref{cB1})
one can immediately see that the scalars $\xi^I$, that are $l$ and $lb^i-c^i$, whose
kinetic term do not involve any $\hat A_4$ or $\hat F_5$,  have no reason to be gauged
under $V^0$. Thus it appears that some additional important ingredient still has to
be found in order to reconcile electric and magnetic fluxes.

Nevertheless, we think that a few lines can be drawn. Recall that in the
definition of the NS fluxes

\beq
H_3 \sim m^A\ax_A -e_A\bx^A,
\eeq the electric and magnetic fluxes are treated on the same footing; they are
coefficients of an expansion on a basis $(\ax_A\, ,\,\bx^B)$ of harmonic forms on $H^3$.
In type IIB, when only electric fluxes are turned on, the spectrum contains the usual
massless RR 2-form $C_2$. However, when only magnetic fluxes are present, $C_2$ becomes
massive, with a mass proportional to the fluxes (\ref{cBNS24}). Since the notion of
"electric" or "magnetic" is just a matter of choice of basis, how is it possible that
magnetic and electric fluxes lead to so different results? Actually the difference only
lies in the way one distributes the degrees of freedom. In \cite{LM2}, it was showed in a
similar situation\footnote{The massive form was the NS 2-form $B_2$ and the fluxes were
coming from the RR sector, but the results can be easily extended to our case.} that the
massive 2-form is dual to a scalar and a massive vector. The scalar has the exact same
couplings as would have the dual of the massless 2-form, and with the emergence of the
extra vector comes a symmetry which allows to eliminate one vector. Thus the number of
physical degrees of freedom is indeed unchanged. Following this idea, on can note that
$C_2$ becomes massive when magnetic fluxes are present only because, in the process of
discarding half of the fields due to self-duality of $\hat F_5$, the independent vectors
in the expansion of $\hat A_4$ are chosen to be the $V^A$. If one chooses instead to keep
the $U_A$, then $C_2$ is no longer massive, and one can check that the magnetic fluxes are
mapped correctly, provided the rules for mapping are slightly modified.

The problem arises when both kinds of fluxes are turned on. If one keeps $V^A$, then $C_2$
acquires a mass proportional to $m^A$, but if one keeps $U_A$, $C_2$ is also massive with
a mass proportional to $e_A$. This means that the obstruction to having both fluxes at
the same time seems to be related to a matter of dualization. Since the electric fluxes
parameterize the failure of $\Ox^+$ to be closed (\ref{F22}), the magnetic fluxes may
naturally be involved in its failure to be co-closed. One may then want to impose

\beq
d^{\dagger}\Ox^+ = m^i\ox_i.
\eeq Recall that $\Ox^+$ and $\Ox^-$ are related by $*\Ox^+\sim \Ox^-$. This would
suggest that magnetic fluxes require a further generalization of the half-flat geometry
allowing the possibility $d\Ox^-\neq 0$.

\chapter{Equations of motion for Nicolai-Townsend multiplet}

\section{Introduction}

The action for $N=4$ supergravity theory containing an antisymmetric tensor was first
given by Nicolai and Townsend \cite{NT81} already in the early eighties. It was derived
from the Lagrangian of \cite{CSF78} after a standard dualization of the pseudo-scalar.
These theories are best understood as coming from compactification of $N=1$ $d=10$ pure
supergravity on a six-dimensional torus $T^6$ \cite{Cha81a}. A consistent truncation of
type IIA supergravity in 10 dimensions is realized by turning off the RR fields $A_1$ and
$C_3$ in the bosonic sector, and by keeping only half of the fermions, one gravitino and
one dilatino \cite{S78,Cha81a}. The resulting action describes the dynamics of the $N=1$
gravity multiplet, composed of the metric $\hat g_{MN}$, the dilaton $\hat\phi$ and the
antisymmetric NS tensor $\hat B_{MN}$ as bosonic fields, one gravitino and one dilatino as
fermionic fields. With a simple counting of the bosonic degrees of freedom, one can deduce
how the reduced fields arrange in $N=4$ multiplets. The reduction of the metric leads to
the metric in 4 dimensions $g_{\mu\nu}$, 6 vectors $A_{\mu m}$ and 21 scalars $A_{mn}$.
The antisymmetric tensor gives one antisymmetric tensor $B_{\mu\nu}$, 6 vectors $B_{\mu
m}$ and 15 scalars $B_{mn}$. In 4 dimensions, the $N=4$ gravity multiplet contains the
metric, 6 vectors, one scalar and one pseudo-scalar dual to an antisymmetric tensor.
Substracting these degrees of freedom from the full spectrum, one is left with 6 vectors
and 36 scalars, which lie in 6 vector multiplets\footnote{A $N=4$ vector multiplet
contains one vector and 6 scalars as bosonic fields.}.

More precisely, apart from the metric, the fields belonging to
the gravity multiplet are the following : 6 vectors
(graviphotons) $v_{\mu}{}^{\au}$, $\au = 1..6$, corresponding to
a particular linear combination\footnote{The other independent
linear combination of the vectors $A_{\mu m}$ and $B_{\mu m}$
corresponds to the vectors, which, together with the 36 scalars
$g_{mn}$ and $B_{mn}$, form the 6 vector multiplets.} of $A_{\mu
m}$ and $B_{\mu m}$, the antisymmetric tensor and the dilaton.
The superspace formulation of this multiplet, which we call the
N-T multiplet in the following, encountered a number of problems
identified in \cite{Gat83} and overcome in \cite{GD89} by
introducing external Chern-Simons forms for the graviphotons.
Recently, a concise geometric formulation was given for this
supergravity theory in central charge superspace \cite{GHK01}.

The geometric approach adopted and described in detail in \cite{GHK01} was based on
the superspace soldering mechanism involving gravity and 2--form geometries in
central charge superspace \cite{AGHH99}. This soldering procedure allowed to identify
various gauge component fields of the one and the same multiplet in two distinct
geometric structures: graviton, gravitini and graviphotons in the gravity sector and
the antisymmetric tensor in the 2--form sector. Supersymmetry and central charge
transformations of the component fields were deduced using the fact that in the
geometric approach these transformations are identified on the same footing with
general space-time coordinate transformations as superspace diffeomorphisms on the
central charge superspace. Moreover, the presence of graviphoton Chern-Simons forms
in the theory was interpreted as an intrinsic property of central charge superspace
and a consequence of the superspace soldering mechanism.

The aim of the present work is to emphasize that the geometric description in
\cite{GHK01} is on-shell, that is the constraints used to identify the component
fields of the N-T supergravity multiplet imply also the equations of motion for these
fields. Therefore, we begin with recalling briefly the basics of extended supergravity
in superspace formalism. Then we identify the component fields \cite{GHK01} and
specify the constraints we use. In section 3, we derive the equations of motion
directly from constraints and Bianchi identities, without any knowledge about a
Lagrangian. Finally, we compare these equations of motion with those found from the
component Lagrangian given in the original article by Nicolai and Townsend
\cite{NT81}.

\section{Extended supergravities in superspace}

In this section we briefly describe the geometry of extended superspace. We only give
the notions that will be relevant to the remainder of this chapter. Very detailed accounts
of superspace formalism can be found for example in \cite{WB83} \cite{BGG01} for $N=1$,
and in \cite{Kis00}\cite{Loy02} for extended supergravities. We use the exact same
conventions as in \cite{BGG01,Kis00}. The superspace is made of the space-time coordinates
$x^m$, with additional fermionic coordinates $\tet_\ta^\a,\tet^\ta_\da$ and bosonic
central charges coordinates $z^\au$, where the index $\ta$ counts the number of
supercharges and $\au$ the number of central charges. As usual, we define the vielbein
one-form

\bea E^\ca & = & d z^\cm E_\cm{}^\ca\label{vielbein1}\eea where
$z^\cm = \left( x^m,\tet_\ta^\a,\tet^\ta_\da,z^\au\right)$ is the
generalized coordinate on the superspace. The generic structure
group is $SL(2,C)\times U(N)$, with connection
$\Phi_{\ca}{}^{{\cal B}}$. The covariant derivatives act on
tensor fields in the following way

\bea Du^\ca & = & d u^\ca + u^{\cal B}\Phi_{\cal B}{}^\ca\label{algebra1}\\[2mm]
Dv_\ca & = & d v_\ca -(-)^{\mbox{deg}(v)}\Phi_\ca{}^{\cal B}
v_{\cal B}\label{algebra2}
  \eea
and lead to the algebra

\bea \left( D_\cc\, ,\, D_{\cal B}\right) u^\ca & = &
-T_{\cc{\cal B}}{}^\cf D_\cf u^\ca + R_{\cc{\cal B}\cf}
{}^\ca u^\cf\label{algebra3}\\[2mm]
\left( D_\cc\, ,\, D_{\cal B}\right) v_\ca & = & -T_{\cc{\cal
B}}{}^\cf D_\cf v_\ca - R_{\cc{\cal B}\ca} {}^\cf
v_\cf\label{algebra4}  \eea where the torsion and the Riemann
tensor are

 \bea T^\ca & = & \vd E^\ca + E^{\cal B}\Phi_{\cal B}{}^\ca\label{torsion1}\\[2mm]
 R_\ca{}^{\cal B} & = & \vd \Phi_\ca{}^{\cal B} +
\Phi_\ca{}^\cc\Phi_\cc{}^{\cal B}.\label{riemann1}\eea
 These tensors are subject to
consistency conditions expressed by the Bianchi Identities

\bea DT^\ca & = & E^{\cal B} R_{\cal B}{}^\ca\label{bianchi1}\\[2mm]
DR_{\cal B}{}^\ca & = & 0. \label{bianchi2} \eea In the case of
the superspace without central charges, Dragon's theorem
\cite{Dra79} states that using (\ref{bianchi1}), one can express
all the components of the Riemann tensor in terms of the
components of the torsion. Moreover, once (\ref{bianchi1}) is
solved, (\ref{bianchi2}) is identically satisfied. Under very
mild assumptions \cite{Kis00}\cite{Loy02}, this theorem can be
extended to the superspace with central charges. This is why in
our case we will only consider (\ref{bianchi1}). In components it
reads

\bea \left(_{\cd\cc{\cal B}}{}^\ca\right)_T\quad&:&\quad E^{\cal
B} E^\cc E^\cd \left(\cd_\cd T_{\cc{\cal
B}}{}^\ca+T_{\cd\cc}{}^\cf T_{\cf{\cal B}}{}^\ca-R_{\cd\cc{\cal
B}}{}^\ca\right) =0. \label{bianchi3}\eea

\section{Identification of the fields}

In this section we recall the essential results of \cite{GHK01} concerning the
identification of the components of the N-T multiplet. Recall that in geometrical
formulation of supergravity theories the basic dynamic variables are chosen to be the
vielbein and the connection. Considering central charge superspace this framework
provides a unified geometric identification of graviton, gravitini and graviphotons
in the frame $E^\ca=(E^a,E_\ta^\a,E^\ta_\da, E^\au)$.

 \be E^a\doubar\ =\ d x^{\mu}
e_{\mu}{}^a~,\quad E_\ta^\a\doubar\ =\ \frac{1}{2}\, d x^{\mu}
\psi_{\mu}{}_\ta^\a~, \quad E^\ta_\da\doubar\ =\ \frac{1}{2}\, d x^{\mu}
\bar{\psi}_{\mu}{}^\ta_\da~, \quad E^\au\doubar\ =\ d x^{\mu}
v_{\mu}{}^\au~, \eqn{frame} while the antisymmetric tensor can be
identified in a superspace 2--form $B$: \be B\doubar\ =\
\frac{1}{2}\, d x^{\mu} d x^{\nu} b_{\nu\mu}. \eqn{B} The remaining
component fields, a real scalar and 4 helicity 1/2 fields, are
identified in the supersymmetry transforms of the vielbein and
2--form, that is in torsion ($T^\ca=DE^\ca$) and 3--form ($H=dB$)
components. The Bianchi identities satisfied by these objects are
\be DT^\ca\ =\ E^{\cal B} R_{\cal B}{}^\ca\,,\qquad\qquad d H\ =\
0\,, \end{equation} and, displaying the 4--form coefficients,

\bea \left(_{\cd\cc{\cal B}\ca}\right)_H\quad&:&\quad E^\ca
E^{\cal B} E^\cc E^\cd \left(2\cd_\cd H_{\cc{\cal
B}\ca}+3T_{\cd\cc}{}^\cf H_{\cf{\cal B}\ca}\right)=0. \eea

By putting constraints on torsion and 3--form we have to solve
two problems at the same time: first, we have to reduce the huge
number of superfluous independent fields contained in these
geometrical objects, and second, we have to make sure that the
antisymmetric tensor takes part of the {\it same} multiplet as
$e_{\mu}{}^a$, $\psi_{\mu}{}_\ta^\a$, $\bar{\psi}_{\mu}{}^\ta_\da$,
$v_{\mu}{}^\au$ (soldering mechanism).

Indeed, the biggest problem in finding a geometrical description of an off-shell
supersymmetric theory is to find suitable covariant constraints which do reduce this
number but do not imply equations of motion for the remaining fields. There are
several approaches to this question. One of them is based on conventional
constraints, which resume to suitable redefinitions of the vielbein and connection
and which do not imply equations of motion \cite{GGMW84b}. However, such
redefinitions leave intact torsion components with 0 canonical dimension and there is
no general recipes to indicate how these torsion components have to be constrained. A
simpler manner of constraining 0 dimensional torsion components together with
conventional constraints give rise to the so-called {\it natural constraints}, which
were analyzed in a systematic way both in ordinary extended superspace \cite{Mul86b}
and in central charge superspace \cite{Kis00}.

The geometrical description of the N-T multiplet is based on a set of natural
constraints in central charge superspace with structure group
$SL(2,\mathbb{C})\otimes U(4)$. The generalizations of the canonical dimension 0
``trivial constraints" \cite{Mul86b} to central charge superspace are

\be
 T{^\tc_\g}{^\tb_\b}{^a_{}}\ =\ 0~, \qquad
 T{^\tc_\g}{^{\dot{\beta}}_\tb}{^a}\ =\ -2i\d{^\tc_\tb}(\s{^a}\eps){_\g}{^{\dot{\beta}}}~,
 \qquad
 T{^{\dot{\gamma}}_\tc}{^{\dot{\beta}}_\tb}{^a}\ =\ 0~,
\eqn{T01} \be T{^\tc_\g}{^\tb_\b}{^\au}\ =\
\eps_{\g\b}T^{[\tc\tb]\au}~,\qquad
T^\tc_\g{}^{\dot{\beta}}_\tb{}^\au\ =\ 0~, \qquad
T{^{\dot{\gamma}}_\tc}{^{\dot{\beta}}_\tb}{^\au}\ =\
\eps^{{\dot{\gamma}}{\dot{\beta}}}T{_{[\tc\tb]}}{}^\au~. \eqn{T02}

As explained in detail in the article \cite{GHK01}, the soldering
is achieved by requiring some analogous, ``mirror''-constraints
for the 2--form sector. Besides the -1/2 dimensional constraints
$H^\tc_\g{}^\tb_\b{}^\ta_\a = H^\tc_\g{}^\tb_\b{}^\da_\ta =
H^\tc_\g{}^{\dot{\beta}}_\tb{}^\da_\ta =
H^{\dot{\gamma}}_\tc{}^{\dot{\beta}}_\tb{}^\da_\ta=0$, we impose
\be
 H{^\tc_\g}{^\tb_\b}{_a}\ =\ 0~, \qquad
 H{^\tc_\g}{^{\dot{\beta}}_\tb}{_a}\ =\
 -2i\d{^\tc_\tb}(\s{_a}\eps){_\g}{^{\dot{\beta}}}L~, \qquad
 H{^{\dot{\gamma}}_\tc}{^{\dot{\beta}}_\tb}{_a}\ =\ 0~,
\eqn{H01} \be
H{^\tb_\b}{^\ta_\a}_\au=\eps_{\b\a}H_\au{}^{[\tb\ta]} \,,\qquad
H^\tc_\g{}^{\dot{\beta}}_\tb{}_\au =0\,,\qquad
H^{\dot{\beta}}_\tb{}^\da_\ta{}_\au=\eps^{{\dot{\beta}}\da}H_{\au[\tb\ta]}\,,
\eqn{H02} with $L$ a real superfield. The physical scalar $\f$ of
the multiplet, called also graviscalar, is identified in this
superfield, parameterized as $L=e^{2\f}$. In turn, the helicity
1/2 fields, called also gravigini fields, are identified as usual
\cite{How82}, \cite{GG83}, \cite{Gat83} in the 1/2--dimensional
torsion component \be \eps^{\b\g}T^\tc_\g{}^\tb_\b{}^\ta_\da\ =\
2T^{[\tc\tb\ta]}{}_\da,\qquad
\eps_{{\dot{\beta}}{\dot{\gamma}}}T_\tc^{\dot{\gamma}}{}_\tb^{\dot{\beta}}{}_\ta^\a\
=\ 2T_{[\tc\tb\ta]}{}^\a. \end{equation}

The scalar, the four helicity 1/2 fields, together with the
gauge--fields defined in \equ{frame} and \equ{B} constitute the
N-T on-shell $N=4$ supergravity multiplet. However, the 0
dimensional natural constraints listed above are not sufficient to
insure that these are the {\it only} fields transforming into
each-other by supergravity transformations. The elimination of a
big number of superfluous fields is achieved by assuming the
constraints \be \cd^{\td\a} T_{[\tc\tb\ta]\a}\ =\
0\,,\qquad\qquad \cd_{\td\da} T^{[\tc\tb\ta]\da}\ =\ 0\,,
\eqn{dt} and \be T_{\az{\cal B}}{}^\ca=0, \eqn{tz} as well as all
possible compatible conventional constraints\footnote{see
equations \equ{conv007} in the appendix} \cite{Mul86b},
\cite{Kis00}.

It is worthwhile to note that even at this stage the assumptions
are not sufficient to constrain the geometry to the N-T
multiplet. This setup allows to give a geometrical description at
least of the coupling of $N=4$ supergravity with antisymmetric
tensor to six copies of $N=4$ Yang-Mills multiplets\cite{Cha81a}.
Nevertheless, they are strong enough to put the underlying
multiplet on-shell. In order to see this, one can easily verify
that the dimension 1 Bianchi identities
$\left(^{\dot{\delta}}_\td{}^{\dot{\gamma}}_\tc{}^\tb_\b{}^\a_\ta\right)_T$
and
$\left(^{\dot{\delta}}_\td{}^{\dot{\gamma}}_\tc{}_\tb^{\dot{\beta}}{}_\da^\ta\right)_T$
for the torsion as well as their complex conjugates imply \be
\begin{array}{lcl} \cd^\td_{\d} T_{[\tc\tb\ta]\a}\ =\ -i
\d_{\tc\tb\ta}^{\td\te\tf}G_{(\d\a)[\te\tf]}\,,&\qquad&
\cd_\td^{{\dot{\delta}}} T^{[\tc\tb\ta]\da}\ =\ -i
 \d^{\tc\tb\ta}_{\td\te\tf}G^{({\dot{\delta}}\da)[\te\tf]}\,,\\[4mm]
\cd_\td^{\dot{\delta}} T_{[\tc\tb\ta]\a}\ =\
P_\a{}^{\dot{\delta}}{}_{[\td\tc\tb\ta]}\,,&& \cd^\td_\d
T^{[\tc\tb\ta]\da}\ =\ P_\d{}^\da{}^{[\td\tc\tb\ta]}\,,
\end{array} \label{dim1}
\end{equation} with $G$ and $P$ a priori some arbitrary
superfields. Let us write one of the last relations as \be
\sum_{\td\tc}\cd_\td^{\dot{\delta}} T_{[\tc\tb\ta]\a}\ =\ 0,
\end{equation} take its spinorial derivative
$\cd^\te_\varepsilon$ \be
\sum_{\td\tc}\left(\left\{\cd^\te_\varepsilon,\cd_\td^{\dot{\delta}}\right\}
T_{[\tc\tb\ta]\a} -\cd_\td^{\dot{\delta}}\left(\cd^\te_\varepsilon
T_{[\tc\tb\ta]\a}\right)\right)\ =\ 0,
\label{dirac1}\end{equation} and observe that the antisymmetric
part of this relation in the indices $\varepsilon$ and $\a$ gives
rise to Dirac equation for the helicity 1/2 fields, that is
$\partial^{\a{\dot{\delta}}}T_{[\tc\tb\ta]\a}=0$ in the linear
approach.

It turns out, that there is a simple solution of both the Bianchi identities of the
torsion and 3--form, which satisfies the above mentioned constraints and reproduce
the N-T multiplet. The non-zero torsion and 3--form components for this solution are
listed in the appendix, we will concentrate here on its properties which are
essential for the identification of the multiplet and the derivation of the equations
of motion for the component fields.

Recall that the particularity of this solution is based on the
identification of the scalar superfield $\f$ in the 0 dimensional
torsion and 3--form components containing a central charge index
\be T^{[\tb\ta]\au}\ =\ 4e^\f{\mathfrak{t}}^{[\tb\ta]\au}\,,
\qquad T_{[\tb\ta]}{}^\au\ =\
4e^\f{\mathfrak{t}}_{[\tb\ta]}{}^\au\,,
\end{equation} \be H_\au{}^{[\tb\ta]}\ =\
4e^\f\hh_\au{}^{[\tb\ta]}\,, \qquad H_{\au[\tb\ta]}\ =\
4e^\f\hh_{\au[\tb\ta]}\,, \end{equation} with
${\mathfrak{t}}^{[\tc\tb]\au}$, ${\mathfrak{t}}_{[\tc\tb]}{}^\au$,
$\hh_\au{}^{[\tb\ta]}$, $\hh_{\au[\tb\ta]}$ constant matrix
elements satisfying the self--duality relations \be
{\mathfrak{t}}^{[\td\tc]\au} \ =\  \frac{q}{4}\,
\varepsilon^{\td\tc\tb\ta}{\mathfrak{t}}_{[\tb\ta]}{}^\au\,,\qquad\qquad
\hh_\au{}^{[\tb\ta]} \ =\
\frac{q}{2}\,\hh_{\au[\td\tc]}\varepsilon^{\td\tc\tb\ta}\qquad\textrm{with}\quad
q=\pm1. \eqn{du0}

Note, that these relations look similar to {\it some} of the properties of the 6
real, antisymmetric $4\times 4$ matrices $\a^n$, $\b^n$, $(n=1,2,3)$ of $SU(2)\otimes
SU(2)$ 
\cite{CSF78}, \cite{FS78}, which appear in the component
formulation of $N=4$ supergravity theories. Indeed, if we define
the matrices \be {\mathfrak{t}}\doteq\left(
\begin{array}{c}{\mathfrak{t}}^{[\td\tc]}{}^\au\\
{\mathfrak{t}}_{[\td\tc]}{}^\au\end{array}\right)\,, \qquad
\hh\doteq\left(
\begin{array}{cc}\hh_\au{}_{[\td\tc]}&
\hh_\au{}^{[\td\tc]}\end{array}\right)\qquad\textrm{and}
\end{equation} \be \S\doteq\left(
\begin{array}{cc}0&\frac{q}{2}\varepsilon^{\td\tc\tb\ta}\\
\frac{q}{2}\varepsilon_{\td\tc\tb\ta}&0\end{array}\right)\,,
\qquad \mathbf{1}\doteq\left(
\begin{array}{cc}\frac{1}{0}\d^{\td\tc}_{\tb\ta}&0\\
0&\frac{1}{2}\d_{\td\tc}^{\tb\ta}\end{array}\right)\qquad\textrm{satisfying}
\qquad \S^2=\mathbf{1}\,, \end{equation} then the properties of
the matrix elements ${\mathfrak{t}}^{[\tc\tb]\au}$,
${\mathfrak{t}}_{[\tc\tb]}{}^\au$, $\hh_\au{}^{[\tb\ta]}$,
$\hh_{\au[\tb\ta]}$ can be resumed in a compact way as follows:
\be \S{\mathfrak{t}}\ =\ {\mathfrak{t}}\,,\qquad\qquad\hh\S\ =\
\hh\,, \eqn{du} \be {\mathfrak{t}}\hh\ =\
\mathbf{1}+\S\,,\qquad\qquad(\hh{\mathfrak{t}})_\au{}^\av\ =\
2\d^\av_\au\,. \eqn{rel_ht}

Recall, however that we didn't fix a priori the number of the central charge
coordinates in the superspace. The interesting feature of the above properties is
that taking the trace of relations in \equ{rel_ht} one finds $\d^\au_\au=6$, that is
the number of central charge indices - and thus, the number of the vector
gauge--fields $v_m{}^\au$ - is determined to be 6.

These matrices serve as converters between the central charge basis (indices \au) and
the $SU(4)$ basis in the antisymmetric representation (indices $[\td\tc]$). In
particular, for the 6 vector gauge fields $v_m{}^\au$ of the N-T multiplet there is
an alternative basis, called the $SU(4)$ basis, defined by \be \left(
\begin{array}{cc}V_{\mu}{}_{[\td\tc]}&V_{\mu}{}^{[\td\tc]}\end{array} \right)\ \doteq\
v_{\mu}{}^\au\left(
\begin{array}{cc}\hh_\au{}_{[\td\tc]}&\hh_\au{}^{[\td\tc]}\end{array}
\right)\,, \end{equation} where the two components are connected
by the self--duality relations

\be V_{\mu}{}^{[\td\tc]} \ =\  \frac{q}{2}\,
\varepsilon^{\td\tc\tb\ta}V_{\mu}{}_{[\tb\ta]}. \end{equation}

Moreover, if we look at self--duality properties \equ{du0} as the
lifting and lowering of $SU(4)$ indices with metric
$\frac{q}{2}\varepsilon_{\td\tc\tb\ta}$, then a corresponding
metric in the central charge basis can be defined by \be
\gg_{\av\au} \ =\
\frac{q}{2}\,\varepsilon_{\td\tc\tb\ta}\,\hh_\av{}^{[\td\tc]}\,\hh_\au{}^{[\tb\ta]},
\qquad\quad \gg^{\av\au}\ =\
\frac{q}{2}\,\varepsilon_{\td\tc\tb\ta}\,{\mathfrak{t}}^{[\td\tc]}{}^\av\,{\mathfrak{t}}^{
[\tb\ta]}{}^\au,
\eqn{metric} satisfying \be \gg_{\au\aw}\gg^{\aw\av} \ = \
\delta_\au^\av~. \end{equation} These are the objects which are
found to connect torsion and 3--form components containing at
least one central charge index \be H_{\cd\cc\au}\ =\
T_{\cd\cc}{}^\az\gg_{\az\au},\qquad \quad T_{\cd\cc}{}^\au\ =\
H_{\cd\cc\az} \gg^{\az\au}, \end{equation} insuring the soldering
of the two geometries.

The four helicity 1/2 fields $T_{[\tc\tb\ta]\a}$,
$T^{[\tc\tb\ta]\da}$ turn out to be equivalent to the fermionic
partner of the graviscalar $\f$ \be \la^\ta_\a\ =
2\cd^\ta_\a\f\,,\qquad\qquad\bla_\ta^\da\ = 2\cd_\ta^\da\f\,,
\end{equation} since the following duality relation holds in this
$N=4$ case: \be T_{[\tc\tb\ta]\a}\ =\
q\varepsilon_{\tc\tb\ta\tf}\bla_\a^\tf\, \qquad\qquad
T^{[\tc\tb\ta]\da}\ =\ q\varepsilon^{\tc\tb\ta\tf}\la^\da_\tf\,.
\eqn{Tla}

It is the soldering mechanism between the geometry of supergravity and the geometry
of the 2--form, that determines how the superfields $G$ and $P$ in the spinorial
derivatives of this helicity 1/2 fields \equ{dim1} are related to the component
fields of the multiplet. In particular, we find that the superfields $G$ are related
to the covariant field strength of the graviphotons $F_{ba}{}^\au$ \be
G_{(\b\a)}{}_{[\tb\ta]}\ = \
-2ie^{-\f}F_{(\b\a)}{}^\au\hh_\au{}_{[\tb\ta]}\,,\label{G-F}
\qquad\qquad
G^{({\dot{\beta}}\da)}{}^{[\tb\ta]}\ =\
 -2ie^{-\f}F^{({\dot{\beta}}\da)}{}^\au\hh_\au{}^{[\tb\ta]}\,,
\end{equation} whereas the superfields $P$ contain the dual field strength
of the antisymmetric tensor and the derivative of the scalar:
\bea \cd^\td_\d
T^{[\tc\tb\ta]\da}&=&q\varepsilon^{\td\tc\tb\ta}P_\d{}^\da\,,\
\textrm{ with }\
P_a\ =\ 2i\cd_a\f+e^{-2\f}H^*_a-\frac{3}{4}\la^\ta\s_a\bla_\ta\,,\label{P}\\
\cd_\td^\da
T_{[\tc\tb\ta]\d}&=&q\varepsilon_{\td\tc\tb\ta}\bP_\d{}^\da\,,\
\textrm{ with }\ \bP_a\ =\
2i\cd_a\f-e^{-2\f}H^*_a+\frac{3}{4}\la^\ta\s_a\bla_\ta\,,\label{barP}
\eea where we can note that the relations \be P_a+\bP_a\ =\
4i\cd_a\f\,,\qquad\qquad P_a-\bP_a\ =\
2e^{-2\f}H^*_a-\frac{3}{2}\la^\ta\s_a\bla_\ta \eqn{PP} allow to
separate the dual field strength of the antisymmetric tensor and
the derivative of the scalar (as "real" and "imaginary" part of
$P$).

Finally, let us precise that the representation of the structure
group in the central charge sector is trivial, $\F_\au{}^\az=0$,
while the $U(4)$ part $\F^\tb{}_\ta$ of the
$SL(2,\mathbb{C})\otimes U(4)$ connection \be \F^\tb_\b{}^\a_\ta\
=\ \d^\tb_\ta\F_\b{}^\a+\d_\b^\a\F^\tb{}_\ta
\qquad\quad\F_\tb^{\dot{\beta}}{}_\da^\ta\ =\
\d_\tb^\ta\F^{\dot{\beta}}{}_\da-\d^{\dot{\beta}}_\da\F^\ta{}_\tb
\end{equation} is determined to be \be \F^\tb{}_\ta\ =\
a^\tb{}_\ta\ +\ \ki^\tb{}_\ta\,, \end{equation} with
$a^\tb{}_\ta$ pure gauge and $\ki^\tb{}_\ta$ a supercovariant
1--form on the superspace with components \bea
\ki_c{}^\tb{}_\ta&=&\frac{1}{4}\d^\tb_\ta\left(ie^{-2\f}H^*_c-
\frac{i}{4}\la^\tf\s_c
\bar{\la}_{\tf}\right)
-\frac{i}{8}(\la^\tb\s_c\bar{\la}_{\ta})\,,\nn\\[2mm]
\ki^\tc_\g{}^\tb{}_\ta&=&\frac{1}{4}\d^\tb_\ta\la^\tc_\g\,,\qquad\qquad
\ki_\tc^{\dot{\gamma}}{}^\tb{}_\ta\ =\
-\frac{1}{4}\d^\tb_\ta\bar{\la}_\tc^{\dot{\gamma}}\,,\qquad\qquad
\ki_\au{}^\tb{}_\ta\ =\ 0\,. \label{connect} \eea This situation
is analogous to the case of the 16+16 $N=1$ supergravity
multiplet which is obtained from the reducible 20+20 multiplet,
described on superspace with structure group
$SL(2,\mathbb{C})\otimes U(1)$, by ``breaking'' the $U(1)$
symmetry \cite{Mul86c}. By eliminating this $U(4)$ part from the
$SL(2,\mathbb{C})\otimes U(4)$ connection and putting the pure
gauge part $a$ to zero, one can define covariant derivatives for
$SL(2,\mathbb{C})$ \bea \hat{D}u^\ca\ =\ Du^\ca-\ki_{\cal
B}{}^\ca u^{\cal B}\qquad\quad \hat{D}u_\ca\ =\
Du_\ca+\ki_\ca{}^{\cal B} u_{\cal B} \eea used in the articles
\cite{GHK01} and \cite{NT81}. Here of course $\ki_{\cal B}{}^\ca$
is defined in such a way that its only non-zero components are
$\ki^\tb_\b{}^\a_\ta=\d_\b^\a\ki^\tb{}_\ta$ and
$\ki_\tb^{\dot{\beta}}{}_\da^\ta=-\d^{\dot{\beta}}_\da\ki^\ta{}_\tb$.
Recall that this redefinition of the connection affects torsion
and curvature components in the following way: \bea
\hat{T}_{\cc{\cal B}}{}^\ca & =& T_{\cc{\cal B}}{}^\ca-\ki_{\cc{\cal
 B}}{}^\ca+(-)^{cb}\ki_{{\cal B}\cc}{}^\ca,\\
\hat{R}_\cd{}_\cc{}^\tb{}_\ta &=&0\,. \eea

In the next section we derive the equations of motion for all the component fields of
the N-T multiplet using its geometrical description presented above.

\section{Equations of motion in terms of supercovariant quantities}

The problem of the derivation of field equations of motion without the knowledge of a
Lagrangian, using considerations on representations of the symmetry group, was
considered a long time ago \cite{Bel74}, \cite{Wei95}. The question is particularly
interesting for supersymmetric theories and there are various approaches which have
been developed. Let us mention for example the procedure based on projection
operators selecting irreducible representations out of superfield with arbitrary
external spin \cite{OS77}. About the same period Wess and Zumino suggested the use of
differential geometry in superspace to reach better understanding of supersymmetric
Yang-Mills and supergravity theories. The techniques used in this approach allowed to
work out a new method for deriving equations of motion, namely looking at consequences
of covariant constraints, which correspond to on-shell field content of a
representation of the supersymmetry algebra.

In order to illustrate the method let us recall as briefly as
possible the simplest example, the $N=1$ Yang-Mills theory
described on superspace considering the geometry of a Lie algebra
valued 1--form $\ca$ \cite{Soh78a}, \cite{BGG01}. Under a gauge
transformation, parameterized by $g$, the gauge potential
transforms as $\ca\mapsto g^{-1}\ca g-g^{-1}d g$ and its field
strength $\cf= d\ca+\ca\ca$ satisfies the Bianchi identity
$D\cf=0$. In order to describe the on-shell multiplet one
constrains the geometry by putting $\cf_{\a\b} =
\cf_\a{}^{\dot{\beta}} = \cf^{\da{\dot{\beta}}} = 0$. Then the
Bianchi identities are satisfied if and only if all the
components of the field strength $\cf$ can be expressed in terms
of two spinor superfields $\cw_\a$, $\bar{\cw}^\da$ and their
spinor derivatives: \be \cf_{\b a}=i(\s_a\bar{\cw})_\b\,,\qquad
\cf^{\dot{\beta}}{}_a=-i(\bs_a\cw)^{\dot{\beta}}\,,
\end{equation} \be
\cf_{(\b\a)}=-\frac{1}{2}\cd_{(\b}\cw_{\a)}\,,\qquad
\cf^{({\dot{\beta}}\da)}=\frac{1}{2}\cd^{({\dot{\beta}}}\bar{\cw}^{\da)}\,,
\end{equation} and the gaugino superfields $\cw_\a$,
$\bar{\cw}^\da$ satisfy \be \cd_\a\bar{\cw}^\da\ =\ 0\,,\qquad
\cd^\da\cw_\a\ =\ 0\,, \eqn{w1} \be \cd^\a\cw_\a\ =\
\cd_\da\bar{\cw}^\da\,. \end{equation} The components of the
multiplet are thus identified as follows: the vector gauge field
in the super 1--form $\ca\doubar=i\vd x^ma_m$, the gaugino
component field as lowest component of the gaugino superfield
$\cw_\a\loco=-i\la_\a$, $\bar{\cw}^\da\loco=i\la^\da$, and the
auxiliary field in their derivatives
$\cd^\a\cw_\a\loco=\cd_\da\bar{\cw}^\da\loco=-2D$.

Note that the supplementary constraint \be \cd^\a\cw_\a\ =\ \cd_\da\bar{\cw}^\da\ =\ 0
\eqn{w2}
 puts this multiplet on-shell. It is a superfield equation and contains all the component field equations of motion. First of all it eliminates the auxiliary field $D$ and we can derive the equations of motion for the remaining fields by successively differentiating it. We obtain the Dirac equation for the  gaugino
\bea
\cd^\da\left(\cd^\a\cw_\a\right)&=& -2i\cd^{\a\da}\cw_\a\ =\ 0\,,\\[2mm]
\cd_\a\left(\cd_\da\bar{\cw}^\da\right)&=& -2i\cd_{\a\da}\bar{\cw}^\da\ =\ 0\,, \eea
and from this we derive the relations \bea
\cd_\b\left(\cd^{\a\da}\cw_\a\right)&=&-2\cd^{\a\da}\cf_{(\b\a)}
+2i\{\cw_\b,\bar{\cw}^\da\}\ =\ 0\,,\\[2mm]
\cd^{\dot{\beta}}\left(\cd_{\a\da}\bar{\cw}^\da\right)&=&2\cd_{\a\da}\cf^{({\dot{\beta}}\da)}
-2i\{\bar{\cw}^{\dot{\beta}},\cw_\a\}\ =\ 0\,, \eea which
correspond to the well-known Bianchi identities
$\cd_{\a{\dot{\beta}}}\cf^{({\dot{\beta}}\da)} -
\cd^{\b\da}\cf_{(\b\a)} = 0$ and equations of motion
$\cd_{\a{\dot{\beta}}}\cf^{({\dot{\beta}}\da)} +
\cd^{\b\da}\cf_{(\b\a)} = 2i\{\cw_\a,\bar{\cw}^{\dot{\beta}}\}$
for the vector gauge field.

The case of supergravity is similar to this, the gravigino superfields
$T_{[\tc\tb\ta]\a}$, $T^{[\tc\tb\ta]\da}$ (or $\la_\a^\ta$, $\bla^\da_\ta$ in
\equ{Tla}) play an analogous r\^ole to the gaugino superfields $\cw_\a$,
$\bar{\cw}^\da$. In order to derive the free equations of motion of component fields
in a supergravity theory it is sufficient to consider only the linearized version
\cite{HL81}, \cite{Sie81b}, \cite{GG83} and the calculations are simple. Considering
the full theory one obtains all the nonlinear terms which arise in equations of
motion derived from a Lagrangian in component formalism.

Recall that the dimension 1 Bianchi identities in the
supergravity sector imply the relations \equ{dim1} for the spinor
derivatives of the gravigino superfields. These properties can be
written equivalently as \be \begin{array}{lcl}
\sum_{\td\tc}\cd_\td^{\dot{\delta}} T_{[\tc\tb\ta]\a}\ =\ 0\,,&&
\sum^{\td\tc}\cd^\td_\d T^{[\tc\tb\ta]\da}\ =\ 0\,,
\\[2mm]
\cd^{\td\a} T_{[\tc\tb\ta]\a}\ =\ 0\,,&& \cd_{\td\da} T^{[\tc\tb\ta]\da}\ =\ 0\,,
\\[2mm]
\cd^\td_{(\d}T_{\a)[\tc\tb\ta]}
-\frac{1}{4}\d^{\td\te\tf}_{\tc\tb\ta}\cd^\tg_{(\d}T_{\a)[\tg\te\tf]}\
=\ 0\,, && \cd_\td^{({\dot{\delta}}}T^{\da)[\tc\tb\ta]}
-\frac{1}{4}\d_{\td\te\tf}^{\tc\tb\ta}\cd_\tg^{({\dot{\delta}}}T^{\da)[\tg\te\tf]}\
=\ 0\,,
\end{array} \end{equation} and they are the $N=4$ analogues of the relations \equ{w1} and
\equ{w2} satisfied by the gaugino superfield corresponding to the on-shell Yang-Mills
multiplet.

Therefore, by analogy to the Yang--Mills case, the equations of motion for the
gravigini, the graviphoton, the scalar and the antisymmetric tensor can be deduced
from the superfield relations \equ{dim1} by taking successive covariant spinorial
derivatives. Let us take the example of Dirac's equation, to clarify ideas. We start
from \equ{dirac1}. The anticommutator can be expressed using \equ{algebra3} and
\equ{algebra4}

\bea \{\cd_\epsilon^\te\, ,\, \cd_\td^{\dot{\delta}}
\}T_{[\tc\tb\ta]\a} & = & -T_{\epsilon\td}^{\te{\dot{\delta}}
f}\cd_f T_{[\tc\tb\ta]\a} -
R_{\epsilon\td\a}^{\te{\dot{\delta}}\;\;\b}T_{[\tc\tb\ta]\b}\label{dirac2}\eea
where we have used the results displayed in the appendix
\ref{torcurv} for the torsion and curvature components, which
make it possible to evaluate this expression explicitly. When we
take the antisymmetric part in $(\a\, , \,\epsilon)$, the second
term in \equ{dirac1} drops out because of the symmetry of
$G_{(\d\a)[\te\td]}$ in \equ{dim1}. Summing on $\td$ and $\te$,
we obtain \equ{Dirac2} below.

 Consider now all possible spinorial derivatives of relations \equ{dim1}. They are
satisfied if and only if in addition to the dimension 1 results
the following relations hold: \bea \cd_\g^\tc G_{(\b\a)[\tb\ta]}&
=& \frac{1}{3}\d^{\tc\tf}_{\tb\ta}\left[
\frac{1}{3}\oint_{\g\b\a}\cd_\g^\te G_{(\b\a)[\te\tf]}
+\frac{i}{2}\sum_{\b\a}\eps_{\g\b}\la_{\da\tf} \bP_\a{}^\da\right]\\
\cd^{\dot{\gamma}}_\tc G^{({\dot{\beta}}\da)[\tb\ta]}& =&
\frac{1}{3}\d_{\tc\tf}^{\tb\ta}\left[
\frac{1}{3}\oint^{{\dot{\gamma}}{\dot{\beta}}\da}\cd^{\dot{\gamma}}_\te
G^{({\dot{\beta}}\da)[\te\tf]}
+\frac{i}{2}\sum^{{\dot{\beta}}\da}\eps^{{\dot{\gamma}}{\dot{\beta}}}\la^{\a\tf}
P_\a{}^\da\right] \eea \bea \cd^\tc_\d
G^{({\dot{\beta}}\da)[\tb\ta]}&=&\cd_\d{}^{({\dot{\beta}}}T^{\da)[\tc\tb\ta]}
+U^\tc_\d{}^{(\da}_\tf T^{{\dot{\beta}})[\tf\tb\ta]}
-U^\tb_\d{}^{(\da}_\tf T^{{\dot{\beta}})[\tf\ta\tc]}
-U^\ta_\d{}^{(\da}_\tf T^{{\dot{\beta}})[\tf\tc\tb]}\\
\cd_\tc^{\dot{\delta}}
G_{(\b\a)[\tb\ta]}&=&\cd_{(\b}{}^{{\dot{\delta}}}T_{\a)[\tc\tb\ta]}
-U^\tf_{(\a}{}^{{\dot{\delta}}}_\tc T_{\b)[\tf\tb\ta]}
+U^\tf_{(\a}{}^{{\dot{\delta}}}_\tb T_{\b)[\tf\ta\tc]}
+U^\tf_{(\a}{}^{{\dot{\delta}}}_\ta T_{\b)[\tf\tc\tb]} \eea and
\bea
\cd_{\b\da}T^{[\tc\tb\ta]\da}&=&\frac{3}{2}U_\b^\tf{}^\da_\tf T^{[\tc\tb\ta]}_\da\label{Dirac1}\\
\cd^{\a{\dot{\beta}}}T_{[\tc\tb\ta]\a}&=&-\frac{3}{2}U_\a^\tf{}^{\dot{\beta}}_\tf
T_{[\tc\tb\ta]}^\a\label{Dirac2} \eea with $U^\tb_\b{}^\da_\ta =
\frac{i}{4}(\la^\tb_\b\bla^\da_\ta-\frac{1}{2}\d^\tb_\ta\la^\tf_\b\bla^\da_\tf)$.

\vspace{0.5cm} \noindent{\bf Equations of motion for the helicity 1/2 fields.}

Note first, that all these relations are implied also by Bianchi identities at dim
3/2. Secondly, note that the last equations, \equ{Dirac1} and \equ{Dirac2}, are the
Dirac equations for the spin 1/2 fields, which may be written in terms of the fields
$\la$ in the following way: \bea
\cd_{\b\da}\bla^\da_\ta&=&ie^{-2\f}H^*_{\b\da}\bla^\da_\ta+\frac{9i}{8}(\bla_\ta\bla_\tf)\la^\tf_\b\,,\label{Diracc1}\\
\cd^{\a{\dot{\beta}}}\la_\a^\ta&=&-ie^{-2\f}H^*{}^{\a{\dot{\beta}}}\la_\a^\ta+\frac{9i}{8}(\la^\ta\la^\tf)\bla_\tf^{\dot{\beta}}\,.\label{Diracc2}
\eea

\begin{enumerate}[a.]
\item Consider the spinorial derivative $\cd^\td_\d$ of the Dirac equation \equ{Dirac1}. The derived identity is satisfied if and only if in addition to the results obtained till dimension 3/2 the following relations take place:
\be
\cd_{\a\da}P^{\a\da}-i\left(e^{-2\f}H^*_{\a\da}+\frac{1}{2}\la^\tf_\a\bla_{\da\tf}\right)P^{\a\da}
+\frac{iq}{2}\varepsilon_{\td\tc\tb\ta}G^{({\dot{\beta}}\da)[\td\tc]}G_{({\dot{\beta}}\da)}{}^{[\tb\ta]}\
=\ 0 \eqn{dL1} \be
\sum_{\b\a}\left[\cd_{\b\da}P_\a{}^{\da}-i\left(e^{-2\f}H^*_{\b\da}+\la^\tf_\b\bla_{\da\tf}\right)P_\a{}^{\da}\right]\
=\ 0. \eqn{dH1}
\item Consider the spinorial derivative $\cd_\td^{\dot{\delta}}$ of \equ{Dirac2}. The identity is satisfied if and only if in addition to the results obtained till dimension 3/2 the following relations take place:
\be
\cd_{\a\da}\bP^{\a\da}+i\left(e^{-2\f}H^*_{\a\da}+\frac{1}{2}\la^\tf_\a\bla_{\da\tf}\right)\bP^{\a\da}
+\frac{iq}{2}\varepsilon^{\td\tc\tb\ta}G_{(\b\a)[\td\tc]}G^{(\b\a)}{}_{[\tb\ta]}\
=\ 0 \eqn{dL2} \be
\sum^{{\dot{\beta}}\da}\left[\cd^{\a{\dot{\beta}}}\bP_\a{}^{\da}+i\left(e^{-2\f}H^*{}^{\a{\dot{\beta}}}+\la^{\tf\a}\bla_{\tf}^{\dot{\beta}}\right)\bP_\a{}^{\da}\right]\
=\ 0. \eqn{dH2} \end{enumerate}

\vspace{0.5cm} \noindent{\bf Equations of motion for the scalar.}

Using properties \equ{PP} the equations of motion for the scalar can be deduced from
the sum of the relations \equ{dL1} and \equ{dL2}: \bea
2\cd_a\left(\cd^a \f\right)&=&e^{-4\f}H^*_aH^{*a}-e^{-2\f}H^*_a(\la^\ta\s^a\bla_\ta)\nn\\
&&-\frac{3}{8}(\la^\tb\la^\ta)(\bla_\tb\bla_\ta)
-\frac{1}{2}e^{-2\f}F_{ba[\tb\ta]}F^{ba[\tb\ta]}\,.\label{scal} \eea This equation
already shows that in the Lagrangian corresponding to these equations of motion the
kinetic terms of the antisymmetric tensor and of the graviphotons are accompanied by
exponentials in the scalar field.

By the way, the difference of relations \equ{dL1} and \equ{dL2} looks as \be
\cd_aH^{*a}\ =\
\frac{1}{2}e^{2\f}(\la^\ta\s_a\bla_\ta)\cd^a\f+\frac{i}{2}F^{*ba[\tb\ta]}F_{ba[\tb\ta]},
\end{equation} and it corresponds of course to the Bianchi identity satisfied by the
antisymmetric tensor gauge field. The topological term
$F^{*ba[\tb\ta]}F_{ba[\tb\ta]}$ is an indication of the intrinsic presence of
Chern-Simons forms in the geometry. This feature is analogous to the case of the
off-shell $N=2$ minimal supergravity multiplet containing an antisymmetric tensor
\cite{AGHH99}. It arises naturally in extended supergravity using the soldering
mechanism with the geometry of a 2--form in central charge superspace.

\vspace{0.5cm} \noindent{\bf Equations of motion for the antisymmetric tensor.}

Note that relations \equ{dH1} and \equ{dH2} are the selfdual and
respectively the anti-selfdual part of the equation of motion for
the antisymmetric tensor. Putting these relations together, we
obtain the equation of motion for the antisymmetric tensor: \bea
\varepsilon_{dcba}\cd^bH^{*a}&=&\left[T_{dc}{}^\a_\ta\la^\ta_\a+T_{dc}{}_\da^\ta\bla_\ta^\da\right]e^{2\f}
-\frac{1}{2}H^*_{[d}(\la^\tf\s_{c]}\bla_\tf)\nn\\
&&+\varepsilon_{dcba}\left[\frac{3}{4}\cd^b(\la^\tf\s^a\bla_\tf)
-(\cd^b\f)(\la^\tf\s^a\bla_\tf) +4e^{-2\f}(\cd^b
\f)H^*{}^a\right]e^{2\f}\label{tens} \eea

Consider the spinorial derivative $\cd_\td^{\dot{\delta}}$ of the
Dirac equation \equ{Dirac1} and the spinorial derivative
$\cd^\td_\d$ of \equ{Dirac2}. The identities obtained this way
are satisfied if and only if in addition to the results obtained
till dimension 3/2 the following relations hold: \bea
4i\cd_{\b\da}G^{({\dot{\delta}}\da)[\tb\ta]}
&=&q\varepsilon^{\tb\ta\td\tc}\left(G_{(\b\a)[\td\tc]}P^{\a{\dot{\delta}}}+i\bla^\da_\td\bla^{\dot{\delta}}_\tc\bP_{\b\da}\right)\nn\\
&&-G^{({\dot{\delta}}\da)[\tb\tf]}\la^\ta_\b\bla_{\tf\da}
-G^{({\dot{\delta}}\da)[\tf\ta]}\la^\tb_\b\bla_{\tf\da}\label{G1}\\[4mm]
4i\cd^{\a{\dot{\beta}}}G_{(\d\a)[\tb\ta]}
&=&q\varepsilon_{\tb\ta\td\tc}\left(G^{({\dot{\beta}}\da)[\td\tc]}\bP_{\d\da}+i\la_\a^\td\la_\d^\tc P^{\a{\dot{\beta}}}\right)\nn\\
&&+G_{(\d\a)[\tb\tf]}\la^{\tf\a}\bla_\ta^{\dot{\beta}}
+G_{(\d\a)[\tf\ta]}\la^{\tf\a}\bla_\tb^{\dot{\beta}}.\label{G2}
\eea

\vspace{0.5cm} \noindent{\bf Equations of motion for the graviphotons.}

Recall that the geometric soldering mechanism between
supergravity and the geometry of the 3-form implies that the
fields $G_{(\b\a)[\tb\ta]}$ and $G^{({\dot{\beta}}\da)[\tb\ta]}$
are related to the covariant field strength of the graviphotons,
$F^\au$, by \equ{G-F}. Then the previous lemma determines both
the equations of motion and the Bianchi identities satisfied by
the vector gauge fields of the multiplet: \bea
\cd_b F^{ba\au}&=&-\frac{i}{2}\left[(P_b+\bP_b)F^{ba\au}+(P_b-\bP_b)F^*{}^{ba\au}\right]\nn\\[2mm]
&&
+\frac{i}{4}\left[P_b(\la^\tb\s^{ba}\bla^\ta){\mathfrak{t}}_{[\tb\ta]}{}^\au
+\bP_b(\la_\tb\bs^{ba}\bla_\ta){\mathfrak{t}}^{[\tb\ta]}{}^\au\right]e^\f\nn\\[2mm]
&&-\frac{i}{4}\left[(\la^\ta\s^{dc}\s^a\bla_\tb)F_{dc}{}^\av\hh_\av{}^{[\tb\tf]}{\mathfrak{t}}_{[\tf\ta]}{}^\au
+(\bla_\ta\bs^{dc}\bs^a\la^\tb)F_{dc}{}^\av\hh_\av{}_{[\tb\tf]}{\mathfrak{t}}^{[\tf\ta]}{}^\au\right]\,,\label{F}\\[2mm]
\cd_b F^*{}^{ba\au}&=&
\frac{i}{4}\left[P_b(\la^\tb\s^{ba}\bla^\ta){\mathfrak{t}}_{[\tb\ta]}{}^\au
-\bP_b(\la_\tb\bs^{ba}\bla_\ta){\mathfrak{t}}^{[\tb\ta]}{}^\au\right]e^\f\nn\\[2mm]
&&-\frac{i}{4}\left[(\la^\ta\s^{dc}\s^a\bla_\tb)F_{dc}{}^\av\hh_\av{}^{[\tb\tf]}{\mathfrak{t}}_{[\tf\ta]}{}^\au
-(\bla_\ta\bs^{dc}\bs^a\la^\tb)F_{dc}{}^\av\hh_\av{}_{[\tb\tf]}{\mathfrak{t}}^{[\tf\ta]}{}^\au\right]\,.\label{F*}
\eea

Further differentiating \equ{G1} and \equ{G2} one can obtain Bianchi identities for
the gravitini and graviton, but here we would like to derive their equations of
motion instead.

\vspace{0.5cm} \noindent{\bf Equations of motion for the gravitini.}

Unlike the equations of motion presented above, the equations of
motion for the gravitini and the graviton are directly given by
the superspace Bianchi identities, once the component fields are
identified. For example, the Bianchi identities at dim 3/2
determinate the torsion components \be T_{(\b\a)}{}^\a_\ta\ =\
\frac{1}{16}\bP_{\b\da}\bla^\da_\ta \qquad
T^{({\dot{\delta}}{\dot{\gamma}})}{}_{\b\td}\ =\
\frac{1}{8}\bP_{\b}{}^{({\dot{\delta}}}\bla^{{\dot{\gamma}})}_\td
+\frac{iq}{8}\varepsilon_{\td\tc\tb\ta}G^{({\dot{\delta}}{\dot{\gamma}})[\tc\tb]}\la^\ta_\b\,,
\end{equation} \be T^{({\dot{\beta}}\da)}{}_\da^\ta\ =\
\frac{1}{16}P^{\a{\dot{\beta}}}\la_\a^\ta \qquad
T_{(\d\g)}{}^{{\dot{\beta}}\td}\ =\
\frac{1}{8}P_{(\d}{}^{{\dot{\beta}}}\la_{\g)}^\td
+\frac{iq}{8}\varepsilon^{\td\tc\tb\ta}G_{(\d\g)[\tc\tb]}\bla_\ta^{\dot{\beta}}\,,
\end{equation} and these components are sufficient to give the equations of
motion for the gravitini: \be
\varepsilon^{dcba}(\bs_cT_{ba\ta})^\da\ =\
\frac{i}{4}(\bla_\ta\bs^d\s^c\eps)^\da\bP_c
+\frac{i}{2}(\bs^{ba}\bs^d\la^\tf)^\da
F_{ba}{}_{[\ta\tf]}e^{-\f}\,, \eqn{gravitini1} \be
\varepsilon^{dcba}(\s_cT_{ba}{}^\ta)_\a\ =\
-\frac{i}{4}(\la^\ta\s^d\bs^c\eps)_\a P_c
-\frac{i}{2}(\s^{ba}\s^d\bla_\tf)_\a F_{ba}{}^{[\ta\tf]}
e^{-\f}\,. \eqn{gravitini2}

\vspace{0.5cm} \noindent{\bf Equations of motion for the graviton.}

In order to give the equations of motion for the graviton we need
the expression of the supercovariant Ricci tensor,
$R_{db}=R_{dcba}\eta^{ca}$, which is given by the superspace
Bianchi identities at canonical dimension 2 \equ{Riccitensor}. The
corresponding Ricci scalar, $R=R_{db}\eta^{db}$, is then \be
R=-2\cd^a\f\cd_a\f-\frac{1}{2}H^*{}^aH^*{}_ae^{-4\f}
+\frac{3}{4}e^{-2\f}H^*{}^a(\la^\ta\s_a\bla_\ta)
+\frac{3}{8}(\la^\tb\la^\ta)(\bla_\tb\bla_\ta). \end{equation}
The knowledge of these ingredients allows us to write down the
Einstein equation \bea
R_{db}-\frac{1}{2}\eta_{db}R&=&-2\left[\cd_d\f\,\cd_b\f-\frac{1}{2}\eta_{db}\cd^a\f\cd_a\f\right]\nn\\[2mm]
&&-\frac{1}{2}e^{-4\f}\left[H^*_d\,H^*_b-\frac{1}{2}\eta_{db}H^*{}^aH^*{}_a\right]\nn\\[2mm]
&&-e^{-2\f}\left[F_{df[\tb\ta]}F_b{}^f{}^{[\tb\ta]}-\frac{1}{4}\eta_{db}F_{ef[\tb\ta]}F^{ef[\tb\ta]}\right]\nn\\[2mm]
&&-\frac{i}{8}\sum_{db}\left[\la^\tf\s_d\cd_b\bla_\tf-(\cd_b\la^\tf)\s_d\bla_\tf\right]\nn\\[2mm]
&&-\frac{1}{8}\left[\frac{1}{4}(\la^\tf\s_d\bla_\tf)\,(\la^\ta\s_b\bla_\ta)+\eta_{db}(\la^\tb\la^\ta)(\bla_\tb\bla_\ta)\right]\nn\\[2mm]
&&-\frac{1}{8}e^{-2\f}\left[H_d^*(\la^\tf\s_b\bla_\tf)-3\eta_{db}H^*{}^a(\la^\ta\s_a\bla_\ta)\right]\,,
\label{gr} \eea where one may recognize on the right-hand-side the usual terms of the
energy-momentum tensor corresponding to matter fields: scalar fields, antisymmetric
tensor, photon fields and spinor fields respectively. As it will be shown by
\equ{Rdbbar}, the contribution of the gravitini is hidden in $R_{db}$.

\section{Equations of motion in terms of component fields}

In the previous section we calculated the equations of motion for
all component fields of the N-T multiplet (graviton \equ{gr},
gravitini \equ{gravitini1}, \equ{gravitini2},  graviphotons
\equ{F}, 1/2-spin fields \equ{Diracc1}, \equ{Diracc2}, scalar
\equ{scal} and the antisymmetric tensor \equ{tens}) in terms of
supercovariant objects, which have only flat (Lorentz) indices.
In order to write these equations of motion in terms of component
fields, one passes to curved (Einstein) indices by the standard
way \cite{WB83}. General formulae are easily written using the
notation $E^\ca\doubar=e^\ca= d x^{\mu} e_{\mu}{}^\ca$
\cite{BGG01}.

\subsection{Supercovariant$\rightarrow$component toolkit}

Recall that the graviton, gravitini and graviphotons are
identified in the super-vielbein. Thus, their field strengths can
be found in their covariant counterparts using \be T^\ca\doubar\
=\  \frac{1}{2} d x^{\mu} d x^{\nu}
\left(\cd_{\nu}e_{\mu}{}^\ca-\cd_{\mu}e_{\nu}{}^\ca\right) \  =\
\frac{1}{2}e^{\cal B} e^\cc T_{\cc{\cal B}}{}^\ca\loco\,.
\end{equation} For $\ca=a$ one finds the relation \be
\cd_{\nu}e_{\mu}{}^a-\cd_{\mu} e_{\nu}{}^a\ =\
i\psi_{[{\nu}\ta}\s^a\bar{\psi}_{{\mu}]}{}^\ta\,, \eqn{Tcba} which
determinates the Lorentz connection in terms of the vierbein, its
derivatives and gravitini fields. For $\ca=_\a^\ta$ and
$\ca=^\da_\ta$ we have the expression of the covariant field
strength of the gravitini \bea
T_{cb}{}_\ta^\a\loco &=& e_b{}^{\mu}e_c{}^{\nu}\cd_{[{\nu}}\psi_{{\mu}]}{}^\a_\ta
-e_b{}^{\mu}e_c{}^{\nu}\frac{q}{4}\varepsilon_{\td\tc\tb\ta}
\bar{\psi}_{\nu}{}^{\td}\bar{\psi}_{\mu}{}^{\tc}\la^{\a}{}^\tb
-\frac{i}{2}e_{[c}{}^{\nu}(\bar{\psi}_{\nu}{}^{\tb}\bar{\s}_{b]}\s^{da})^{\a}F_{da}{}_{[\tb\ta]}\loco e^{-\f}\nn\\[2mm]
&& +\frac{i}{4}(\psi_{{\nu}\tf}\s_{f[b})^{\a}e_{c]}{}^{\nu}
\left[\la^{\tf}\s^f\bar{\la}_{\ta}
-\frac{1}{2}\d_{\ta}^{\tf}\la^{\tb}\s^f\bar{\la}_{\tb}\right]\,,
\eea \bea
T_{cb}{}^\ta_\da\loco &=& e_b{}^{\mu}e_c{}^{\nu}\cd_{[{\nu}}\bp_{{\mu}]}{}_\da^\ta
-e_b{}^{\mu}e_c{}^{\nu}\frac{q}{4}\varepsilon^{\td\tc\tb\ta}
{\psi}_{\nu}{}_{\td}{\psi}_{\mu}{}_{\tc}\bar{\la}_{\da}{}_\tb
-\frac{i}{2}e_{[c}{}^{\nu}({\psi}_{\nu}{}_{\tb}{\s}_{b]}\bs^{da})_{\da}F_{da}{}^{[\tb\ta]}\loco e^{-\f}\nn\\[2mm]
&&
-\frac{i}{4}(\bar{\psi}_{\nu}{}^\tf\bar{\s}_{f[b})_{\da}e_{c]}{}^{\nu}
\left[{\la}^{\ta}\s^f\bar{\la}_{\tf}
-\frac{1}{2}\d^{\ta}_{\tf}\la^{\tb}\s^f\bar{\la}_{\tb}\right]\,.
\eea As for $\ca=\au$, the central charge indices, we obtain the
covariant field strength of the graviphotons \bea
F_{ba}{}^\au\loco &=& e_b{}^{\nu} e_a{}^{\mu}\cf_{{\nu}{\mu}}{}^\au
+e_b{}^{\nu}
e_a{}^{\mu}\left[\bar{\psi}_{{\nu}}{}^{\tc}\bar{\psi}_{{\mu}}{}^{\tb}
+i\bar{\psi}_{[{\nu}}{}^{\tc}\bar{\s}_{{\mu}]}\la^{\tb}\right]e^\f{\mathfrak{t}}_{[\tc\tb]}{}^{\au}\nn\\[2mm]
&& +e_b{}^{\nu}
e_a{}^{\mu}\left[\psi_{{\nu}\tc}\psi_{{\mu}\tb}+i\psi_{[{\nu}\tc}\s_{{\mu}]}\bar{\la}_{\tb}\right]e^\f{\mathfrak{t}}^{[\tc\tb]\au}\,,
\eea with $\cf_{{\nu}{\mu}}{}^\au$ the field strength of the
graviphotons
$\cf_{{\nu}{\mu}}{}^\au=\partial_{{\nu}}v_{{\mu}}{}^\au
-\partial_{{\mu}}v_{{\nu}}{}^\au$. In the $SU(4)$ basis this
becomes \bea
F_{ba}{}^{[\tb\ta]}\loco &=& e_b{}^{\nu} e_a{}^{\mu}\cf_{{\nu}{\mu}}{}^{[\tb\ta]} 
+e_b{}^{\nu}
e_a{}^{\mu}\left[\bar{\psi}_{{\nu}}{}^{[\tb}\bar{\psi}_{{\mu}}{}^{\ta]}
+i\bar{\psi}_{[{\nu}}{}^{[\tb}\bar{\s}_{{\mu}]}\la^{\ta]}\right]e^\f\nn\\[2mm]
&& +e_b{}^{\nu}
e_a{}^{\mu}\frac{q}{2}\varepsilon^{\td\tc\tb\ta}\left[\psi_{{\nu}\td}\psi_{{\mu}\tc}
+i\psi_{[{\nu}\td}\s_{{\mu}]}\bar{\la}_{\tc}\right]e^\f\,, \eea
with the field strength
$\cf_{{\nu}{\mu}}{}^{[\tb\ta]}=\cf_{{\nu}{\mu}}{}^\au\hh_\au{}^{[\tb\ta]}=\partial_{{\nu}}V_{{\mu}}{}^{[\tb\ta]}
-\partial_{{\mu}}V_{{\nu}}{}^{[\tb\ta]}$.

Since the antisymmetric tensor is identified in the 2--form, the
development of its covariant field strength on component fields
is deduced using \be H\doubar\ =\ \frac{1}{2} d x^{\mu} d x^{\nu}
d x^{\rho}
\partial_{\rho}b_{{\nu}{\mu}}\ =\ \frac{1}{3!}e^\ca e^{\cal B} e^\cc
H_{\cc{\cal B}\ca}\loco \end{equation} and one finds \bea
H^*{}^a\loco &=& e_{\lambda}{}^a\cg^{\lambda}
+ie_{\lambda}{}^a\left[\psi_{{\rho}\tf}\s^{{\lambda}{\rho}}\la^{\tf}
-\bar{\psi}_{\rho}{}^{\tf}\bar{\s}^{{\lambda}{\rho}}\bar{\la}_{\tf}
+\frac{1}{2}\varepsilon^{{\lambda}{\rho}{\nu}{\mu}}\psi_{{\rho}\tf}\s_{\nu}\bar{\psi}_{{\mu}}{}^{\tf}\right]e^{2\f}\,,
\label{compH} \eea with \be \cg^{\lambda}\ =\
\frac{1}{2}\varepsilon^{{\lambda}{\rho}{\nu}{\mu}}\left[\partial_{\rho}b_{{\nu}{\mu}}-v_{\rho}{}^\au\gg_{\au\av}\cf_{{\nu}{\mu}}{}^\av\right]
\ =\
\frac{1}{2}\varepsilon^{{\lambda}{\rho}{\nu}{\mu}}\left[\partial_{\rho}b_{{\nu}{\mu}}-V_{{\rho}[\tb\ta]}\cf_{{\nu}{\mu}}{}^{[\tb\ta]}\right]\,.
\end{equation} Note, that the dual field strength,
$\frac{1}{2}\varepsilon^{\la{\rho}{\nu}{\mu}}\partial_{\rho}b_{{\nu}{\mu}}$,
of the antisymmetric tensor appears in company with the
Chern-Simons term
$\frac{1}{2}\varepsilon^{\la{\rho}{\nu}{\mu}}v_{\rho}{}^\au\gg_{\au\av}\cf_{{\nu}{\mu}}{}^\av
$. We use the notation $\cg^{\la}$ in order to accentuate this
feature. Recall also, that one of the fundamental aims of the
article \cite{GHK01} was to explain in detail that this
phenomenon is quite general and arises as an intrinsic property
of soldering in superspace with central charge coordinates.

The lowest component of the derivative of the scalar can be
calculated using $D\f\doubar = d x^{\mu}\cd_{\mu}\f = e^\ca\cd_\ca
\f\loco$, and it is \be \cd_a\f\loco =
e_a{}^{\mu}\left(\cd_{\mu}\f
-\frac{1}{4}\psi_{{\mu}\tf}\la^{\tf}-\frac{1}{4}\bar{\psi}_{\mu}{}^{\tf}\bar{\la}_{\tf}\right)\,,
\end{equation} while the lowest component of the double derivative
$\cd_a\cd^a\f\loco$, needed for the expansion of the equation of
motion for the scalar \equ{scal}, becomes \bea 2\cd_a\cd^a\f\loco
&=& 2\Box \f +e_a{}^{\mu}\cd_{\mu}e^{a{\nu}}\left[2\cd_{\nu}\f
-\frac{1}{2}\psi_{{\nu}\tf}\la^{\tf}-\frac{1}{2}\bar{\psi}_{\nu}{}^{\tf}\bar{\la}_{\tf}\right]
-\frac{1}{2}H^{*a}\loco\psi_{{\mu}\tf}\s_a\bar{\psi}^{{\mu}\tf}\, e^{-2\f}\nn\\[2mm]
&&
-\frac{1}{2}\cd^{\mu}(\psi_{{\mu}\tf}\la^{\tf}+\bar{\psi}_{\mu}{}^{\tf}\bar{\la}_{\tf})
-\frac{1}{2}(\psi_{{\mu}\tf}\cd^{\mu}\la^{\tf}+\bar{\psi}_{\mu}{}^{\tf}\cd^{\mu}\bar{\la}_{\tf})\nn\\[2mm]
&&
-\frac{3i}{32}(\la^{\tc}\s^{\nu}\bar{\la}_{\tc})\left(\psi_{{\nu}\tf}\la^{\tf}-\bar{\psi}_{\nu}{}^{\tf}\bar{\la}_{\tf}\right)
-\frac{3}{4}(\psi_{{\mu}\tf}\la^{\ta})(\bar{\psi}_{\mu}{}^{\tf}\bar{\la}_{\ta})\nn\\[2mm]
&&
-\frac{1}{4}(\psi_{{\mu}\tf}\la^{\tf})(\psi^{\mu}_{\tc}\la^{\tc})
+\frac{1}{2}(\psi^{\mu}_{\tf}\la^{\tf})(\bar{\psi}_{\mu}{}^{\tc}\bar{\la}_{\tc})
-\frac{1}{4}(\bar{\psi}_{\mu}{}^{\tf}\bar{\la}_{\tf})(\bar{\psi}^{{\mu}\tc}\bar{\la}_{\tc})\nn\\[2mm]
&&
+F_{ba[\tf\tc]}\loco\left[\frac{1}{2}\bar{\psi}_{\mu}{}^{\tf}\bar{\s}^{ba}\bar{\psi}^{{\mu}\tc}
+\frac{i}{4}\bar{\psi}_{\nu}{}^{\tf}\bar{\s}^{\nu}\s^{ba}\la^{\tc}\right]e^{-\f}\nn\\[2mm]
&&
+F_{ba}{}^{[\tf\tc]}\loco\left[\frac{1}{2}\psi_{{\mu}\tf}\s^{ba}\psi^{\mu}_{\tc}
+\frac{i}{4}\psi_{\nu}{}_{\tf}\s^{\nu}\bar{\s}^{ba}\bar{\la}_{\tc}\right]e^{-\f}\,.
\eea In order to compare our results with the component
expression of the scalar's equation of motion derived from
\cite{NT81}, we have to replace in this expression
$e_a{}^{\mu}\cd_{\mu}e^{a{\nu}}$ with \[
e_a{}^{\mu}\cd_{\mu}e^{a{\nu}}\ =\
V^{-1}\partial_{\mu}(Vg^{{\mu}{\nu}})-ig^{{\mu}{\nu}}\psi_{[{\mu}\ta}\s^k\bp_{k]}{}^\ta\,,
\] as a consequence of \equ{Tcba}.

Finally, using $R_b{}^a\doubar = \frac{1}{2} d x^{\mu} d x^{\nu}
\cR_{{\nu}{\mu}b}{}^a = \frac{1}{2}e^{\cal B} e^\cc R_{\cc{\cal
B}}{}_b{}^a\loco $, one obtains for the lowest component of the
covariant Ricci tensor $ R_{db}$ the expression

\bea R_{db}\loco &=&
\frac{1}{2}\sum_{db}\left\{e_d{}^{\nu}e^{{\mu}a}{\cal
R}_{{\nu}{\mu}ba}\loco
+\frac{1}{2}\varepsilon_b{}^{{\mu}ef}\psi_{{\mu}\td}\s_dT_{ef}{}^\td\loco
-\frac{1}{2}\varepsilon_b{}^{{\mu}ef}\bp_{\mu}{}^{\td}\bs_d T_{ef\,\td}\loco\right.\nn\\[2mm]
&&+\frac{1}{4}\left(i\psi_\td^{\mu}\s_d\bs^f\s_{b{\mu}}
\la^\td-ie_d{}^{\nu}\d_b^f\psi_{{\nu}\td}\la^\td\right)P_f\loco\nn\\[2mm]
&&+\frac{1}{4}\left(i\bp^{{\mu}\td}\bs_d\s^f\bs_{b{\mu}}
\bla_\td-ie_d{}^{\nu}\d_b^f\bp_{\nu}^\td\bla_\td\right)\bar{P}_f\loco\nn\\[2mm]
&&-\frac{1}{2}e^{-\f}F^{ef[\td\tf]}\loco\left(i\textrm{tr}(\s_{b{\mu}}\s_{ef})\psi_\td^{\mu}\s_d\bla_\tf+\frac{i}{2}
e_d{}^{\nu}\psi_{{\nu}\td}\s_{ef}\s_b\bla_\tf\right)
\nn\\[2mm]
&&-\frac{1}{2}e^{-\f}F^{ef}{}_{[\td\tf]}\loco\left(i\textrm{tr}(\bs_{b{\mu}}\bs_{ef})\bp^{{\mu}\td}\bs_d\la^\tf+\frac{i}{2}
e_d{}^{\nu}\bp_{\nu}^\td\bs_{ef}\bs_b\la^\tf\right)
\nn\\[2mm]
&&+\frac{1}{2}e_d{}^{\nu}e^{-\f}\left(\textrm{tr}(\bs_b{}^{\mu}\bs_{ef})\psi_{{\nu}\td}\psi_{{\mu}\tc}F^{ef[\td\tc]}\loco
+\textrm{tr}(\s_b{}^{\mu}\s_{ef})\bp_{\nu}^\td\bp_{\mu}^\tc F^{ef}{}_{[\td\tc]}\loco\right)\nn\\[2mm]
&&-\left.\frac{1}{2}e_d{}^{\mu}(\d^\td_\tb\d^\tc_\ta-\frac{1}{2}\d^\td_\ta\d^\tc_\tb)
\left[(\psi_{[{\mu}\td}\s_b{}^{\nu}\la^\tb)(\bp_{{\nu}]}{}^\ta\bla_\tb)-(\psi_{[{\mu}\td}\la^\tb)(\bp_{{\nu}]}{}^\ta\bs_b{}^{\nu}\bla_\tb)
\right]\right\}. \label{Rdbbar} \eea

\subsection{The equations of motion}

In the last subsection we deduced the expression of all quantities appearing in the
supercovariant equations of motion in terms of component fields. We are therefore
ready now to replace these expressions in \equ{gr}, \equ{gravitini1},
\equ{gravitini2}, \equ{F}, \equ{Diracc1}, \equ{Diracc2}, \equ{scal}, \equ{tens}
 and give the equations of motion in terms
of component fields.

It turns out that the expressions \bea \tilde{H}_{{\rho}} &=&
e_{\rho}{}^a H^*_a\loco
-\frac{i}{2}e^{2\f}\psi_{{\rho}\ta}\la^{\ta}
+\frac{i}{2}e^{2\f}\bp_{\rho}{}^{\ta} \bla_{\ta}
-\frac{3}{4}e^{2\f}\la^{\ta}\s_{\rho}\bla_{\ta}\nn\\[2mm]
&=&\cg_{\rho}
+\frac{i}{2}e^{2\f}\left[\psi_{{\kappa}\tf}\s_{\rho}\bs^{{\kappa}}\la^{\tf}
-\bar{\psi}_{\kappa}{}^{\tf}\bs_{\rho}\s^{{\kappa}}\bar{\la}_{\tf}
+\varepsilon_{\rho}{}^{{\kappa}{\nu}{\mu}}\psi_{{\kappa}\tf}\s_{\nu}\bar{\psi}_{{\mu}}{}^{\tf}\right]
-\frac{3}{4}e^{2\f}\la^{\ta}\s_{\rho}\bla_{\ta} \eea and \bea
\tilde{F}^{{\nu}{\kappa}\az} &=&e^{\nu}{}^be^{\kappa}{}^a
F_{ba}{}^\az\loco
+\frac{i}{2}\varepsilon^{{\nu}{\kappa}{\mu}{\rho}}
\left[\bp_{\mu}{}^{\td}\bp_{\rho}{}^{\tc}
-i\frac{q}{2}\varepsilon^{\td\tc\tb\ta}\bla_{\tb}\bs_{\mu}\psi_{{\rho}\ta}\right]e^{\f}t_{[\td\tc]}{}^{\az}\nn\\[2mm]
&&-\frac{i}{2}\varepsilon^{{\nu}{\kappa}{\mu}{\rho}}
\left[\psi_{\mu}{}_{\td}\psi_{\rho}{}_{\tc}
-i\frac{q}{2}\varepsilon_{\td\tc\tb\ta}\la^{\tb}\s_{\mu}\bp_{\rho}{}^\ta\right]e^{\f}t^{[\td\tc]}{}^{\az}\nn\\[2mm]
&=&
\cf^{{\nu}{\kappa}\az}-\textrm{tr}(\s^{{\nu}{\kappa}}\s^{{\mu}{\rho}})
\left[\bp_{\mu}{}^{\td}\bp_{\rho}{}^{\tc}
-i\frac{q}{2}\varepsilon^{\td\tc\tb\ta}\bla_{\tb}\bs_{\mu}\psi_{{\rho}\ta}\right]e^{\f}t_{[\td\tc]}{}^{\az}\nn\\[2mm]
&&-\textrm{tr}(\bs^{{\nu}{\kappa}}\bs^{{\mu}{\rho}})
\left[\psi_{\mu}{}_{\td}\psi_{\rho}{}_{\tc}
-i\frac{q}{2}\varepsilon_{\td\tc\tb\ta}\la^{\tb}\s_{\mu}\bp_{\rho}{}^\ta\right]e^{\f}t^{[\td\tc]}{}^{\az}
\eea appear systematically, and using them, the equations take a
quite simple form. Let us also denote the quantity
$\hat{F}_{{\nu}{\kappa}}{}^\az=e_{\nu}{}^be_{\kappa}{}^a
F_{ba}{}^\az\loco$, which is called the supercovariant field
strength of the graviphotons in the component approach
\cite{CS77b}, \cite{NT81}.

\vspace{0.5cm} \noindent{\bf Equations of motion for the helicity
1/2 fields.} \bea
\left(\s^{\mu}\hat{\cd}_{\mu}\bla_{\ta}\right)_{\b} &=&
-ie^{-2\f}\tilde{H}^{\mu}
\left[\frac{i}{2}(\s^{\nu}\bar{\s}_{\mu}\psi_{{\nu}\ta})_{\b}-\frac{3}{4}(\s_{\mu}\bla_{\ta})_{\b}\right]
-\frac{i}{2}(\bar{\psi}_{\nu}{}^{\tf}\bla_\tf)(\s^{\mu}\bs^{\nu}\psi_{{\mu}\ta})_\b\nn\\[2mm]
&&+i\partial_{\nu}\f(\s^{\mu}\bar{\s}^{\nu}\psi_{{\mu}\ta})_{\b}-e^{-\f}\hat{F}_{{\kappa}{\rho}[\tf\ta]}(\s^{\mu}\bar{\s}^{{\kappa}{\rho}}\bar{\psi}_{\mu}{}^{\tf})_{\b}
-\frac{3i}{8}(\bar{\la}_{\ta}\bar{\la}_{\tf}) \la_{\b}^{\tf} \eea

\vspace{0.5cm} \noindent{\bf Equations of motion for the
gravitini.} \bea
\varepsilon^{{\rho}{\kappa}{\nu}{\mu}}(\bs_{\kappa}{}\hat{\cd}_{\nu}\psi_{{\mu}\ta})^\da
&=&
-\frac{i}{4}e^{-2\f}\tilde{H}_{\nu}\left[\varepsilon^{{\rho}{\kappa}{\nu}{\mu}}(\bs_{\kappa}\psi_{{\mu}\ta})^{\da}
+(\bla_{\ta}\bs^{\rho}\s^{\nu}\varepsilon)^{\da}\right]
-\frac{1}{2}\partial_{\nu}\f(\bla_{\ta}\bs^{\rho}\s^{\nu}\varepsilon)^{\da}\nn\\[2mm]
&&-e^{-\f}\hat{F}_{{\mu}{\nu}[\ta\tf]}\left[\textrm{tr}(\s^{{\rho}{\kappa}}\s^{{\mu}{\nu}})\bp_{\kappa}{}^{\tf\da}
+\frac{i}{2}\textrm{tr}(\bs^{{\rho}{\kappa}}\bs^{{\mu}{\nu}})(\s_{\kappa}\la^\tf)^\da\right]\nn\\[2mm]
&&+\frac{1}{8}\psi_{{\nu}\ta}\la^\tf(\bs^{{\rho}{\nu}}\bla_\tf)^\da+\frac{3}{8}(\psi_{{\nu}\ta}\s^{{\rho}{\nu}}\la^\tf)
\bla_\tf^\da-\frac{1}{4}(\psi_{{\nu}\tf}\s^{\rho}\bs^{\nu}\la^\tf)\bla_\ta^\da
\nn\\[2mm]
&&+\frac{q}{4}\varepsilon^{{\rho}{\kappa}{\nu}{\mu}}\varepsilon_{\tc\tb\tf\ta}
\bp_{\nu}{}^{\tc}\bp_{\mu}{}^{\tb}(\bs_{\kappa}\la^{\tf})^{\da}
\eea

\vspace{0.5cm} \noindent{\bf Equations of motion for the scalar.}
\bea 0 &=& 2V^{-1}\partial_{\mu}(Vg^{{\mu}{\nu}}\partial_{\nu}\f)
+\frac{1}{2}V^{-1}\partial_{\mu}\left(V\la^{\ta}\s^{\nu}\bs^{\mu}\psi_{{\nu}\ta}+V\bla_{\ta}\bs^{\nu}\s^{\mu}\bp_{\nu}^{\ta}
\right)\nn\\[2mm]
&&-e^{-4\f}\cg^{\mu}\tilde{H}_{\mu}
+\frac{1}{2}e^{-2\f}\cf_{{\nu}{\mu}[\tb\ta]}\tilde{F}^{{\nu}{\mu}[\tb\ta]}
\eea

\vspace{0.5cm} \noindent{\bf Equations of motion for the antisymmetric tensor.} \bea
\partial_{{\kappa}}\left(e^{-4\f}V\varepsilon^{{\mu}{\nu}{\kappa}{\rho}}\tilde{H}_{{\rho}}\right) &=& 0
\eea

\vspace{0.5cm} \noindent{\bf Equations of motion for the graviphotons.} \bea
\partial_{\nu}\left(Ve^{-2\f}\tilde{F}^{{\nu}{\kappa}\au}\right) &=&
\frac{1}{2}Ve^{-4\f}\varepsilon^{{\rho}{\nu}{\mu}{\kappa}}\tilde{H}_{\rho}\cf_{{\nu}{\mu}}{}^{\au}
\eea

\vspace{0.5cm} \noindent{\bf Equations of motion for the graviton.}

The Einstein equation in terms of component fields is also
deduced in a straightforward manner from \equ{gr} and
\equ{Rdbbar} with the usual Ricci tensor
$\cR_{{\mu}{\nu}}=\frac{1}{2}\sum_{{\mu}{\nu}}e_{\nu}{}^be^{{\kappa}a}\cR_{{\mu}{\kappa}ba}$.
Here we give the expression of the Ricci scalar: \bea \cR &=&
\frac{1}{2}\varepsilon^{{\rho}{\kappa}{\nu}{\mu}}\psi_{{\rho}\ta}\s_{\kappa}\hat{\cd}_{\nu}\bp_{\mu}{}^\ta-\frac{1}{2}\varepsilon^{{\rho}{\kappa}{\nu}{\mu}}
\bp_{\rho}{}^{\ta}\bs_{\kappa}\hat{\cd}_{\nu}\psi_{{\mu}\ta}\nn\\[2mm]
&&-\frac{i}{4}\la^\ta\s^{\mu}\hat{\cd}_{\mu}\bla_\ta-\frac{i}{4}\bla_\ta\bs^{\mu}\hat{\cd}_{\mu}\la^\ta
-2\partial^{\mu}\f\partial_{\mu}\f\nn\\[2mm]
&&-e^{-\f}\hat{F}_{{\kappa}{\rho}}{}^{[\tb\ta]}\left(\textrm{tr}(\bs^{{\kappa}{\rho}}\bs^{{\mu}{\nu}})\psi_{{\mu}\tb}\psi_{{\nu}\ta}
+\frac{i}{2}\textrm{tr}(\s^{{\kappa}{\rho}}\s^{{\mu}{\nu}})\psi_{{\mu}\tb}\s_{\nu}\bla_\ta\right)\nn\\[2mm]
&&-e^{-\f}\hat{F}_{{\kappa}{\rho}[\tb\ta]}\left(\textrm{tr}(\s^{{\kappa}{\rho}}\s^{{\mu}{\nu}})\bp_{\mu}{}^\tb\bp_{\nu}{}^\ta
+\frac{i}{2}\textrm{tr}(\bs^{{\kappa}{\rho}}\bs^{{\mu}{\nu}})\bp_{\mu}{}^{\tb}\bs_{\nu}\la^\ta\right)\nn\\[2mm]
&&-\frac{1}{2}e^{-4\f}\tilde{H}^{{\rho}}\left(\cg_{\rho}+\frac{i}{2}e^{2\f}(\psi_{{\kappa}\tf}\s_{\rho}\bs^{{\kappa}}\la^{\tf}
-\bar{\psi}_{\kappa}{}^{\tf}\bs_{\rho}\s^{{\kappa}}\bar{\la}_{\tf}
+2\varepsilon_{\rho}{}^{{\kappa}{\nu}{\mu}}\psi_{{\kappa}\tf}\s_{\nu}\bar{\psi}_{{\mu}}{}^{\tf}\right)\nn\\[2mm]
&&+\frac{1}{2}(\psi_{{\rho}\ta}\la^\ta)(\bp^{{\rho}\ta}\bla_\ta)+\frac{i}{2}(\la^\tf\s^{\rho}\bla_\tf)
(\psi_{{\rho}\ta}\la^\ta-\bp_{\rho}{}^{\ta}\bla_\ta)\nn\\[2mm]
&&-\frac{3i}{16}\varepsilon^{{\rho}{\mu}{\nu}{\kappa}}(\psi_{{\rho}\tf}\s_{\mu}\bp_{\nu}{}^\tf)(\la^\ta\s_{\kappa}\bla_\ta)
-\frac{i}{2}\varepsilon^{{\rho}{\mu}{\nu}{\kappa}}(\psi_{{\rho}\tf}\s_{\mu}\bp_{\nu}{}^\ta)(\la^\tf\s_{\kappa}\bla_\ta)
\eea

\section{Conclusion}

The aim of this work was to deduce the equations of motion for
the components of the N-T multiplet from its geometrical
description in central charge superspace, and compare these
equations with those deduced from the Lagrangian of the component
formulation of the theory with the same field content \cite{NT81}.

We showed that the constraints on the superspace which allow to
identify the components in the geometry imply equations of motion
in terms of supercovariant quantities. Moreover, we succeeded in
writing these equations of motion in terms of component fields in
an elegant way, using the objects $\tilde{H}_{\mu}$ and
$\tilde{F}_{\mu\nu}{}^\au$. The equations found this way are in
perfect concordance with the ones deduced from the Lagrangian of
Nicolai and Townsend \cite{NT81}. This result resolves all
remaining doubt about the equivalence of the geometric
description on central charge superspace of the N-T multiplet and
the Lagrangian formulation of the theory with the same field
content.

As a completion of this work one may ask oneself about an
interpretation of the objects $\tilde{H}_{\mu}$ and
$\tilde{F}_{\mu\nu}{}^\au$, which seem to be some natural building
blocks of the Lagrangian. Concerning this question let us just
remark the simplicity of the relation \be -i\ki_{\mu}{}^\ta{}_\ta\
=\ e^{-2\f}\tilde{H}_{\mu}+\frac{3}{8}\la^\ta\s_{\mu}\bla_\ta
\end{equation} between $\tilde{H}_{\mu}$ and the $U(1)$ part of the
initial connection \equ{connect} of the central charge superspace
with structure group $SL(2,\mathbb{C})\otimes U(4)$.

Having at our disposal now a well-defined and elegant formalism to
describe $N=4$ supergravity, it would be interesting to try to
incorporate matter multiplets in this framework. In particular,
according to \cite{Cha81a}, 6 vector multiplets naturally couple
to gravity in the process of dimensional reduction. In the central charge superspace we have used here, before putting all constraints, the maximal number of central charges is 12. Once the constraints have been taken into account, this number reduces to 6 (\ref{rel_ht}), which is exactly the number of vectors contained in the $N=4$ gravity multiplet. However, if one relaxes some of the constraints listed in this chapter, one may expect that the full number of supercharges would be available. This would mean that 6 new vectors appear, and it seems natural to think that these vectors would correspond to those arising in the reduction of the gravity multiplet of $N=1$ $d=10$ supergravity \cite{Cha81a}. We hope to report on this issue in a close futur.


\appendix

\chapter{Notions of differential geometry}\label{AppA}

%
\section{Einstein's Gravity}\label{eg}
%

We take the signature $(-, +, +,\ldots)$. The space-time has $d$
dimensions and indices are running from $0$ to $d-1$. The
Christoffel connection is the unique torsion-free connection for
which the metric is covariantly constant

\bea \Gamma_{\mu\nu}{}^{\rho} & = &
\frac{1}{2}g^{\rho\lambda}(\del_{\mu}
g_{\nu\lambda}+\del_{\nu} g_{\mu\lambda}-\del_{\lambda} g_{\mu\nu})\label{eg1}\\[2mm]
\nabla_{\mu} g_{\nu\rho} & = & \del_{\mu}
g_{\nu\rho}-\Gamma_{\mu\nu}{}^{\lambda}
g_{\lambda\rho}-\Gamma_{\mu\rho}{}^{\lambda} g_{\nu\lambda} = 0.
\label{eg2}\eea Here is a useful relation :

\bea \del_{\mu}\left(\sqrt{-g}V^{\mu}\right) &=&
\sqrt{-g}\nabla_{\mu} V^{\mu} \label{eg3}\eea that can be checked
using the formula for the determinant of a matrix $A$

\bea \det(A) & = & exp\left(\textrm{Tr}\ln A\right).\label{eg4}
\eea Since there is no torsion, the commutator of two covariant
derivatives is given by the Riemann tensor as follows

\bea [\nabla_{\mu}\, ,\, \nabla_{\nu}] V_{\rho} &=&
-\cR_{\mu\nu\rho}{}^{\lambda} V_{\lambda} \label{comm1}\eea and
its expression is

\bea \cR_{\mu\nu\rho}{}^{\lambda} &=& \del_{\mu}
\Gamma_{\nu\rho}{}^{\lambda}-\del_{\nu}\Gamma_{\mu\rho}{}^{\lambda}+\Gamma_{\mu\sigma}{}^{
\lambda}
\Gamma_{\nu\rho}{}^{\sigma}-\Gamma_{\nu\sigma}{}^{\lambda}\Gamma_{\mu\rho}{}^{\sigma}.
\eea The Ricci tensor and scalar are :

\bea
\cR_{\mu\nu} &=& \cR_{\mu\lambda\nu}{}^{\lambda}\nn\\[2mm]
\cR &=& g^{\mu\nu}\cR_{\mu\nu}. \eea Under a Weyl rescaling

\bea
g_{\mu\nu} &=& \Omega^{-2}\,\hg_{\mu\nu}\nn\\[2mm]
g^{\mu\nu} &=& \Omega^2\,\hg^{\mu\nu}\label{eg6}\\[2mm]
\sqrt{-g} &=& \Omega^{-d}\,\sqrt{-\hg} \eea it behaves like

\bea \int \rd x^d\,\sqrt{-g}\, \Omega^{d-2}\,\cR &=& \int \rd
x^d\,\sqrt{-\hg}\, \left(\cR
+(d-1)(d-2)(\frac{\del\Omega}{\Omega})^2\right). \eea Finally,
the covariant derivative acts on a Majorana spinor $\eta_{\ax}$
with the connection $\Gamma_{mab}$ as

\beq \nabla_m\eta_{\ax} =
\del_m\eta_{\ax}-\frac14\Gamma_{mab}(\Gamma^{ab})_{\ax}{}^{\bx}\eta_{\bx}
\label{conn1} \eeq where $\Gamma^{ab}$ is the antisymmetrized
product of two $\Gamma$-matrices.

\section{Forms}\label{forms}

A p-form is defined by :

\bea F_p &=& \frac{1}{p!}F_{M_1M_2\ldots M_p}\rd y^{M_1}\rd
y^{M_2}\ldots \rd y^{M_p}. \eea The Hodge dual is

\bea *F_p &=& \frac{1}{\sqrt{-g}}\frac{1}{p!(d-p)!}F_{M_1M_2\ldots
M_p}\epsilon^{M_1M_2\ldots M_p}{}_{M_{p+1}\ldots M_d}\rd
y^{M_{p+1}}\ldots \rd y^{M_d}.\label{forms02} \eea where we use
for the epsilon tensor the convention $\epsilon^{12\ldots d}=+1$
and

\bea \epsilon_{M_1M_2\ldots M_d} &=& g_{M_1N_1}\ldots g_{M_d
N_d}\epsilon^{N_1\ldots N_d}. \label{eps1}\eea This tensor enjoys
the following properties

\bea
\epsilon^{M_1\ldots M_d}\epsilon_{M_1\ldots M_d} &=& g\cdot\, d!\nn\\[2mm]
\epsilon^{M_1M_2\ldots M_d}\epsilon_{N_1M_2\ldots M_d} &=& g\cdot\,
 (d-1)!\delta^{M_1}_{N_1}\nn\\[2mm]
\epsilon^{M_1M_2M_3\ldots M_d}\epsilon_{N_1N_2M_3\ldots M_d} &=& g\cdot\,
 (d-2)!\delta^{M_1M_2}_{N_1N_2}\nn\\[2mm]
\epsilon^{M_1\ldots M_p\ldots M_d}\epsilon_{N_1\ldots
N_pM_{p+1}\ldots M_d} &=& g\cdot\, (d-p)!\delta^{M_1\ldots
M_p}_{N_1\ldots N_p}. \label{eps2}\eea where $g=\det(g_{MN})$ and
the $\delta$-symbols satisfy

\bea \delta^{M_1\ldots M_p}_{N_1\ldots N_p}\:F_{M_1\ldots M_p}
&=& p!\,F_{N_1\ldots N_p}. \eea Then we have that

\bea
H_p*F_p &=& -\frac{1}{p!}H^{M_1\ldots M_p}F_{M_1\ldots M_p}\sqrt{-g}d^dy\nn\\[2mm]
**F_p &=& -(-)^{p(d-p)}F_p, \eea the extra minus sign being due
to the negative signature of the minkowsky space (absent for
Euclidean manifolds). Suppose the whole space-time splits into
$\cM_{10} = \cM_4\times {\cal I}_6$, where ${\cal I}_6$ is an
internal manifold. Then the coordinates of both spaces do not
mix, and it is possible to define the Hodge dual on each space.
In this particular case, the 10-dimensional Hodge dual splits in
the product of 2 forms, one belonging to each space, in the
following way

\beq *_{10}(E_n\wg I_p) = (-)^{np} *_4 E_n\wg *_6
I_p,\label{forms01}\eeq where $*_4$ and $*_6$ are defined as in
(\ref{forms02}), $E_n$ and $I_p$ are an external $n$-form and an
internal $p$-form.

Finally, the action of the differential operator $d$, which
brings a p-form to a (p+1)-form, reads

\bea d F_p &=& \frac{1}{p!}\del_N F_{M_1\ldots M_p}\, d y^N d
y^{M_1}\ldots d y^{M_p} \label{forms03}\eea and satisfies

\bea d (F_pG_q) &=& d F.G +(-)^pF. d G. \label{forms7}\eea It is a
nilpotent operator

\beq dd F_p = 0\label{forms8}\eeq and satisfies Stoke's theorem

\beq \int_{M} d F_p = \int_{\del M} F_p.\label{forms9}\eeq Its
conjugate $d{\dagger}$ brings a p-form to a (p-1)-form. The
generalized Laplacian is

\beq \Delta = dd^{\dagger} + d^{\dagger}d,\label{forms10}\eeq and
the component expression of these operators\footnote{For Euclidean
spaces, signs are opposite.} is

\bea d^{\dagger} F_p & = &
-\frac{1}{(p-1)!}(-)^{(p-1)(d-p)}\nabla^{\nu}(F_p)_{\nu\nu_1\ldots\nu_{p-1}}\rd
x^{\nu_1}\ldots\rd x^{\nu_{p-1}}\label{forms11}\\[2mm]
(\Delta F_p)_{\nu_1\ldots\nu_p} & = &
-\nabla^{\nu}\nabla_{\nu}(F_p)_{\nu_1\ldots\nu_p}+
\sum_i[\nabla^{\nu}\, ,\,
\nabla_{\nu_i}](F_p)_{\nu_1\ldots\nu_{(i)}\ldots\nu_p}.\label{forms12}\eea
Using the positivity of the scalar product (for Euclidean spaces)

\beq <F_p|H_p> = \int F_p\wg * H_p\label{forms13}\eeq and
considering the product $<F_p|\Delta F_p>$, one can show that a
form is harmonic if and only if it is closed and co-closed

\beq \Delta F_p = 0 \Longleftrightarrow d F_p = 0 \quad {
\rm{and}}\quad d^{\dagger} F_p =0.\label{forms14}\eeq

\section{Clifford algebra in 6 Euclidean dimensions}\label{cad}

For a review of these results, see \cite{AvP}. We consider
Majorana spinors. There are six Gamma matrices, of dimension $8$
obeying Clifford algebra

\beq \{ \Gamma_m\, ,\, \Gamma_n\} = 2\eta_{mn}.\label{cad1}\eeq
Their conjugation relation is $\Gamma_m^{\dagger} = \Gamma_m$. The
chirality operator $\Gx_7$ is defined as

\begin{equation}
  \Gx_7 = i \Gx_1 \ldots \Gx_6 \label{cad01}\end{equation} and satisfies the same
conjugation relation $(\Gamma_7)^{\dagger} = \Gamma_7$. Majorana
spinors on the 6 dimensional internal space can be defined if we
adopt the following conventions for the charge conjugation matrix
$\cC$

\begin{equation}
  \cC^T = \cC \ , \qquad \qquad \Gx_m^T = - \cC \Gx_m \cC^{-1} \ ,
\label{cad02}\end{equation} while the Majorana condition on a
spinor $\eta$ reads
\begin{equation}
  \eta^\dagger = \eta^T \cC.\label{cad03}
\end{equation} From the single $\Gamma$-matrices, we can build the antisymmetrized
product of k matrices

\beq (\Gamma^{(k)})_{m_1m_2\ldots m_k} =
\frac{1}{k!}\sum_{\sigma\in
S^k}\epsilon(\sigma)\Gamma_{m_{\sigma(1)}}\Gamma_{m_{\sigma(2)}}\ldots\Gamma_{m_{\sigma(k)
}}, \label{cad2}\eeq which is a basis for all 8-dimensional
matrices for k from 0 to 6. For $k=0$ and $k=3$ (modulo 4),
$\Gamma^{(k)}$ is symmetric\footnote{More precisely, the symmetry
properties are true for $\Gamma^{(k)}{\cal C}$.}, and for $k=1$
or $k=2$ it is antisymmetric. These symmetry properties of the
gamma matrices and $\cC$ with the above conventions imply that
for a commuting Majorana spinor $\eta$ the following quantities
vanish

\begin{equation} \eta^\dagger \Gx_{(1)} \eta = \eta^\dagger \Gx_{(2)} \eta =
\eta^\dagger \Gx_{(5)} \eta = \eta^\dagger \Gx_{(6)} \eta = 0\ ,
\label{cad05}\end{equation} We will meet several times the
product of two such matrices. As an 8-dimensional matrix, this
product can be expanded on the $\Gamma^{(k)}$. The (heuristic)
rule is the following. Take the product $\Gamma_{m_1\ldots
m_k}\Gamma^{n_1\ldots n_p}$. It will be a linear combination of
$\Gamma^{l}$ for l from 0 to $k+p$. Since the only tensor we can
use, other than the matrices, is the $\delta$ which has two
indices, the only terms which survive are the ones of order
$k+p$, $k+p-2$ and so on, until there are no ways to take indices
from the lowest order matrix in the product. The expansion reads
 \beq \Gamma_{m_1\ldots
m_k}\Gamma^{n_1\ldots n_p} = \Gamma_{m_1\ldots m_k}{}^{n_1\ldots
n_p}+ A\dx_{[m_1}{}^{[n_1}\Gamma_{m_2\ldots m_k]}{}^{n_2\ldots
n_p]}+... \label{cad3}\eeq For the sign of $A$, look at the first
term $\Gamma_{m_1\ldots m_k}{}^{n_1\ldots n_p}$ and imagine you
take $n_1$ to the right of $m_1$. Then you have to go through
$m_k , m_{k-1}\ldots m_2$. Each time you get a minus sign, so the
sign of $A$ is $(-)^{k-1}$. Its absolute value is $C_k^1C_p^1$,
where the 1 stands for the number of indices you take in the
$\Gamma$, and $C$ is the combination $C_n^l =
\frac{n!}{l!(n-l)!}$. The next term will have an order 2
$\delta$, its sign will be the previous sign times the sign
obtained when going with $n_2$ through $m_k , m_{k-1}\ldots m_3$
to go to the right of $m_2$, so$(-)^{k-1}(-)^{k-2}$, and its
coefficient will be $C_k^2C_p^2$, and so on. Suppose $k<p$, then
the last term in the expansion will be proportional to
$\dx_{m_1m_2\ldots m_k}^{[n_1n_2\ldots n_k}\Gamma^{n_{k+1}\ldots
n_p]}$. The absolute value of its coefficient will be
$C_k^kC_p^k$. Let's take an explicit example to make things
clearer. Following the above rules, it can be checked that

\bea \Gamma_{m_1m_2m_3}\Gamma^{n_1n_2n_3n_4} & = &
\Gamma_{m_1m_2m_3}{}^{n_1n_2n_3n_4}
+3*4\delta_{[m_1}^{[n_1}\Gamma_{m_2m_3]}{}^{n_2n_3n_4]}\nn\\[2mm]
&&-3*6\delta_{[m_1m_2}^{[n_1n_2}\Gamma_
{m_3]}{}^{n_3n_4]}-1*4\delta_{m_1m_2m_3}^{[n_1n_2n_3}\Gamma^{n_4]}.\label{cad4}\eea

\section{Cohomology and homology classes}\label{coho}

A p-form $F_p$ is said to be \emph{closed} iff $\rd F_p = 0$. It
is \emph{exact} iff there exists a (p-1)-form $G_{p-1}$ such that
$F_p = \rd G_{p-1}$. Obviously, using the property of the
derivation (\ref{forms8}), an exact form is closed. But the
converse needs not be true.

Let's define the group $Z^p$ of closed p-forms, with additive
law, and the group $B^p$ of exact p-forms. Then the group $H^p =
Z^p/B^p$ contains the classes of closed forms which are equal up
to an exact form. The harmonic forms are examples of closed but
not exact forms. A classical result is that there is actually a
unique harmonic form in each cohomology class. Thus the number of
harmonic p-forms is exactly the dimension of $H^p$. Finally we
notice that, when $F$ is harmonic, so is $*F$. This implies
Poincar\'e duality $H^p\sim H^{n-p}$. Let $h^p$ be the dimension
of $H^p$ (Betti number). Then the Euler characteristic $\chi$ is

\beq \chi = \sum_p (-)^ph^p.\label{coho1}\eeq

A p-dimensional submanifold $\gamma_p$ is a \emph{p-cycle} iff it
has no boundary $\del \gamma_p =0$. Since a boundary has no
boundary, $\del$ is nilpotent. Thus the boundaries are cycles.
But again, the converse needs not be true, there can be cycles
which are not boundaries of some submanifold.

Let's define the group $Z_p$ of p-cycles, and the group $B_p$ of
p-boundaries. Then the group $H_p = B_p/Z_p$ contains the classes
of p-cycles which are equal up to a boundary.

The analogy between homology and cohomology is striking. Indeed,
De Rham's theorem states that $H^p\sim H_p$. Since we expand all
our forms on harmonic forms, the Betti numbers, the numbers of
such forms, are extremely important to us. These numbers are
topological invariants, which can be found by looking for
independent p-cycles.

\section{Almost complex, complex and K\"ahler manifolds}\label{ack}

For a detailed review of the notion of (complex) manifolds, see
\cite{Nakahara:1990th,candelasTS}. Here we briefly recall the main
results. A differentiable manifold is described by a set of
patches, that can overlap. On the overlap of two patches, the two
sets of coordinates are related by a $C_{\infty}$ diffeomorphism.
A manifold of even dimension 2n is locally diffeomorph to
$R^{2n}=C^n$. If moreover, the changes of coordinates are
holomorphic, then the manifold is complex. An other way to define
the notion of complex manifold is the following. If there is a
globally defined 2-tensor $J_m{}^n$ squaring to $-1$, the
manifold is called almost complex, and $J$ is the almost complex
structure. This does not mean that $J$ can be used to define
complex coordinates globally. To do so, the almost complex
structure must be integrable. This is quantified by its
"torsion", the Nijenhuis tensor \cite{Green:1987mn}

\beq N_{mn}{}^k = J_m{}^l(\nabla_lJ_n^k -
\nabla_nJ_l{}^k)-J_n{}^l(\nabla_lJ_m^k - \nabla_mJ_l{}^k).
\label{ack1}\eeq If $N$ vanishes, then the manifold is complex.
This means that complex coordinates can be used safely. Let's
denote them $z^{\alpha}, \bar z^{\bar\alpha}$, $\alpha,\ab$
taking values in $1,2,3$ and $\bar 1,\bar 2,\bar 3$. The complex
structure takes the form

\beq J_{\ax}{}^{\bx} = +i\dx_{\ax}{}^{\bx}\quad ,\quad
J_{\ab}{}^{\bb} = -i\dx_{\ab}{}^{\bb}\quad ,\quad J_{\ax}{}^{\bb}
= J_{\ab}{}^{\bx} = 0.\label{ack2}\eeq It is always possible to
choose a hermitian metric for which only the mixed components are
non-vanishing. Lowering the indices of the complex structure with
this hermitian metric, one can check that $J_{mn} = -J_{nm}$. It
is thus natural to define a 2-form out of $J$

\beq J = \frac{1}{2!}J_{mn}\rd y^m\rd y^n. \label{ack3}\eeq If
$J$ is closed, it is called the K\"ahler class and the manifold
is called K\"ahler. Some very important properties come from this
assumption

\beq \rd J = 0. \label{ack4}\eeq First of all, in complex
component, (\ref{ack4}) can be rewritten

\beq \del_{\alpha}g_{\beta\bb} =
\del_{\bx}g_{\ax\bb}\label{ack5}\eeq which means that the metric
can be written in terms of a (K\"ahler) potential $K$

\beq g_{\ax\ab} = \del_{\ax}\del_{\ab} K.\label{ack6}\eeq This in
turns implies that the only non-vanishing Christoffell symbols
are completely pure in their indices

\beq \Gamma_{\ax\bx}{}^{\cx} =
g^{\cx\cb}\del_{\ax}g_{\bx\cb}.\label{ack7}\eeq For the Riemann
tensor, it can be checked that the only non-vanishing component is
$R_{\ax\ab\bx\bb}$ and the ones obtained by complex conjugation
or using the symmetry property

\beq  R_{[mnp]}{}^q = 0.\label{ack8}\eeq

\section{Homogeneous functions of degree 2}

Let $F(X)$ be a homogeneous function of degree 2 of the scalars
$X^1, X^2, \ldots X^n$. This means that $F$ is a polynomial of
the $X$'s, with integer powers, such that the sum of all powers
is 2. $F$ can be written as

\bea F(X) & = & \sum
a_{\lambda_1\lambda_2\ldots\lambda_n}(X^1)^{\lambda_1}(X^2)^{\lambda_2}\ldots
(X^n)^{\lambda_n} \eea where the sum is other all n-uplets
$(\lambda_1\ldots\lambda_n)$ in $Z^n$ such that
$\lambda_1+\ldots+\lambda_n = 2$ (with only a finite number of
non-zero n-uplets). Obviously such a function has the important
property $F(\alpha X^1,\alpha X^2\ldots \alpha X^n) = \alpha^2
F(X^1,X^2\ldots X^n)$, which is precisely the reason why it is
used in $N=2$ supergravities.

Let $F_I$ be the derivative of $F$ with respect to $X^I$. Then

\begin{equation}\begin{array}{lllll} F_1 & = & \frac{\del}{\del X^1} F & = & \sum
\lambda_1
a_{\lambda_1\lambda_2\ldots\lambda_n}(X^1)^{\lambda_1-1}(X^2)^{\lambda_2}\ldots
(X^n)^{\lambda_n}\nn\\[2mm]
\cdot&&&&\nn\\[2mm]
\cdot&&&&\nn\\[2mm]
\cdot&&&&\nn\\[2mm]
F_n & = & \frac{\del}{\del X^n} F & = & \sum \lambda_n
a_{\lambda_1\lambda_2\ldots\lambda_n}(X^1)^{\lambda_1}(X^2)^{\lambda_2}\ldots
(X^n)^{\lambda_n-1}. \end{array}\end{equation} From this we deduce

\bea X^IF_I & = & \sum (\lambda_1+\ldots +\lambda_n)
a_{\lambda_1\lambda_2\ldots\lambda_n}(X^1)^{\lambda_1}(X^2)^{\lambda_2}\ldots
(X^n)^{\lambda_n} = 2F.\label{homo4}\eea Now we successively
differentiate (\ref{homo4}) with respect to $X^J,X^K$ and we
obtain

\bea X^IF_I & = & 2F\label{homo5}\\[2mm]
X^IF_{IJ} & = & F_J\label{homo6}\\[2mm]
X^IF_{IJK} & = & 0.\label{homo7} \eea

\chapter{Calabi-Yau manifolds}\label{AppB}

\section{Main properties of $CY_3$}\label{cyp}

We will start from the definition involving a spinor. \emph{A
Calabi-Yau manifold admits exactly one covariantly constant
spinor}. Using this spinor $\eta$, we build the tensor

\beq J_m{}^n = -i\eta^{\dagger}\Gamma_m{}^n\Gamma_7\eta
\label{cyp1}\eeq and we want to show that it squares to $-1$. We
evaluate the expression

\bea J_m{}^nJ_n{}^p & = & -\eta^{\dagger\, a
}\eta^b\eta^{\dagger\, c }\eta^d
(\Gamma_m{}^n\Gamma_7)_{ab}(\Gamma_n{}^p\Gamma_7)_{cd}\nn\\[2mm]
& = & -\eta^{\dagger\, a }\eta^b\eta^{\dagger\, c }\eta^d
(M_m{}^p)_{abcd}.\label{squ1}\eea We rearrange the spinor indices
with Fierz method

\bea (M_m{}^p)_{abcd} & = & (k_m{}^p)_{ad}\dx_{cb} +
(k_{nm}{}^p)_{ad}(\Gamma^n)_{cb} +
(k_{nqm}{}^p)_{ad}(\Gamma^{nq})_{cb}\nn\\[2mm]
&& + (k_{lnqm}{}^p)_{ad}(\Gamma^{lnq})_{cb}+ (\tilde
k_{nqm}{}^p)_{ad}(\Gamma^{nq}\Gamma_7)_{cb}\nn\\[2mm]
&& + (\tilde k_{nm}{}^p)_{ad}(\Gamma^{n}\Gamma_7)_{cb}+ (\tilde
k_{m}{}^p)_{ad}(\Gamma_7)_{cb}.\label{squ2} \eea Considering the
symmetry of spinor indices in (\ref{squ1}), we only compute the
coefficients $(k_m{}^p)_{ad}$, $(k_{lnqm}{}^p)_{ad}$ and $(\tilde
k_{nqm}{}^p)_{ad}$. Multiplying (\ref{squ2}) respectively by
$\dx^{bc}$, $(\Gamma_{lnq})^{bc}$ and
$(\Gamma_{nq}\Gamma_7)^{bc}$, we obtain

\bea k_m{}^p & = & \frac58 \dx_m^p\nn\\[2mm]
k_{lnqm}{}^p & = & -\frac{1}{48}\dx_m^p\Gamma_{lnq}\nn\\[2mm]
k_{nqm}{}^p & = & -\frac{1}{16}\dx_m^p\Gamma_{nq}\Gamma_7-\frac14
g_{m[n}\Gamma_{q]}{}^p\Gamma_7-\frac14\dx_{[n}^p\Gamma_{q]m}\Gamma_7.
\eea This leads to

\beq \frac32 J_m{}^nJ_n{}^p +\frac{1}{16}\dx_m^p J_{nq}J^{nq} =
-\frac98\dx_m^p \label{squ3}\eeq and the final conclusion

\beq J_m{}^nJ_n{}^q = -\dx_m^p.\label{squ4}\eeq

$J$ is thus an almost complex structure. Since $\eta$ is
covariantly constant (with Christoffell connection), so is $J$,
and the Nijenhuis tensor (\ref{ack1}) vanishes : the complex
structure is integrable and the manifold is complex. Moreover,
since $J$ is covariantly constant, it is obviously K\"ahler. The
last defining property, Ricci-flatness, comes from a consistency
condition on the spinor. The fact that $\eta$ is covariantly
constant is written, according to (\ref{conn1})

\beq \nabla_m\eta_{\ax} =
\del_m\eta_{\ax}-\frac14\Gamma_{mab}(\Gamma^{ab})_{\ax}{}^{\bx}\eta_{\bx}
=0.\label{cov1}\eeq Applying an other covariant derivative and
taking the commutator, we find

\beq [\nabla_m\, ,\,\nabla_n]\eta = -R_{mnpq}\Gamma^{pq}\eta =
0.\label{cov2}\eeq We want a constraint on $R$, so we multiply
successively by $\Gamma^n$, $\Gamma^l$ and $\eta^{\dagger}$.
Using the identities

\bea \Gamma^n\Gamma^{pq} & = & \Gamma^{npq} +
2g^{n[p}\Gamma^{q]}\label{cov3}\\[2mm]
\Gamma_l\Gamma^{npq} & = & \Gamma_l{}^{npq} +
3\dx_l^{[n}\Gamma^{pq]}\label{cov4}\\[2mm]
\Gamma_l\Gamma^q & = & \Gamma_l{}^q +\dx_l^q\label{cov5} \eea and
the symmetry properties (\ref{cad05}), we obtain

\beq \eta^{\dagger}\Gamma_l{}^{npq}\eta R_{mnpq} - 2R_{ml} =
0.\eeq Since $R$ is subject to the Bianchi identity

\beq R_{m[npq]} = 0,\eeq the only surviving term expresses
Ricci-flatness

\beq R_{mn} = 0.\label{Ric1}\eeq For a proof of the structure of
the Hodge diamond (\ref{cyr5}), see \cite{candelasTS}.

\section{Integrals on $CY_3$}\label{icy}

On the Calabi-Yau manifold one can define complex coordinates
$\xi^i$

\bea \xi^1 = \frac{y^1+iy^2}{\sqrt{2}} ;\xi^2 =
\frac{y^3+iy^4}{\sqrt{2}} ;\xi^3 = \frac{y^5+iy^6}{\sqrt{2}}.
\label{icy1}\eea We define the three-dimensional epsilon tensors
such that

\bea d\xi^{\ax}d\xi^{\bx}d\xi^{\cx}d\xi^{\ab} d\xi^{\bb}
d\xi^{\cb} &=& \epsilon^{\ax\bx\cx}\epsilon^{\ab\bb\cb}d^6\xi
\label{icy2}\eea that is to say
$\epsilon^{123}=\epsilon^{\bar{1}\bar{2}\bar{3}}=+1$ and
$d^6\xi=d\xi^1d\xi^2d\xi^3d\xi^{\bar{1}}d\xi^{\bar{2}}d\xi^{\bar{3}}=-id^6y$.
The indices are lowered with the metric as in (\ref{eps1}), and,
keeping in mind that $g_{ij}=g_{\bi\bj}=0$, one has the
properties similar to (\ref{eps2})

\bea \epsilon^{\ax\bx\cx}\epsilon_{\ax\bx\cx} &=& \sqrt{g}3!
\;\ldots \label{icy3}\eea

The relation with the 6-dimensional epsilon tensor is the
following. We define the real $\epsilon$-symbol by
$\epsilon^{123456} = +1$. The indices
  are lowered with the metric. It follows that in terms of complex
  indices  one has

  \begin{equation}
    \epsilon^{\ax \bx \cx \ab \bb \cb} = - i \epsilon^{\ax \bx \cx} \,
    \epsilon^{\ab \bb \cb} \ .
  \label{icy4}\end{equation}

\subsection{(1,1)-form sector} Harmonic $(1,1)$-forms are denoted by $(\ox_i)_{\ax\bb}$,
 $i$ running from $1$ to
$h^{(1,1)}$. The integrals we abbreviate as

\begin{eqnarray}
  \label{K}
  \cK & = & \frac16 \int_{Y} J \wg J \wg J  \, , \quad  \cK_i = \int_{Y}
  \ox_i \wg J \wg J \ , \\
  \cK_{ij} & = & \int_{Y} \ox^i \wg \ox ^j \wg J \ , \quad \cK_{ijk} =
  \int_{Y} \ox^i \wg \ox^j \wg \ox^k  \ ,\nn
\end{eqnarray} where $\cK$ is the volume and $J$ is the K\"ahler form which can be
expanded in terms of the basis $\ox_i$  as
\begin{equation}\label{Jexp}
  J= v^i \ox_i \ .
\end{equation} This implies the following identities

\bea \cK_{ijk}v^k = \cK_{ij}\quad & ; & \quad \cK_{ij}v^j = \cK_{i}\nn\\[2mm]
 \cK_{i}v^i = 6\cK\label{icy04} \eea We also define the metric on the
 complexified K\"ahler cone

\begin{equation}
  \label{gH11}
  g_{i j}  = \frac{1}{4 \cK} \int_{Y} \ox_i \wg * \ox_j \ , \end{equation} which, using
  \cite{AS3}

\beq *\ox_i = -J\wg \ox_i + \frac{\cK_i}{4\cK}J\wg J,
\label{icy5}\eeq can be rewritten

\begin{equation}
  \label{icy6}
  g_{i j}  = -\frac{1}{4\cK}\left(\cK_{ij}-\frac{1}{4\cK}\cK_i\cK_j\right). \end{equation}

On a \CY{} threefold $H^{2,2} (Y)$ is dual to $H^{(1,1)} (Y)$ and
it is useful to introduce the dual basis $\tilde \ox^i$
normalized by \begin{equation}
  \label{normH2}
  \int_{Y} \ox_i \wg \tilde \ox^j\ =\ \dx_i^j\ .
\end{equation} With this normalization the following relations hold \begin{equation}
  \label{oxstar}
  g^{ij} = 4 \cK \int_{Y} \tilde \ox^i \wg * \tilde \ox^j \, , \quad
  * \ox_i = 4 \cK g_{ij} \tilde \ox^j \, , \quad
  * \tilde \ox^i = \frac{1}{4 \cK}\, g^{ij} \ox_j \, ,
\quad \ox_i \wg \ox_j \sim \cK_{ijk} \tilde \ox^k \, ,
\end{equation} where the symbol $\sim$ denotes the fact that the
quantities are in the same cohomology class.

\subsection{3-form sector}

There are two standard choices of basis for the 3-form sector.
One is obviously complex, with $(\eta_a)_{\ax\bx\cb}$, $a$
running from $1$ to $h^{(2,1)}$, a basis for the $(2,1)$-forms,
and $\Omega_{\ax\bx\cx}$ is the unique holomorphic $(3,0)$-form.
The other choice is a complete set of $2(h^{(2,1)}+1)$ real forms
$\ax_A\, ,\,\bx^A$, $A$ running from $0$ to $h^{(2,1)}$. This
basis is orthonormal in the following sense

\bea \int_Y\ax_A\wg\ax_B = \int_Y\bx^A\wg\bx^B = 0\label{icy30}\\[2mm]
\int_Y\ax_A\wg\bx^B = -\int_Y\bx^B\wg\ax_A =
\dx_A^B.\label{icy31} \eea The $(3,0)$-form $\Ox$ can be expanded
on this basis with coefficients to be interpreted later

\beq \Ox = z^A\ax_A - \cF_A\bx^A. \label{icy32}\eeq Since $\Ox$
is covariantly constant, $||\Ox||^2$ defined by

\beq ||\Ox||^2 \equiv \frac{1}{3!} \Ox_{\ax\bx\cx}\bar
\Ox^{\ax\bx\cx}\label{icy33}\eeq is a constant on the Calabi-Yau
manifold.

\section{Lichnerowicz's equation}\label{lse}

Consider a deformation of the metric $g_{mn} = g_{mn}^0 +
\dg_{mn}$ such that $g_{mn}^0$ is hermitian with vanishing Ricci
tensor. Using invariance under diffeomorphisms and tracelessness
of the metric, we are allowed to impose the following constraints
on $\dg$

\bea g^{0\, mn}\dg_{mn} & = & 0\label{lse1}\\[2mm]
\nabla^m\dg_{mn} & = & 0. \label{lse2}\eea The first order
variation of the Christoffel symbols can be written as

\beq \dx\Gamma_{mn}{}^p =  \frac{1}{2}g^{0\, pl}(\nabla_{m}
\dg_{nl}+\nabla_{n} \dg_{ml}-\nabla_{l} \dg_{mn})\label{lse3}\eeq
and leads to the variation of the Riemann tensor

\beq \dx R_{mnp}{}^q =  \nabla_{m}\dx
\Gamma_{np}{}^q-\nabla_{n}\dx \Gamma_{mp}{}^q.\label{lse4}\eeq
Since we want our manifold to remain Calabi-Yau, we have to impose

\beq \dx R_{mn} = 0.\label{lse5}\eeq From (\ref{lse1}), one can
immediately see that

\beq \dx\Gamma_{lm}{}^l = 0.\label{lse6}\eeq Plugging this in
(\ref{lse4}) and using (\ref{lse2}), we find Lichnerowicz equation

\beq \nabla^l\nabla_l\dg_{mn}-[\nabla^l\, ,\,
\nabla_m]\dg_{ln}-[\nabla^l\, ,\, \nabla_n]\dg_{lm} = 0
.\label{lse8}\eeq This splits into two equations, one on the
mixed variations and one on the pure variations. Taking

\beq \dx g_{\ax\ab} = -iv^i(\ox_i)_{\ax\ab}\eeq obviously solves
(\ref{lse8}), but for pure variations, the problem is more
tricky. First of all there are no (2,0)-forms on the Calabi-Yau,
and second of all, $\dx g_{\ax\bx}$ is symmetric. We show now that

\beq \dg_{\ax\bx} = \frac{i}{||\Ox||^2}\bar
z^a(\bar\eta_a)_{\ax\bb\cb}\Ox^{\bb\cb}{}_{\bx}\eeq is indeed
symmetric and solution to (\ref{lse8}). Consider

\beq L_{\ax\bx} \equiv
(\bar\eta_a)_{\ax\bb\cb}\Ox^{\bb\cb}{}_{\bx} -
(\bar\eta_a)_{\ax\bb\cb}\Ox^{\bb\cb}{}_{\bx}\label{syme1}\eeq and
multiply this equation by $\bar\Ox^{\bx}{}_{\bar\lambda\bar\mu}$.
We find

\beq L_{\ax\bx}\bar\Ox^{\bx}{}_{\bar\lambda\bar\mu} =
-2(\bar\eta_a)^{\bar\rho}{}_{\bar\rho\bar\lambda}g_{\ax\bar\mu}
-2(\bar\eta_a)^{\bar\rho}{}_{\bar\mu\bar\rho}g_{\ax\bar\lambda}
.\label{syme2}\eeq From this we can see that if
$(\bar\eta_a)^{\bar\rho}{}_{\bar\rho\bar\lambda}=0$,
$L_{\ax\bx}=0$. Conversely, by contracting with
$g^{\ax\bar\lambda}$ we obtain

\beq L_{\ax\bx}\bar\Ox^{\bx\ax}{}_{\bar\mu} =
4(\bar\eta_a)^{\bar\rho}{}_{\bar\rho\bar\mu}\label{syme4}\eeq
such that if $L_{\ax\bx}=0$, then
$(\bar\eta_a)^{\bar\rho}{}_{\bar\rho\bar\mu}=0$. Finally, $\dx
g_{\ax\bx}$ is symmetric iff
$(\bar\eta_a)^{\bar\rho}{}_{\bar\rho\bar\mu}=0$. Let us write the
fact that $\bar\eta_a$ is harmonic (\ref{forms12})

\bea \nabla^{m}\nabla_{m}(\bar\eta_a)_{\ax\bb\cb} & - &
[\nabla^{m}\,
,\, \nabla_{\ax}](\bar\eta_a)_{m\bb\cb}\nn\\[2mm]
& - & [\nabla^{m}\, ,\, \nabla_{\bb}](\bar\eta_a)_{\ax m \cb}\nn\\[2mm]
& - & [\nabla^{m}\, ,\, \nabla_{\cb}](\bar\eta_a)_{\ax\bb m} = 0
.\label{syme5}\eea We want to compute

\bea L_{\cb} \equiv
\nabla^{m}\nabla_{m}(\bar\eta_a)^{\bb}{}_{\bb\cb} & - &
[\nabla^{m}\, ,\, \nabla_{\cb}](\bar\eta_a)^{\bb}{}_{\bb\cb}
,\label{syme6}\eea therefore we contract (\ref{syme5}) with
$g^{\ax\bb}$. To evaluate the commutators, we use (\ref{comm1})
and the Ricci flatness. We find

\beq L_{\cb} = -R^{\bx}{}_{\ax\cb\dx}(\bar\eta_a)_{\bx}{}^{\ax\dx}
-R^{\db}{}_{\bb\cb\dx}(\bar\eta_a)^{\bb}{}_{\db}{}^{\dx} = 0.
\eeq This means that $(\bar\eta_a)^{\bb}{}_{\bb\cb}$ is a
harmonic 1-form. Since there are no such forms on a Calabi-Yau
manifold, $(\bar\eta_a)^{\bb}{}_{\bb\cb}$ is zero which is the
final proof for the symmetry of $\dx g_{\ax\bx}$.

Lichnerowicz equation on $\dx g_{\ax\bx}$ reads

\beq \nabla^l\nabla_l\dg_{\ax\bx}-[\nabla^{\cx}\, ,\,
\nabla_{\ax}]\dg_{\cx\bx}-[\nabla^{\cx}\, ,\,
\nabla_{\bx}]\dg_{\cx\ax} = 0.\label{lse20}\eeq Plugging the
expression (\ref{ms3}) for $\dx g_{\ax\bx}$, and contracting with
$\bar\Ox_{\bb\cb}{}^{\bx}$, this is equivalent to

\bea \nabla^{m}\nabla_{m}(\bar\eta_a)_{\ax\bb\cb} & + &
2R^{\cx}{}_{\ax}{}^{\bar\rho}{}_{\bb}
(\bar\eta_a)_{\cx\bar\rho\cb}-2R^{\cx}{}_{\ax}{}^{\bar\rho}{}_{\cb}
(\bar\eta_a)_{\cx\bar\rho\bb}= 0 .\label{syme8}\eea This is
exactly the equation of harmonicity of $\bar\eta_a$
(\ref{syme5}). Thus (\ref{ms3}) is indeed solution to
Lichnerowicz equation.

\section{Compactification of the Ricci scalar}\label{crs}

We perform an expansion of the Ricci scalar up to order 2 in the
moduli. The components of the metric and its inverse are

\bea g_{\ax\bx} & = & 0 + \bar z^a (\bar b_a)_{\ax\bx}\label{crs1}\\[2mm]
g_{\ax\ab} & = & g_{\ax\ab}^0 -i v^i (\ox_i)_{\ax\ab}\label{crs2}\\[2mm]
g^{\ax\bx} & = & 0 - z^a (b_a)_{\ab\bb}g^{0\,\ax\ab}g^{0\,\bx\bb}\label{crs3}\\[2mm]
g^{\ax\ab} & = & g^{0\,\ax\ab} + i v^i
(\ox_i)_{\bx\bb}g^{0\,\ax\bb}g^{0\,\bx\ab}\label{crs4}\eea with

\beq (\bar b_a)_{\ax\bx} =
\frac{i}{||\Ox||^2}(\bar\eta_a)_{\ax\bb\cb}\Ox^{\bb\cb}{}_{\bx}.
\label{crs5}\eeq Apart from the purely space-time ones, the
non-vanishing Christoffell symbols are

\bea \Gamma_{\mu\ax}{}^{\bx} & = &
\frac{1}{2}(\ox_i)_{\cx\cb}(\ox_j)_{\ax\bb}
g^{0\,\cx\bb}g^{0\,\bx\cb}v^i\del_{\mu}v^j-\frac{i}{2}(\ox_i)_{\ax\bb}g^{0\,
\bx\bb}
\del_{\mu}v^i\nn\\[2mm]
&&-\frac{1}{2}(b_a)_{\bb\cb}(\bar
b_b)_{\ax\cx}g^{0\,\cx\bb}g^{0\,\bx\cb}
z^a\del_{\mu}\bar z^b \label{crs6}\\[2mm]
\Gamma_{\mu\ax}{}^{\bb} & = & \frac{1}{2}(\bar
b_a)_{\ax\bx}g^{0\,\bx\bb}\del_{\mu}\bar z^a +
\frac{i}{2}(\ox_i)_{\cx\cb}(\bar
b_a)_{\ax\bx}g^{0\, \bx\cb}g^{0\, \cx\bb}v^i\del_{\mu}\bar z^a\nn\\[2mm]
&&+\frac{i}{2}(\ox_i)_{\ax\cb}(\bar b_a)_{\cx\bx}g^{0\,
\bx\cb}g^{0\,
\cx\bb}\del_{\mu}v^i\bar z^a \label{crs7}\\[2mm]
\Gamma_{\ax\bx}{}^{\mu} & = & -\frac{1}{2}(\bar
b_a)_{\ax\bx}\del^{\mu}\bar z^a
\label{crs8}\\[2mm]
\Gamma_{\ax\bb}{}^{\mu} & = &
+\frac{i}{2}(\ox_i)_{\ax\bb}\del^{\mu}v^i. \label{crs9}
 \eea
The 10-dimensional Ricci scalar has the following decomposition

\bea R_{10} & = & R_4 + g^{\mu\nu}R_{\mu\ax\nu}{}^{\ax} +
g^{\ax\bx}\left( R_{\ax\mu\bx}{}^{\mu} + R_{\ax\cx\bx}{}^{\cx} +
R_{\ax\cb\bx}{}^{\cb}\right)\label{crs10}\\[2mm]
&& + g^{\ax\bb}\left( R_{\ax\mu\bb}{}^{\mu} +
R_{\ax\cx\bb}{}^{\cx} + R_{\ax\cb\bb}{}^{\cb}\right) + \rm c .\rm
c .\label{crs11}\eea where $\rm c .\rm c.$ means complex
conjugate of all terms except $R_4$. Before going further in this
calculation, let's look at a trick that we will use several
times. There will appear terms like

 \beq \sqrt{-g_4}\sqrt{g_6}\nabla_{\mu}V^{\mu}.
\label{crs12}\eeq Recalling (\ref{eg3}), we see that this is not
exactly a total derivative, because the summation is not on all
indices, but only on the space-time ones. A generic expression
like (\ref{crs12}) will be transformed into

\beq \sqrt{-g_4}\sqrt{g_6}\nabla_{\mu}V^{\mu}\sim
-\sqrt{-g_4}V^{\mu}\del_{\mu}\sqrt{g_6}. \label{crs13}\eeq where
$\sim$ means equal up to a total space-time derivative. For the
derivative of the determinant of the metric, we use (\ref{eg4})
and we find

\beq \del_{\mu}\sqrt{g_6} =
\frac{1}{2}\sqrt{g_6}\left(g^{\ax\bx}\del_{\mu}g_{\ax\bx}+g^{\ab\bb}\del_{\mu}g_{\ab\bb}
+2g^{\ax\bb}\del_{\mu}g_{\ax\bb}\right), \label{crs14}\eeq which
leads to the following expressions for the Ricci scalar

\bea g^{\mu\nu}R_{\mu\ax\nu}{}^{\ax} & \sim &
\frac{1}{2}\left((\ox_ig)(\ox_jg)
-\frac{1}{2}\ox_i\ox_j\right)\del_{\mu}v^i\del^{\mu}v^j\nn\\[2mm]
&& + \frac{1}{4}b_a\bar
b_b\del_{\mu}z^a\del^{\mu}\bar z^b\label{crs015}\\[2mm]
g^{\ax\bx}R_{\ax\mu\bx}{}^{\mu} & \sim & \frac{1}{2}b_a\bar
b_b\del_{\mu}z^a\del^{\mu}\bar z^b\label{crs016}\\[2mm]
g^{\ax\bx}\left( R_{\ax\cx\bx}{}^{\cx} +
R_{\ax\cb\bx}{}^{\cb}\right) & = & O(3)\label{crs017}\\[2mm]
g^{\ax\bb}R_{\ax\mu\bb}{}^{\mu} & \sim &
\frac{1}{2}\left((\ox_ig)(\ox_jg)
-\frac{1}{2}\ox_i\ox_j\right)\del_{\mu}v^i\del^{\mu}v^j\nn\\[2mm]
&& - \frac{1}{4}b_a\bar
b_b\del_{\mu}z^a\del^{\mu}\bar z^b\label{crs018}\\[2mm]
g^{\ax\bb}R_{\ax\cx\bb}{}^{\cx} & = &
-\frac{1}{4}\left((\ox_ig)(\ox_jg)
-\ox_i\ox_j\right)\del_{\mu}v^i\del^{\mu}v^j\label{crs019}\\[2mm]
g^{\ax\bb}R_{\ax\cb\bb}{}^{\cb} & = & -\frac{1}{4}(\ox_ig)(\ox_jg)
\del_{\mu}v^i\del^{\mu}v^j\nn\\[2mm]
&& - \frac{1}{4}b_a\bar b_b\del_{\mu}z^a\del^{\mu}\bar
z^b\label{crs020}\eea where $O(3)$ means of order in the moduli
greater or equal to 3. Finally we obtain

\bea \int\rd^{10}\hat x\sqrt{-\hat g}\hat R  = &
\int\rd^4x\sqrt{-g_4}& \left(\cK
R_4+P_{ij}\del_{\mu}v^i\del^{\mu}v^j\right.
\nn\\[2mm]
&&\left. +Q_{ab}\del_{\mu}z^a\del^{\mu}\bar z^b\right),
 \label{crs15}\eea with

\bea P_{ij} & = & \int_{Y}(\ox_ig)(\ox_jg) -\frac{1}{2}\ox_i\ox_j\label{crs16}\\[2mm]
Q_{ab} & = & \frac12\int_Y b_a\bar b_b\label{crs0017}\\[2mm]
(\ox_ig) & = & (\ox_i)_{\ax\bb}g^{0\,\ax\bb}\label{crs17}\\[2mm]
\ox_i\ox_j & = & (\ox_i)_{\ax\ab}(\ox_j)_{\bx\bb}g^{0\,\ax\bb}g^{0\,\bx\ab}\label{crs18}\\[2mm]
b_a\bar b_b & = & (b_a)_{\ab\bb}(\bar
b_b)_{\ax\bx}g^{0\,\ax\bb}g^{0\,\bx\ab}.\label{crs19} \eea
Observe that in components, we have

\beq \cK_{ij} = \int_{Y}
\ox_i\ox_j-(\ox_ig)(\ox_jg)\label{crs20}\eeq which leads to

\beq P_{ij} = -\cK_{ij} +\frac{1}{2}\int_{Y}
\ox_i\ox_j.\label{crs21}\eeq Writing (\ref{gH11}) in components,
we find

\bea g_{ij} & = & -\frac{1}{4\cK}\int_Y \ox_i\ox_j\label{crs22}\\[2mm]
P_{ij} & = & -\cK_{ij} -2\cK g_{ij} = +2\cK
g_{ij}-\frac{1}{4\cK}\cK_i\cK_j\label{crs23}.
  \eea where we have used (\ref{icy6}) for the last equation.

\section{Moduli space}\label{msa}

For a detailed account of the moduli space of Calabi-Yau
threefold, see \cite{CdO}. One of the main results is that the
moduli space $\cM$ splits into the product of two spaces :
$\cM_{1,1}$ corresponds to the deformations of the K\"ahler class
, which are parameterized by the harmonic $(1,1)$-forms, and
$\cM_{2,1}$ corresponds to the deformations of the complex
structure, parameterized by the harmonic $(2,1)$-forms. Both
spaces are special K\"ahler \cite{CRTvP}. The whole moduli space
is thus

\beq \cM = \cM_{1,1}\times\cM_{2,1}. \label{msa1}\eeq

\subsection{K\"ahler class moduli space}

$\cM_{1,1}$ describes the sector of the scalars of the vector
multiplets for type IIA, and the sector of the scalars of
hypermultiplets for type IIB. Recall that the NS-NS 2-form $B_2$
is present in type IIA and type IIB, and that it combines with
the K\"ahler class in the following way (\ref{S4A0})

\beq B_2+iJ\quad\longrightarrow\quad t^i = b^i + iv^i
\label{msa2}\eeq The metric for this sector is

\begin{equation}
  \label{msa3}
  g_{i j}  = \frac{1}{4\cK}\int\,\ox_i\wg *\ox_j = -\frac{1}{4\cK}
  \left(\cK_{ij}-\frac{1}{4\cK}\cK_i\cK_j\right) ; \end{equation} from this we deduce
  that it is K\"ahler with K\"ahler potential $K$

  \begin{equation}
  \label{msa4}
  e^{-K} = 8 \cK \ .
\end{equation} This potential can be written in terms of a
prepotential $\cF$

\beq e^{-K} = i\left(\bar X^I\cF_I- X^I\bar\cF_I\right),\qquad
\cF_I\equiv \frac{\del}{\del X^I}\cF \label{msa5}\eeq with

\beq \cF = -\frac{1}{3!}\frac{\cK_{ijk}X^iX^jX^k}{X^0}.
\label{msa6}\eeq Here the index $I$ is running from 0 to
$h^{1,1}$, and we have defined the $X^I$ in terms of the special
coordinates $t^i$ as $X^I = (1, t^i)$. The potential (\ref{msa5})
has a symplectic invariance and the corresponding manifold is
called Special K\"ahler.

The vectors of the vector multiplets in type IIA couple through a
matrix $\cN$ given below. This matrix is also the one appearing
in the hypermultiplet sector of type IIB. It is defined by

\beq \cN_{IJ} = \bar\cF_{IJ}+\frac{2i}{X^P\IM\cF_{PQ} X^Q
}\IM\cF_{IK}X^K\IM\cF_{JL}X^L \label{msa0021}\eeq and its real and
imaginary parts are

\begin{eqnarray}
  \label{eq:N}
  \RE \cN_{00} = - \frac13 \cK_{ijk} b^i b^j b^k \, , & & \IM \cN_{00} = -
  \cK + \left(\cK_{ij} - \frac14 \, \frac{\cK_i \cK_j}{\cK} \right) b^i
  b^j\ ,
  \nn \\
  \RE \cN_{i0} = \frac12 \cK_{ijk} b^j b^k \ , & & \IM \cN_{i0} = -
  \left(\cK_{ij} - \frac14 \, \frac{\cK_i \cK_j}{\cK} \right) b^j \ ,\\
  \RE \cN_{ij} = - \cK_{ijk} b^k \ , & & \IM \cN_{ij} = \left(\cK_{ij} - \frac14
  \, \frac{\cK_i \cK_j}{\cK} \right) \ ,\nn
\end{eqnarray} We will also need the inverse of the imaginary part of $\cN$

  \begin{equation}
  \label{ImN-1e}
  \left(\IM \cN\right)^{-1} = - \frac{1}{\cK}\left(
    \begin{array}{cc}
      1 & b^i \\
      & \\
      b^i \quad & \frac{g^{ij}}{4} + b^i b^j
    \end{array} \right) \,
\end{equation}
and the inverse metric has the explicit form

\beq
g^{ij} = -4\cK\left( \cK^{ij}-\frac{v^iv^j}{2\cK}\right)
\eeq
in terms of $\cK^{ij}$ defined by

\beq
\cK^{ij}\cK_{jk} = \dx^i_k\label{metric01}.
\eeq
Multiplying \eqref{metric01} by $v^k$, we also obtain the following useful relation

\beq
\cK^{ij}\cK_j = v^i.
\eeq

\subsection{Complex structure moduli space}

According to Kodaira's formula \cite{Tian:1987g},
$\frac{\del}{\del z^a}\Ox$ is in $H^{3,0}+H^{2,1}$

\beq \frac{\del}{\del z^a}\Ox = k_a\Ox + i\eta_a. \label{msa7}\eeq
where $\eta_a$ is the basis for $(2,1)$-forms used in
(\ref{crs5}). The metric for the scalars $z^a$ was found in
(\ref{crd13}), and can be rewritten

\beq g_{a\bar b} =
-\frac{i}{\cK||\Ox||^2}\int_Y\eta_a\wg\bar\eta_b.
\label{msa08}\eeq From this and (\ref{msa7}) we deduce that the
metric for complex structure deformations $g_{ab}$ is also
K\"ahler with K\"ahler potential given by

\begin{equation}
  \label{Kpotcs}
  e^{-K}\ =\ i \int_Y \Ox \wg \bar \Ox\ =\  \cK\, ||\Ox||^2.
\end{equation} It is argued that $\Ox$ can be taken homogeneous of degree one
\cite{CdO}, so the coordinates $z^A$ in the expansion
(\ref{icy32}) are actually projective. Again this potential can
be written in terms of a prepotential $\cF$ as

\beq e^{-K} = i\left(\bar z^A\cF_A- z^A\bar\cF_A\right),\qquad
\cF_A\equiv \frac{\del}{\del z^A}\cF \label{msa8}\eeq where
$\cF_A$, the one appearing in (\ref{icy32}), is a function of the
$z^A$. The moduli space for the K\"ahler class deformations is
thus Special K\"ahler.

In order to evaluate the integrals in the reduction we need to
recall that the Hodge-dual basis $(* \ax_A,* \bx^A)$ is related
to $(\ax_A, \bx^A)$ via
\begin{equation}
  \label{star}
  * \ax_A =  {A_A}^B \, \ax_B + B_{AB} \, \bx^B \ , \qquad
  * \bx^A = C^{AB} \, \ax_B + {D^A}_B \, \bx^B\ ,
\end{equation} where $A, \ B, \ C,\ D$ are some unknown matrices. The relation

\beq \int\ax_A\wg *\bx^B = \int\bx^B\wg *\ax_A\label{msa9} \eeq
implies

\beq A_A{}^B = -D^B{}_A,\label{msa10} \eeq and similarly it can
be obtained that $B$ and $C$ are symmetric. Following
\cite{CeDF,CaDFsk,CaDFsg,Suz}, we will show that these matrices
can be expressed in terms of the moduli. To this end, we start by
noticing the identities

\bea *\Ox & = & -i\Ox \label{msa11}\\[2mm]
*\pi & = & +i\pi \label{msa011} \eea for the $(3,0)$-form $\Ox$
and any $(2,1)$-form $\pi$. (\ref{msa11}) can be expressed
directly using the expansion (\ref{icy32}) and the definitions
(\ref{star})

\bea z^AA_A{}^B-\cF_AC^{AB} & = & -iz^B\label{msa12}\\[2mm]
z^AB_{AB}+\cF_AA_B{}^A & = & i\cF_B.\label{msa13}\eea Since the
forms $\eta_a$ are $(2,1)$, they do not contribute to the integral

\beq \int_Y \del_a\Ox\wg\bar\Ox = k_a
\int_Y\Ox\wg\bar\Ox\label{msa012}\eeq which enables us to
evaluate $k_a$

\beq k_a = \frac{1}{<z|\bar z>}\IM\cF_{aB}\bar z^B = -\del_a K
\label{msa013}\eeq where we defined the inner product

\beq <F,\bar G> = \IM\cF_{AB}F^A\bar G^B \label{msa14}\eeq and we
have used formula (\ref{homo6}) for homogeneous functions.
Adapting the argument in \cite{CdO} to the coordinate $z^0$, we
know that $\frac{\del}{\del z^0}\Ox\in H^{3,0}+H^{2,1}$, and we
define the $(2,1)$ piece by

\beq \frac{\del}{\del z^0}\Ox = \frac{1}{<z|\bar
z>}\IM\cF_{0B}\bar z^B\Ox + i\eta_0.\label{msa014}\eeq With the
general expression

\beq \frac{\del}{\del z^A}\Ox =
\ax_A-\cF_{AB}\bx^B\label{msa015}\eeq one can expand $\eta_a$ on the $(\ax_A\, ,
\,\bx^B)$ basis, and imposing (\ref{msa011}) for $\eta_A$, we
find the equations

\bea A_A{}^B-\cF_{AC}C^{CB} & = & i\dx_A^B
-\frac{2i}{<z|\bar z>}\IM\cF_{AC}\bar z^Cz^B\label{msa016}\\[2mm]
B_{AB}+\cF_{AC}A_B{}^C & = & -i\cF_{AB}+\frac{2i}{<z|\bar
z>}\IM\cF_{AC}\bar z^C\cF_B.\label{msa017}\eea Remark that
multiplying (\ref{msa016}) and (\ref{msa017}) by $z^A$ one
recovers (\ref{msa12}) and (\ref{msa13}). Separating the real and
imaginary parts of (\ref{msa016}) and (\ref{msa017}) we find the
expressions of the matrices

\bea A_A{}^B & = &
-\RE\cF_{AC}\left(\IM\cF^{-1}\right)^{CB}+\frac{\bar
z^B\cF_A+z^B\bar \cF_A}{<z|\bar z>}\label{msa18}\\[2mm]
B_{AB} & = &
\IM\cF_{AB}+\RE\cF_{AC}\left(\IM\cF^{-1}\right)^{CD}\RE\cF_{DB}-\frac{\bar\cF_A\cF_B+\cF_A
\bar \cF_B}{<z|\bar z>}\label{msa19}\\[2mm]
C^{AB} & = & -\left(\IM\cF^{-1}\right)^{AB} + \frac{\bar z^A
z^B+z^A \bar z^B}{<z|\bar z>}.\label{msa20}\eea We introduce the
matrix $\cM$

\beq \cM_{AB} =
\bar\cF_{AB}+\frac{2i}{<z|z>}\IM\cF_{AC}z^C\IM\cF_{BD}z^D
\label{msa21}\eeq and we give the expression of $A, B, C$ in
terms of $\cM$

\begin{eqnarray}
  \label{A-N}
  A & = & \left(\RE \cM \right) \left(\IM \cM \right)^{-1}\ , \nn \\
  B & = & - \left(\IM \cM \right) - \left(\RE \cM \right) \left(\IM
  \cM\right)^{-1} \left(\RE \cM \right)\ , \nn \\
  C & = & \left(\IM \cM \right)^{-1}\ .
\end{eqnarray}

\chapter{Type II supergravities on {CY}$_3$ with NS-form fluxes}\label{typeIINS}

\section{Type IIA with NS fluxes}\label{IIANS}

In this section we briefly recall the results of \cite{LM2} for
the compactification of type IIA supergravity on Calabi-Yau
three-folds $Y$ when background NS fluxes are turned on.

The bosonic spectrum of type IIA supergravity in ten dimensions
features the following fields: the graviton $\hat g_{MN}$, a
two-form $\hat B_2$ and the dilaton $\hat \phi$ in the NS-NS
sector and a one form $\hat A_1$ and a three-form $\hat C_3$ in
the RR sector. The action governing the interactions of these
fields can be written as \cite{JP}
\begin{eqnarray}
  S & = & \int \, e^{-2\hat\phi} \left( -\frac12 \hat R *\! {\bf 1} + 2
  d \hat\phi \wg * d \hat\phi - \frac14  \hat H_3\wg * \hat H_3 \right) \nn \\
  & & - \frac12  \int \, \left(\hat  F_2 \wg * \hat F_2 + \hat F_4 \wg * \hat
  F_4 \right) + \frac12 \int \hat H_3 \wedge \hat C_3 \wedge d \hat C_3 \, ,
\end{eqnarray}
where
\begin{equation}
   \hat H_3 = d\hat B_2\ ,\qquad \hat F_2 = d\hat A_1\ ,\qquad \hat F_4 = d
   \hat C_3 - \hat A_1 \wedge\hat H_3.
\end{equation}
Upon compactification on a \CY three-fold the four-dimensional
spectrum can be read from the expansion of the ten-dimensional
fields in the \CY harmonic forms
\begin{eqnarray}
  \hat A_1 & = & A^0 \ ,\nn \\
  \hat C_3 & = & C_3 + A^i \wg \ox_i + \xi^A \ax_A + \txi_A \bx^A \ ,\\
  \hat B_2 & = & B_2 + b^i \ox_i \ .\nn
\end{eqnarray}
Correspondingly, in $D=4$ we find a three-form $C_3$, a two-form
$B_2$, the vector fields $(A^0,A^i)$ and the scalars $b^i, \xi^A,
\txi_A$. Together with the K\"ahler class and complex structure
deformations $v^i$ and $z^a$ these fields combine into a gravity
multiplet $(G_{\mu \nu}, A^0)$, $h^{(1,1)}$ vector multiplets
$(A^i, v^i, b^i), \ i = 1,\ldots , h^{(1,1)}$,  $h^{(1,2)}$
hyper-multiplets $(z^a, \xi^a, \txi_a), \ a = 1, \ldots ,
h^{(1,2)}$ and a tensor multiplet $(B_2, \phi, \xi^0, \txi_0)$.

We assume that turning on background fluxes does not change the
light spectrum and thus the only modification in the KK Ansatz is
a shift in the field strength of $\hat B_2$
\begin{equation}
  \label{cAns1}
  \hat H_3 = H_3 + d  b^i\wg\ox_i +p^A\ax_A +q_A\bx^A \ .
\end{equation}
This leads to the following expressions for the different terms
appearing in the ten-dimensional action (\ref{SIIA10})
\begin{eqnarray}
  \label{cAns2}
   - \frac14 \int_Y\hat H_3 \wg * \hat H_3
  & = & - \frac{\cK}{4} \, H_3 \wg * H_3 - \cK
  g_{ij} db^i \wg * db^j - V*{\bf 1} \ , \nn \\ [5mm]
  - \frac12 \int_Y \hat F_2 \wg * \hat F_2 & = &
  - \frac{\cK}{2} \, d A^0 \wg * dA^0 \ , \nn \\ [5mm]
  - \frac12 \int_Y {\hat {F}_4} \wg * {\hat {F}_4} & = & -
  \frac{\cK}{2}\, (d C_3 - A^0\wedge H_3) \wg * (d C_3 - A^0\wg H_3) \\
  & & - 2 \cK g_{ij} (d A^i - A^0 d b^i) \wg * (d A^j - A^0 d b^j) \nn \\
  & & + \frac{1}{2}\left(\IM \cM ^{-1} \right)^{AB}
  \Big[ D\tilde\xi_A +  \cM_{AC} D\xi^C \Big] \wg * \Big[ D\tilde\xi_B +
  \bar \cM_{BD} D\xi^D \Big] \ , \nn \\[5mm]
  \frac12 \int_Y \hat H_3 \wedge \hat C_3 \wedge d \hat C_3
  & = &  - \frac12 H_3 \wg (\xi^A  d \txi_A - \txi_A  d  \xi^A) +\frac12 d b^i
  \wg A^j \wg d A^k \cK_{ijk} \nn \\
  && + d  C_3\wg\left( p^A\tilde\xi_A-q_A\xi^A\right) \nn  \, .
\end{eqnarray}
Even from this stage one can notice that some of the fields
effectively became charged
\begin{equation}
  \label{cAns07}
   D \xi^A =  d  \xi^A - p^A A^0 \quad , \quad D \tilde \xi_A =  d  \tilde
   \xi_A - q_A A^0  \ ,
\end{equation}
and a potential term is induced
\begin{equation}
  \label{cAns06}
  V = -\frac{1}{4}\rme^{-\hat\phi}\left(q+\cM p\right) \IM \cM^{-1} \left(q
  +
  \bar \cM p \right) \ .
\end{equation}
\footnote{%
  For a systematic study of the \CY\!\!\! moduli space we refer the reader to
  the literature \cite{BCF,CdO}.}

Next, the compactification proceeds as usually by dualizing the
fields $C_3$ and $B_2$ to a constant and to a scalar
respectively. We do not perform these steps here, but we just
recall the final results. (for more details see \cite{LM2,BGG}).
First the dualization of $C_3$ to a constant $e$ results in
\begin{equation}
  \label{cAns10}
  \cL_e = \cL_{C_3} = -\frac{\rme^{4\phi}}{2\cK}
  \left(p^A \tilde \xi_A - q_A \xi^A + e \right)^2 *{\bf 1}
  + \left(p^A \tilde \xi_A - q_A \xi^A + e \right) A^0 \wg H_3 \ .
\end{equation}
It was shown in \cite{LM2} that the constant $e$ plays a special
role in the case of RR fluxes. however, it is irrelevant for the
analysis in this paper and thus we will set it to zero. Dualizing
now the two-form $B_2$, one obtains an axion, which due to the
Green-Schwarz term in \eqref{cAns10} becomes charged and its
covariant derivative reads
\begin{equation}
  \label{cAns18}
  Da =  d  a - \left( p^A \tilde \xi_A - q_A \xi^A \right) A^0.
\end{equation}

Collecting all terms one can write the final form of the
action\footnote{ We have further redefined the gauge fields as
$A^i \longrightarrow A^i - b^i A^0$ and also appropriately
rescaled the metric in order to go to the Einstein frame.}
\begin{eqnarray}
  \label{cAns19}
  S_{IIA} & = & \int \Big[ -\frac12 R ^* {\bf 1} - g_{ij} dt^i \wg * d
  {\bar t}^j - h_{uv} D q^u \wg * D q ^v - V_{IIA}*{\bf 1} \nn \\
  & & \qquad + \frac{1}{2}\, \IM \cN_{IJ} F^I\wg * F^{J}
  + \frac{1}{2} \, \RE \cN_{IJ} F^I \wg F^J \Big] \ ,
\end{eqnarray}
where the potential can be read from \eqref{cAns06} and
\eqref{cAns10}
\begin{equation}
  \label{cAns20}
  V_{IIA} = -\frac{1}{4\cK} \rme^{2 \phi} \left(q + \cM p \right) \IM \cM^{-1}
  \left(q + \bar \cM p \right) + \frac{1}{2 \cK} \rme^{4 \phi} \left(p^A
  \tilde \xi_A - q_A \xi^A \right)^2 \ ,
\end{equation}
while the metric for the hyper-scalars $h_{uv}$ has the standard
form of \cite{FeS}
\begin{eqnarray}
  \label{cAns21}
  h_{uv} Dq^u \wg * Dq ^v & = &  d \phi \wg * d\phi + g_{ab} dz^a
  \wg * d \bar z^b \\
  & & + \frac{e^{4\phi}}{4} \, \Big[ Da +
  (\tilde\xi_A D \xi^A-\xi^A D\tilde\xi_A) \Big] \wg *
  \Big[Da + (\tilde\xi_A D \xi^A-\xi^A D \tilde\xi_A) \Big] \nn \\
  & & - \frac{e^{2\phi}}{2}\left(\IM \cM^{-1} \right)^{AB}
  \Big[ D\tilde\xi_A + \cM_{AC} D\xi^C \Big]
  \wg * \Big[D \tilde \xi_B + \bar \cM_{BD} D \xi^D \Big] \ .
   \nn
\end{eqnarray}

%
\section{Type IIB with NS fluxes}\label{cBNS}
%

In this section we recall the compactification of type IIB
supergravity on a \CY{}3-folds with NS 2-form fluxes. The only
differences with section \ref{cB} appear in the expansions of the
field strengths which have to take into account $H_3$ fluxes

\bea
   \hat H_3 & = & H_3 + d b^i \wg \ox_i +\p^A\ax_A - \q_A
  \bx^A .\label{cBNS5}\eea Consequently, $\hat F_5$ is also
  modified, according to

\bea
  \hat F_5 & = & \tilde F^A\wg\ax_A-\tilde G_A\bx^A + \left(
 d D_2^i - d b^i\wg C_2 - c^iH_3\right)\wg\ox_i\nn\\[2mm]
&& + d\rho_i\wg\tox^i-c^i d b^j\wg\ox_i\wg\ox_j\label{cBNS6} \eea
with

\beq \tilde F^A = F^A -\tilde m^A C_2 \quad ;\quad \tilde G_A =
G_A -\tilde e_A C_2.\label{cBNS7} \eeq  The straightforward
expansion reads

\begin{eqnarray}
   - \frac14\rme^{-\hat\phi} \int_Y\hat H_3 \wg * \hat H_3
  & = & - \frac{\cK}{4}\rme^{-\hat\phi} \, H_3 \wg * H_3 - \cK\rme^{-\hat\phi}
  g_{ij} db^i \wg * db^j \nn\\
  && +\frac{1}{4}\rme^{-\hat\phi}\left(\tilde e-\cM \tilde m
\right)Im\cM^{-1}\left(\tilde e-\bar\cM \tilde m\right)*{\bf
1}\,\label{cBNS8}\\[5mm]
 - \frac12\rme^{2\hat\phi} \int_Y  d \hat l \wg
* d \hat l & = &
  - \frac{\cK}{2}\rme^{2\hat\phi} \, d l \wg * d l \ ,\label{cBNS9}\\ [5mm]
  - \frac12 \rme^{\hat\phi}\int_Y {\hat {F}_3} \wg * {\hat {F}_3} & = & -
  \frac{\cK}{2}\rme^{\hat\phi}\, (d C_2 - l H_3) \wg * (d C_2 - l H_3)
  \nn \\
  & & - 2 \cK\rme^{\hat\phi} g_{ij} (d c^i - l d b^i) \wg * (d c^j - l d b^j)
  \label{cBNS10} \\
  & & + \frac{1}{2}\rme^{\hat\phi}l^2\left(\tilde e-\cM \tilde m
\right)Im\cM^{-1}\left(\tilde e-\bar\cM \tilde m\right)*{\bf 1} \ , \nn \\[5mm]
 -\frac{1}{4}\int \hat F_5\wg * \hat F_5
  & = &  + \frac14 Im\cM^{-1}\left(\tilde G-\cM \tilde F
\right)\wg *\left(\tilde G-\bar\cM \tilde F\right) \nn\\
  && -\cK g_{ij}d \tilde D_2^i \wg *d\tilde D_2^j-\frac{1}{16\cK} g^{ij}d \tilde
  \rho_i \wg *d\tilde\rho_j\label{cBNS11}\\[5mm]
  -\frac12\int\hat A_4\wg \hat H_3\wg d\hat C_2 & = &
  -\frac12\cK_{ijk}D_2^i\wg d b^j\wg d c^k-\frac12\rho_i\left(d
  B_2\wg d c^i + d b^i\wg d C_2\right)\nn\\
  && +\frac12 C_2\wg\left( F^A\tilde e_A - G_A\tilde
  m^A\right)\label{cBNS012}
\end{eqnarray} where $d\tilde D_2^i$ and $d\tilde\rho_i$ have been defined in
 (\ref{cB12})
and (\ref{cB13}). Since the sector of $D_2^i$ and $\rho_i$ is not
modified by the fluxes, the elimination of $D_2^i$ is made in
exactly the same way as in section {\ref{cB}. However, the sector
of $C_2$ gets modified. We still add the duality Lagrangian

\beq +\frac12 F^A\wg G_A \eeq which imposes the new self-duality
condition

\bea *\tilde G & = & Re\cM *\tilde F - Im\cM \tilde F
\label{cBNS20}\\[2mm]
\tilde G & = & Re\cM \tilde F + Im\cM *\tilde F \label{cBNS21}\eea
as an equation of motion for $G_A$. After elimination of $G_A$,
we obtain

\bea \cL_{F^A} & = & + \frac12 Im\cM_{AB}\tilde F^A \wg *\tilde
F^B +\frac12 Re\cM_{AB} \tilde F^A\wg \tilde F^B\nn\\[2mm]
&&+\frac12 \tilde e_AC_2\wg\left( F^A+\tilde F^A\right)
.\label{cBNS23} \eea It can be checked that this is compatible
with the new component $(2,6)$ of the equation of motion of $\hat
C_2$ (\ref{cB15}) which reads now

\bea \rme^{\hat\phi}\cK d * d C_2 & = & -\left(F^A\tilde
e_A-G_A\tilde m^A\right)+d\rho_i\wg d b^i.\label{cBNS18}\eea After
the Weyl rescaling of the volume and the rotation of $v^i$, the
whole action is

\begin{eqnarray}
  S_{IIB}^{(4)} &=& \int - \frac{1}{2} R *\! {\bf 1} - g_{ab} dz^a \wg
  *d\bar{z}^{b} - g_{ij} dt^i \wg *d\bar{t}^j - d\phi \wg *d \phi
  \nonumber\\[2mm]
  && -\frac{1}{4} e^{-4\phi} dB_2 \wg * dB_2 - \frac{1}{2} e^{-2\phi} \cK
  \left( dC_2 - l dB_2 \right) \wg *\left( dC_2 - l dB_2 \right)
  \nonumber\\[2mm]
  && - \frac{1}{2} \cK e^{2\phi} dl \wg * dl - 2 \cK e^{2\phi} g_{ij} \left(
    dc^i - l db^i \right)\wg * \left( dc^j - l db^j \right)\nonumber\\[2mm]
  && - \frac{e^{2\phi}}{8 \cK} g^{-1\,ij} \left( d\rho_i -
    \cK_{ikl} c^k db^l \right) \wg *\left( d\rho_j -
    \cK_{jmn} c^m db^n \right) \nonumber \\[2mm]
  && +  \left( db^i \wg C_2 + c^i dB_2 \right)\wg
  \left( d\rho_i -  \cK_{ijk} c^j db^k \right) + \frac{1}{2}
  \cK_{ijk} c^i c^j dB_2 \wg db^k \nonumber \\[2mm]
  && + \frac{1}{2} Re\cM_{AB} \tilde{F}^A \wg \tilde{F}^B + \frac{1}{2} Im
  \cM_{AB} \tilde{F}^A \wg * \tilde{F}^B + \frac{1}{2} \q_A \left( F^A +
    \tilde{F}^A \right) \wg C_2\nn \\[2mm]
  && + \frac{1}{2} e^{4\phi} \left(l^2 + \frac{e^{-2\phi}}{2 \cK }\right)
  \left(\q - \cM \p \right)_A Im \cM^{-1AB} \left(\q - \bar{\cM}
  \p\right)_B *\! {\bf 1} \ .\label{cBNS24}
\end{eqnarray}

We again want to dualize the 2-form $C_2$ but it has become
massive with a mass involving only the magnetic fluxes. This is
one of the reasons why the study of the magnetic fluxes is more
involved and will not be addressed in this thesis. From now on we
consider only the electric fluxes. $C_2$ and $B_2$ are thus
massless and can be dualized to the scalars $h_1$ and $h_2$. We
add first

\beq +d C_2\wg d h_1 \label{cBNS25}\eeq and the Lagrangian for
$C_2$ is

\bea \cL_{C_2} & = & - \frac{1}{2} e^{-2\phi} \cK
  \left( dC_2 - l dB_2 \right) \wg *\left( dC_2 - l dB_2 \right)\nn\\[2mm]
  && -  b^i d C_2 \wg d\rho_i +\q_A F^A \wg C_2+d C_2\wg d h_1.\label{cBNS26}
    \eea We eliminate $d C_2$ with its equation of motion and we
    find

\bea \cL_{C_2} & = & - \frac{1}{2\cK} e^{2\phi}
  \left( d h_1 - b^id\rho_i -\q_AV^A \right) \wg *\left( d h_1 - b^jd\rho_j
  -\q_AV^A \right)\nn\\[2mm]
  && +l d B_2 \wg\left( d h_1-b^id\rho_i-\q_AV^A\right).\label{cBNS27}
    \eea Repeating the same procedure with $B_2$, we obtain the action for
    type IIB supergravity on a Calabi-Yau manifold with electric
    NS fluxes

\begin{eqnarray}
  S_{IIB}^{(4)} &=& \int - \frac{1}{2} R *\! {\bf 1} - g_{ab} dz^a \wg
  *d\bar{z}^{b} - g_{ij} dt^i \wg *d\bar{t}^j - d\phi \wg *d \phi
  \nonumber\\[2mm]
  && - \frac{e^{2\phi}}{8 \cK} g^{-1\,ij} \left( d\rho_i -
    \cK_{ikl} c^k db^l \right) \wg *\left( d\rho_j -
    \cK_{jmn} c^m db^n \right) \nonumber \\[2mm]
  &&  - 2 \cK e^{2\phi} g_{ij} \left( dc^i - l db^i \right)\wg * \left( dc^j - l db^j
 \right)
  - \frac{1}{2} \cK e^{2\phi} dl \wg * dl\nonumber\\[2mm]
  &&  - \frac{1}{2\cK} e^{2\phi}\left( d h_1 - b^id\rho_i -\q_A V^A \right)
  \wg *\left( d h_1 - b^jd\rho_j -\q_A V^A\right) \nonumber \\[2mm]
  && -\rme^{4\phi} D\tilde h\wg * D\tilde h\nn\\[2mm]
  && + \frac{1}{2} Re\cM_{AB} F^A \wg F^B + \frac{1}{2} Im
  \cM_{AB} F^A \wg * F^B\nn\\[2mm]
  && + \frac{1}{2} e^{4\phi} \left(l^2 + \frac{e^{-2\phi}}{2 \cK }\right)
  \q_A Im \cM^{-1AB}\q_B *\! {\bf 1}  \label{cBNS28}
\end{eqnarray} with

\beq D\tilde h = d h_2 +ld h_1 +
  (c^i-lb^i)d\rho_i-l\q_AV^A-\frac12\cK_{ijk}c^ic^jd
  b^k.\label{cBNS29}\eeq Applying the map described in section
  \ref{msy}, we find

\begin{eqnarray}
  S_{IIB}^{(4)} & = & \int -\frac{1}{2}R * \! {\mathbf 1} - g_{a b} dz^a \wg
  *d \bar{z}^{b} - h_{uv} D q^u \wg * D q ^v - V_{IIB} * \! {\mathbf 1} \nonumber\\
  && + \frac{1}{2}Re \cM_{AB} F^A \wg F^B + \frac{1}{2} Im \cM_{AB}F^A \wg *
  F^B \ ,
  \label{action3}
\end{eqnarray}
where the quaternionic metric is given by
\begin{eqnarray}
h_{uv} D q^u \wg * D q ^v &=&
g_{ij} dt^i \wg *d \bar{t}^j + d\phi \wg *d \phi \\[2mm]
  && - \frac{1}{2} e^{2\phi} Im\cN^{-1 \, IJ} \left(D \tilde{\xi}_I + \cN_{IK}
    D \xi^K \right) \wg * \left( D \tilde{\xi}_J + \bar{\cN}_{JL} D \xi^L
  \right) \nonumber\\[2mm]
  && + \frac{1}{4} e^{4\phi} \left(D a + (\tilde{\xi}_I D \xi^I -
\xi^I D \tilde{\xi}_I) \right) \wg * \left(D a + (\tilde{\xi}_I D
\xi^I - \xi^I D \tilde{\xi}_I) \right) \ , \nonumber
\end{eqnarray}
while the potential reads
\begin{equation}
  V_{IIB} = -\frac{1}{2}\, e^{4 \phi} \Big(l^2 + \frac{e^{-2\phi}}{2\cK} \Big)\,
  \q_A  \left[(Im \cM)^{-1}\right]^{AB}\q_B .
  \label{potIIB}
\end{equation}
The presence of the electric fluxes has gauged some of the
isometries of the hyperscalars as can be seen from the covariant
derivatives
\begin{equation}
 D a  =  da - \xi^0 \q_A V^A\ ,\qquad
  D\tilde{\xi}_0 =  d \tilde{\xi}_0 + \q_A V^A \ , \qquad
D\tilde{\xi}_i =  d \tilde{\xi}_i \ , \qquad D \xi^I =  d \xi^I \
.
  \label{action5}
\end{equation}

\chapter{$G$-structures} \label{acs}

In this section we assemble a few facts about $G$-structures as
taken from the mathematical literature where one also finds the
proofs omitted here. (See, for example,
\cite{salamon,CS,yano,candelasTS,joyce,FFS,salamonb}.) We
concentrate on the example of manifolds with $\SU(3)$-structure.

\subsection{Almost Hermitian manifolds}
\label{ahm}

Before discussing $G$-structures in general, let us recall the
definition of an almost Hermitian manifold. This allows us to
introduce useful concepts, and, as we subsequently will see,
provides us with a classic example of a $G$-structure.

A manifold of real dimension $2n$ is called \emph{almost complex}
if it admits a globally defined tensor field $J_m{}^n$ which obeys
\begin{equation}\label{J2}
  J_m{}^p J_p{}^n = -\dx_m{}^n \ .
\end{equation}
A metric $g_{mn}$ on such a manifold is called Hermitian if it
satisfies
\begin{equation}\label{hm}
  J_m{}^p J_n{}^r g_{pr} = g_{mn}\ .
\end{equation}
An almost complex  manifold endowed with a Hermitian metric is
called an \emph{almost Hermitian manifold}. The relation
(\ref{hm}) implies that $J_{mn} = J_m{}^p g_{pn}$ is a
non-degenerate 2-form which is called \emph{the fundamental form}.

On any even dimensional manifold one can locally introduce complex
coordinates. However, complex manifolds have to satisfy in
addition that, first, the introduction of complex coordinates on
different patches is consistent, and second that the transition
functions between different patches are holomorphic functions of
the complex coordinates. The first condition corresponds to the
existence of an almost complex structure. The second condition is
an integrability condition, implying that there are coordinates
such that the almost complex structure takes the form
\begin{equation}\label{diagJ}
  J = \left(
  \begin{array}{cc}
    i \bf{1}_{n\times n} & 0 \\
    0 & -i \bf{1}_{n\times n}
  \end{array}\right) \ .
\end{equation}
The integrability condition is satisfied if and only if the
Nijenhuis tensor $N_{mn}{}^p$ vanishes. It is defined as
\begin{equation}
\label{Ntens}
\begin{split}
   N_{mn}{}^p &= J_m{}^q \left (\partial_q J_n{}^p
      - \partial_n J_q{}^p \right)
      - J_n{}^q \left (\partial_q J_m{}^p - \partial_m J_q{}^p \right)
   \\
      &= J_m{}^q \left (\nabla_q J_n{}^p - \nabla_n J_q{}^p \right)
      - J_n{}^q \left (\nabla_q J_m{}^p - \nabla_m J_q{}^p \right) \ ,
\end{split}
\end{equation}
where $\nabla$ denotes the covariant derivative with respect to
the Levi--Civita connection.

One can also consider an even stronger condition where
$\nabla_mJ_{np}=0$. This implies $N_{mn}{}^p=0$ but in addition
that $\dd J=0$ and means we have a {\em K\"ahler manifold}. In
particular, it implies that the holonomy of the Levi--Civita
connection $\nabla$ is $U(n)$.

Even if there is no coordinate system where it can be put in the
form~\eqref{diagJ}, any almost complex structure obeying
(\ref{J2}) has eigenvalues $\pm i$. Thus even for non-integrable
almost complex structures one can define the projection operators
\begin{equation}\label{Pdef}
  (P^\pm)_m{}^n = \frac12(\dx_m^n \mp i J_m{}^n)\ ,
\end{equation}
which project onto the two eigenspaces, and satisfy
\begin{equation}\label{Pprop}
P^\pm P^\pm = P^\pm \ , \qquad P^+ P^- =0\ .
\end{equation}
On an almost complex manifold one can define $(p,q)$ projected
components $\ox^{p,q}$ of a real $(p+q)$-form $\ox^{p+q}$ by
using (\ref{Pdef})
\begin{equation}
  \ox^{p,q}_{m_1\ldots m_{p+q}} =
    (P^+)_{m_1}{}^{n_1} \ldots (P^+)_{m_p}{}^{n_p}
    (P^-)_{m_{p+1}}{}^{n_{p+1}} \ldots (P^-)_{m_{p+q}}{}^{n_{p+q}}
\ox_{n_1  \ldots n_{p+q}}^{p+q} \ .
\end{equation}
Furthermore, a real $(p+q)$-form is of the type $(p,q)$ if it
satisfies
\begin{equation}
  \ox_{m_1 \ldots m_p n_1 \ldots n_q} =
     (P^+)_{m_1}{}^{r_1} \ldots (P^+)_{m_q}{}^{r_p}
     (P^-)_{n_1}{}^{s_1} \ldots (P^-)_{n_q}{}^{s_q} \ox_{r_1
  \ldots s_q} \ .
\end{equation}

In analogy with complex manifolds  we denote  the projections on
the subspace of eigenvalue $+i$ with an unbarred  index $\ax$ and
the projection on the subspace of eigenvalue $-i$ with a barred
index $\ab$. For example the hermitian metric of an almost
Hermitian manifold is of type $(1,1)$ and has one barred and one
unbarred index. Thus, raising and lowering indices using this
hermitian metric converts holomorphic indices into
anti-holomorphic ones and vice versa. Moreover the contraction of
a holomorphic and an anti-holomorphic index vanishes, i.e.\ given
$V_m$ which is of type $(1,0)$ and $W^n$ which is of type
$(0,1)$, the product $V_m W^m$ is zero. Similarly, on an almost
hermitian manifold of real dimension $2n$ forms of type $(p,0)$
vanish for $p>n$. Finally, derivatives of $(p,q)$-forms pick up
extra pieces compared to complex manifolds precisely because $J$
is not constant. One finds \cite{candelasTS}
\begin{equation}
\label{ddecomp}
  d \ox^{(p,q)} = (d \ox)^{(p-1, q+2)} + (d \ox)^{(p, q+1)} + (d
  \ox)^{(p+1,q)} + (d \ox)^{(p+2, q-1)} \ .
\end{equation}

\subsection{$G$-structures and $G$-invariant tensors}
\label{Gstruc}

An orthonormal frame on a $d$-dimensional Riemannian manifold $M$
is given by a basis of vectors $e_i$, with $i=1,\dots,d$,
satisfying $e_i^me_j^ng_{mn}=\delta_{ij}$. The set of all
orthonormal frames is known as the frame bundle. In general, the
structure group of the frame bundle is the group of rotations
$O(d)$ (or $\SO(d)$ is $M$ is orientable). The manifold has a
$G$-structure if the structure group of the frame bundle is not
completely general but can be reduced to $G\subset O(d)$. For
example, in the case of an almost Hermitian manifold of dimension
$d=2n$, it turns out one can always introduce a complex frame and
as a result the structure group reduces to $U(n)$.

An alternative and sometimes more convenient way to define
$G$-structures is via $G$-invariant tensors, or, if $M$ is spin,
$G$-invariant spinors. A non-vanishing, globally defined tensor or
spinor $\xi$ is $G$-invariant if it is invariant under $G\subset
O(d)$ rotations of the orthonormal frame. In the case of almost
Hermitian structure, the two-form $J$ is an $U(n)$-invariant
tensor. Since the invariant tensor $\xi$ is globally defined, by
considering the set of frames for which $\xi$ takes the same
fixed form, one can see that the structure group of the frame
bundle must then reduce to $G$ (or a subgroup of $G$). Thus the
existence of $\xi$ implies we have a $G$-structure. Typically,
the converse is also true. Recall that, relative to an
orthonormal frame, tensors of a given type form the vector space
for a given representation of $O(d)$ (or $\Spin(d)$ for spinors).
If the structure group of the frame bundle is reduced to
$G\subset O(d)$, this representation can be decomposed into
irreducible representations of $G$. In the case of almost complex
manifolds, this corresponds to the decomposition under the $P^\pm$
projections~\eqref{Pdef}.  Typically there will be some tensor or
spinor that will have a component in this decomposition which is
invariant under $G$. The corresponding vector bundle of this
component must be trivial, and thus will admit a globally defined
non-vanishing section $\xi$. In other words, we have a globally
defined non-vanishing $G$-invariant tensor or spinor.

To see this in more detail in the almost complex structure
example, recall that we had a globally defined fundamental
two-form $J$. Let us specialize for definiteness to a
six-manifold, though the argument is quite general. Two-forms are
in the adjoint representation $\rep{15}$ of $\SO(6)$ which
decomposes under $U(3)$ as
\begin{equation}
   \rep{15} = \rep{1} + \rep{8} + (\rep{3} + \bar{\rep{3}})\  .
\end{equation}
There is indeed a singlet in the decomposition and so given a
$U(3)$-structure we necessarily have a globally defined invariant
two-form, which is precisely the fundamental two-form $J$.
Conversely, given a metric and a non-degenerate two-form $J$, we
have an almost Hermitian manifold and consequently a
$U(3)$-structure.

In this paper we are interested in $\SU(3)$-structure. In this
case we find two invariant tensors. First we have the fundamental
form $J$ as above. In addition, we find an invariant complex
three-form $\Omega$. Three-forms are in the $\rep{20}$
representation of $\SO(6)$, giving two singlets in the
decomposition under $\SU(3)$,
\begin{equation}
\begin{aligned}
   \rep{15} &= \rep{1} + \rep{8} + \rep{3} + \bar{\rep{3}}
       \quad \Rightarrow \quad J \  , \\
   \rep{20} &= \rep{1} + \rep{1} + \rep{3} + \bar{\rep{3}}
       + \rep{6} + \bar{\rep{6}}
       \quad \Rightarrow \quad \Omega = \Omega^+ +\ii\Omega^- \ .
 \end{aligned}
 \end{equation}
In addition, since there is no singlet in the decomposition of a
five-form, one finds that
\begin{equation}\label{JOcond}
   J \wedge \Omega = 0 \ .
\end{equation}
Similarly, a six-form is a singlet of $\SU(3)$, so we also must
have that $J\wedge J\wedge J$ is proportional to
$\Omega\wedge\bar\Omega$. The usual convention is to set
\begin{equation}
\label{JOcond2}
   J\wedge J \wedge J
      = \frac{3\ii}{4}\, \Omega \wedge \bar{\Omega} \ , \\
\end{equation}
Conversely, a non-degenerate $J$ and $\Omega$
satisfying~\eqref{JOcond} and~\eqref{JOcond2} implies that $M$ has
$\SU(3)$-structure. Note that, unlike the $U(n)$ case, the metric
need not be specified in addition; the existence of $J$ and
$\Omega$ is sufficient~\cite{H}. Essentially this is because,
without the presence of a metric, $\Omega$ defines an almost
complex structure, and $J$ an almost symplectic structure.
Treating $J$ as the fundamental form, it is then a familiar
result on almost Hermitian manifolds that the existence of an
almost complex structure and a fundamental form allow one to
construct a Hermitian metric.

We can similarly ask what happens to spinors for a structure
group $SU(3)$. In this case we have the isomorphism
$\Spin(6)\cong\SU(4)$  and the four-dimensional spinor
representation decomposes as
\begin{equation}
   \rep{4}=\rep{1}+\rep{3} \quad \Rightarrow \quad \eta \ .
\end{equation}
We find one singlet in the decomposition, implying the existence
of a globally defined invariant spinor $\eta$. Again, the
converse is also true. A metric and a globally defined spinor
$\eta$ implies that $M$ has $\SU(3)$-structure.

\subsection{Intrinsic torsion}
\label{app:IT}

One would like to have some classification of $G$-structures. In
particular, one would like a generalization of the notion of a
K\"ahler manifold where the holonomy of the Levi--Civita
connection reduces to $U(n)$. Such a classification exists in
terms of the {\em
  intrinsic torsion}. Let us start by recalling the definition of
torsion and contorsion on a Riemannian manifold $(M,g)$.

Given any metric compatible connection $\nabla'$ on $(M,g)$, i.e.
one satisfying $\nabla'_mg_{np}=0$, one can define the Riemann
curvature tensor and the torsion tensor as follows
\begin{equation}
\label{RT}
  [\nabla'_m, \nabla'_n] V_p =
     - R_{mnp}{}^q V_q - 2 T_{mn}{}^r \nabla'_r V_p \ ,
\end{equation}
where $V$ is an arbitrary vector field. The Levi-Civita
connection is the unique torsionless connection compatible with
the metric and is given by the usual expression in terms of
Christoffel symbols $\Gx_{mn}{}^p = \Gx_{nm}{}^p$. Let us denote
by $\nabla$ the covariant derivative with respect to the
Levi-Civita connection while a connection with torsion is denoted
by $\nabla^{(T)}$. Any metric compatible connection can be
written in terms of the Levi-Civita connection
\begin{equation}
  \label{cont}
  \nabla^{(T)} = \nabla - \CT \ ,
\end{equation}
where $\CT_{mn}{}^p$ is the contorsion tensor. Metric
compatibility implies
\begin{equation}\label{Kprop}
  \CT_{mnp} = - \CT_{mpn}\ , \quad \textrm{where} \quad
\CT_{mnp} = \CT_{mn}{}^r g_{rp} \ .
\end{equation}
Inserting (\ref{Kprop}) into (\ref{RT}) one finds a one-to-one
correspondence between the torsion and the contorsion
\begin{equation}
\begin{aligned}
  \label{T-K}
  T_{mn}{}^p &=
    \frac12 (\CT_{mn}{}^p - \CT_{nm}{}^p) \equiv \CT_{[mn]}{}^p \ , \\
  \CT_{mnp} &=  T_{mnp} + T_{pmn} + T_{pnm} \ .
\end{aligned}
\end{equation}
These relations tell us that given a torsion tensor $T$ there
exist a unique connection $\nabla^{(T)}$ whose torsion is
precisely $T$.

Now suppose $M$ has a $G$-structure. In general the Levi-Civita
connection does not preserve the $G$-invariant tensors (or
spinor) $\xi$. In other words, $\nabla\xi\neq 0$. However, one can
show~\cite{joyce}, that there always exist some other connection
$\nabla^{(T)}$ which is compatible with the $G$ structure so that
\begin{equation}
   \nabla^{(T)}\xi = 0\ .
\end{equation}
Thus for instance, on an almost Hermitian manifold one can always
find $\nabla^{(T)}$ such that $\nabla^{(T)} J=0$. On a manifold
with $\SU(3)$-structure, it means we can always find
$\nabla^{(T)}$ such that both $\nabla^{(T)} J=0$ and
$\nabla^{(T)}\Omega=0$. Since the existence of an
$\SU(3)$-structure is also equivalent to the existence of an
invariant spinor $\eta$, this is equivalent to the condition
$\nabla^{(T)}\eta=0$.

Let $\CT$ be the contorsion tensor corresponding to
$\nabla^{(T)}$. From the symmetries~\eqref{Kprop}, we see that
$\CT$ is an element of $\Lambda^1\otimes\Lambda^2$ where
$\Lambda^n$ is the space of $n$-forms. Alternatively, since
$\Lambda^2\cong\so(d)$, it is more natural to think of
$\CT_{mn}{}^p$ as one-form with values in the Lie-algebra
$\so(d)$ that is $\Lambda^1\otimes\so(d)$. Given the existence of
a $G$-structure, we can decompose $\so(d)$ into a part in the Lie
algebra $g$ of $G\subset\SO(d)$ and an orthogonal piece
$g^\perp=\so(d)/g$. The contorsion $\CT$ splits accordingly into
\begin{equation}
   \CT = \CT^0 + \CT^g \ ,
\end{equation}
where $\CT^0$ is the part in $\Lambda^1\otimes g^\perp$. Since an
invariant tensor (or spinor) $\xi$ is fixed under $G$ rotations,
that action of $g$ on $\xi$ vanishes and we have, by definition,
\begin{equation}
\label{ICTdef}
   \nabla^{(T)}\xi = \left(\nabla - \CT^0 - \CT^g\right)\xi
       = \left(\nabla - \CT^0\right)\xi = 0 \ .
\end{equation}
Thus, any two $G$-compatible connections must differ by a piece
proportional to $\CT^g$  and they have a common term $\CT^0$ in
$\Lambda^1\otimes g^\perp$ called the ``intrinsic contorsion''.
Recall that there is an isomorphism~\eqref{T-K} between $\CT$ and
$T$. It is more conventional in the mathematics literature to
define the corresponding torsion
\begin{equation}
\label{IT}
   T^0_{mn}{}^p = \CT^0_{[mn]}{}^p \in \Lambda^1\otimes g^\perp \ ,
\end{equation}
known as the {\em intrinsic torsion}.

From the relation~\eqref{ICTdef} it is clear that the intrinsic
contorsion, or equivalently torsion, is independent of the choice
of $G$-compatible connection. Basically it is a measure of the
degree to which $\nabla\xi$ fails to vanish and as such is a
measure solely of the $G$-structure itself. Furthermore, one can
decompose $\CT^0$ into irreducible $G$ representations. This
provides a classification of $G$-structures in terms of which
representations appear in the decomposition. In particular, in
the special case where $\CT^0$ vanishes so that $\nabla\xi=0$,
one says that the structure is ``torsion-free''. For an almost
Hermitian structure this is equivalent to requiring that the
manifold is complex and K\"ahler. In particular, it implies that
the holonomy of the Levi--Civita connection is contained in $G$.

Let us consider the decomposition of $T^0$ in the case of
$SU(3)$-structure. The relevant representations are
\begin{equation}
\Lambda^1 \sim \rep 3\oplus \rep{\bar 3}\ ,\qquad g \sim \rep 8\
, \qquad g^\perp \sim \rep 1 \oplus \rep 3\oplus \rep{\bar 3}\ .
\end{equation}
Thus the intrinsic torsion, which is an element of
$\Lambda^1\otimes\su(3)^\perp$, can be decomposed into the
following $SU(3)$ representations
\begin{equation}
\label{ITdecomp}
\begin{split}
   \Lambda^1 \otimes \su(3)^\perp &=
      (\rep 3\oplus \rep{\bar 3}) \otimes
         (\rep 1 \oplus \rep 3\oplus\rep{\bar 3 }) \\
      &= (\rep 1 \oplus \rep 1) \oplus (\rep 8 \oplus \rep 8)
         \oplus (\rep 6 \oplus \rep{\bar  6})
         \oplus (\rep 3 \oplus \rep{\bar  3})
         \oplus (\rep 3 \oplus \rep{\bar  3})' \ .
\end{split}
\end{equation}
The terms in parentheses on the second line correspond precisely
to the five classes $\W_1,\ldots,\W_5$ presented in
table~\ref{tabW}. We label the component of $T^0$ in each class
by $T_1,\dots,T_5$.

In the case of $\SU(3)$-structure, each component $T_i$ can be
related to a particular component in the $\SU(3)$ decomposition of
$\dd J$ and  $\dd\Omega$. From~\eqref{ICTdef}, we have
\begin{equation}
\label{dJdO}
\begin{aligned}
    d  J_{mnp} &= 6 T^0_{[mn}{}^r J_{r|p]} \ , \\
    d  \Ox_{mnpq} &= 12 T^0_{[mn}{}^r \Ox_{r|pq]}\ .
\end{aligned}
\end{equation}
Since $J$ and $\Omega$ are $\SU(3)$ singlets, $ d  J$ and $ d
\Omega$ are both elements of $\Lambda^1\otimes\su(3)^\perp$. Put
another way, the contractions with $J$ and $\Omega$
in~\eqref{dJdO} simply project onto different $\SU(3)$
representations of $T^0$. We can see which representations appear
simply by decomposing the real three-form $ d  J$ and complex
four-form $ d \Omega$ under $\SU(3)$. We have,
\begin{equation}
\label{dJdecomp}
\begin{aligned}
    d  J &= \big[( d  J)^{3,0} + ( d  J)^{0,3} \big]
       + \big[( d  J)^{2,1}_0 + ( d  J)^{1,2}_0 \big]
       + \big[( d  J)^{1,0} + ( d  J)^{0,1} \big] \ , \\
   \rep{20} &= (\rep 1 \oplus \rep 1)
       \oplus (\rep 6 \oplus \rep{\bar  6})
       \oplus (\rep 3 \oplus \rep{\bar  3}) \ ,
\end{aligned}
\end{equation}
and
\begin{equation}
\label{dOdecomp}
\begin{aligned}
    d  \Omega &= ( d  \Omega)^{3,1} + ( d \Omega)^{2,2}_0  +
      ( d \Omega)^{0,0} \ , \\
   \rep{24} &= (\rep 3 \oplus \rep{\bar  3})'
      \oplus (\rep 8 \oplus \rep 8)
      \oplus (\rep 1 \oplus \rep 1) \ .
\end{aligned}
\end{equation}
The superscripts in the decomposition of $ d  J$ and $ d \Omega$
refer to the $(p,q)$-type of the form. The $0$ subscript refers
to the irreducible $\SU(3)$ representation where the trace part,
proportional to $J^n$ has been removed. Thus in particular, the
traceless parts $( d  J)^{2,1}_0$ and $( d \Omega)^{2,2}_0$
satisfy $J\wedge( d  J)^{2,1}_0=0$ and $J\wedge( d
\Omega)^{2,2}_0=0$ respectively. The trace parts on the other
hand, have the form $( d  J)^{1,0}=\alpha\wedge J$ and $( d
\Omega)^{0,0}=\beta J\wedge J$, with $\alpha\sim *(J\wedge d  J)$
and $\beta\sim *(J\wedge d \Omega)$ respectively. Note that a
generic complex four-form has 30 components. However, since
$\Omega$ is a $(3,0)$-form, from~\eqref{ddecomp} we see that $ d
\Omega$ has no $(1,3)$ part, and so only has 24 components.
Comparing~\eqref{dJdecomp} and~\eqref{dOdecomp}
with~\eqref{ITdecomp} we see that
\begin{equation}
    d  J \in \W_1 \oplus \W_3 \oplus \W_4 \ , \quad
    d \Omega \in \W_1 \oplus \W_2 \oplus \W_5 \ ,
\end{equation}
and as advertised, $ d  J$ and $ d \Omega$ together include all
the components $T_i$. Explicit expressions for some of these
relations are given above in~\eqref{cxT12} and \eqref{T3}. Note
that the singlet component $T_1$ can be expressed either in terms
of $( d  J)^{0,3}$, corresponding to $\Omega\wedge d  J$ or in
terms of $( d \Omega)^{0,0}$ corresponding to $J\wedge d \Omega$.
This is simply a result of the relation~\eqref{JOcond} which
implies that $\Omega\wedge d  J=J\wedge d \Omega$.


\chapter{The Ricci scalar of half-flat manifolds} \label{Rhf}

 The Riemann curvature tensor is defined as
  \begin{equation}
    \label{RTexp}
    R_{mnp}{}^q = \partial_m \phi_{np}{}^q - \partial_n \phi_{mp}{}^q
    - \phi_{mp}{}^r \phi_{nr}{}^q + \phi_{np}{}^r \phi_{mr}{}^q,
  \end{equation}
  where $\phi$ denotes a general connection that contains two contributions
$\phi_{mn}{}^p = \Gx_{mn}{}^p + \CT_{mn}{}^p$ where
$\Gx_{mn}{}^p=\Gx_{nm}{}^p$ denote the Christoffel symbols and
$\CT_{mn}{}^p$ is the contorsion, out of which we define the
torsion

\bea T_{mn}{}^p & = & \frac12
(\CT_{mn}{}^p-\CT_{nm}{}^p)\label{tors1}\\[2mm]
\CT_{mnp} & = & T_{mnp} + T_{pmn} + T_{pnm}.\label{tors2}
 \eea
For the Ricci tensor we use $R_{np} = R_{nmp}{}^m$. The simplest
way to derive the Ricci scalar for the manifold considered in
section \ref{Yhat} is by using the integrability condition one
can derive from the Killing spinor equation (\ref{torKS}).
\begin{equation}
  \label{intKS1}
  R^{(T)}_{mnpq} \Gx^{pq} \eta = 0,
\end{equation}
where the Riemann tensor of the connection with torsion is given
by (\ref{RTexp})
\begin{equation}
  \label{RT1}
  R_{mnpq}^{(T)} = R(\Gx)_{mnpq} + \nabla_m \CT_{npq} - \nabla_n \CT_{mpq} -
  \CT_{mp}{}^r \CT_{nrq} + \CT_{np}{}^r \CT_{mrq} \ .
\end{equation}
Here $R(\Gx)_{mnpq}$ represents the usual Riemann tensor for the
Levi-Civita connection and the covariant derivatives are again
with respect to the Levi-Civita connection. For definiteness we
choose the solution of the Killing spinor equation (\ref{torKS})
to be a Majorana spinor.\footnote{The results are independent of
the choice of the spinor, but the derivations may be more
involved.} Multiplying (\ref{intKS1}) by $\Gx^n$ and summing over
$n$ one obtains
\begin{equation}
  \label{intKS2}
  R^{(T)}_{mnpq} \Gx^{npq} \eta - 2 R^{(T)}_{mn} \Gx^{n} \eta = 0\ .
\end{equation}
Contracting from the left with $\eta^\dagger \Gx^m$ and using the
conventions for the Majorana spinors (\ref{cad05}) one derives
\begin{equation}
  \label{intKS3}
  2 R^{(T)} = R^{(T)}_{mnpq} \eta^\dagger \Gx^{mnpq} \eta\ .
\end{equation}
where $R^{(T)}$ represents the Ricci scalar which can be defined
from the Riemann tensor (\ref{RT1}). Expressing $ R^{(T)}_{mnpq}$
in terms of $R(\Gx)_{mnpq}$ from (\ref{RT1}), using the Bianchi
identity $R(\Gx)_{m[npq]} = 0$ and the fact that the contorsion
is traceless $\CT_{mn}{}^m = \CT^m{}_{mn} = 0$ which holds for
half flat manifolds one can derive the formula for the Ricci
scalar of the Levi-Civita connection
\begin{equation}
  \label{Rapp}
  R = - \CT_{mnp} \CT^{npm} - \frac{1}{2} \epsilon^{mnpqrs} (\nabla_m \CT_{npq} -
  \CT_{mp}{}^l \CT_{nlq}) J_{rs} \ .
\end{equation}

In order to simplify the formulas we evaluate (\ref{Rapp}) term
by term. The strategy will be to express first the contorsion
$\CT$ in terms of the torsion $T$ (\ref{tors2}) and then go to
complex indices splitting the torsion in its component parts
$T_{1\oplus2}$ and $T_3$ which are of definite type with respect
to the almost complex structure $J$.

The first term can be written as
\begin{equation}
  \label{A1}
  A \equiv - \CT_{mnp} \CT^{npm} = - (T_{mnp} + T_{pmn} + T_{pnm}) T^{npm} = T_{mnp}
  T^{mnp} - 2 T_{mnp} T^{npm}.
\end{equation}
Using (\ref{su3Nt}) and (\ref{T3}) one sees that the first two
indices of $T$ are of the same type and thus one has
\begin{equation}
  \label{Af}
  A = (T_{1\oplus 2})_{\ax \bx \cx} (T_{1\oplus 2})^{\ax \bx \cx} - 2
  (T_{1\oplus 2})_{\ax \bx \cx} (T_{1\oplus 2})^{\bx \cx \ax}  + (T_3)_{\ax
  \bx \cb} (T_3)^{\ax \bx \cb} + c.c. \ ,
\end{equation}
where $c.c.$ denotes complex conjugation.

The second term can be computed if one takes into account that
the 4 dimensional effective action appears after one integrates
the 10 dimensional action over the internal space, in this case
$\hat Y$. Thus the second term in (\ref{Rapp}) can be integrated
by parts to give\footnote{ Strictly speaking in 10 dimensions the
Ricci scalar comes multiplied with a dilaton factor (\ref{SIIA}).
However in all that we are doing we consider that the dilaton is
constant over the internal space so it still make sense to speak
about integration by parts without introducing additional factors
with derivatives of the dilaton. }
\begin{equation}
  \label{B1}
  B \equiv - \frac12 \epsilon^{mnpqrs} (\nabla_m \CT_{npq}) J_{rs} \sim
  \frac12 \epsilon^{mnpqrs} \CT_{npq} \nabla_m J_{rs}.
\end{equation}
Using (\ref{torJO}) and (\ref{T-K}) we obtain after going to
complex indices
\begin{eqnarray}
  \label{B2}
  B & =& - \epsilon^{mnpqrs} T_{mnp} T_{qr}{}^t J_{ts} \\
  &=& -\epsilon^{\ax \bx \cx
  \ab \bb \cb} (T_{1\oplus 2})_{\ax \bx \cx} (T_{1\oplus 2})_{\ab \bb}{}^\dx
  J_{\dx \cb} -\epsilon^{\ax \bx \cb \ab \bb \cx} (T_3)_{\ax \bx \cb}
  (T_3)_{\ab \bb}{}^{\bar \dx} J_{\bar \dx \cx} + c.c. \ . \nonumber
\end{eqnarray}
The 6 dimensional $\epsilon$ symbol splits as
\begin{equation}
  \epsilon^{\ax \bx \cx \ab \bb \cb} = -i \epsilon^{\ax \bx \cx} \epsilon^{\ab
  \bb \cb} \ ,
\end{equation}
and after some algebra involving the 3 dimensional $\epsilon$
symbol one finds
\begin{equation}
  \label{Bf}
  B = -2 (T_{1\oplus 2})_{\ax \bx \cx} (T_{1\oplus 2})^{\ax \bx \cx} -
  4(T_{1\oplus 2})_{\ax \bx \cx} (T_{1\oplus 2})^{\bx \cx \ax} - 2 (T_3)_{\ax
  \bx \cb} (T_3)^{\ax \bx \cb} + c.c. \ .
\end{equation}

In the same way one obtains for the last term
\begin{equation}
  \label{Cf}
  C \equiv \frac12 \epsilon^{mnpqrs} \CT_{mp}{}^t \CT_{ntq} J_{rs} = 2
  (T_{1\oplus 2})_{\ax \bx \cx} (T_{1\oplus 2})^{\ax \bx \cx} + 2  (T_3)_{\ax
  \bx \cb} (T_3)^{\ax \bx \cb} + c.c. \ .
\end{equation}
Collecting the results from (\ref{Af}), (\ref{Bf}) and (\ref{Cf})
the formula for the Ricci scalar (\ref{Rapp}) becomes
\begin{equation}
  \label{Rint}
  R = (T_{1\oplus 2})_{\ax \bx \cx} (T_{1\oplus 2})^{\ax \bx \cx} - 6
  (T_{1\oplus 2})_{\ax \bx \cx} (T_{1\oplus 2})^{\bx \cx \ax}  + (T_3)_{\ax
  \bx \cb} (T_3)^{\ax \bx \cb} + c.c. \ .
\end{equation}
The first two terms in the above expression can be
straightforwardly computed using (\ref{su3Nt}), (\ref{F4}) and

\beq (T_{1\oplus 2})_{\ax \bx \cx} =
\frac{e_i}{4||\Ox||^2}(\tox^i)_{\ax\bx\ab\bb}\Ox^{\ab\bb}{}_{\cx}.
\label{exptors}\eeq After a little algebra we find
\begin{eqnarray}
  \label{AB}
  (T_{1\oplus 2})_{\ax \bx \cx} (T_{1\oplus 2})^{\ax \bx \cx} & = & \frac{e_i e_j}{8 ||\Ox||^2}
  (\tox^i)_{\ax \bx \ab \bb} (\tox^j)^{\ax \bx \ab \bb}  \\
  (T_{1\oplus 2})_{\ax \bx \cx} (T_{1\oplus 2})^{\bx \cx \ax} & = & - \frac{e_i e_j}{8 ||\Ox||^2}
  (\tox^i)_{\ax \bx \ab \bb} (\tox^j)^{\ax \bx \ab \bb} - \frac{e_i e_j}{4
  ||\Ox||^2} (*\tox^i)_{\ax \bb} (*\tox^j)^{\bb\ax} + \frac{(e_i v^i)^2}{4
  ||\Ox ||^2 \cK^2} \ . \nn
\end{eqnarray}
In order to obtain the above expressions we have used (\ref{Jexp})
and  \cite{AS3}

\bea (\tox^i)_{\ax\bx\bb}{}^{\bx} & = &
  -ig_{\ax\bb}(*\tox^i)_{\cx\cb}g^{\cx\cb}+i(*\tox^i)_{\ax\bb}\\[2mm]
\label{D.15}
  (\tox^i)_{\ax \bx}{}^{\ax \bx} & = & \frac{2 v^i}{\cK}.
\eea

Integrating (\ref{AB}) over $\hat Y$ we obtain
\begin{eqnarray}
  \label{ABf}
  \int_{\hat Y} (T_{1\oplus 2})_{\ax \bx \cx} (T_{1\oplus 2})^{\ax \bx \cx} & = &
  \frac{e_i e_j g^{ij}}{8 ||\Ox||^2 \cK}\ , \\
  \int_{\hat Y} (T_{1\oplus 2})_{\ax \bx \cx} (T_{1\oplus 2})^{\bx \cx \ax} & = &
  - \frac{e_i e_j g^{ij}}{16 ||\Ox||^2 \cK} + \frac{(e_i v^i)^2}{4 ||\Ox ||^2
  \cK} \ .\nn
\end{eqnarray}

Finally, we have to compute the third term in (\ref{Rint}). For
this we note that, since $dJ$ is in $W_1^-+W_2^-+W_3$, the
torsions $T_{12}$ and $T_3$ both appear in \eqref{eq:dJOe}.
Projecting \eqref{eq:dJOe} on the $(3,0)$ component, we find a
relation between $T_{12}$ and $\bx^0|_{(3,0)}$, which using the
explicit expression for the torsion \eqref{exptors} can be written

\begin{equation}
  \label{bt}
  \bx^0_{\ax\bx\cx} = -\frac{i}{\cK||\Ox||^2}\Ox_{\ax\bx\cx}.
\end{equation} The $(2,1)$ component gives $T_3$ in terms of the $(2,1)$ part of
$\bx^0$

\begin{equation} (T_3)_{\ax\bx\cb} =
-\frac{i}{2}v^ie_i\,\bx^0_{\ax\bx\cb}.\label{expr2}
\end{equation}

To obtain an explicit expression for the third term in
(\ref{Rint}), we compute the following quantity

\begin{equation}
  \label{chi2}
   \int_{\hat Y} \bx^0 \wg * \bx^0 = \frac{2}{\cK||\Ox||^2}+ \int_{\hat Y} \bx^0_{\ax\bx\cb}
   \bx^{0\, \ax\bx\cb}.
\end{equation}
On the other hand, the same integral appears in

\begin{equation}
  \label{imM}
  \int \bx^0 \wg * \bx^0 = - \left[(\IM \cM)^{-1} \right]^{00}=
  \frac{8}{||\Ox||^2 \cK}.
\end{equation}
The simplest way to see this is by using a mirror symmetry
argument. We know that under mirror symmetry the gauge couplings
$\cM$ and $\cN$ are mapped into one another. This also means that
$(\IM \cM)^{-1}$ is mapped into $(\IM \cN)^{-1}$ and this matrix
is given in (\ref{ImN-1e}) for a Calabi-Yau space. From here one
sees that the element $\left[(\IM \cN)^{-1}\right]^{00}$ is just
the inverse volume of the mirror Calabi-Yau space. Using again
mirror symmetry and the fact that the K\"ahler potential of the
K\"ahler moduli (\ref{msa4}) is mapped into the K\"ahler
potential of the complex structure moduli (\ref{Kpotcs}) we end
up with the RHS of the above equation.

Now we obtain

\begin{equation}
   \label{C'f}
  \int_{\hat Y} (T_3)_{\ax \bx \cb} (T_3)^{\ax \bx \cb} = \frac32 \; \frac{(e_i
  v^i)^2}{||\Ox||^2 \cK} \ .
\end{equation}

Inserting (\ref{ABf}) and (\ref{C'f}) into (\ref{Rint}) and
taking into account that all the terms in (\ref{ABf}) and
(\ref{C'f}) are explicitly real such that the term '$c.c.$' in
(\ref{Rint}) just introduces one more factor of $2$, we obtain the
final form of the Ricci scalar
\begin{equation}
  \label{Rfin}
  R = - \frac{1}{8} \, e_i e_j g^{ij} \left[(\IM \cM)^{-1} \right]^{00} \ ,
\end{equation}
where we have used again (\ref{imM}).

\chapter{Display of torsion and curvature
components}\label{torcurv}

The conventional constraints compatible with the assumptions
(\ref{T01} -- \ref{T02}) and \equ{tz} are the following: \be
\begin{array}{lcl} T^\tc_\g{}_b{}^a\ =\
0&\qquad\quad&T_\tc^{\dot{\gamma}}{}_b{}^a\ =\ 0 \\ [2mm]
T^\tc_\g{}^\tb_\b{}_{\ta}^\a\ =\
0&&T_\tc^{\dot{\gamma}}{}_\tb^{\dot{\beta}}{}^{\ta}_\da\ =\ 0 \\
[2mm] T^\tc_\g{}^{\dot{\beta}}_\tb{}_\da^\ta\ =\
0&&T_\tc^\g{}_{\dot{\beta}}^\tb{}^\da_\ta\ =\ 0 \\ [2mm]
T_c{}^\tb_\a{}^\a_\ta\ =\ 0&&T_c{}_\tb^\da{}_\da^\ta\ =\ 0 \\
[2mm] &T_{cb}{}^a\ =\ 0\,.&
\end{array} \eqn{conv007}

There is a particular solution of the Bianchi identities for the
torsion and 3--form subject to the constraints (\ref{T01} --
\ref{H02}), (\ref{dt} -- \ref{tz}) and \equ{conv007}, which
describes the N-T supergravity multiplet. Besides the constant
$T^{\dot{\gamma}}_\tc{}^\tb_\b{}^a$ and the supercovariant field
strength of the graviphotons, $T_{cb}{}^\au\doteq F_{cb}{}^\au$,
the non-zero torsion components corresponding to this solution
are then the following: \be \begin{array}{lcl}
T_\g^\tc{}_\b^\tb{}^\au\ =\
4\eps_{\g\b}{\mathfrak{t}}^{[\tc\tb]}{}^\au e^\f&&
T^{\dot{\gamma}}_\tc{}^{\dot{\beta}}_\tb{}^\au\ =\
4\eps^{{\dot{\gamma}}{\dot{\beta}}}{\mathfrak{t}}_{[\tc\tb]}{}^\au
e^\f
\\[2mm]
T^\tc_\g{}^\tb_\b{}^\ta_\da\ =\
q\eps_{\g\b}\varepsilon^{\tc\tb\ta\tf}\bla_\tf{}_\da &\qquad&
T_\tc^{\dot{\gamma}}{}_\tb^{\dot{\beta}}{}_\ta^\a\ =\
q\eps^{{\dot{\gamma}}{\dot{\beta}}}\varepsilon_{\tc\tb\ta\tf}\la^\tf{}^\a \\
[2mm] T^\tc_\g{}_b{}^\au\ =\
ie^\f(\s_b\bla_\ta)_\g{\mathfrak{t}}^{[\ta\tc]\au}&&
T_\tc^{\dot{\gamma}}{}_b{}^\au \ =\
ie^\f(\bs_b\la^\ta)^{\dot{\gamma}}{\mathfrak{t}}_{[\ta\tc]}{}^\au
\\ [2mm] T^\tc_\g{}_b{}^\a_\ta\ =\ -2(\s_{ba})_\g{}^\a
U^a{}^\tc{}_\ta&&T_\tc^{\dot{\gamma}}{}_b{}_\da^\ta \ =\
2(\bs_{ba})^{\dot{\gamma}}{}_\da U^a{}^\ta{}_\tc \\ [2mm]
T^\tc_\g{}_b{}_\da^\ta\ =\
\frac{i}{2}(\s_b\bs^{dc})_{\g\da}F_{dc}{}^{[\tc\ta]}e^{-\f}
&&T_\tc^{\dot{\gamma}}{}_b{}^\a_\ta\ =\
\frac{i}{2}(\bs_b\s^{dc})^{{\dot{\gamma}}\a}F_{dc}{}_{[\tc\ta]}e^{-\f} \\
[2mm] T_{cb}{}_{\ta\a}\ =\ -(\eps\s_{cb})^{\g\b}\S_{(\g\b\a)\ta}
&& T_{cb}{}^{\ta\da}\ =\
-(\eps\bs_{cb})_{{\dot{\gamma}}{\dot{\beta}}}\S^{({\dot{\gamma}}{\dot{\beta}}\da)\ta}
\\ [2mm]
\quad-\frac{1}{4}\textrm{tr}(\bs_{cb}\bs_{af})F^{af}{}_{[\ta\tf]}\la^\tf_\a
&&
\quad-\frac{1}{4}\textrm{tr}(\s_{cb}\s_{af})F^{af}{}^{[\ta\tf]}\bla_\tf^\da
\\ [2mm]
\quad-\frac{1}{12}\left(\d^{af}_{cb}-\frac{i}{2}\varepsilon_{cb}{}^{af}\right)\bP_a(\s_f\bla_\ta)_\a&&
\quad-\frac{1}{12}\left(\d^{af}_{cb}+\frac{i}{2}\varepsilon_{cb}{}^{af}\right)P_a(\bs_f\la^\ta)^\a
\end{array} \end{equation} with
$U_a{}^\tb{}_\ta=-\frac{i}{8}(\la^\tb\s_a\bla_\ta-\frac{1}{2}\d^\tb_\ta\la^\tf\s_a\bla_\tf)$,
$P$ and $\bP$ are given in equations \equ{P} and \equ{barP},
while $\S_{(\g\b\a)\ta}$ and
$\S^{({\dot{\gamma}}{\dot{\beta}}\da)\ta}$ are the gravitino
"Weyl" tensors.

Furthermore, the Lorentz curvature has components \be
\begin{array}{lcl} R^{\td\,\tc}_{\d\,\g\, ba}\ =\
2\eps_{\d\g}\textrm{tr}(\s_{dc}\s_{ba})F^{dc[\td\tc]}e^{-\f}&&
R_{\td\,\tc}^{{\dot{\delta}}\,{\dot{\gamma}}}{}_{ ba}\ =\
2\eps^{{\dot{\delta}}{\dot{\gamma}}}\textrm{tr}(\bs_{dc}\bs_{ba})F^{dc}{}_{[\td\tc]}e^{-\f}\\[2mm]
R_{\td\,\g\, ba}^{{\dot{\delta}}\,\tc}\ =\ -4\eps_{dcba}U^d{}^\td{}_\tc(\s^c\varepsilon)_\g{}^{\dot{\delta}}&& R_{\cc\au\,ba}\ =\ 0\\[2mm]
R^\td_\d{}_c{}_{ba}\ =\
-2i(\s_c)_{\d\da}(\eps\bs_{ba})_{{\dot{\gamma}}{\dot{\beta}}}\S^{({\dot{\gamma}}{\dot{\beta}}\da)\td}
&&R_\td^{\dot{\delta}}{}_c{}_{ba}\ =\
-2i(\bs_c)^{\a{\dot{\delta}}}(\eps\s_{ba})^{\g\b}\S_{(\g\b\a)\td}
\\[2mm]
\quad-\frac{i}{2}\textrm{tr}(\s_{ba}\s_{ef})(\s_c\bla_\ta)_\d
F^{ef}{}^{[\td\ta]}e^{-\f}
&&\quad-\frac{i}{2}\textrm{tr}(\bs_{ba}\bs_{ef})(\bs_c\la^\ta)^{\dot{\delta}}
F^{ef}{}_{[\td\ta]}e^{-\f}
\\[2mm]
\quad+\frac{i}{4}(\s_c\bs_e\s_{ba}\la^\td)_\d P^e
&&\quad+\frac{i}{4}(\bs_c\s_e\bs_{ba}\bla_\td)^{\dot{\delta}}
\bP^e
\end{array} \end{equation} and \bea
R_{dc}{}_{ba}&=&(\eps\s_{dc})^{\d\g}(\eps\s_{ba})^{\b\a}V_{(\d\g\b\a)}
+(\eps\bs_{dc})_{{\dot{\delta}}{\dot{\gamma}}}(\eps\bs_{ba})_{{\dot{\beta}}\da}V^{({\dot{\delta}}{\dot{\gamma}}{\dot{\beta}}\da)}\nn\\[2mm]
&&+\frac{1}{2}\left(\eta_{db}R_{ca}-\eta_{da}R_{cb}+\eta_{ca}R_{db}-\eta_{cb}R_{da}\right)
-\frac{1}{6}(\eta_{db}\eta_{ca}-\eta_{da}\eta_{cb})R \eea with
the supercovariant Ricci tensor, $R_{db}=R_{dcba}\eta^{ca}$,
given by \bea R_{db}\  &=&
-2\cd_d\f\,\cd_b\f-\frac{1}{2}e^{-4\f}H^*_d\,H^*_b
-e^{-2\f}F_{df[\tb\ta]}F_b{}^f{}^{[\tb\ta]}+\frac{1}{4}\eta_{db}e^{-2\f}F_{ef[\tb\ta]}F^{ef[\tb\ta]}\nn\\[2mm]
&&+\frac{1}{8}\sum_{db}\left\{i(\cd_b\la^\tf)\s_d\bla_\tf-i\la^\tf\s_d\cd_b\bla_\tf
+e^{-2\f}H_d^*(\la^\tf\s_b\bla_\tf)\right\}\nn\\[2mm]
&&-\frac{1}{32}(\la^\tf\s_d\bla_\tf)\,(\la^\ta\s_b\bla_\ta)-\frac{1}{16}\eta_{db}(\la^\ta\la^\tf)(\bla_\ta\bla_\tf)
\label{Riccitensor} \eea and the corresponding Ricci scalar,
$R=R_{db}\eta^{db}$, which is then \be
R=-2\cd^a\f\cd_a\f-\frac{1}{2}H^*{}^aH^*{}_ae^{-4\f}
+\frac{3}{4}e^{-2\f}H^*{}^a(\la^\ta\s_a\bla_\ta)
+\frac{3}{8}(\la^\tb\la^\ta)(\bla_\tb\bla_\ta). \end{equation}
The tensors $V_{(\d\g\b\a)}$ and
$V^{({\dot{\delta}}{\dot{\gamma}}{\dot{\beta}}\da)}$ are
components of the usual Weyl tensor. Like the gravitino Weyl
tensors, $\S_{(\g\b\a)\ta}$ and
$\S^{({\dot{\gamma}}{\dot{\beta}}\da)\ta}$, their lowest
components do not participate in the equations of motion.

As for the 2--form sector, besides the supercovariant field
strength of the antisymmetric tensor, $H_{cba}$, the non-zero
components of the 3--form $H$, which do not have central charge
indices, are \be
H^{\dot{\gamma}}_\tc{}^\tb_\b{}_a=-2i\d^\tb_\tc(\s_a\eps)_\b{}^{\dot{\gamma}}
e^{2\f} \qquad H^\tc_\g{}_{ba}\ =\ 4(\s_{ba}\la^\tc)_\g e^{2\f}
\qquad H_\tc^{\dot{\gamma}}{}_{ba}\ =\
4(\bs_{ba}\bla_\tc)^{\dot{\gamma}} e^{2\f}\ .
\end{equation} The components with at least one central charge
index, are related to the torsion components by \be
H_{\cd\cc\au}\ =\ T_{\cd\cc}{}^\az\gg_{\az\au}, \end{equation}
with the metric $\gg_{\az\au}$ defined in \equ{metric}.

\newpage

\subsection*{Acknowledgments}

This thesis was supported by the CNRS -- the French Center for
National Scientific Research.

I thank R. Grimm for advising me during these three years, for
helping me finding a way through scientific problems as well as
administrative issues, and for making me benefit from his
numerous acquaintance in the scientific community. I thank the
CPT for providing suitable facilities necessary to this work.

I would like to thank A. Van Proeyen for welcoming me at K. U.
Leuven and the Erasmus Exchange Program for supporting this visit.

I am very grateful to J. Louis for inviting me several times at
Martin-Luther University, Halle, as well as at Desy, Hamburg. A
several month stay in Halle was supported by the DAAD -- the
German Academic Exchange Service.

I also had the opportunity to stay at Kyoto University, with
financial help from the summer program of the MEXT, the japanese
Ministry of Education, Research and Science, and the JSPS, the
Japanese Society for Promotion of Science. I could appreciate the
welcoming of all members of the University, especially T. Kugo.

 Finally I thank R. Grimm, E. Loyer, A. M. Kiss, A. Van Proeyen, J. Louis,
A. Micu, D. Waldram for interesting discussions and fruitful
collaboration.

\bibliography{CalabiRef,newREF}

\providecommand{\href}[2]{#2}\begingroup\raggedright\begin{thebibliography}{10%
0}

\bibitem{GSW}
M.~B. Green, J.~H. Schwarz  and E.~Witten, \emph{Superstring theory},
Cambridge, Uk: Univ. Pr. ( 1987) 596 P. ( Cambridge Monographs On Mathematical
  Physics)

\bibitem{LT}
D.~L{\"u}st and S.~Theisen, \emph{Lectures on string theory}, Springer-Verlag,
  Berlin
(1989)

\bibitem{JP}
J.~Polchinski, \emph{String theory},
Cambridge University Presse

\bibitem{AHDD}
N.~Arkani-Hamed, S.~Dimopoulos  and G.~Dvali, \emph{The Hierarchy Problem and
  New Dimensions at a Millimeter}, Phys. Rev. Lett. {\bf B429} (1998) 263,
\href{http://www.arXiv.org/abs/hep-th/9803315}{{\tt hep-th/9803315}}

\bibitem{AAHDD}
I.~Antoniadis, N.~Arkani-Hamed, S.~Dimopoulos  and G.~Dvali, \emph{New
  Dimensions at a Millimeter to a Fermi and superstrings at a TeV}, Phys. Lett.
  {\bf B436} (1998) 257,
\href{http://www.arXiv.org/abs/hep-th/9804398}{{\tt hep-th/9804398}}

\bibitem{CKLT}
G.~Curio, A.~Klemm, D.~L{\"u}st  and S.~Theisen, \emph{On the vacuum structure
  of type II string compactifications on Calabi-Yau spaces with H-fluxes},
  Nucl. Phys. {\bf B609} (2001) 3,
\href{http://www.arXiv.org/abs/hep-th/0012213}{{\tt hep-th/0012213}}

\bibitem{LM2}
J.~Louis and A.~Micu, \emph{Type II Theories Compactified on Calabi-Yau
  Threefolds in the Presence of Background Fluxes}, Nucl. Phys. {\bf B635}
  (2002) 395,
\href{http://www.arXiv.org/abs/hep-th/0202168}{{\tt hep-th/0202168}}

\bibitem{Vafa}
T.~Taylor and C.~Vafa, \emph{Superstrings and topological strings at large N},
  J. Math. Phys. {\bf 42} (2001) 2798,
\href{http://www.arXiv.org/abs/hep-th/0008142}{{\tt hep-th/0008142}}

\bibitem{GLMW}
S.~Gurrieri, J.~Louis, A.~Micu  and D.~Waldram, \emph{Mirror symmetry in
  generalized Calabi-Yau compactifications}, Nucl. Phys. {\bf B654} (2003) 61,
\href{http://www.arXiv.org/abs/hep-th/0211102}{{\tt hep-th/0211102}}

\bibitem{GM}
S.~Gurrieri and A.~Micu, \emph{Type IIB on Half-flat manifolds}, Class. Quant.
  Grav. {\bf 20} (2003) 2181,
\href{http://www.arXiv.org/abs/hep-th/0212278}{{\tt hep-th/0212278}}

\bibitem{Cha81a}
A.~Chamseddine, \emph{{{N}=4 Supergravity Coupled to {N}=4 Matter and Hidden
  Symmetries}}, Nucl.Phys. {\bf B185} (1981)
403--415

\bibitem{NT81}
H.~Nicolai and P.~K. Townsend, \emph{{N} = 1 supersymmetry multiplets with
  vanishing trace anomaly : building blocks of the {N} $<$ 3 supergravities},
  Phys. Lett. {\bf 98B} (1981)
257--260

\bibitem{GHK01}
R.~Grimm, C.~Herrmann  and A.~Kiss, \emph{N=4 supergravity with antisymmetric
  tensor in central charge superspace}, Class. Quant. Grav. {\bf 18} (2001)
  1027--1038,
hep-th/0009201

\bibitem{GK02}
S.~Gurrieri and A.~Kiss, \emph{Equations of motion for $N=4$ supergravity with
  antisymmetric tensor from its geometrical descritption in central charge
  superspace map}, JHEP {\bf 02040} (2002)
\href{http://www.arXiv.org/abs/hep-th/0201234}{{\tt hep-th/0201234}}

\bibitem{BCF}
M.~Bodner, A.~Cadavid  and S.~Ferrara, \emph{$(2,2)$ vacuum configurations for
  type IIA superstrings: $N=2$ supergravity Lagrangians and algebraic
  geometry}, Class. Quant. Grav. {\bf 8} (1991)
789

\bibitem{Bodner:1990ca}
M.~Bodner and A.~Cadavid, \emph{Dimensional Reduction Of Type IIb Supergravity
  And Exceptional Quaternionic Manifolds}, Class. Quant. Grav. {\bf 7} (1990)
829

\bibitem{BGHL}
R.~B{\"o}hm, H.~G{\"u}nther, C.~Herrmann  and J.~Louis, \emph{Compactification
  of type IIB string theory on Calabi-Yau threefolds}, Nucl. Phys. {\bf B569}
  (2000) 229--246,
\href{http://www.arXiv.org/abs/hep-th/9908007}{{\tt hep-th/9908007}}

\bibitem{Haack:2001ha}
M.~Haack, {\em Calabi-Yau Fourfold Compactifications in String Theory}.
\newblock PhD thesis, Martin-Luther Universitat, Halle-Wittenberg,
2001.
\newblock

\bibitem{Duff:1986np}
M.~Duff, B.~Nilsson  and C.~Pope, \emph{Kaluza-Klein supergravity}, Phys. Rept.
  {\bf 130} (1986)
1

\bibitem{Nieu:1985ni}
P.~van Nieuwenhuizen, \emph{An introduction to simple supergravity and the
  Kaluza-Klein program},
Proceedings of Les Houches 1983, 'Relativity, groups and topology II'

\bibitem{Louis:2002am}
J.~Louis and A.~Micu, \emph{Type II Theories Compactified on Calabi-Yau
  Threefolds in the Presence of Background Fluxes}, Nucl. Phys. {\bf B635}
  (2002) 395,
\href{http://www.arXiv.org/abs/hep-th/0202168}{{\tt hep-th/0202168}}

\bibitem{FeS}
S.~Ferrara and S.~Sabharwal, \emph{Quaternionic manifolds for type II
  superstring vacua of Calabi-Yau spaces}, Nucl. Phys. {\bf B332} (1990)
317

\bibitem{deWit:2002am}
B.~de~Wit and A.~V. Proeyen, \emph{Potentials and symmetries of general gauged
  $N=2$ supergravity - Yang-Mills models}, Nucl. Phys. {\bf B245} (1984)
89

\bibitem{Bagger:1983wi}
J.~Bagger and E.~Witten, \emph{Matter couplings in $N=2$ supergravity}, Nucl.
  Phys. {\bf B222} (1983)
1

\bibitem{deWit:1985lp}
B.~de~Wit, P.~Lauwers  and A.~V. Proeyen, \emph{Lagrangians of $N=2$
  supergravity - matter systems}, Nucl. Phys. {\bf B255} (1985)
569

\bibitem{Dauria:2001fp}
R.~D'Auria, S.~Ferrara  and P.~Fre, \emph{Special and quaternionic isometries:
  General couplings in $N=2$ supergravity and the scalar potential}, Nucl.
  Phys. {\bf B359} (1991)
705

\bibitem{N=2}
L.~Andrianopoli, M.~Bertolini, A.~Ceresole, R.D'Auria, S.~Ferrara, P.~Fre  and
  T.~Magri, \emph{$N=2$ supergravity and $\cN = 2$ super Yang-Mills theory on
  general scalar manifolds: Symplectic covariance, gaugings and the momentum
  map}, J. Geom. Phys. {\bf 23} (1997) 111,
\href{http://www.arXiv.org/abs/hep-th/9605032}{{\tt hep-th/9605032}}

\bibitem{mirror}
S.~Hosono, A.~Klemm  and S.~Theisen, \emph{Lectures on mirror symmetry},
\href{http://www.arXiv.org/abs/hep-th/9403096}{{\tt hep-th/9403096}}

\bibitem{PS}
J.~Polchinski and A.~Strominger, \emph{New vacua for type II string theory},
  Phys. Lett. {\bf B388} (1996) 736,
\href{http://www.arXiv.org/abs/hep-th/9510227}{{\tt hep-th/9510227}}

\bibitem{JM}
J.~Michelson, \emph{Compactifications of type IIB strings to four dimensions
  with non-trivial classical potential}, Nucl. Phys. {\bf B495} (1997) 127,
\href{http://www.arXiv.org/abs/hep-th/9610151}{{\tt hep-th/9610151}}

\bibitem{TV}
T.~Taylor and C.~Vafa, \emph{R-R Flux on Calabi-Yau and Partial Supersymmetry
  Breaking}, Phys. Lett. {\bf B474} (2000) 130,
\href{http://www.arXiv.org/abs/hep-th/9912152}{{\tt hep-th/9912152}}

\bibitem{PM1}
P.~Mayr, \emph{On supersymmetry breaking in string theory and its realization
  in brane worlds}, Nucl. Phys. {\bf B593} (2001) 99,
\href{http://www.arXiv.org/abs/hep-th/0003198}{{\tt hep-th/0003198}}

\bibitem{GKP}
S.~B. G.~S. Kachru and J.~Polchinski, \emph{Hierarchies from fluxes in string
  compactifications},
\href{http://www.arXiv.org/abs/hep-th/0105097}{{\tt hep-th/0105097}}

\bibitem{CKKL}
G.~Curio, A.~Klemm, B.~K{\"o}rs  and D.~L{\"u}st, \emph{Fluxes in heterotic and
  type II string compactifications}, Nucl. Phys. {\bf B620} (2002) 237,
\href{http://www.arXiv.org/abs/hep-th/0106155}{{\tt hep-th/0106155}}

\bibitem{GD}
G.~Dall'Agata, \emph{Type IIB supergravity compactified on a Calabi-Yau
  manifold with H-fluxes}, JHEP {\bf 11} (2001) 005,
\href{http://www.arXiv.org/abs/hep-th/0107264}{{\tt hep-th/0107264}}

\bibitem{CKL}
G.~Curio, B.~K{\"o}rs  and D.~L{\"u}st, \emph{Fluxes and Branes in Type II
  Vacua and M-theory Geometry with G(2) and Spin(7) Holonomy}, Nucl. Phys. {\bf
  B636} (2002) 197,
\href{http://www.arXiv.org/abs/hep-th/0111165}{{\tt hep-th/0111165}}

\bibitem{GVW}
S.~Gukov, C.~Vafa  and E.~Witten, \emph{CFT's from Calabi-Yau four-folds},
  Nucl. Phys. {\bf B608} (2001) 477,
\href{http://www.arXiv.org/abs/hep-th/9906070}{{\tt hep-th/9906070}}

\bibitem{gukov}
S.~Gukov, \emph{Solitons, superpotentials and calibrations}, Nucl. Phys. {\bf
  B574} (2000) 169,
\href{http://www.arXiv.org/abs/hep-th/9911011}{{\tt hep-th/9911011}}

\bibitem{FFS}
M.~Falcitelli, A.~Farinola  and S.~Salamon, \emph{Almost-Hermitian Geometry},
  Diff. Geo. {\bf 4} (1994)
259

\bibitem{salamonb}
S.~Salamon, \emph{Riemannian Geometry and Holonomy Groups},
Pitman Research Notes in Mathematics, Longman, Harlow

\bibitem{joyce}
D.~Joyce, \emph{Compact Manifolds with Special Holonomy},
Oxford University Press, Oxford

\bibitem{friedrich}
T.~Friedrich and S.~Ivanov, \emph{Parallel spinors and connections with
  skew-symmetric torsion in string theory},
\href{http://www.arXiv.org/abs/math.dg/0102142}{{\tt math.dg/0102142}}

\bibitem{salamon}
S.~Salamon, \emph{Almost Parallel Structures},
\href{http://www.arXiv.org/abs/math.DG/0107146}{{\tt math.DG/0107146}}

\bibitem{CS}
S.~Chiossi and S.~Salamon, \emph{The Intrinsic Torsion of $SU(3)$ and $G_2$
  Structures},
\href{http://www.arXiv.org/abs/math.DG/0202282}{{\tt math.DG/0202282}}

\bibitem{AS1}
A.~Strominger, \emph{Superstrings with torsion}, Nucl. Phys. {\bf B274} (1986)
253

\bibitem{Hull1}
C.~M. Hull,
\emph{Superstring Compactifications With Torsion And Space-Time Supersymmetry},

\bibitem{rocek}
M.~Rocek,
\emph{Modified Calabi-Yau manifolds with torsion},

\bibitem{rocek2}
S.~Gates, C.~Hull  and M.~Rocek, \emph{Twisted Multiplets And New
  Supersymmetric Nonlinear Sigma Models}, Nucl. Phys. {\bf B248} (1984)
157

\bibitem{papa}
S.~Ivanov and G.~Papadopoulos, \emph{Vanishing theorems and string
  backgrounds}, Class. Quant. Grav. {\bf 18} (2001) 1089,
\href{http://www.arXiv.org/abs/math-DG/0010038}{{\tt math-DG/0010038}}

\bibitem{papa2}
S.~Ivanov and G.~Papadopoulos, \emph{A no-go theorem for string warped
  compactifications}, Phys.\ Lett.\ {\bf B497} (2001) 309,
\href{http://www.arXiv.org/abs/hep-th/0008232}{{\tt hep-th/0008232}}

\bibitem{papa3}
G.~Papadopoulos, \emph{KT and HKT geometries in strings and in black hole
  moduli spaces},
\href{http://www.arXiv.org/abs/hep-th/0201111}{{\tt hep-th/0201111}}

\bibitem{papa4}
J.~Gutowski, S.~Ivanov  and G.~Papadopoulos, \emph{Deformations of generalized
  calibrations and compact non-Kahler manifolds with vanishing first Chern
  class},
\href{http://www.arXiv.org/abs/math-DG/0205012}{{\tt math-DG/0205012}}

\bibitem{waldram}
J.~Gauntlett, N.~Kim, D.~Martelli  and D.~Waldram, \emph{Fivebranes wrapped on
  SLAG three-cycles and related geometry}, JHEP {\bf 0111} (2001) 018,
\href{http://www.arXiv.org/abs/hep-th/0110034}{{\tt hep-th/0110034}}

\bibitem{waldram2}
J.~Gauntlett, D.~Martelli, S.~Pakis  and D.~Waldram, \emph{G-structures and
  wrapped NS5-branes},
\href{http://www.arXiv.org/abs/hep-th/0205050}{{\tt hep-th/0205050}}

\bibitem{KMPT}
P.~Kaste, R.~Minasian, M.~Petrini  and A.~Tomasiello, \emph{Kaluza-Klein
  bundles and manifolds of exceptional holonomy}, JHEP {\bf 0209} (2002) 033,
\href{http://www.arXiv.org/abs/hep-th/0206213}{{\tt hep-th/0206213}}

\bibitem{BD}
K.~Becker and K.~Dasgupta, \emph{Heterotic strings with torsion},
\href{http://www.arXiv.org/abs/hep-th/0209077}{{\tt hep-th/0209077}}

\bibitem{Romans}
L.~J. Romans, \emph{Massive $N=2a$ supergravity in ten-dimensions}, Phys. Lett.
  {\bf B169} (1986)
374

\bibitem{GP}
F.~Giani and M.~Pernici, \emph{$N=2$ Supergravity In Ten-Dimensions}, Phys.
  Rev. {\bf D30} (1984)
325

\bibitem{Green:1987mn}
M.~B. Green, J.~H. Schwarz  and E.~Witten, \emph{Superstring theory. Vol. 2:
  Loop amplitudes, anomalies and phenomenology},
Cambridge, Uk: Univ. Pr. ( 1987) 596 P. ( Cambridge Monographs On Mathematical
  Physics)

\bibitem{dWS}
B.~de~Wit, D.~J. Smit  and N.~D.~H. Dass, \emph{Residual supersymmetry of
  compactified $d=10$ supergravity}, Nucl. Phys. {\bf B283} (1987)
165

\bibitem{H}
N.~Hitchin, \emph{Stable forms and special metrics},
\href{http://www.arXiv.org/abs/math.DG/0107101}{{\tt math.DG/0107101}}

\bibitem{GLMW2}
S.~Gurrieri, J.~Louis, A.~Micu  and D.~Waldram, \emph{?},
in preparation

\bibitem{BGGF}
P.~Binetruy, F.~Pillon, G.~Girardi  and R.~Grimm, \emph{The 3-Form Multiplet in
  Supergravity}, Nucl. Phys. {\bf B477} (1996) 175,
\href{http://www.arXiv.org/abs/hep-th/9603181}{{\tt hep-th/9603181}}

\bibitem{BGG}
P.~Binetruy, G.~Girardi  and R.~Grimm, \emph{Supergravity couplings: A
  geometric formulation}, Phys. Rept. {\bf 343} (2001) 255,
\href{http://www.arXiv.org/abs/hep-th/0005225}{{\tt hep-th/0005225}}

\bibitem{H2}
N.~Hitchin, \emph{Generalized Calabi-Yau manifolds},
\href{http://www.arXiv.org/abs/math.DG/0209099}{{\tt math.DG/0209099}}

\bibitem{CSF78}
E.~Cremmer, J.~Scherk  and S.~Ferrara, \emph{{SU(4)} Invariant Supergravity
  theory}, Phys. Lett. {\bf 74B} (1978)
61

\bibitem{S78}
J.~Scherk, {\em Recent developpments in gravitation}.
\newblock 1978,
Cargese, ed. M. Levy and S. Deser (Plenum Press)

\bibitem{Gat83}
S.~Gates, \emph{{On-shell and conformal {N}=4 supergravity in superspace}},
  Nucl.Phys. {\bf B213} (1983)
409--444

\bibitem{GD89}
S.~Gates and J.~Durachta, \emph{{Gauge two-form in D=4, N=4 supergeometry with
  SU(4) supersymmetry}}, Mod. Phys. Lett. {\bf A4} (1989)
2007

\bibitem{AGHH99}
G.~Akemann, R.~Grimm, M.~Hasler  and C.~Herrmann, \emph{{N=2 central charge
  superspace and a minimal supergravity multiplet}}, Class.Quant.Grav. {\bf 16}
  (1999) 1617--1623,
hep-th/9812026

\bibitem{WB83}
J.~Wess and J.~Bagger, {\em Supersymmetry and Supergravity}.
\newblock Princeton Series in Physics. Princeton University Press, Princeton,
  1983,
2nd edition 1992

\bibitem{BGG01}
P.~Bin\'etruy, G.~Girardi  and R.~Grimm, \emph{{Supergravity Couplings: a
  Geometric Formulation}}, Phys. Rept. {\bf 343} (2001) 255--462,
hep-th/0005225

\bibitem{Kis00}
A.~Kiss, {\em Formulation g\'eom\'etrique des th\'eories de supergravit\'e
  {N}=4 et {N=8} en superespace avec charges centrales}.
\newblock PhD thesis, Universit\'e de la M\'editerran\'ee Aix-Marseille II,
  december, 2000.
\newblock
CPT-2000/P.4106.

\bibitem{Loy02}
E.~Loyer, {\em Formulation g\'eom\'etrique de la supergravit\'e {N=8} en
  superespace avec charges centrales}.
\newblock PhD thesis, Universit\'e de la M\'editerran\'ee Aix-Marseille II,
  october, 2002.
\newblock
CPT-2002/P.XXXX.

\bibitem{Dra79}
N.~Dragon, \emph{Torsion and curvature in extended supergravity}, Z. Physik
  {\bf C2} (1979)
29--32

\bibitem{GGMW84b}
G.~Girardi, R.~Grimm, M.~M{\"{u}}ller  and J.~Wess, \emph{Superspace geometry
  and the minimal, non minimal, and new minimal supergravity multiplets}, Z.
  Phy. {\bf C26} (1984)
123--140

\bibitem{Mul86b}
M.~M{\"{u}}ller, \emph{{Natural Constraints for Extended Superspace}}, Z. Phys.
  {\bf C 31} (1986)
321--325

\bibitem{How82}
P.~Howe, \emph{Supergravity in superspace}, Nucl. Phys. {\bf B199} (1982)
309--364

\bibitem{GG83}
S.~Gates and R.~Grimm, \emph{{Consequences of conformally covariant constraints
  for {N}$>$4 superspace}}, Phys. Lett. {\bf 133B} (1983)
192

\bibitem{FS78}
D.~Freedman and J.~Schwartz, \emph{N=4 supergravity theory with local
  {$SU(4)\otimes SU(4)$} invariance}, Nucl. Phys. {\bf B137} (1977)
225--230

\bibitem{Mul86c}
M.~M{\"{u}}ller, \emph{Supergravity in ${U}(1)$ superspace with a two-form
  gauge potential}, Nucl. Phys. {\bf B264} (1986)
292--316

\bibitem{Bel74}
B.~Belinicher, \emph{Relativistic wave equations and Lagrangian formalism for
  particles of arbitrary spin}, Theor. Math. Phys. {\bf 20} (1974)
849 (320)

\bibitem{Wei95}
S.~Weinberg, {\em The quantum theory of fields}, vol.~1 : Foundations.
\newblock Cambridge University Press,
1995

\bibitem{OS77}
V.~I. Ogievetsky and E.~Sokatchev, \emph{SUPERFIELD EQUATIONS OF MOTION}, J.
  Phys. {\bf A10} (1977)
2021--2030

\bibitem{Soh78a}
M.~Sohnius, \emph{Bianchi identities for supersymmetric gauge theories},
  Nucl.Phys. {\bf B136} (1978)
461--474

\bibitem{HL81}
P.~Howe and U.~Lindstr{\"{o}}m, \emph{Higher Order Invariants in Extended
  Supergravity}, Nucl. Phys. {\bf B181} (1981)
487--501

\bibitem{Sie81b}
W.~Siegel, \emph{On-Shell {O(N)} Supergravity in Superspace}, Nucl. Phys. {\bf
  B177} (1981)
325--332

\bibitem{CS77b}
E.~Cremmer and J.~Scherk, \emph{{Algebraic Simplifications in Supergravity
  Theories}}, Nucl. Phys. {\bf B127} (1977)
259

\bibitem{AvP}
A.~Van~Proeyen, \emph{Tools for supersymmetry}, Annals of the University of
  Craiova, Physics AUC {\bf 9 (part I)} (1999) 1--48,
\href{http://www.arXiv.org/abs/hep-th/9910030}{{\tt hep-th/9910030}}

\bibitem{Nakahara:1990th}
M.~Nakahara, \emph{Geometry, topology and physics},
Bristol, UK: Hilger (1990) 505 p. (Graduate student series in physics)

\bibitem{candelasTS}
P.~Candelas, \emph{Lectures on complex manifolds}, proceedings to the Trieste
  Spring School (1987) 1,
\href{http://www.arXiv.org/abs/`in Superstrings 87'}{{\tt `in Superstrings
  87'}}

\bibitem{AS3}
A.~Strominger, \emph{Yukawa Couplings In Superstring Compactification}, Phys.
  Rev. Lett. {\bf 55} (1985)
286

\bibitem{CdO}
P.~Candelas and X.~de~la Ossa, \emph{Moduli space of Calabi-Yau manifolds},
  Nucl. Phys. {\bf B365} (1991)
455--481

\bibitem{CRTvP}
B.~Craps, F.~Roose, W.~Troost  and A.~van Proeyen, \emph{What is Special
  K{\"a}hler Geometry}, Nucl. Phys. {\bf B503} (1997) 565--613,
\href{http://www.arXiv.org/abs/hep-th/9703082}{{\tt hep-th/9703082}}

\bibitem{Tian:1987g}
G.~Tian, {\em Mathematical aspects of string theory}.
\newblock World Scientific, s.-t yau~ed., 1987,
Singapore

\bibitem{CeDF}
A.~Ceresole, R.~D'Auria  and S.~Ferrara, \emph{The symplectic structure of
  $N=2$ Supergravity and its central extension}, Nucl. Phys. Proc. Suppl. {\bf
  46} (1996) 67,
\href{http://www.arXiv.org/abs/hep-th/9509160}{{\tt hep-th/9509160}}

\bibitem{CaDFsk}
L.~Castellani, R.~D'Auria  and S.~Ferrara, \emph{Special Kahler geometry: An
  intrinsic formulation from $N=2$ space-time supersymmetry}, Phys. Lett. {\bf
  B241} (1990)
57

\bibitem{CaDFsg}
L.~Castellani, R.~D'Auria  and S.~Ferrara, \emph{Special geometry without
  special coordinates}, Class. Quant. Grav. {\bf 7} (1990)
1767

\bibitem{Suz}
H.~Suzuki, \emph{Calabi-Yau compactification of type IIB string and a mass
  formula of the extreme black holes}, Mod. Phys. Lett. {\bf A11} (1996) 1--48,
\href{http://www.arXiv.org/abs/hep-th/9508001}{{\tt hep-th/9508001}}

\bibitem{yano}
K.~Yano, \emph{Differential geometry on complex and almost complex spaces},
Macmillan, New York, 1965

\end{thebibliography}\endgroup

\bibliographystyle{toine}

\end{document}